\documentclass[12pt]{article}
\usepackage{a4wide,graphicx,amsmath,amssymb,hyperref}
\usepackage[numbers,sort&compress]{natbib}
\usepackage{wrapfig,rotating,subfigure}

\def\TeV{{\rm TeV}}

\def\lapproxeq{\lower .7ex\hbox{$\;\stackrel{\textstyle                                                    
<}{\sim}\;$}}                                                    
\def\gapproxeq{\lower .7ex\hbox{$\;\stackrel{\textstyle                                                    
>}{\sim}\;$}} 
\begin{document}

\titlepage

\begin{flushright}
LCTS/2013-24\\
\end{flushright}

\vspace*{0.2cm}

\begin{center}
{\Large \bf  The Effect of LHC Jet Data on MSTW PDFs}

\vspace*{1cm}
\textup{B.J.A. Watt, P. Motylinski and R.S. Thorne \\
Department Of Physics and Astronomy, University College London\\
          Gower Place, London, WC1E 6BT, UK}
\vspace*{0.0cm} 

\end{center}

\vspace*{0.0cm}

\begin{abstract}
We consider the effect on LHC jet cross sections on 
partons distribution functions (PDFs), in particular the MSTW2008 set of PDFs.
We first compare the published inclusive jet data to the predictions using 
MSTW2008, finding a very good description. We also use the parton distribution 
reweighting procedure to estimate the impact of these new data on the PDFs, 
finding that the combined ATLAS 2.76~TeV and 7~TeV data, and CMS 7~TeV data have 
some significant impact. We then also investigate the impact of ATLAS, CMS and D{\O} 
dijet data using the same techniques. In this case we investigate the effect 
of using different scale choices for the NLO cross section calculation. We find 
that the dijet data is generally not completely compatible with the 
corresponding inclusive jet data, often tending to pull PDFs, particularly the 
gluon distribution, away from the default values. However, the effect depends 
on the dijet data set 
used as well as the scale choice. We also note that conclusions 
may be affected by limiting the pull on the data luminosity chosen by the best 
fit, which is sometimes a number of standard deviations. 
Finally we include the inclusive 
jet data in a new PDF fit explicitly. This enables us to check the consistency 
of the exact result with that obtained from 
the reweighting procedure. There is generally good, but 
not full quantitative agreement. Hence, the conclusion remains that MSTW2008 PDFs 
already fit the published jet data well, but the central values and uncertainties 
are altered and improved respectively by significant, but 
not dramatic extent by inclusion of these data.   
\end{abstract}

%\newpage
\vspace*{0.0cm}

\section{Introduction}
\label{Introduction}

When considering hadron collider data for the determination of PDFs, one of 
the most effective and distinguishing sets is the cross section for 
production of high-$p_T$ jets. Indeed, this is one of the few direct probes of
the gluon distribution in PDF fits, with the gluon constraint from fitting DIS 
and Drell-Yan data being overwhelmingly indirect via the quark and antiquark 
evolution.  Until recently, the only hadron collider data on the inclusive
jet cross section which was 
available for PDF fits was that produced at the Tevatron by the CDF 
\cite{cdfjet} and D{\O} 
\cite{d0jet} collaborations. These were shown to have a 
significant constraining effect on the PDFs, and were particularly useful in 
decoupling the correlation between the gluon distribution and the strong 
coupling. 
The introduction of LHC data is expected to have an even larger 
impact on the current modern PDF sets 
\cite{MSTW,nnpdfpaper,herapdf,ct10,abm}
due to the extension in the range of 
$x$ and $Q^2$ probed. We will consider the quality of the fit to LHC 
inclusive jet data, both from ATLAS and CMS, in Section 2 of this article. 
As well as investigating how well the current MSTW PDFs fit the data we will 
examine the impact of the data both by considering the PDF uncertainty 
eigenvectors and checking which improve the fit quality  and which 
cause it to deteriorate, 
and also by using the PDF reweighting procedure. The latter provides a 
quantitative estimate of the genuine effect of a new data set on both the 
central value and uncertainty of a PDF set.  
  
So far the only type of hadron collider jet data included in the 
determination of the MSTW 2008 PDF sets, or indeed any other available PDF 
set, is the inclusive jet production. There is some Tevatron dijet data 
\cite{d0-dijet-paper} spanning the same range in energy and rapidity as the 
inclusive data, but this has not been used in used obtaining PDFs, though
some studies of the fit quality and potential impact have been made
\cite{wattthorne,Alekhin:2012ig}.
This is perhaps largely due to the fact that the dijet data sample has a 
significant overlap with the inclusive jet data sample.  
The inclusive data were chosen due to there being a less reliable 
theoretical understanding of the high rapidity 
dijet production as a function of dijet mass, $M_{JJ}$. This issue will 
be studied in more detail in Section 3 of this article, for both the older 
D{\O} dijet data and the more recent ATLAS and CMS data. As for the inclusive 
data the fit quality using the existing MSTW2008 PDFs, and the potential impact 
of the new data will be studied.

In the next section we will include the ATLAS and CMS inclusive data in 
a new fit explicitly using the MSTW2008 framework. This will provide the 
most detailed results on the impact of these new data sets, also including the
effect on the strong coupling constant $\alpha_S(M_Z^2)$ obtained from the 
fit. It also provides an opportunity to compare the results from including
a new data set explicitly in the fit with the results obtained from PDF 
reweighting, the first time this has been studied for the reweighting 
procedure using the Hessian approach. We find reasonable agreement between
the results obtained using reweighting and from fitting explicitly, but the 
former seems 
to imply a slightly greater reduction in uncertainty than is found from direct 
inclusion 
of data. We also briefly investigate different forms of reweighting and 
the uncertainty estimation for PDFs.

Throughout this 
article we will base our main results on an analysis using PDFs and
jet cross sections at NLO in QCD, removing any ambiguity due to the lack of 
knowledge of the full NNLO jet cross sections. We will comment on NNLO 
corrections at 
the end of the article. There are also electroweak corrections (see 
\cite{ewcorr} for a summary) 
which could potentially be quite large. However, there is still some 
disagreement upon the nature of these corrections, and so they are omitted 
from the analysis.  We also base our study on the framework of the MSTW2008 
PDF fit. Although there have been updates \cite{cheb,mstwhera,robert1} 
including new data sets, PDF parameterisations, deuterium corrections and 
heavy flavour schemes, and even some publicly released PDFs \cite{cheb}, 
in order to 
isolate the singular effect of the inclusion of jet data without potential 
contamination from   
these other updates we present the
impact of the jet data and nothing else on MSTW2008 PDFs. A forth-coming PDF
update will include all these various sources of improvement or update, and
the specific impact of the jet data within this larger set of changes
is known to be very similar to that presented in this article.

\section{Inclusive Jets}

\begin{figure}
\begin{center}
\subfigure[D0]{\includegraphics[width=0.7\textwidth]{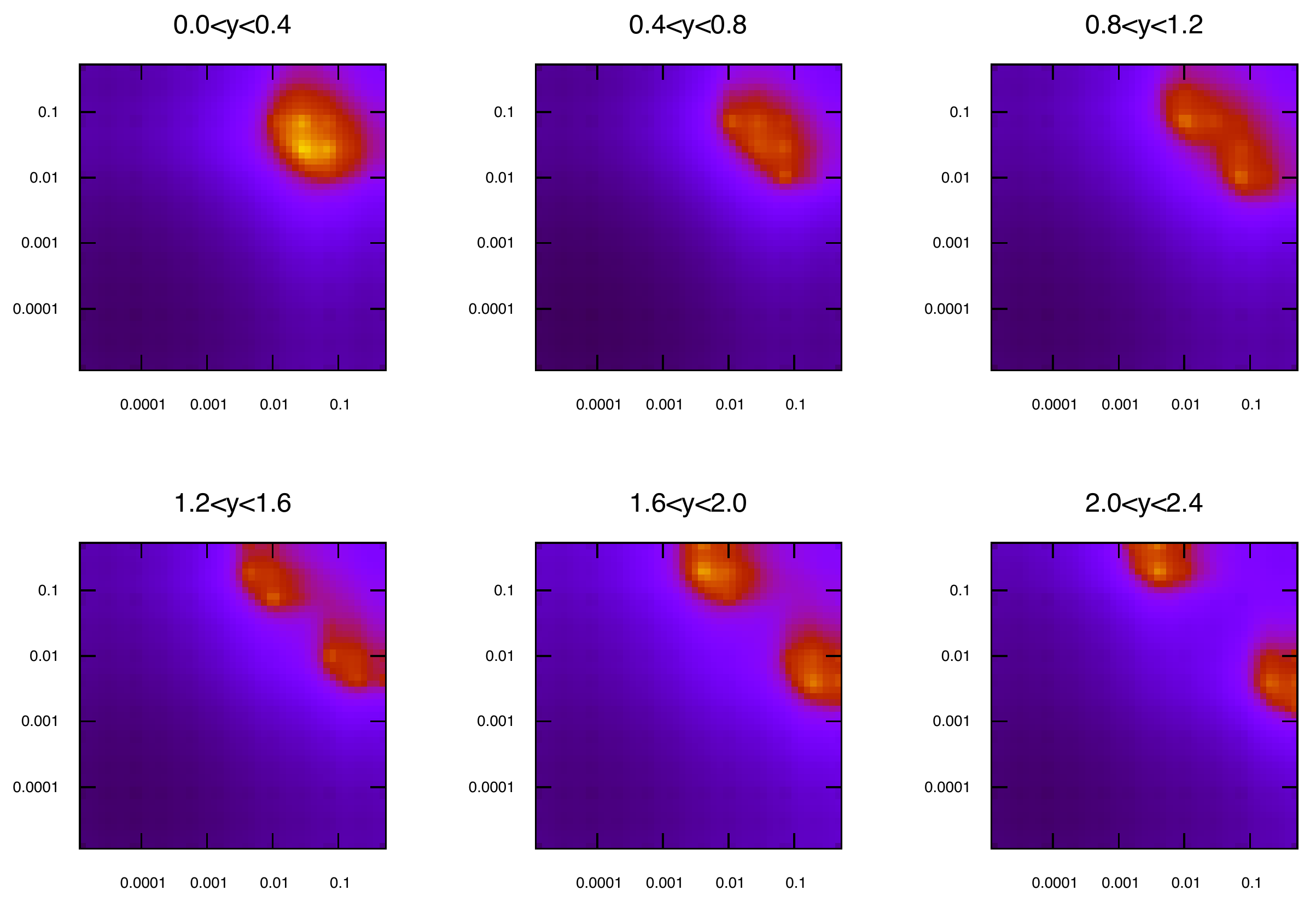}}
\subfigure[ATLAS]{\includegraphics[width=0.8\textwidth]{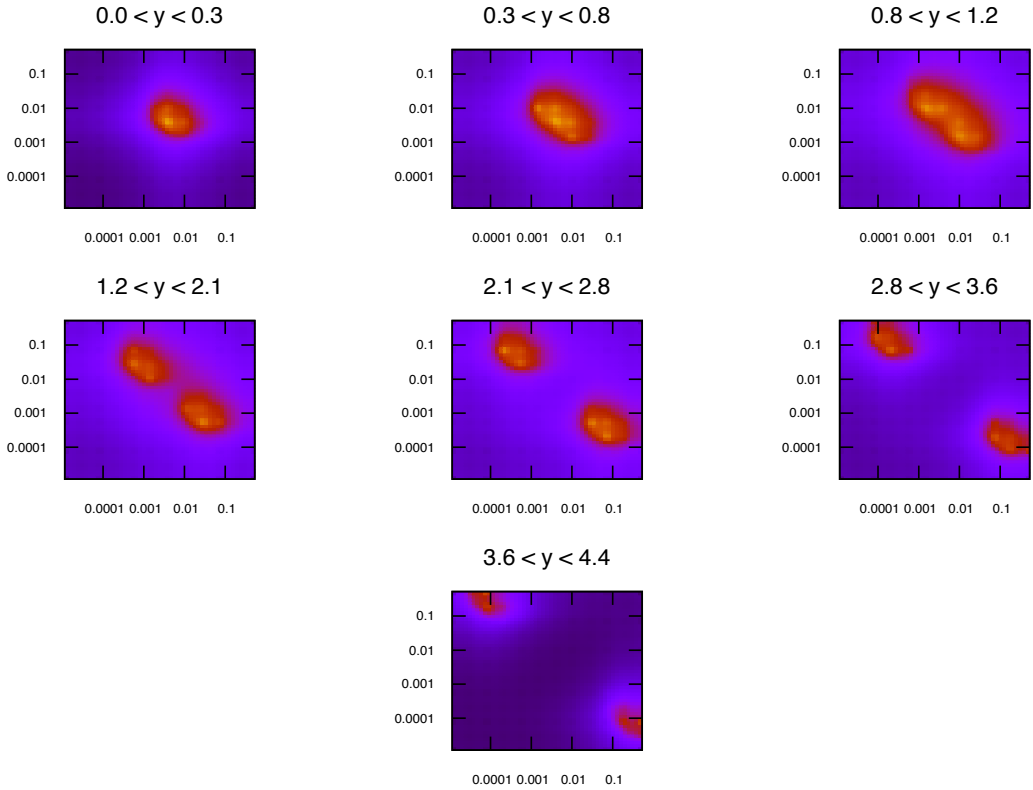}}
\end{center}
\vspace{-0.8cm}
\caption{Values of $x_1$ (higher x) and $x_2$ (lower x) for each event generated in NLOJet++ for inclusive jets at the Tevatron ($\sqrt{s}=1.96$~TeV) and LHC ($\sqrt{s}=7$~TeV). The lowest $p_T$ jets dominate in each rapidity bin, so the higher values of $x$ probed at large $p_T$ do not appear.}
\label{xreach2}
\end{figure}

\begin{figure}
\begin{center}
\subfigure[$|y|<0.3$]{\includegraphics[width=0.32\textwidth]{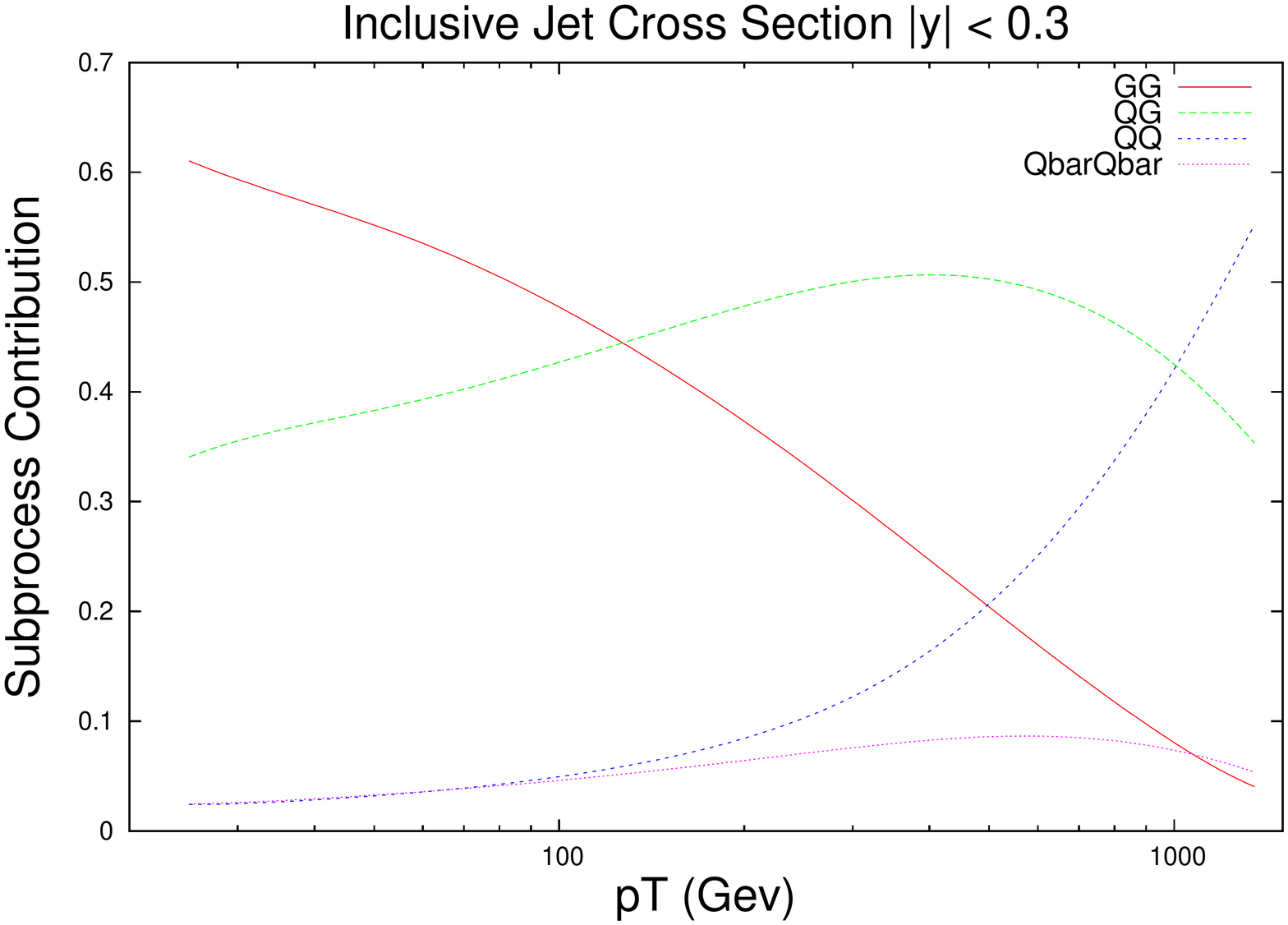}}
\subfigure[$0.3<|y|<0.8$]{\includegraphics[width=0.32\textwidth]{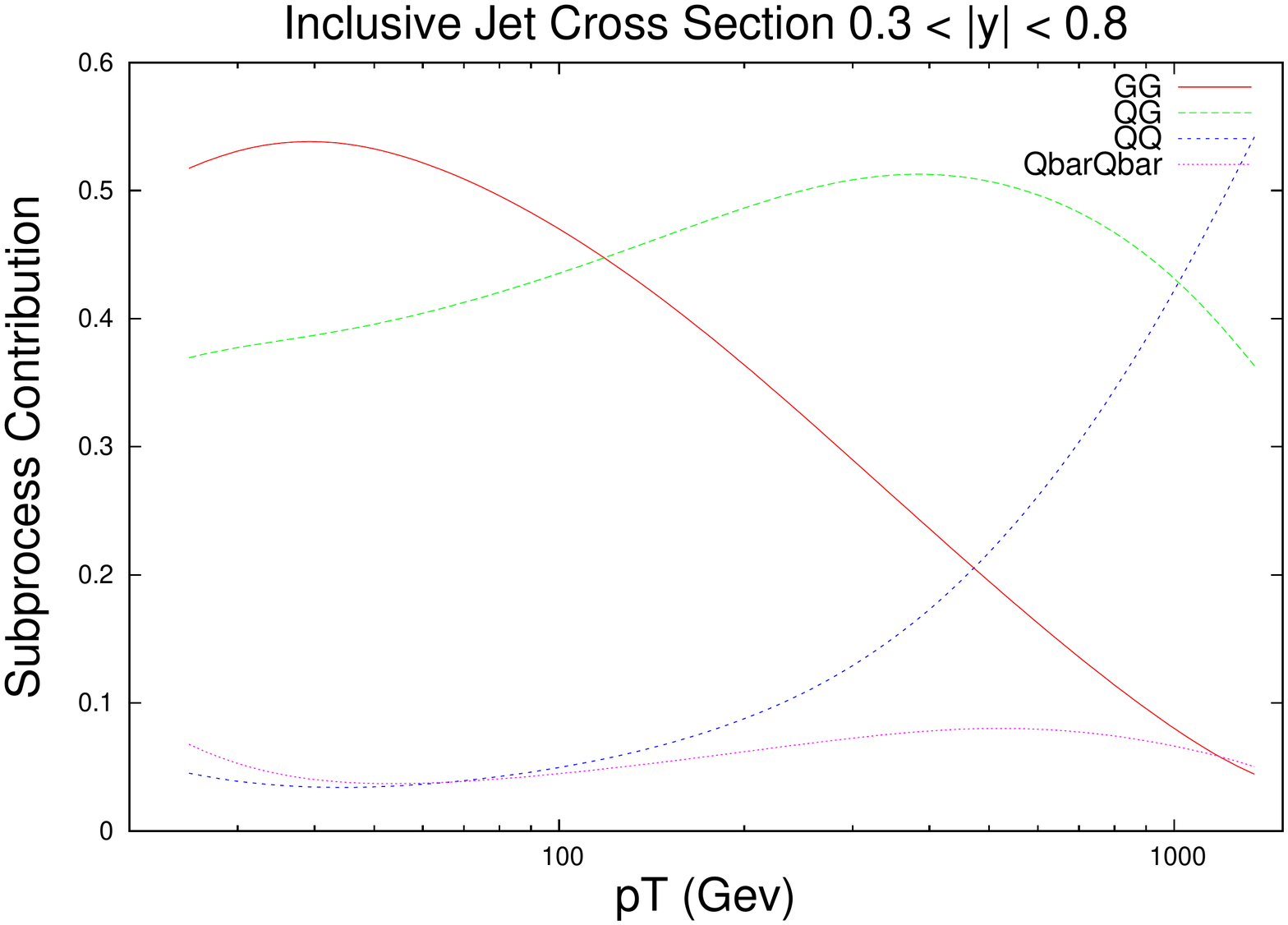}}
\subfigure[$0.8<|y|<1.2$]{\includegraphics[width=0.32\textwidth]{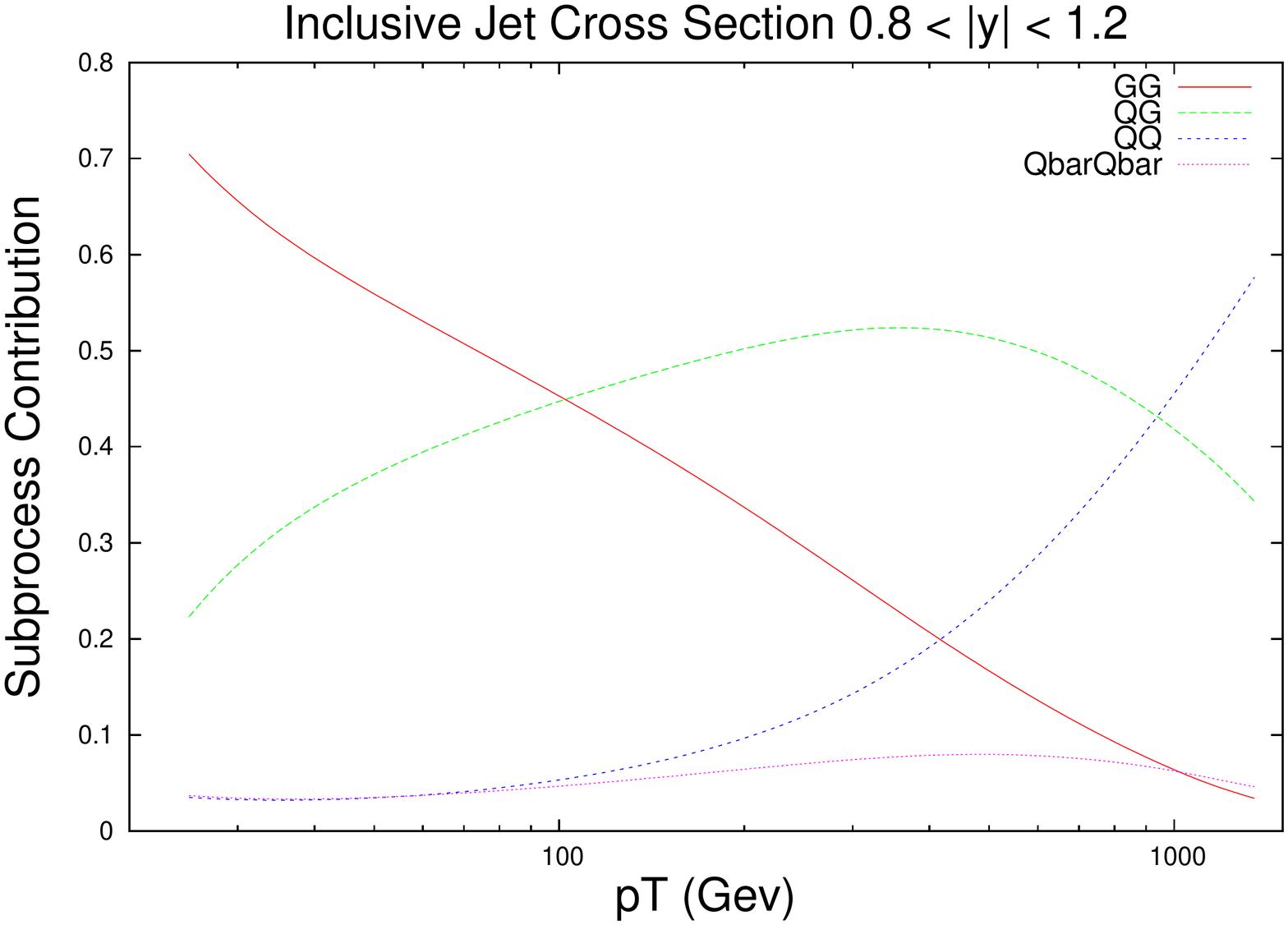}}
\subfigure[$1.2<|y|<2.1$]{\includegraphics[width=0.32\textwidth]{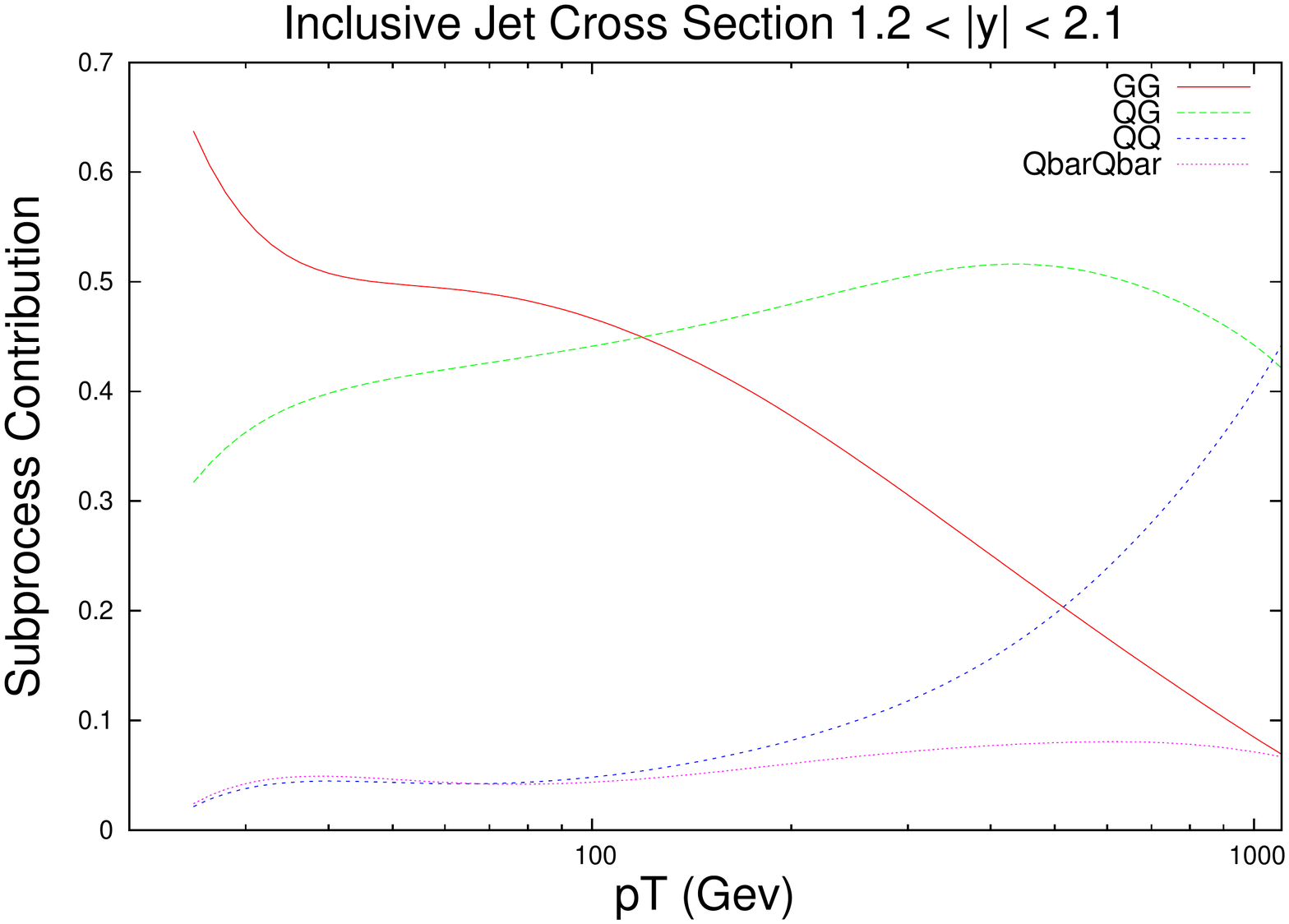}}
\subfigure[$2.1<|y|<2.8$]{\includegraphics[width=0.32\textwidth]{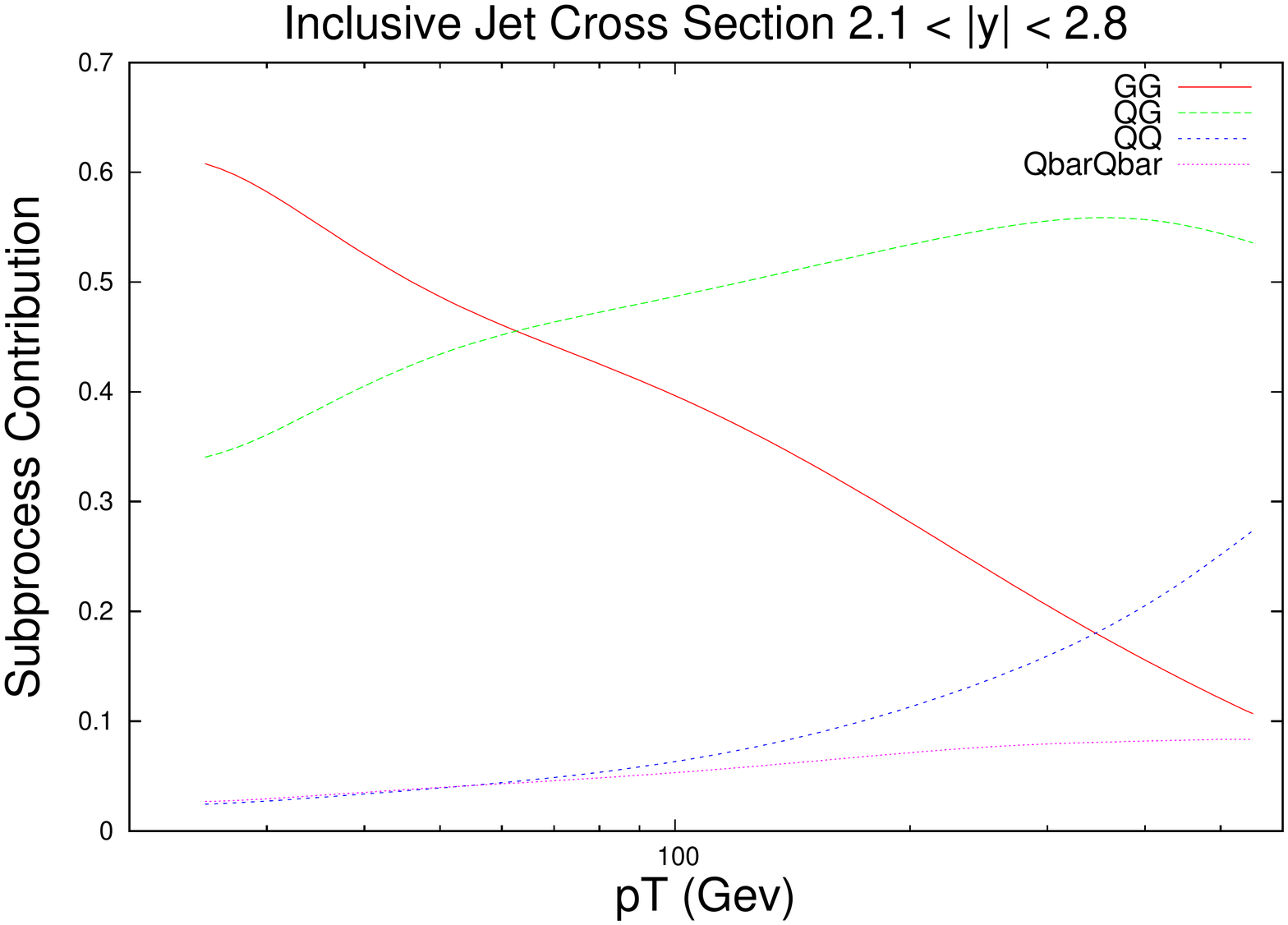}}
\subfigure[$2.8<|y|<3.6$]{\includegraphics[width=0.32\textwidth]{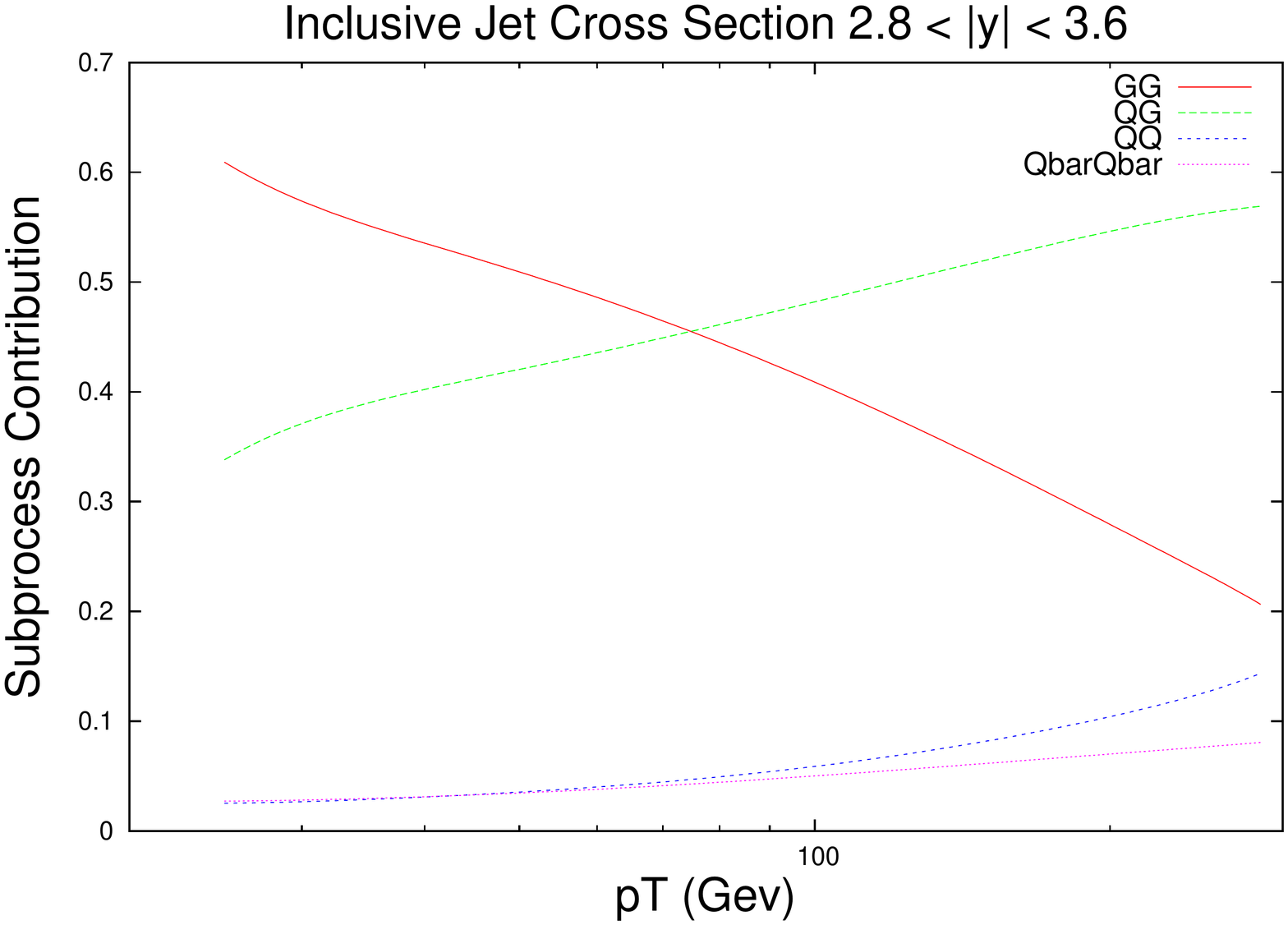}}
\subfigure[$3.6<|y|<4.4$]{\includegraphics[width=0.32\textwidth]{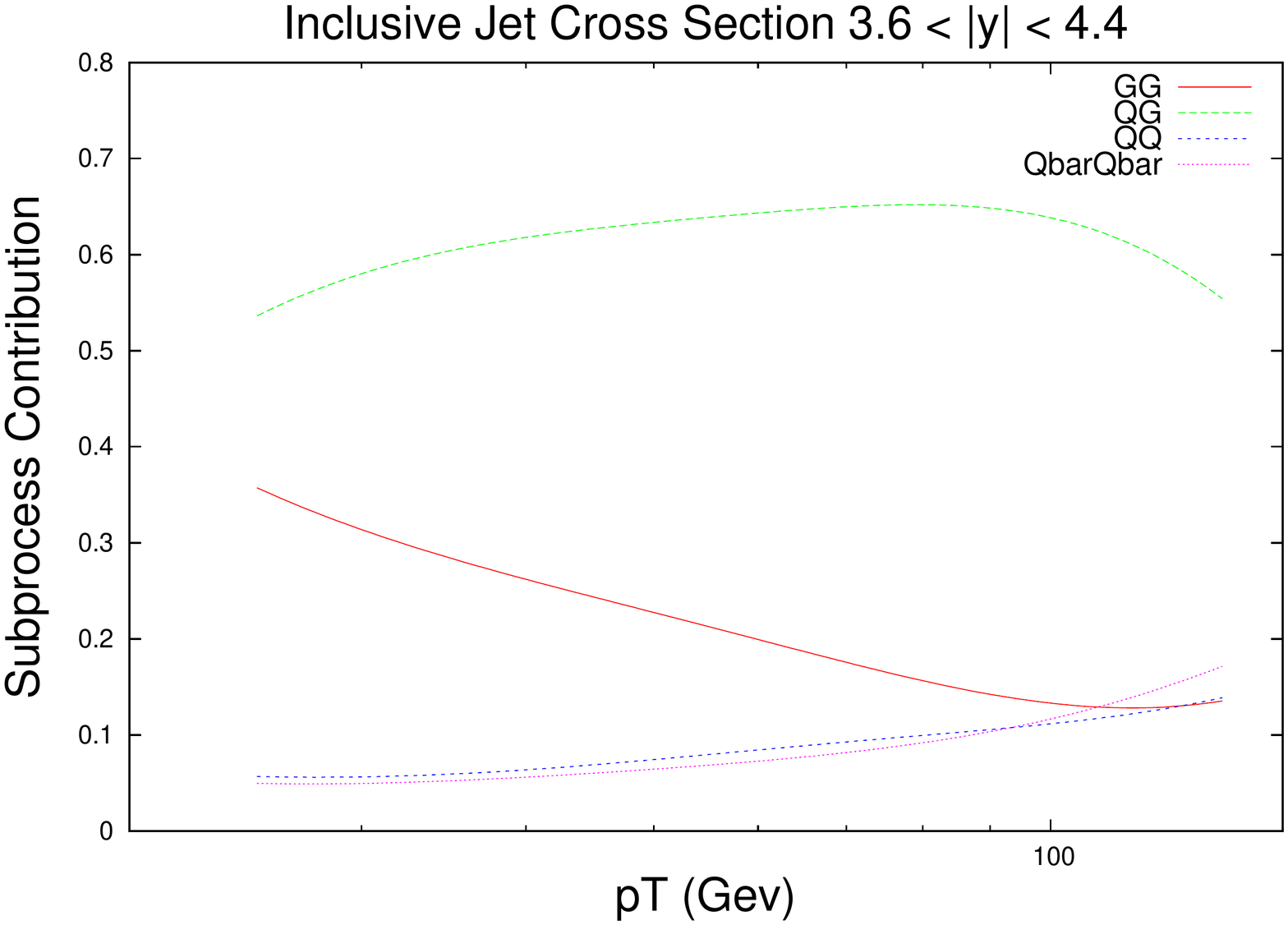}}
\end{center}
\vspace{-0.0cm}
\caption{Contributions of different initial-state parton combinations to the inclusive jet cross section calculation at ATLAS.}
\label{partons}
\end{figure}

In this section the details of the 
theoretical prediction for inclusive jet cross sections at the LHC are 
studied and the effects they have on the PDFs is analysed.
The first LHC data to have a true ability to probe new regions of the $(x,Q^2)$ 
plane for current PDFs was that from the ATLAS collaboration on the inclusive 
jet and dijet cross sections at 7~TeV using $36pb^{-1}$ of data 
\cite{atlas-inc-paper}. To demonstrate this ability, Figure \ref{xreach2} shows 
the distribution of the parton momentum fractions $x_1$ and $x_2$ for NLOJet++ 
\cite{nlojet,nlojet2} events at the Tevatron and the LHC. 
In these, and similar subsequent plots, the points have been generated at NLO
using unweighted events, though the plots would look very similar at LO. 
In the highest 
rapidity bin, the ATLAS data is probing values of  $x\approx10^{-5}$, 2 orders 
of magnitude lower than at D{\O}. These plots are dominated by the low $p_T$ bins 
within each rapidity bin, due to the orders of magnitude greater number of jets 
produced at low $p_T$. The higher $p_T$ jets require higher $x$ values, and the 
spots in Fig. \ref{xreach2} shift along the diagonal line parallel to 
$x_1=x_2$ towards higher $x$ as the $p_T$ of the jets is increased. Comparing 
the plots at LHC and Tevatron energies shows the value of the LHC data.
For inclusive jets, the PDFs can be probed down to $x=10^{-5}$ at high 
rapidities, a factor of 10 better than the Tevatron reach.
The sensitivity of the data to different partons is demonstrated in 
Fig. \ref{partons}, where the cross section calculation is broken down into 4 
partonic subprocesses: gluon-gluon, quark-gluon, quark-quark, antiquark-antiquark.

Clearly, different areas of phase space provide more information about certain 
PDFs than others. In the lowest rapidity bin for instance, the low $p_T$ jets 
are produced predominantly by initial state gluons, whereas the hardest jets are 
dominated by the quark-quark process. By combining this information with that 
obtained from Figure \ref{xreach2}, we can see that the low-$p_T$ central jets 
will provide information on the low x gluon, whereas high-$p_T$ will shed light 
on the high x valence quark distributions.   The fraction probed changes as 
a function of rapidity. As the rapidity of the inclusive jets increases, the 
events are produced predominantly by a combination of one low $x$ and one high 
$x$ parton, which can again be seen in the plots of Fig. \ref{xreach2}. This 
means that the quark-gluon process becomes dominant at high rapidities, 
especially at high $p_T$, and so these bins in the data will simultaneously 
probe the gluon and the quark distributions.

The $\chi^2$ used to compare data to theory is similar to that used 
for jet data in MSTW PDF fits. Each data point is allowed to move with respect 
to the theory prediction due to the many systematic uncertainties in the 
measurement. For each source of systematic uncertainty, a nuisance parameter 
$r_k$ is introduced, such that shifts will only occur if the reduction in 
$\chi^2$ is significant. The exact form of the expression used is:

\begin{equation}
\chi^2=\sum_{i=1}^{N_{pts}}\left(\frac{D_i-\sum_{k=1}^{N_{corr}}r_k\sigma_{k,i}^{corr}-T_i}{\sigma_i^{uncorr}}\right)^2+\sum_{k=1}^{N_{corr}}r_k^2
\label{chi2def}
\end{equation}

\noindent where $i$ labels the individual data points and $k$ labels the 
correlated systematics. For the ATLAS 7~TeV data, the number of correlated 
systematics is 88 when including the hadronisation uncertainty. The uncorrelated 
error is the sum in quadrature of the statistical error and the 3 uncorrelated 
systematic errors. This definition is not identical to the standard MSTW fit due 
to the treatment of normalisations, which here is considered a standard source 
of systematic error. In the actual fits, the normalisations are treated 
separately, and this will be discussed later in the article.

It is possible to solve this equation for $r_k$ analytically, giving the 
optimum systematic shifts directly. By minimising the $\chi^2$ the result is:

\begin{equation}
r_k=\sum_{k'=1}^{N_{corr}}(A^{-1})_{kk'}B_{k'}
\end{equation}

where

\begin{equation}
A_{kk'}=\delta_{kk'}+\sum_{i=1}^{N_{pts}}\frac{\sigma^{corr}_{k,i}\sigma^{corr}_{k',i}}{(\sigma^{uncorr}_i)^2}, B_k=\sum_{i=1}^{N_{pts}}\frac{\sigma^{corr}_{k,i}(D_i-T_i)}{(\sigma^{uncorr}_i)^2}.
\end{equation}

Hence, by calculating and subsequently inverting the $88\times88$ matrix $A$, 
and the vector $B$, the optimal values of the nuisance parameters can be found.

The correlated systematics for both the inclusive and dijet data sets are mostly 
antisymmetric, and so a method of symmetrising to obtain a single error for each 
data point must be employed. Since this is a matter of choice and should not 
affect the results in any meaningful way, three opposing methods were used to 
test the effect. These were:

\begin{align*}
\sigma_{corr} &= |{\sigma_{corr}}^+| \\
\sigma_{corr} &= |{\sigma_{corr}}^-| \\
\sigma_{corr }&= \frac{(|{\sigma_{corr}}^+| + |{\sigma_{corr}}^-|)}{2}.
\end{align*}

\noindent ${\sigma_{corr}}^{+/-}$ denotes the two opposing values of 
the antisymmetric errors, and the convention is that the sign denotes the 
sign of the error for the first point, i.e. lowest rapidity and $p_T$ 
(or $M_{jj}$ for dijets). The sign of the error is not necessarily maintained throughout 
the whole data set. 
The difference in $\chi^2$ obtained from these methods 
varied by no more than $3\%$ across all theory predictions. In all the following 
results, the third definition is used to calculate the $\chi^2$ values.

\subsection{The Effect of the ATLAS Inclusive Jet Data at 7~TeV}

\begin{figure}[h!]
\begin{center}
\includegraphics[width=0.9\textwidth]{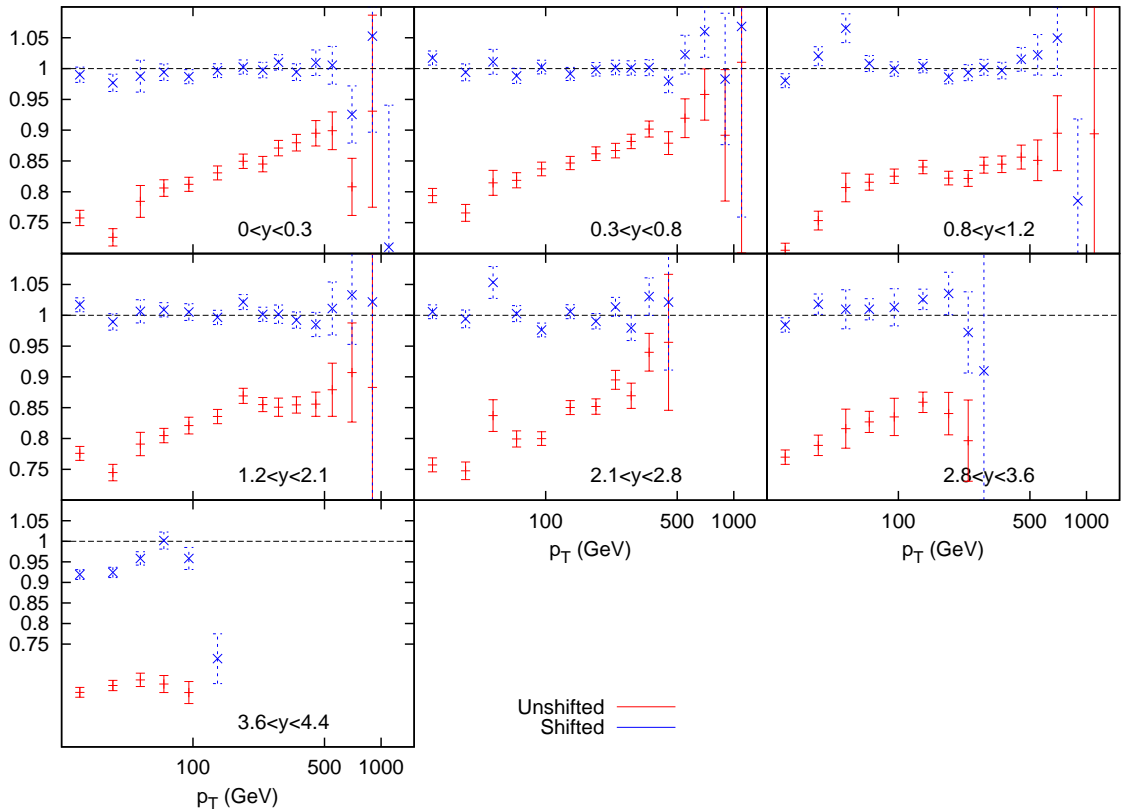}
\caption{Ratio of data to theory for ATLAS inclusive jets (R=0.4)}
\label{inclusiveratio}
\end{center}
\end{figure}

Fig. \ref{inclusiveratio} shows the ratio of data to theory 
(calculated with FastNLO \cite{fastnlo} version 2 \cite{fastnlov2} which uses 
NLOJet++ \cite{nlojet,nlojet2})
using MSTW2008 NLO PDFs for the ATLAS 7~TeV R=0.4 inclusive jet cross section, 
both before and after the correlated systematics are taken into account. The 
former gives a very poor agreement, with all data points above theory by up to 
40\%. The systematics are, however, large and the shifted points, defined as 
$(D_i-\sum_{k=1}^{N_{corr}}r_k\sigma_{k,i}^{corr})/T_i$ are almost all within 
1$\sigma$ of 1. The R=0.4 data set is chosen as default over the R=0.6 due to 
the much smaller hadronisation corrections in the case of the smaller jet 
parameter.
Tables \ref{table:nolin} and \ref{table:nolin2} demonstrate that a $\chi^2$ of 
less than 1 per point is achieved for a variety of scale choices and both R 
parameter choices, whilst the vast majority of the $r_k$ penalty terms are less 
than 0.5. This implies that the fit is a very good one, however the large shifts 
observed in the data alongside the small penalty terms implies that the 
systematic uncertainties are very large, and may be drowning out any underlying 
physics effects.

\begin{table}[h!]
\centering
\begin{tabular}{c c c c}
\hline\hline
Scale & $p_T$/2 & $p_T$ & 2$p_T$ \\
\hline
R=0.4 & 0.75 & 0.78 & 0.70\\
R=0.6 & 0.85 & 0.79 & 0.72\\
\hline
\end{tabular}
\caption{$\chi^2$ per point (90 points)}
\label{table:nolin}
\end{table}

\begin{table}[h!]
\centering
\begin{tabular}{c c c c c c c c}
\hline\hline
$|r_k| $ & $<0.5$ & 0.5--1.5 & 1.5--2.5 & 2.5--3.5 & \\
\hline
R=0.4 & 72 & 15 & 1 & 0 \\
R=0.6 & 74 & 13 & 1 & 0\\
\hline
\end{tabular}
\caption{Distribution of $r_k$'s (Total 88)}
\label{table:nolin2}
\end{table}

\begin{figure}[h!]
\centering
\includegraphics[width=0.8\textwidth]{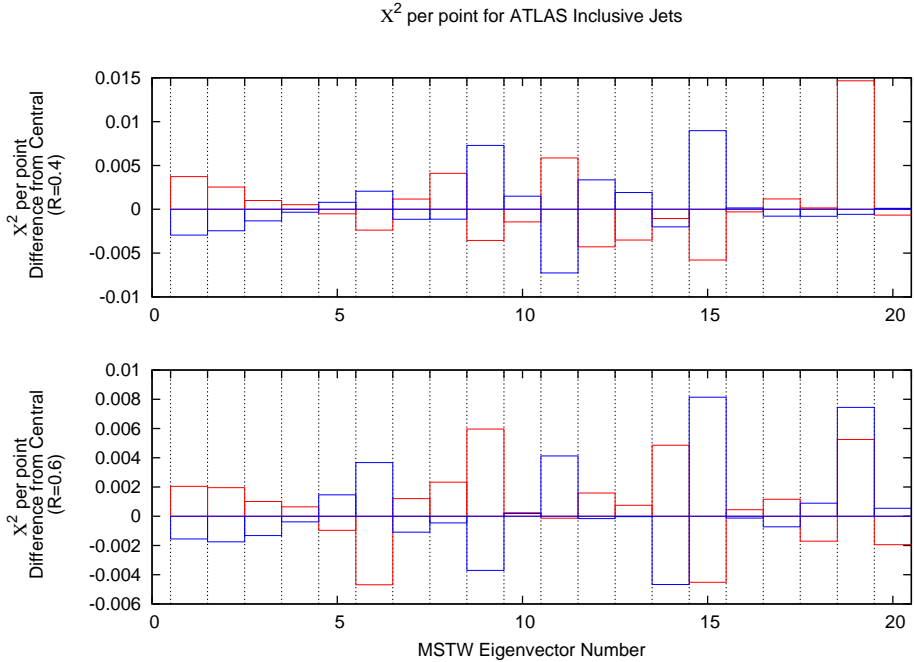}
\vspace{-0.8cm}
\caption{Change in fit quality for each MSTW eigenvector direction for ATLAS inclusive jets for both R-parameters used. The blue (red) bars indicate positive (negative) movement in the eigenvector direction}
\label{eig_atlas}
\end{figure}

A useful tool for extracting information on how the partons are affected by a 
new data set is to analyse the change in fit quality when using the different 
eigenvector sets in a global PDF fit. The global minimum of the PDF set will not 
necessarily give the best fit to any individual data set, due to competing 
influences from other sets used in the global fit.
The overcompensation of systematic effects is further demonstrated in 
Figs. \ref{eig_atlas}. Firstly, the individual eigenvectors are varied, and 
predictions produced corresponding to 1$\sigma$ deviations in each direction. 
The change in $\chi^2$ is negligible for all eigenvectors, with a maximum 
improvement of 0.007 per point in the R=0.4 fit for eigenvector 11. 

In order to best determine if there is any impact on the PDFs from a new dataset, 
another method which can be employed is the reweighting procedure. This was  
first suggested in \cite{geile}, reintroduced and modified in 
\cite{nnrew1,nnrew2} in the context of PDFs obtained from fits to replicas 
of data, 
and then extended for use with replicas obtained from PDF eigenvectors in the 
Hessian approach in \cite{reweight} (see also \cite{dilorenzi}). We briefly 
summarise this last approach. Firstly, 
the prediction for each eigenvector in the MSTW2008 fit is produced. These 
predictions are used to produce 1000 PDFs randomly distributed in 
eigenvector space, using the formula:

\begin{equation}
F(S_k) = F(S_0)+\sum_{j=1}^n [F(S_j^{\pm})-F(S_0)] |R_{jk}|
\end{equation}

\noindent where $R_{jk}$ is a Gaussian-distributed random number. The $F(S_k)$
can be any observable calculated using a PDF eigenvector $S_k$, in 
this case the jet cross sections. By sampling the eigenvector sets directly 
and weighting each PDF equally, an accurate estimate of the Hessian error on 
each data point is obtained. The central prediction is estimated simply by 
taking the average of these unweighted predictions. Although this does not 
exactly reproduce the prediction from the PDF set at the global minimum of 
the fit, the deviations are small and always well within the $1\!-\!\sigma$ error 
band. The main source of these deviations is the nonlinear dependence of the 
parameters on $x$. For example, if a function takes the form $(1-x)^{\eta}$
the average of $(1-x)^{\eta+\delta}$ and $(1-x)^{\eta-\delta}$ is not exactly 
equal to $(1-x)^{\eta}$, even if $\delta$ is small. 

\begin{figure}[h!]
\centering
\includegraphics[width=0.7\textwidth]{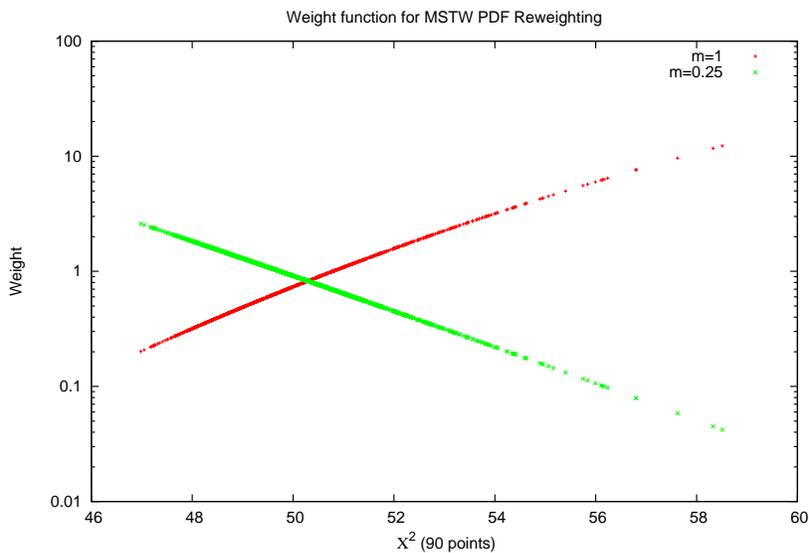}
\vspace{-0.8cm}
\caption{Weights for 1000 random PDFs, each fit to a dataset of 90 points with many PDFs giving $\chi^2$ better than $1$ per point. In this instance the standard reweighting function breaks down, and a value of $m<1$ is needed to properly weight the PDFs.}
\label{weightfunc}
\end{figure}

In order to obtain the effect of a new data set on the PDFs each random PDF 
is weighted according to its $\chi^2$, and then the statistical combination 
provides an updated ideal PDF for the dataset in question. The weighting 
formula advocated in \cite{nnrew1,nnrew2} is

\begin{equation}
w_i(\chi^2_i)=\frac{W_i(\chi^2_i)}{\frac{1}{N_{pdf}}\sum_{j=1}^{N_{pdf}} W_j(\chi^2_j)} , \;\;\;
W_i(\chi^2_i)=[\chi^2_i]^{\frac{m*(N_{pts}-1)}{2}}\exp\left(-\frac{\chi^2_i}{2}\right)
\label{reweight_equation}
\end{equation}

\noindent where $\chi^2_i$ is the fit quality of the $i$th random PDF, 
$N_{pdf}$ is the number of random PDFs generated and $N_{pts}$ is the number 
of points in the fit, and by default $m=1$. There is some active discussion 
about the correct weighting function to use (see e.g. 
\cite{Sato:2013ika,Paukkunen:2014zia}), 
although the most appropriate 
choice is certainly related to the procedure used in the PDF fit to obtain 
the uncertainty. Since in the MSTW fitting procedure the so-called 
``dynamical tolerance'' procedure, based on the confidence level of the fit 
quality to individual data sets is used, the weighting in 
eq.(\ref{reweight_equation}) might seem appropriate. 
The weighting function used here is also modified to include the multiplying 
factor $m$, to account for the case where the fit gives a $\chi^2$ 
significantly better than $1$ per point. In this instance, the weight 
function has a turning point, and so assigns lower weights to the best 
fits than those slightly worse. This is demonstrated in Fig. 
\ref{weightfunc}, where all random PDFs give a better fit than $1$ per 
point (e.g. ATLAS inclusive jet data). In this case, a value of $m<1$ is 
required to ensure the weights are assigned correctly. The actual value of 
$m$ to choose will affect how quickly the weights decrease as the fit 
worsens. For the ATLAS inclusive jet data values of $m \approx 0.5$ 
were used since the $\chi^2$ per point can be not much more than 0.5. 
The shape of the function 
$[\chi^2_i]^{\frac{m*(N_{pts}-1)}{2}}\exp(-\frac{\chi^2_i}{2})$ does not 
change much for variations of $m$ about 0.5 for the quite narrow range of 
$\chi^2$ values produced by the random PDF sets, and hence  the 
effect of this on the final reweighted PDFs is very small. However, in 
practice there is surprisingly little variation with even lower values of
$m$, and in terms of results simply ensuring that the function does not turn 
over appears to be sufficient. This insensitivity is unlikely to be so marked 
if a wider range of $\chi^2$ values is produced, i.e. if the new data set 
strongly constrains some eigenvector directions. 

\begin{figure}[h!]
\centering
\includegraphics[width=0.49\textwidth]{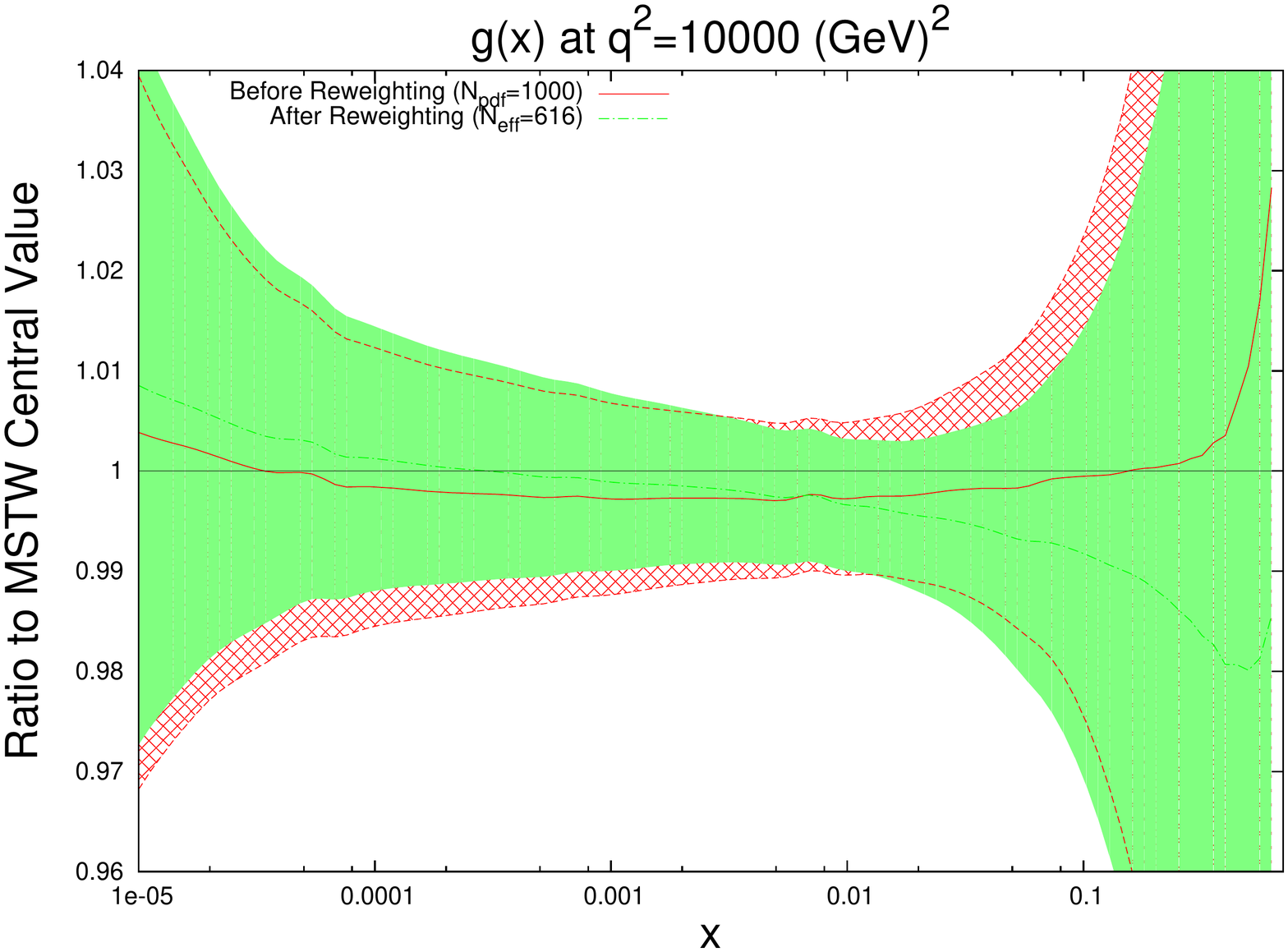}
\includegraphics[width=0.49\textwidth]{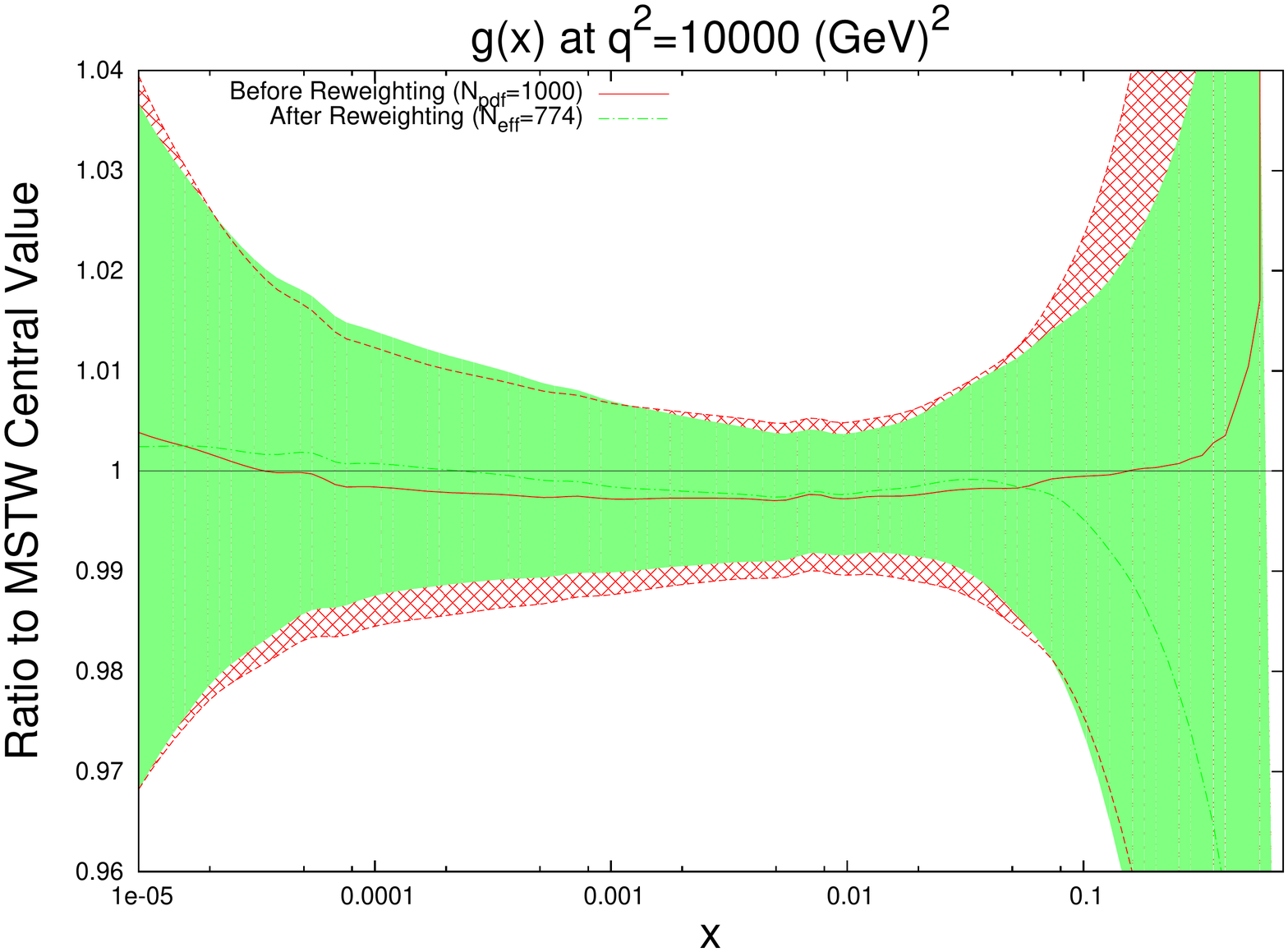}\\
\vspace{-0.8cm}
\caption{The effect of reweighting the MSTW2008 gluon using ATLAS inclusive jet data. Jet size parameter $R=0.4$ (left), and $R=0.6$ (right).}
\label{reweight_inc}
\end{figure}

A number which can provide more information on the reweighting procedure is 
$N_{eff}$, the effective number of PDFs included in the reweighted distribution. 
This is calculated by:

\begin{equation}
N_{eff}=exp\left(\frac{1}{N_{pdf}}\sum_{i=1}^{N_{pdf}}w_i\ln\left(\frac{N_{pdf}}{w_i}\right)\right).
\end{equation}

\noindent If the data set reweighted to has no effect, then all weights are 1 
and $N_{eff}=N_{pdf}$, however as soon as there are some weights larger than 
others, $N_{eff}$ will provide an estimate for the number of random PDFs which 
have contributed.

\begin{table}[h!]
\centering
\begin{tabular}{c c c c}
\hline\hline
Scale & $p_T$/2 & $p_T$ & 2$p_T$ \\
\hline
Multiplicative (R=0.4) & 0.645 & 0.584 & 0.556\\
Multiplicative (R=0.6) & 0.630 & 0.584 & 0.587\\
Additive (R=0.4) & 0.752 & 0.773 & 0.703\\
Additive (R=0.6) & 0.845 & 0.790 & 0.721\\
\hline
\end{tabular}
\caption{$\chi^2$ per point using multiplicative and additive errors}
\label{table:mult}
\end{table}

The results of the reweighting procedure using $\mu = p_T$ 
are shown in Fig. \ref{reweight_inc} for 
the gluon, which is the only PDF noticeably affected by the data. There is a 
very slight trend for the gluon to increase at low-$x$ and decrease at high-$x$, 
but again it is clear that very little can be deduced with the swamping effect 
of the systematics. The reweighted PDF produces a comparison to data with a 
$\chi^2$ of 0.73, slightly down from an unweighted value of 0.78. 

Another issue regarding the treatment of systematics is that of 
whether to use multiplicative or additive definitions. The systematic 
errors  in the data are presented as percentages, and so in order to obtain an 
absolute value of any given error, this percentage can be multiplied either 
by the data values or theory. If the percentage errors are multiplied by the 
data, they are considered additive since they are equivalent to an absolute 
error, whereas if they are multiplied by the theory they are considered 
multiplicative. By the nature of this particular fitting method, the data points 
themselves are significantly shifted in one direction by the systematics before 
the $\chi^2 $ is evaluated (in this case upwards, since the theory lies above 
data in general). Therefore, if the absolute errors are obtained from the raw 
data, they will be proportionally smaller after the shift. The effect of this 
can be seen in Table \ref{table:mult} where the $\chi^2$ for the two separate 
treatments of errors is summarised. The multiplicative treatment shows a 
considerably lower $\chi^2$ than the additive treatment, 
due to the larger absolute size of each error.
The table also demonstrates that the physics being probed depends upon the treatment of the errors. In the multiplicative case with R=0.6, the best fit is obtained with a scale choice of $p_T$, whereas it is $2p_T$ when using additive. Whilst it is a small discrepancy, it shows the importance of the treatment of errors, since everything else in the two fits is identical.

\subsection{ATLAS 2.76~TeV and 7~TeV combined data sets}

\begin{figure}[h!]
\centering
\includegraphics[width=0.49\textwidth]{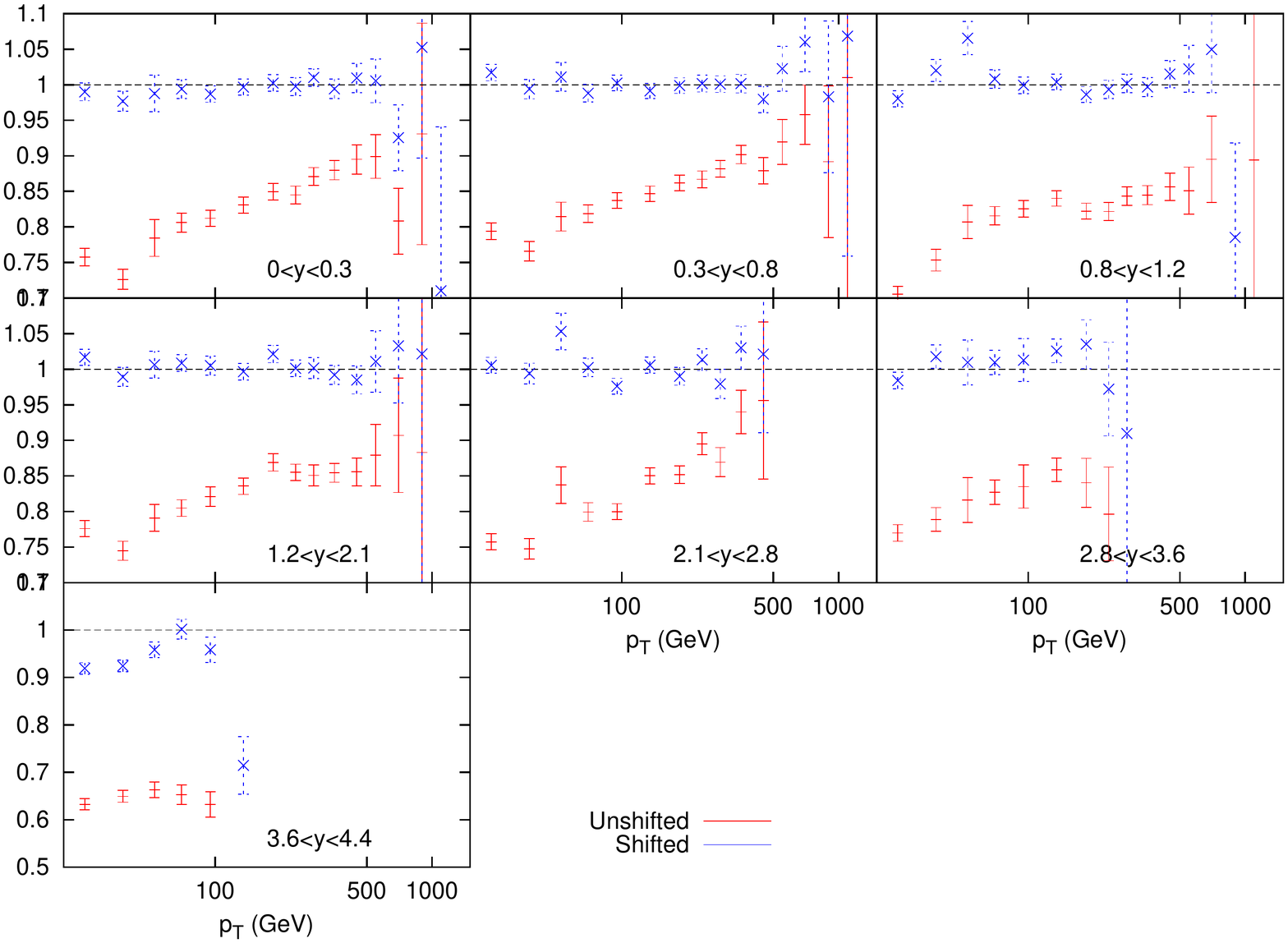}
\includegraphics[width=0.49\textwidth]{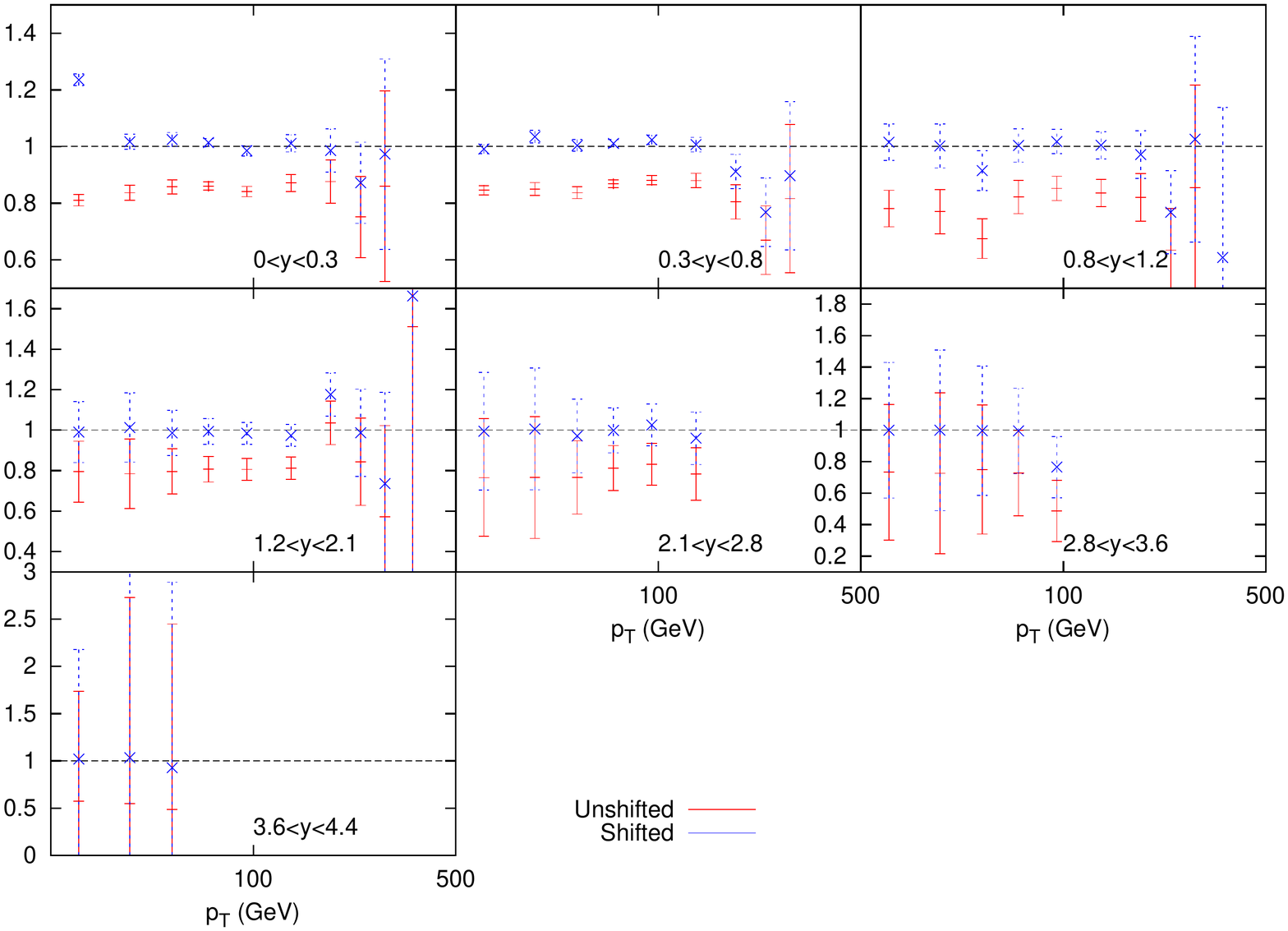}
\caption{Ratio of data over theory for MSTW PDFs convoluted with APPLgrid for the ATLAS inclusive jet combined data. The left hand plots are the 7~TeV data points, whilst the right hand side shows the 2.76~TeV data. There is more fluctuation in the shifted points for 7~TeV with the constraints imposed by concurrent 2.76~TeV fit, than for the pure 7~TeV fit.}
\label{ratio_nocuts}
\end{figure}

A method for possibly reducing the effect of the systematic uncertainties of 
the inclusive jet cross section data is to perform a simultaneous fit of
data taken at two 
different centre of mass energies, as done in \cite{ratiopaper}. The largest 
source of such uncertainties is the Jet Energy Scale (JES), which for ATLAS 
comprises of 14 separate uncertainties correlated across all bins in the 
measurement. Since the source of JES uncertainties is the same at any centre of 
mass energy, performing a PDF fit across two measurements will significantly 
reduce the allowed systematic shift of data points, allowing better constraints 
on PDFs. The prediction for MSTW2008, using NLOjet++ interfaced with APPLgrid
\cite{applgrid} is shown in Fig. \ref{ratio_nocuts}, both 
before and after the systematics shifts in the $\chi^2$ calculation are taken 
into account. The data again must be moved upwards for all points in the 
combined set to match the theory, however, when compared to the equivalent plot 
for the 7~TeV data (Fig. \ref{inclusiveratio}), it can be seen that the 
systematics are having less of an effect on this particular data set, with more 
fluctuations in the shifted points, especially at high rapidity.
Both the measurement of the inclusive jet cross section at 2.76~TeV and that 
at 7~TeV contain 21 sources of correlated systematic uncertainty which translate into 88 individual uncertainties after considering the correlations between rapidity bins. Only 3 of the sources are not correlated between the two datasets, and so the combined measurement contains 91 separate correlated uncertainties, an increase of only 3 whilst increasing the data points from 90 to 149.

\begin{table}[h!]
\centering
\begin{tabular}{|c |c| c|c|}
\hline\hline
 & No Cuts & HERAPDF Cuts & Additive Errors\\
\hline
MSTW 2008 & 1.43 & 0.93 & 1.46 \\
\hline
\end{tabular}
\caption{$\chi^2$ per point for ATLAS combined data, both with and without $p_T$ cuts. The third column uses additive errors and has two additional anomalous points cut.}
\label{table:chi_ratio}
\end{table}

\begin{figure}[h!]
\centering
\includegraphics[width=0.49\textwidth]{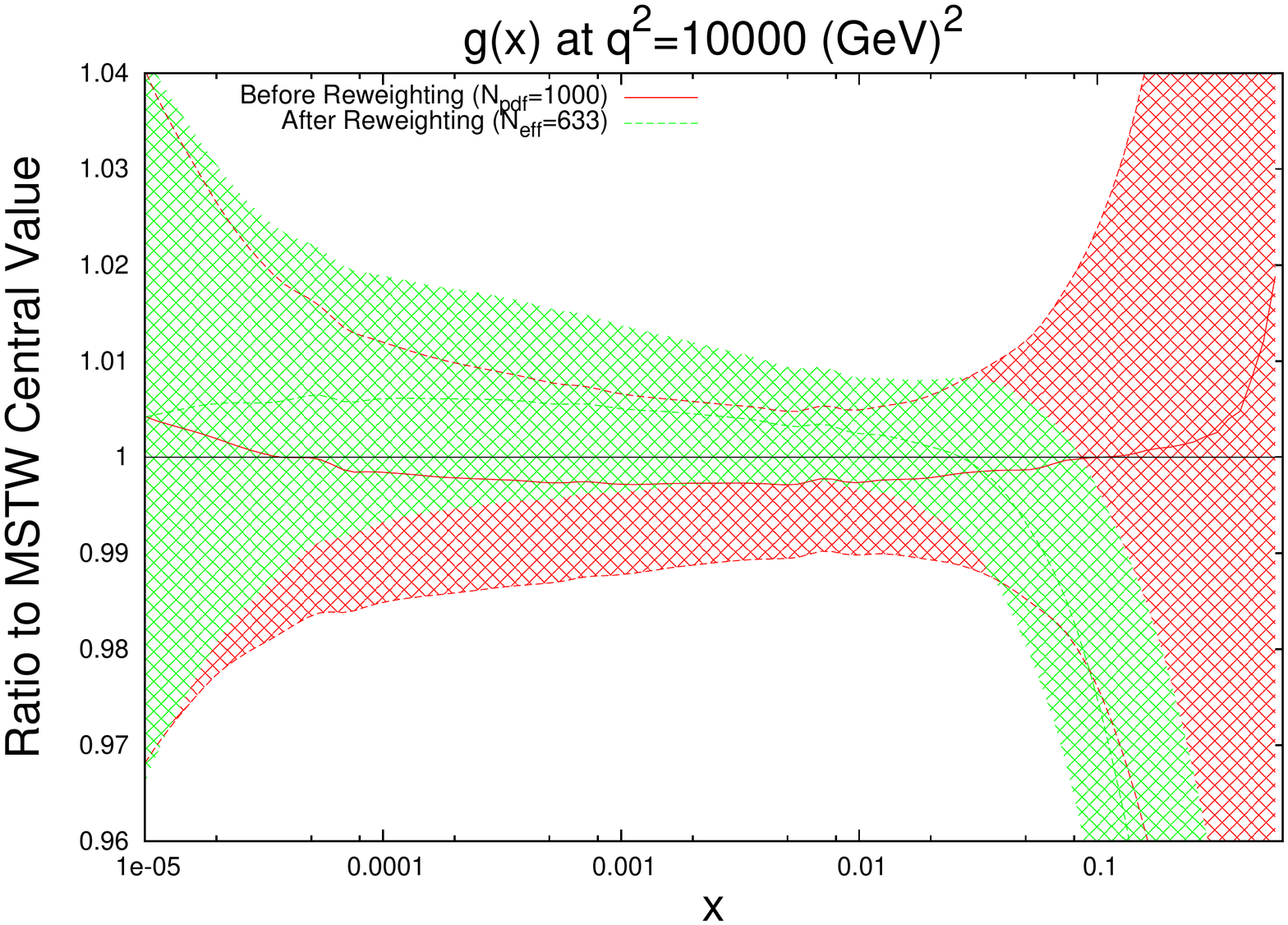}
\includegraphics[width=0.49\textwidth]{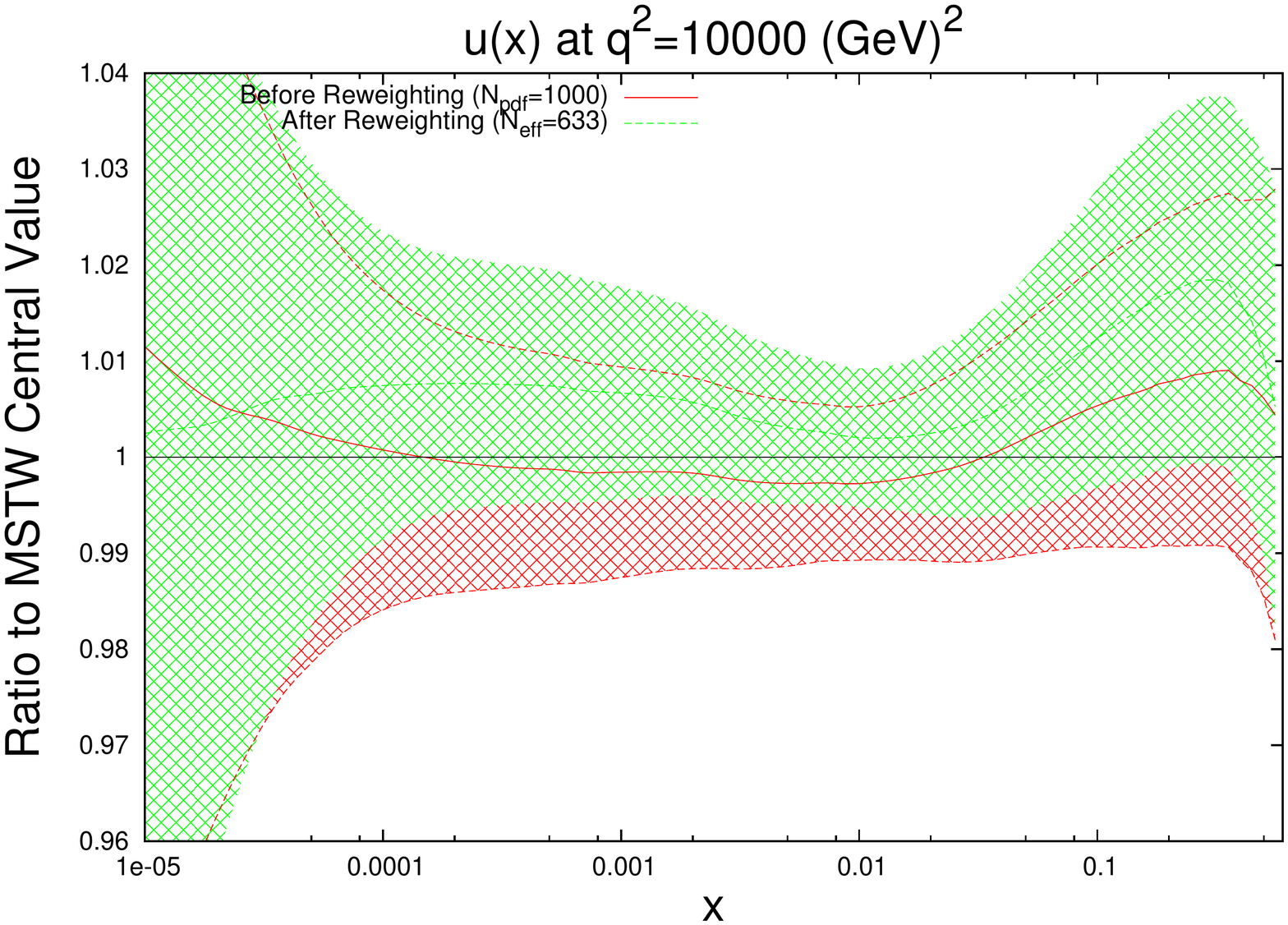}\\	
\includegraphics[width=0.49\textwidth]{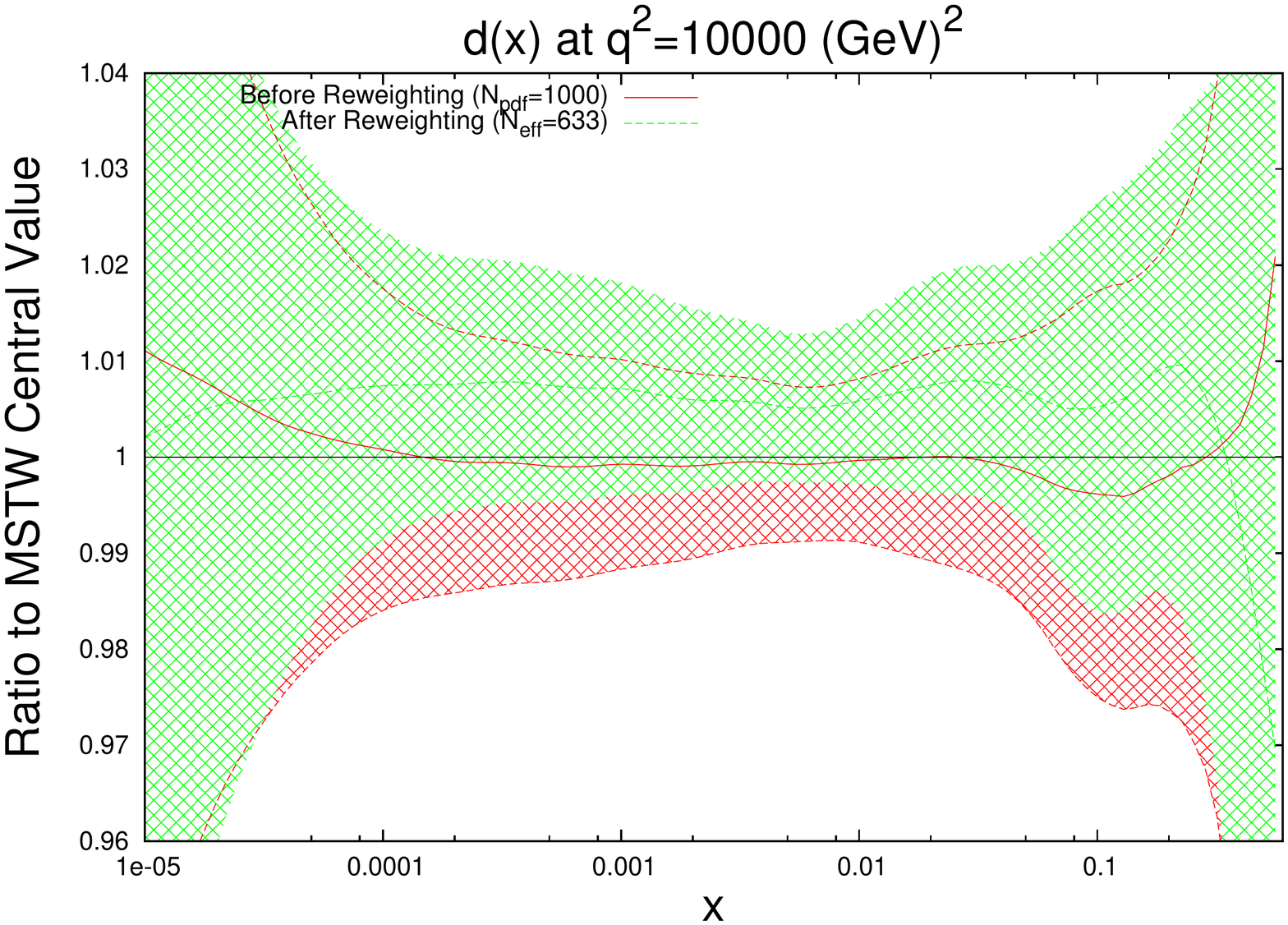}
\includegraphics[width=0.49\textwidth]{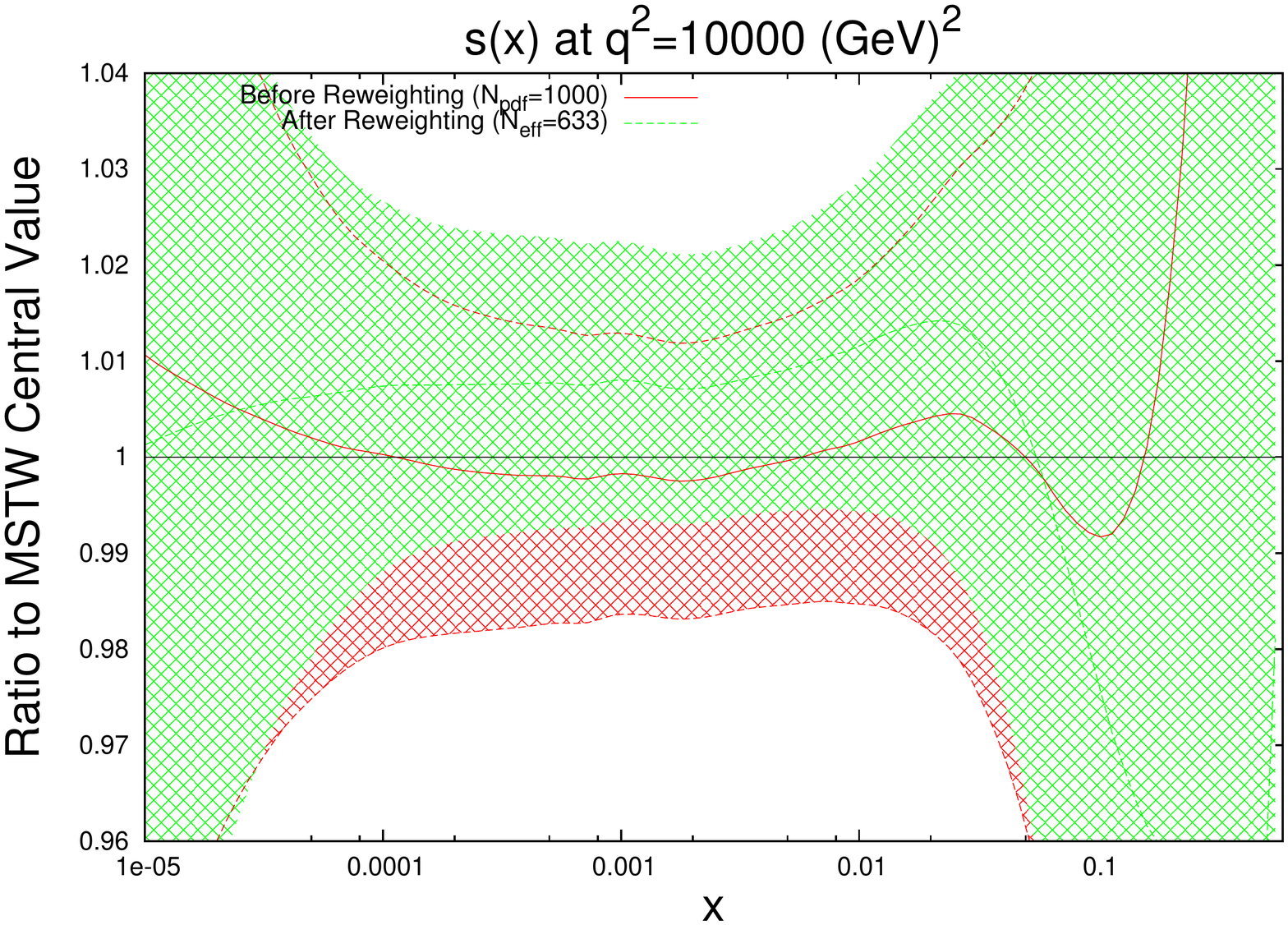}\\
\vspace{-0.8cm}
\caption{Effect of the ATLAS combined inclusive jet data on the gluon and quark PDFs. Here, multiplicative errors are used, and the lowest two bins in $p_T$ in all rapidity bins and the highest $p_T$ bins in the 2.76~TeV rapidity bins are excluded as per the HERAPDF analysis.}
\label{reweight_ratio2}
\end{figure}

The original paper \cite{ratiopaper} to produce such a PDF analysis was 
produced by the HERAPDF collaboration in conjunction with ATLAS. In this 
analysis, a minimum $p_T$ cut is applied of 45~GeV for all bins in both 
datasets, whilst the 2.76~TeV dataset includes a further maximum $p_T$ cut 
of 400~GeV which is applied in all but the $1.2<y<2.1$ rapidity bin. These cuts 
are motivated by the large hadronisation corrections in the stated bins, which 
can be as high as 12\% for some low $p_T$ bins using $R=0.4$ (and higher using 
$R=0.6$). For this analysis, both 
definitions will be tested. The difference in fit quality is shown in 
Table \ref{table:chi_ratio}, where a large improvement is seen when including 
the $p_T$ cuts for various PDFs. The source of this improvement is from the 
low-$p_T$ bins, where the statistical errors are the smallest, and so any 
deviation from the data produces a comparatively large increase in $\chi^2$. 
Indeed, the source of the increase in $\chi^2$ for the data without the cuts 
can be traced 
to one or two points in the set. The lowest $p_T$ bin in the $0<y<0.3$ 
rapidity bin of the $2.76$~TeV dataset contributes over 100 points to the 
total $\chi^2$ when using the MSTW2008 PDF set.

\begin{figure}[h!]
\centering
\includegraphics[width=0.49\textwidth]{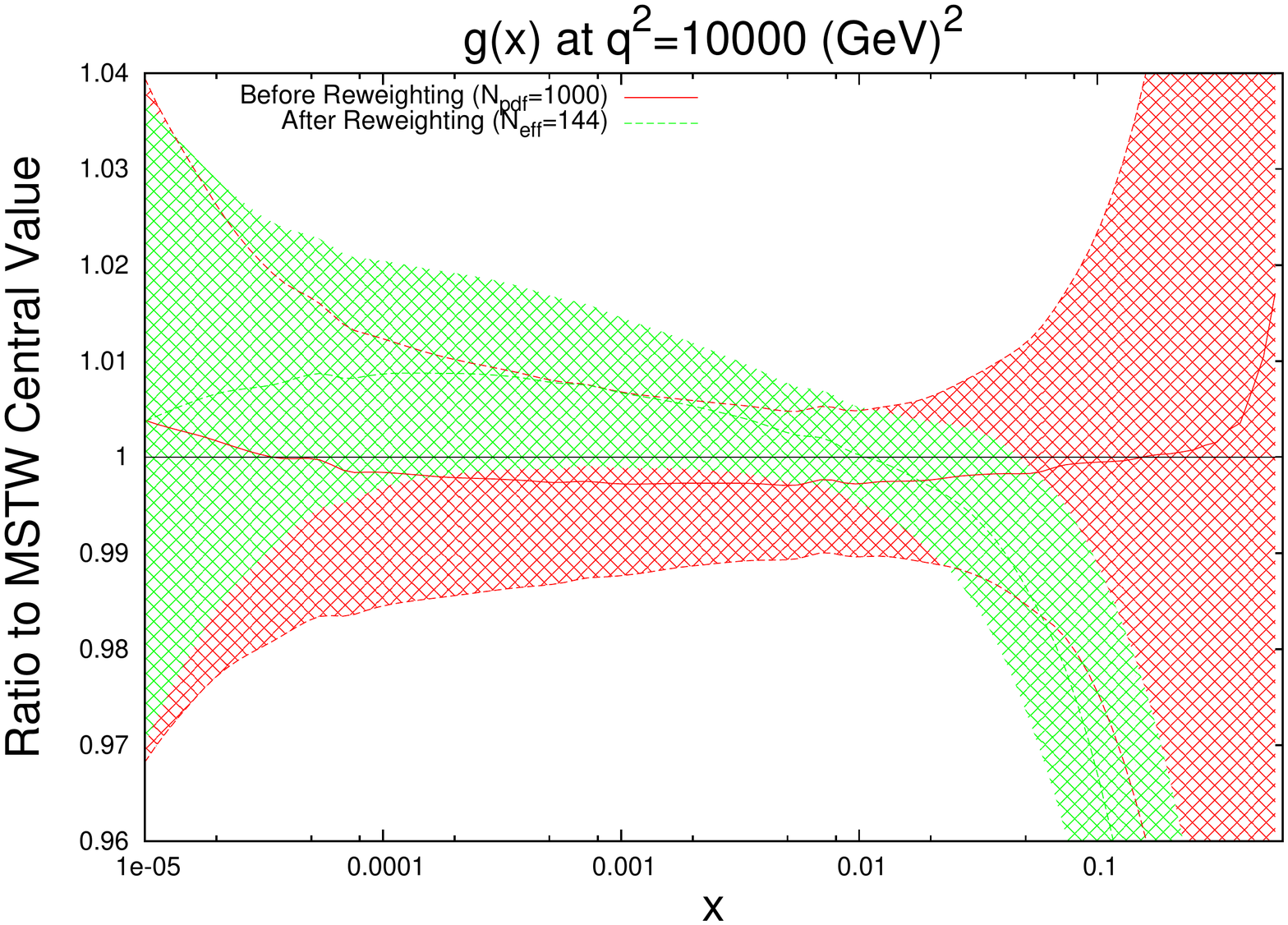}
\includegraphics[width=0.49\textwidth]{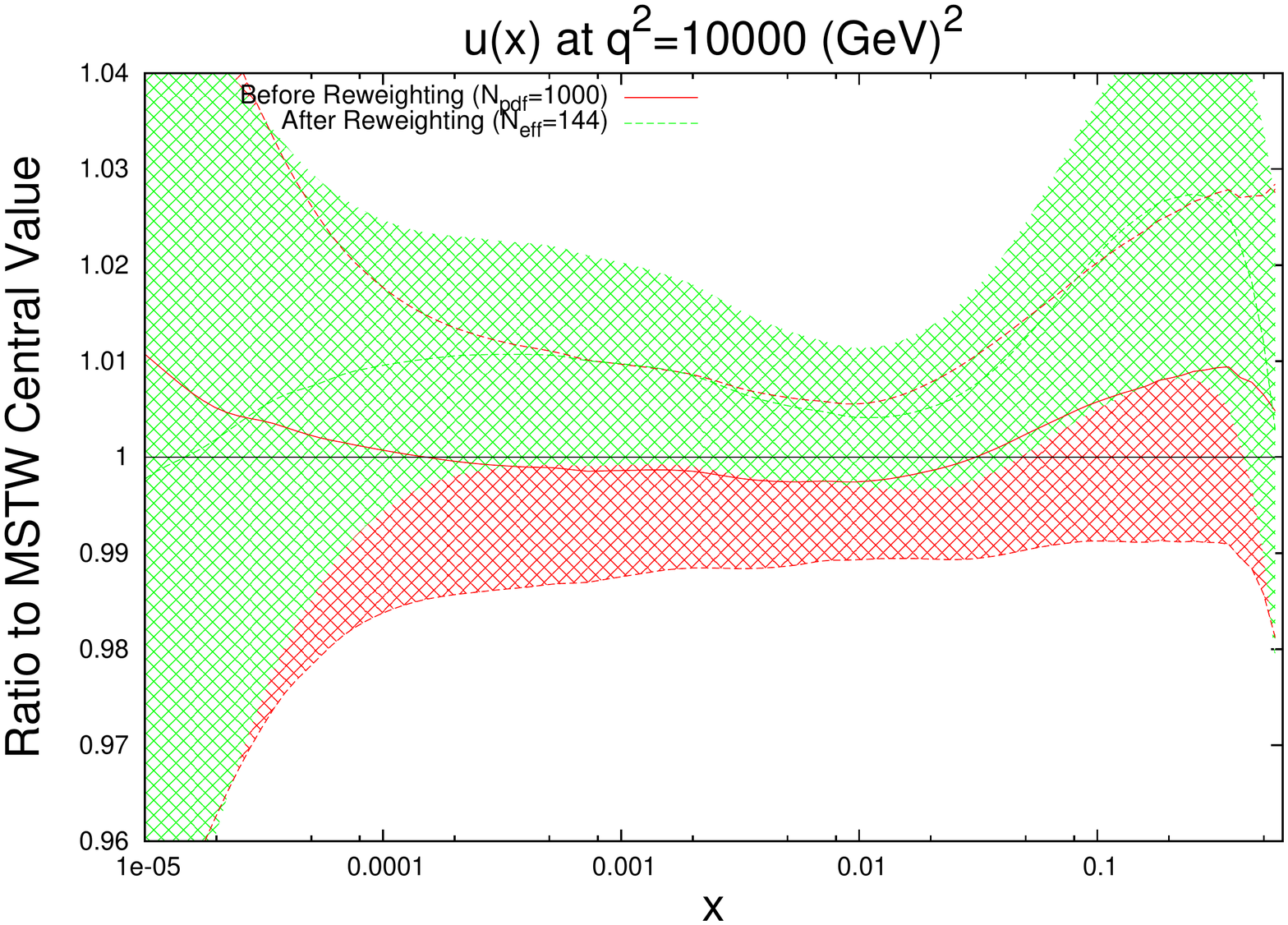}\\
\includegraphics[width=0.49\textwidth]{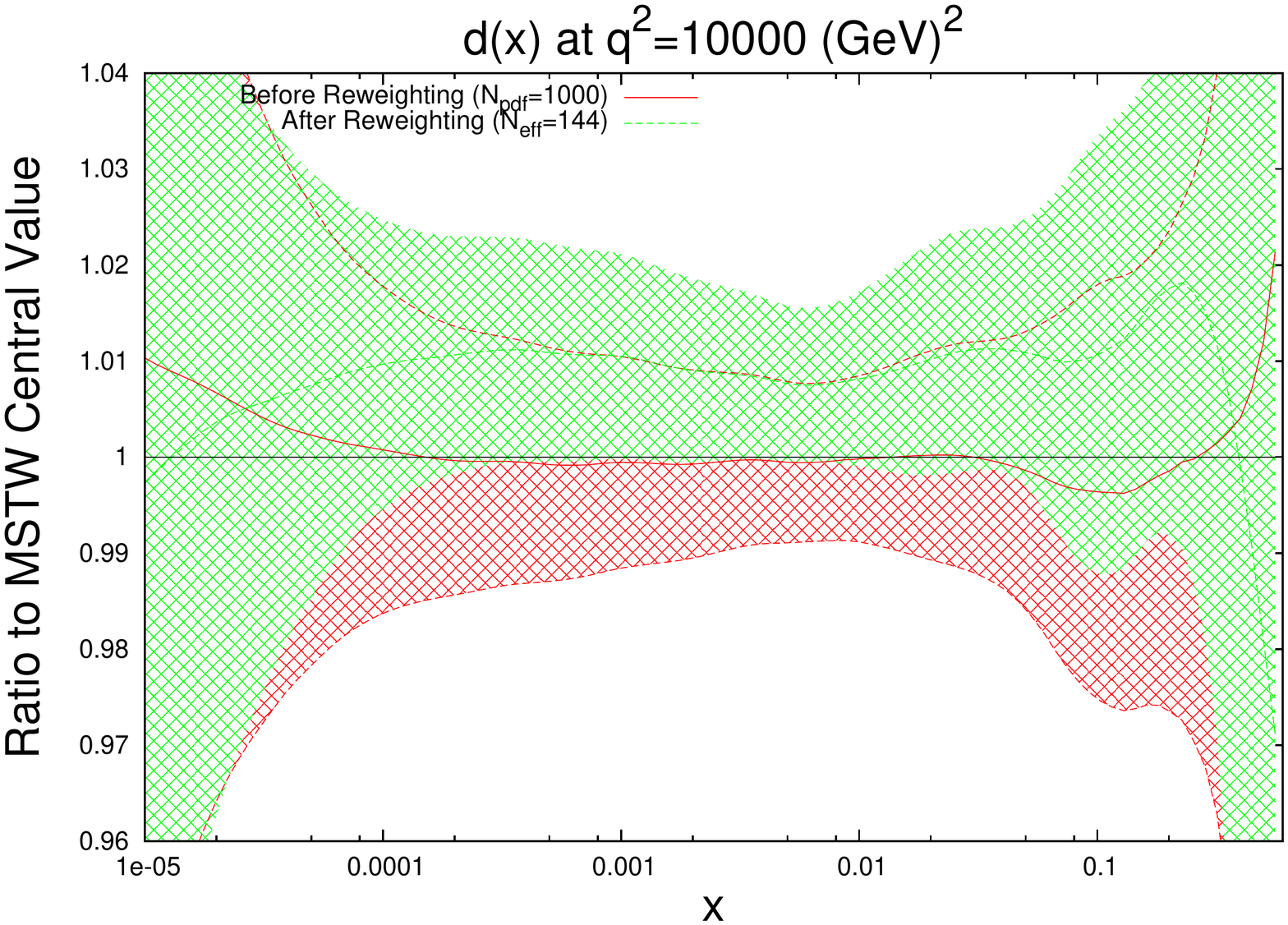}
\includegraphics[width=0.49\textwidth]{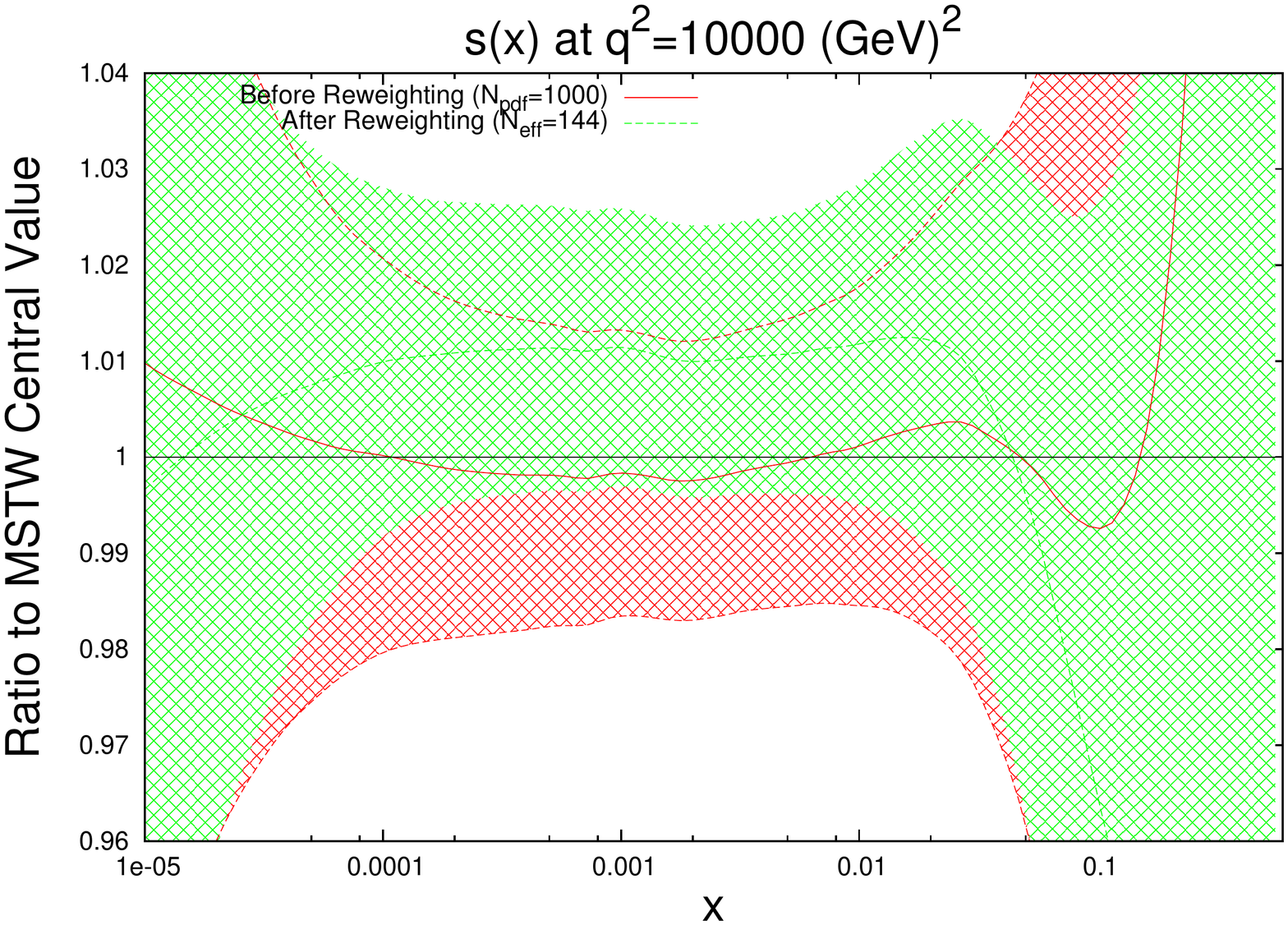}\\
\vspace{-0.8cm}
\caption{Effect of the ATLAS combined inclusive jet data on the gluon and quark PDFs. Here, additive errors are used in the determination of the $\chi^2$, and the two anomalous points discussed in the previous section are additionally cut.}
\label{reweight_ratio}
\end{figure}

As discussed for the pure 7~TeV fit, the way in which the systematic errors are 
treated is important to the quality of fit due to the systematic shift between 
data and theory. So far for this combined data set the multiplicative definition 
has been used since this is the treatment which most closely follows the 
HERAPDF/ATLAS analysis. Now, the additive definition is discussed. Since the 
same shift upwards from the data to the theory is seen in the ATLAS combined 
data set, it is expected to give a worse fit. This is true, and for MSTW2008 NLO 
PDFs, the fit becomes 2.44 per point, more than doubling the $\chi^2$ from the 
multiplicative treatment. However, a large part of this $\chi^2$ is localised to 
two anomalous points, even after the HERAPDF cuts. These are the highest $p_T$ 
bin of the highest rapidity bin of the 7~TeV data, and the lowest $p_T$ bin 
(after cuts) of the third rapidity bin of the 2.76~TeV data. Removing just 
these two additional points reduces the $\chi^2$ to 1.46 per point (the effect 
is rather less pronounced for the $R=0.6$ data). Since the MSTW fitting code 
currently uses additive errors for all data sets, it is proposed to remove 
these points for a PDF fit including this data. The $\chi^2$ value is shown 
in the third column of Table \ref{table:chi_ratio}.

The effect on the PDFs using the reweighting technique is shown for the case of 
multiplicative errors in Fig. \ref{reweight_ratio2} and for additive errors in 
Fig. \ref{reweight_ratio}. In both cases, the central value of the reweighted 
gluon is consistent with standard MSTW 2008 central value across all values of 
$x$ and in the multiplicative case it is very similar to the reweighted pure 
7~TeV gluon. The error bands are reduced in size more significantly than when 
using just the 7~TeV data, and the additive treatment seems to have more of an 
effect in this sense than the multiplicative. The upward shift in the quark PDFs 
and also the error constraints are larger when using the additive treatment. 
Clearly there is more constraint on the gluon with the 2.76~TeV data included, 
and the reduction in systematics is allowing more information on the PDFs to be 
extracted. The quark PDFs are also shown; although the effect is again larger 
than the pure 7~TeV case, there is very little movement from the central MSTW 
value. 
%Another hint to the improvement in PDF extraction from using only 7~TeV 
%is the reduction in the number of effective PDFs, $N_{eff}$, in the reweighting 
%procedure for both of the error treatments. 
When using multiplicative errors, 
the $\chi^2$ is reduced from 0.974 to 0.962 by reweighting and $N_{eff}=633$, 
and for additive 
errors the effect is larger as expected from the reweighted plots, changing from 
1.45 to 1.26, and $N_{eff}=144$, a much lower value. 
(Note that the $\chi^2$ obtained from the average of the random PDF
sets is not identical to that obtained using the best fit PDF set, but is 
always very close.) Whilst the two additional data points cut were deemed 
necessary to provide an acceptable $\chi^2$ value, it was observed that even 
when including these points, the reweighed PDFs for the 
additive treatment were essentially unchanged. Hence, the difference between 
Figs. \ref{reweight_ratio} and \ref{reweight_ratio2} can be 
attributed to the differing error treatments.

\subsection{CMS Inclusive Jets}

\begin{figure}[h!]
\includegraphics[width=1.0\textwidth]{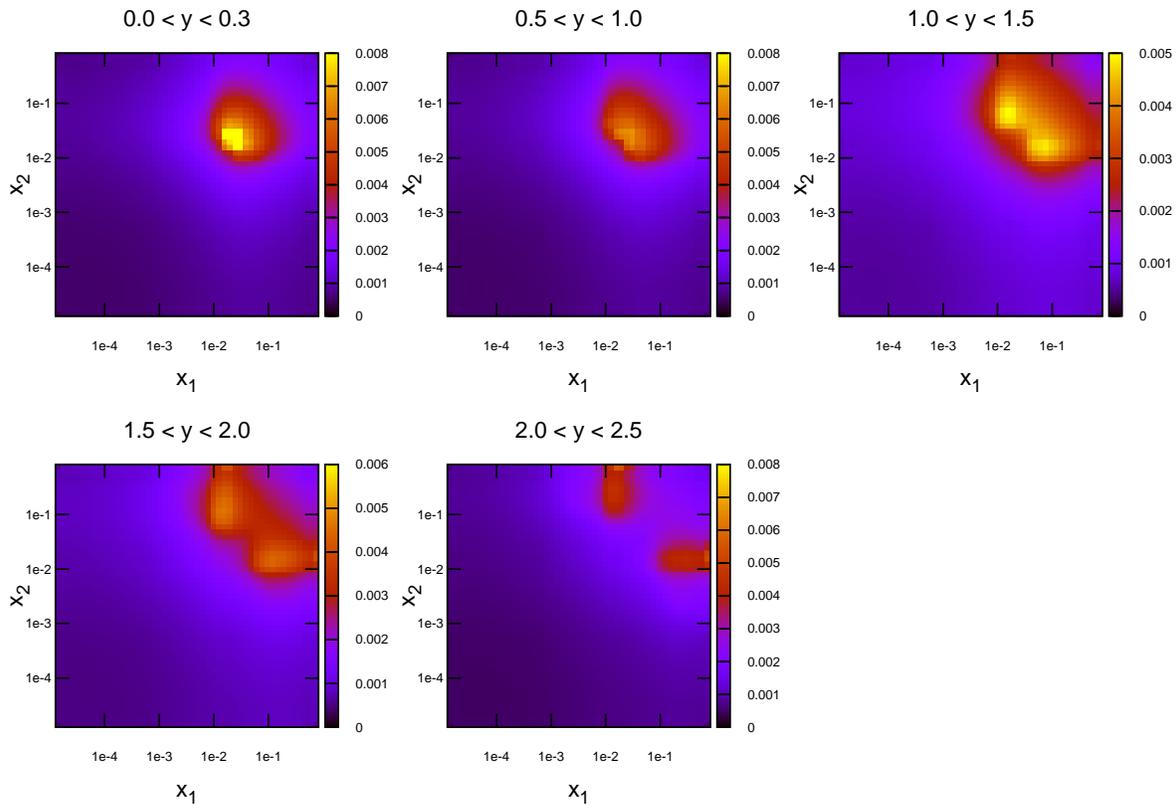}
\vspace{-0.8cm}
\caption{Distribution of $x_{1,2}$ values for NLOjet++ events in the CMS inclusive jet calculation}
\label{partons_cms}
\end{figure}

\begin{figure}
\begin{center}
\subfigure[$y<0.5$]{\includegraphics[width=0.4\textwidth]{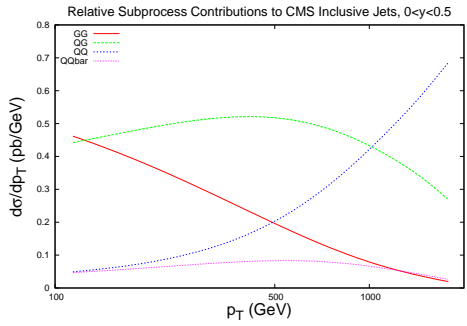}}
\subfigure[$0.5<y<1.0$]{\includegraphics[width=0.4\textwidth]{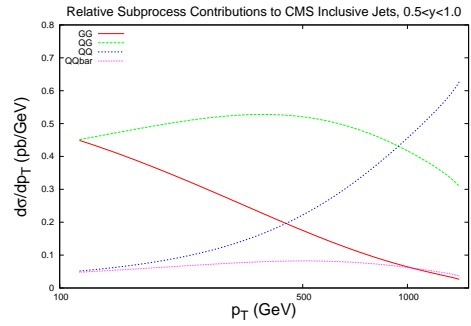}}
\subfigure[$1.0<y<2.5$]{\includegraphics[width=0.4\textwidth]{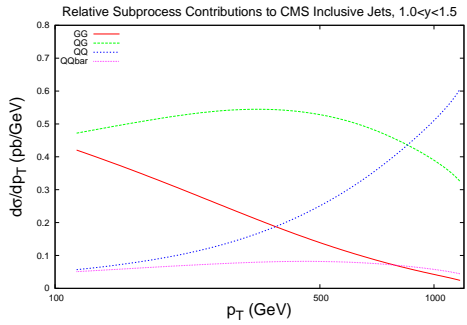}}
\subfigure[$1.5<y<2.0$]{\includegraphics[width=0.4\textwidth]{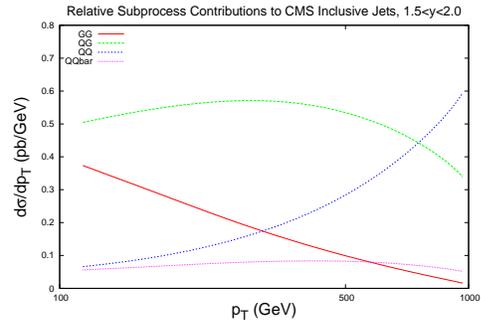}}
\subfigure[$2.0<y<2.5$]{\includegraphics[width=0.4\textwidth]{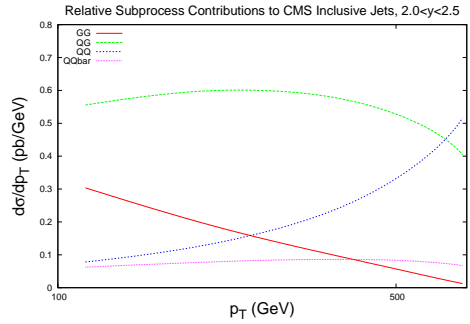}}
\end{center}
\vspace{-0.8cm}
\caption{Contributions of different initial-state parton combinations to the CMS inclusive jet cross section calculation}
\label{partons2_cms}
\end{figure}

To date the LHC dataset with the most resolving power for PDFs is that released 
by the CMS collaboration \cite{cmsinc} in early 2013. This analysis, like the 
earlier ATLAS analysis, was performed at 7~TeV. However, a much higher collected 
luminosity of 5$fb^{-1}$ is included, and so statistical errors are greatly 
reduced across the phase space.
Compared to the ATLAS measurement, the jet $p_T$ spectrum extends much higher to 
2~TeV, whereas it is also cut off higher, only going down to 114~GeV. There is 
also less rapidity span for the CMS jets, which are only measured to a 
rapidity of 2.5. The overall effect is to have more pronounced sensitivity to 
high-x PDFs, and lower sensitivity to low-x PDFs. This can be seen in 
Fig. \ref{partons_cms}, where the $(x_1,x_2)$ distribution for each event 
generated is shown. The reach to low $x$ is limited to $10^{-3}$, but each 
distribution is shifted more towards the high $(x_1,x_2)$ region.
The partons which are probed by the data are therefore naturally different from 
those of ATLAS. The greater emphasis on medium to high $x$ partons means a 
greater relative contribution from quarks. Fig. \ref{partons2_cms} shows the 
partonic composition of the calculation at each point in phase space. Unlike the 
ATLAS jets, the $gg$ subprocess does not dominate anywhere in the phase space, 
with $gq$ contributing maximally everywhere except for the very highest 
$p_T$ jets.

\begin{figure}[h!]
\centering
\includegraphics[width=0.9\textwidth]{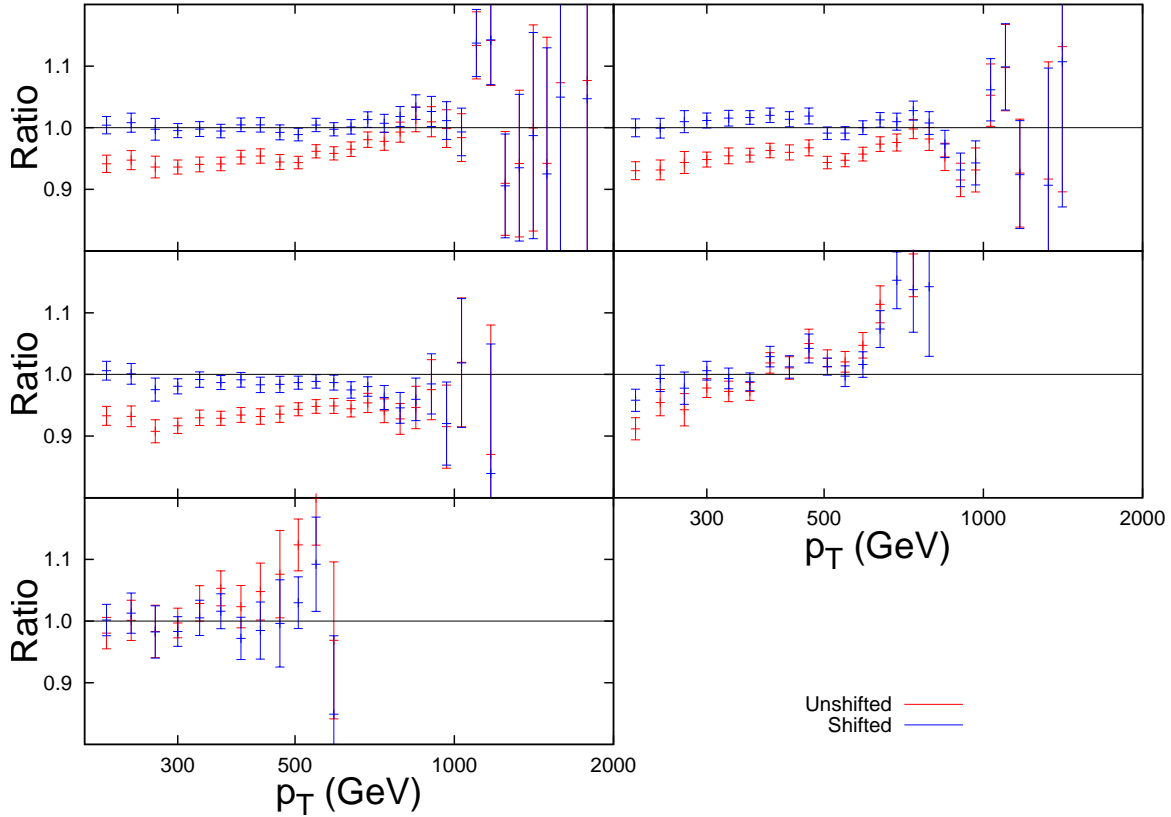}
\caption{Ratio of data to theory using MSTW 2008 NLO for CMS inclusive jets. Both the raw APPLgrid calculation and the calculation after systematic effects are taken into account are shown.}
\label{ratio_cms}
\end{figure}

The raw calculation using NLOjet++ interfaced with APPLgrid is in much better 
agreement with data than the ATLAS inclusive cross section. Whilst the ATLAS jet 
calculation was up to 30\% too high in some bins, the CMS calculation is never 
more than 10\% off. The systematics must again be taken account of in a $\chi^2$ 
fit, and the comparison to data again improves after this consideration. 
However, as Fig. \ref{ratio_cms} shows, the shifted data/theory points reflect 
the statistical fluctuations present in the unshifted points. For the ATLAS 
fit, it was clear that the statistical fluctuations were being washed out by 
the large freedom provided by the systematics.
The table of fits is shown in Table \ref{table:chi_cms}, and the corresponding 
systematic shifts in Table \ref{table:rk_cms}. The $\chi^2$ values are generally 
worse than for the ATLAS data. With fewer $r_k$ values, it is clear that there is 
less freedom to compensate for differences by using the systematic shifts. This 
is reflected by the distribution of $r_k$'s, which for ATLAS produced a majority 
below 0.5, but for CMS has a larger number of higher values.

\begin{table}[h!]
\centering
\begin{tabular}{c c c c}
\hline\hline
Scale & $p_T$/2 & $p_T$ & 2$p_T$ \\
\hline
MSTW 2008 & 1.92 & 1.48 & 1.12\\
\hline
\end{tabular}
\caption{$\chi^2$ per point (133 points) for NLO PDFs for CMS inclusive jet data.}
\label{table:chi_cms}
\end{table}

\begin{table}[h!]
\centering
\begin{tabular}{c c c c c c c c}
\hline\hline
$|r_k|$ & $<0.5$ & 0.5--1.5 & 1.5--2.5 & 2.5--3.5 & 3.5--4.5 &\\
\hline
MSTW 2008 & 8 & 8 & 2 & 1 & 0\\
\hline
\end{tabular}
\caption{Distribution of $r_k$'s (Total 19)}
\label{table:rk_cms}
\end{table}

\begin{figure}[h!]
\centering
\includegraphics[width=0.8\textwidth]{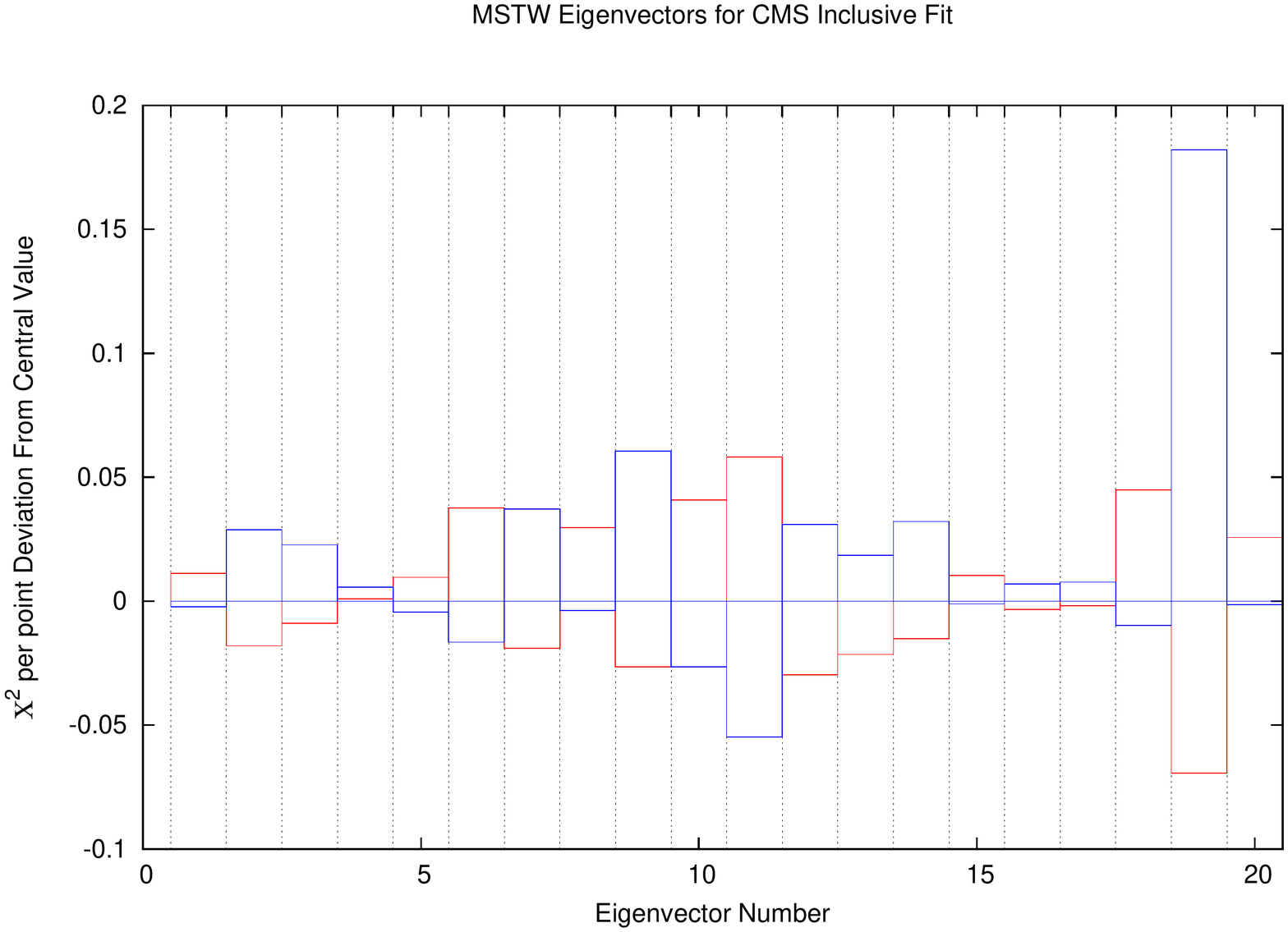}
\vspace{-0.8cm}
\caption{Change in fit quality from the central MSTW2008 PDF for each eigenvector in the set.}
\label{eig_cms}
\end{figure}

The same procedure as described for the ATLAS jets is applied to the CMS 
dataset. The variations of the fit under movements in the eigenvector directions 
are shown in Fig. \ref{eig_cms}. This time, there are significant improvements in 
some directions, with eigenvectors 11 and 19 reducing the $\chi^2$ the most. 
Eigenvectors 19 is most influenced by the gluon 
distribution while 11 contributes significantly to the uncertainty of a wide 
variety of PDFs.

\begin{figure}[h!]
\centering
\includegraphics[width=0.49\textwidth]{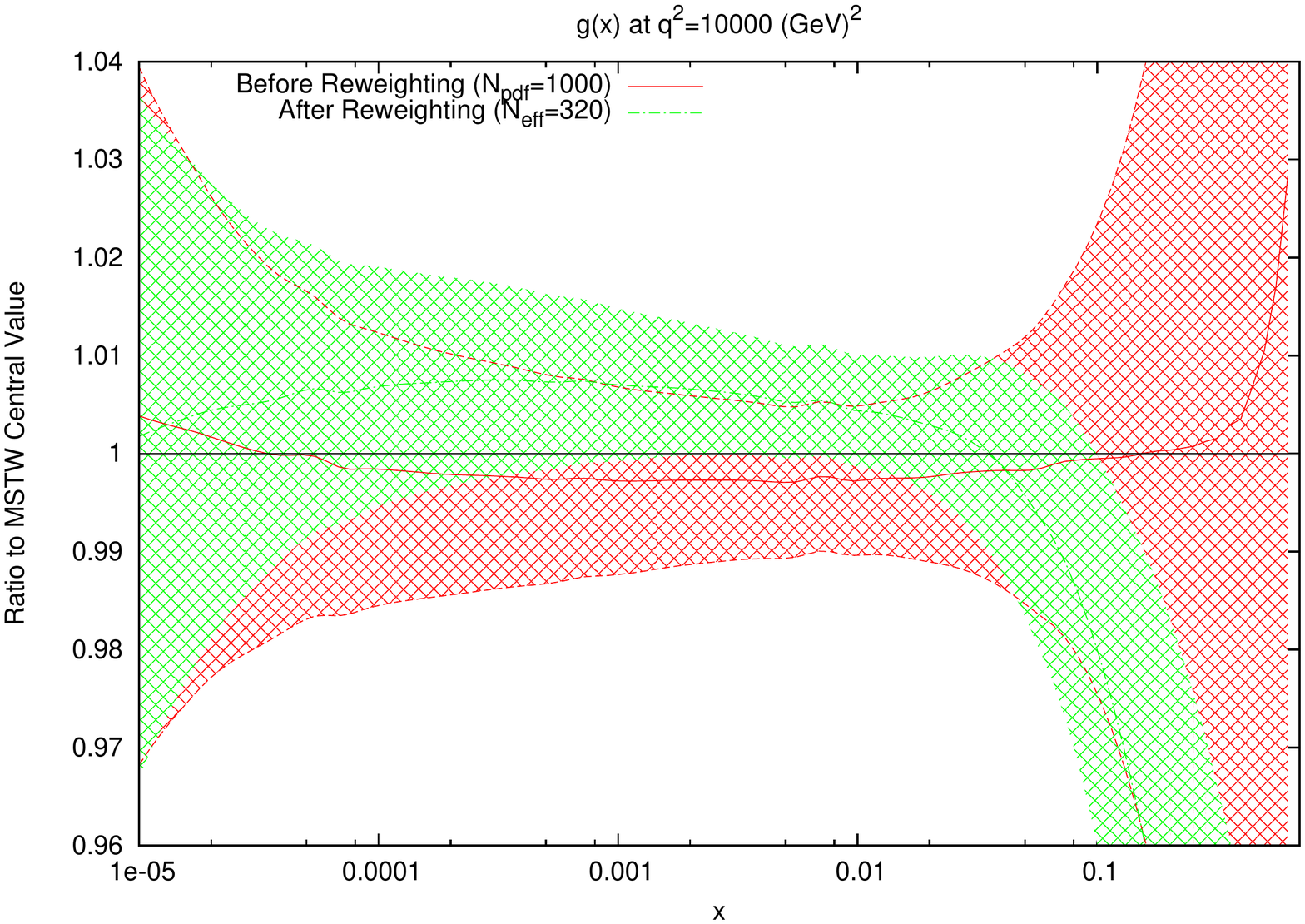}
\includegraphics[width=0.49\textwidth]{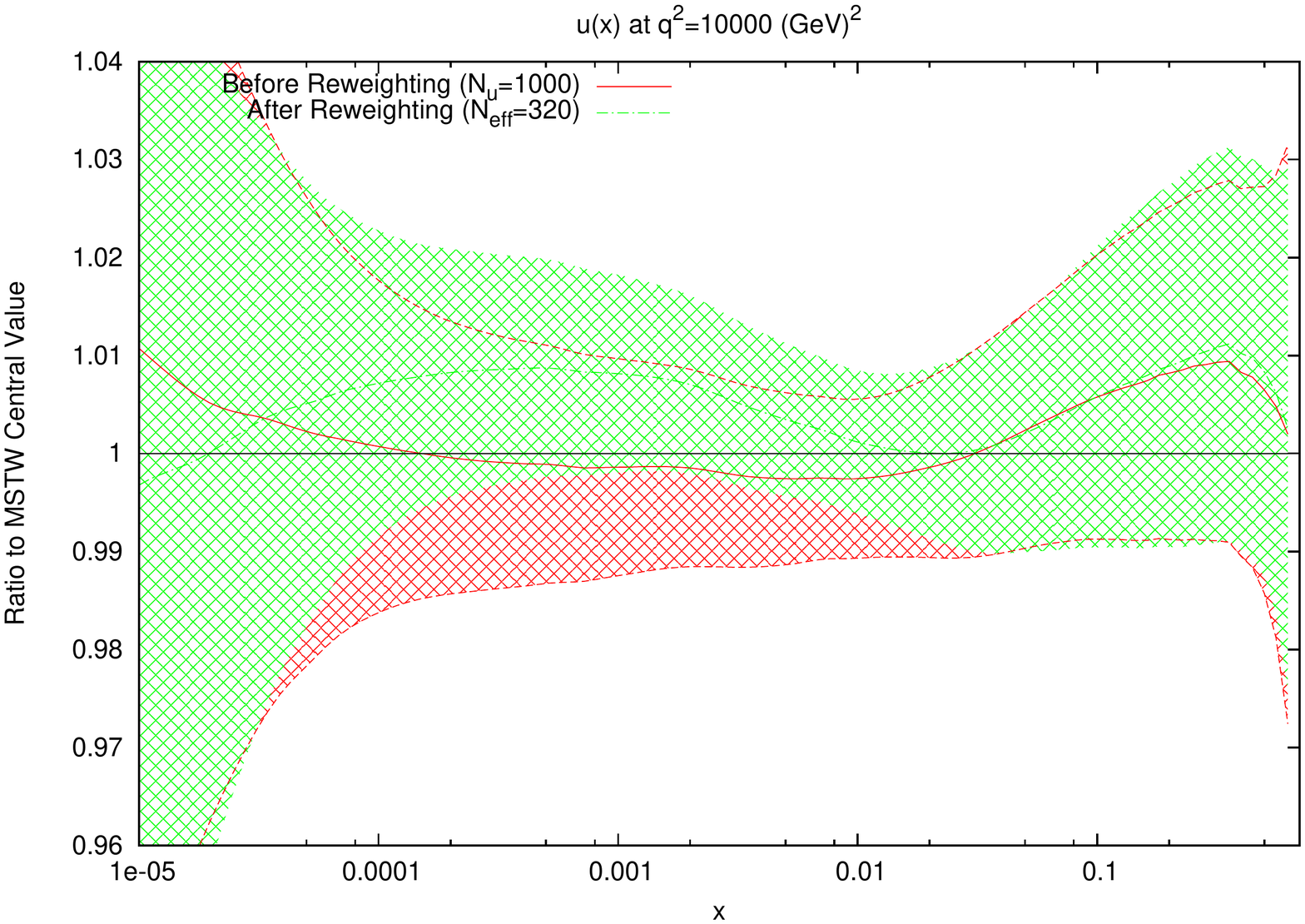}\\
\includegraphics[width=0.49\textwidth]{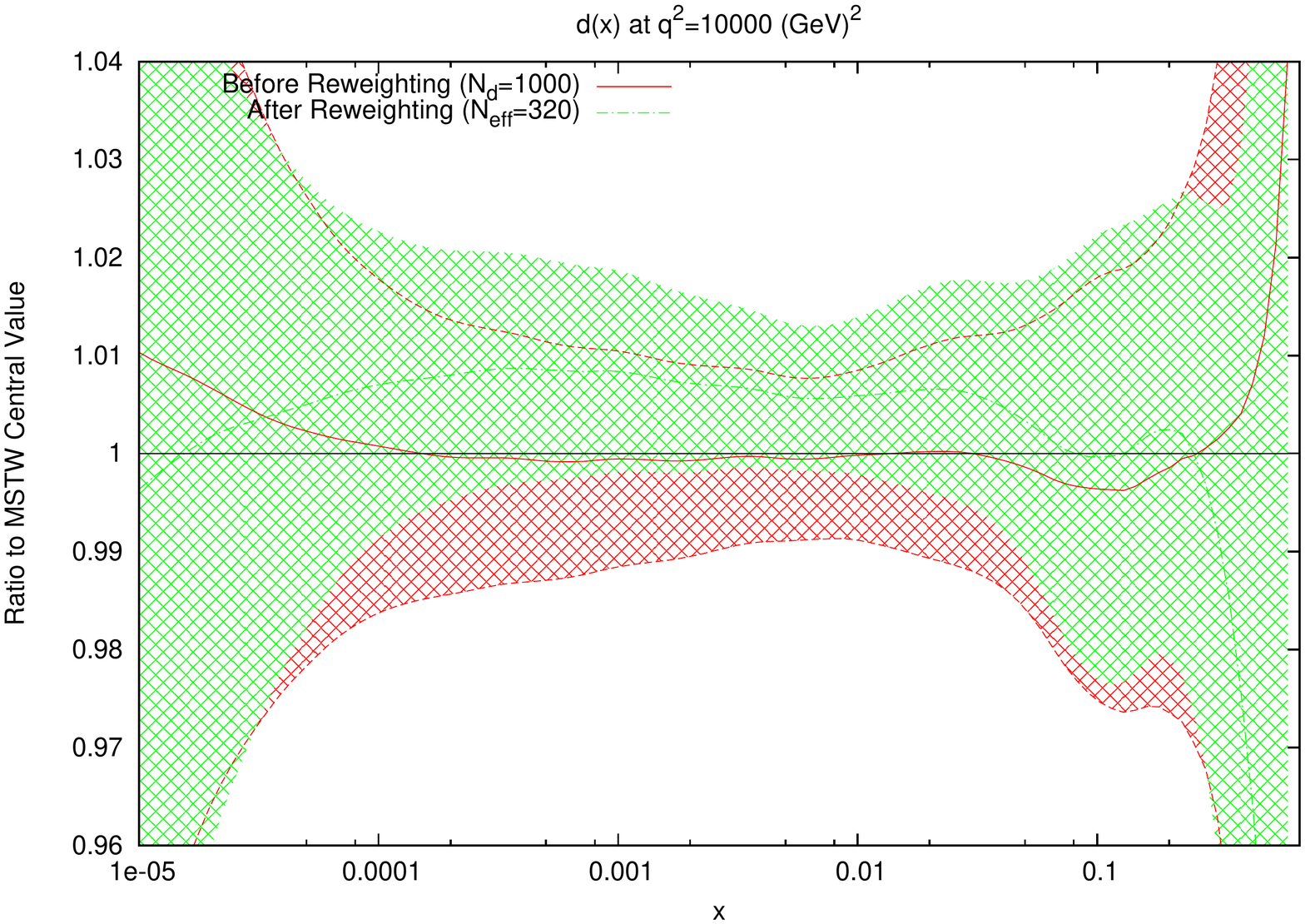}
\includegraphics[width=0.49\textwidth]{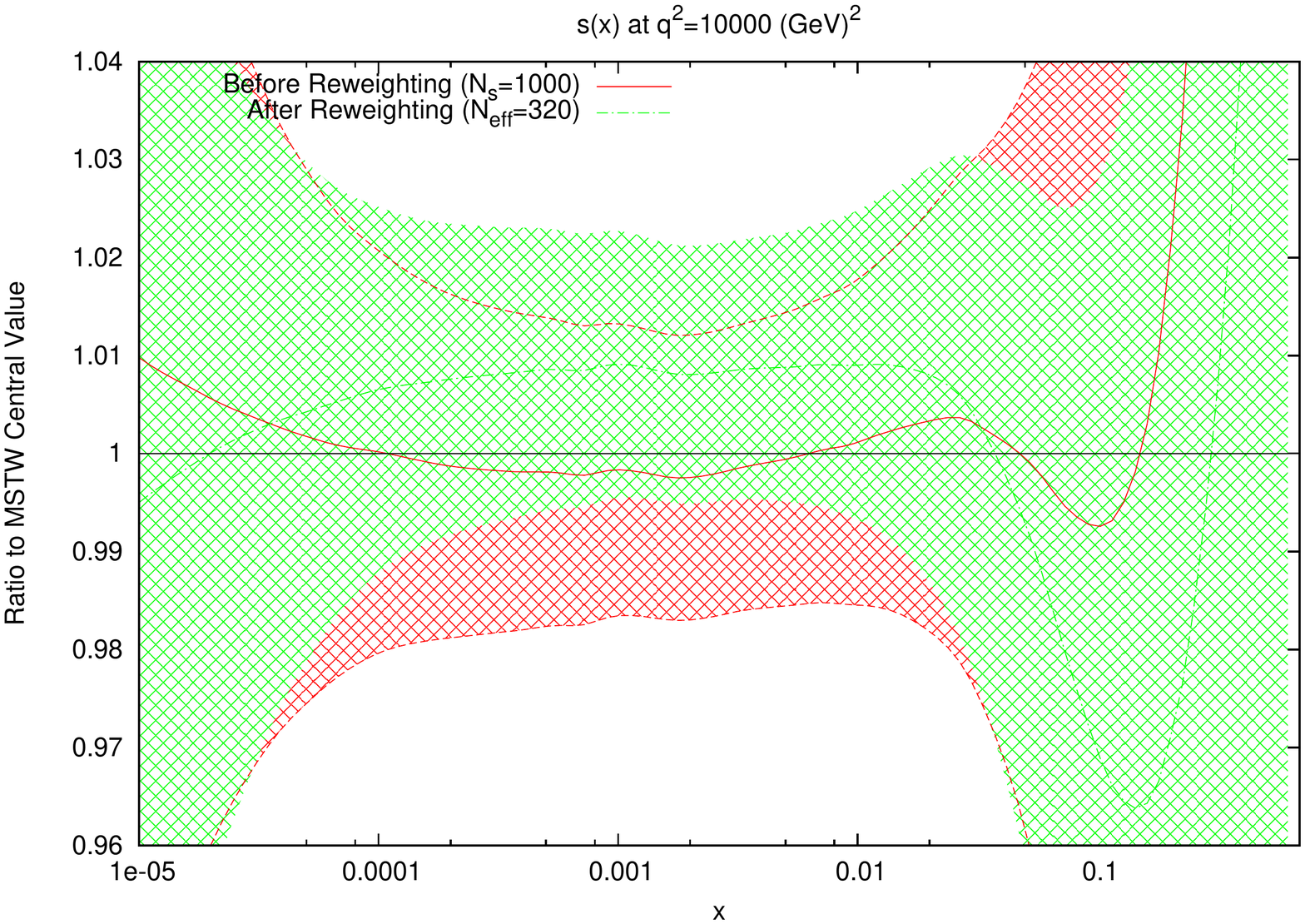}\\
\vspace{-0.8cm}
\caption{Effect of the CMS inclusive jet data on the gluon and quark PDFs.}
\label{reweight_cms}
\end{figure}

When the reweighting procedure is applied, the results of which are shown in 
Fig. \ref{reweight_cms}, the effect is larger than the full ATLAS combined 
dataset. The shape of the reweighted gluon agrees with the ATLAS reweighting, 
with a lower gluon at high $x$. What is significant in this case is the increased 
sensitivity to the quark PDFs. The reduction in error band in the up and 
down distributions is similar to that for the gluon. Even the error in the 
strange distribution is reduced in both directions across almost all values of 
$x$. The focus of the CMS data on higher values of $x$ has lead to a less 
dramatic effect on the gluon, but consistently better constraining of all 
quark PDFs. The reweighting improves the fit quality from 1.47 to 1.29.

\section{Dijet Cross Sections}
\label{chap:dijet}

\begin{figure}[h!]
\centering
\subfigure[$p_T^{av}$]{\includegraphics[width=0.9\textwidth]{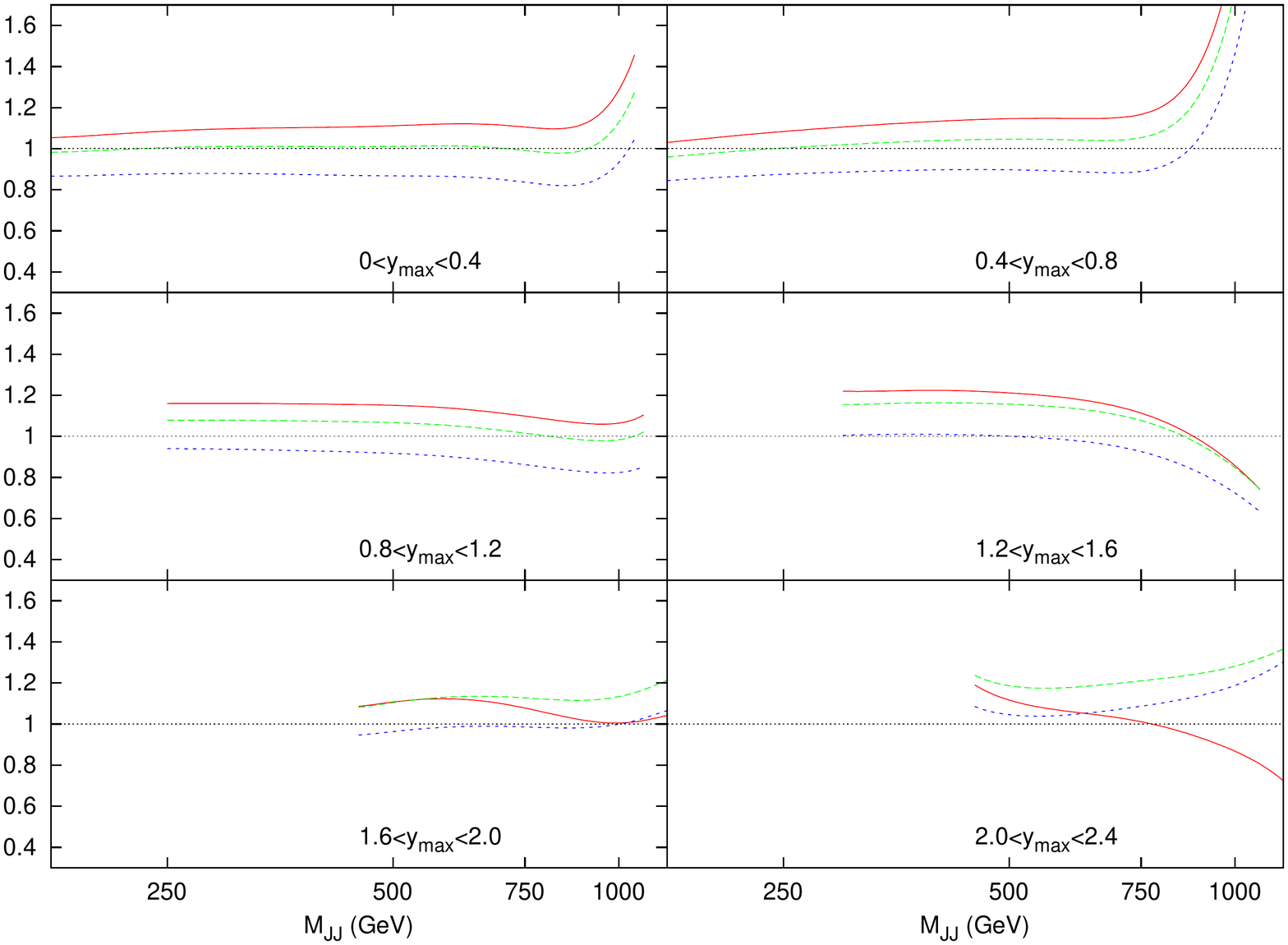}}
\vspace{-0.5cm}
\caption{Theory/Data ratio for D{\O} dijets, using multiples of $p_T^{av}$ as the choice of $\mu_R$ and $\mu_F$. The multiples are $0.5$ (red), $1.0$ (green) \& $2.0$ (blue)}
\label{D0_data_ratios1}
\end{figure}

In all previous MSTW fits, only inclusive jet data has been included into the 
fit. This is due to the overlap with the inclusive jet cross sections, but also 
to the uncertainties in the calculation of dijet cross sections and the scale 
choices therein. Whilst there is very limited scope for changing the kinematic 
choice of scale for inclusive jets, there are many possibilities when 
considering dijet cross sections. As such, this section presents a thorough 
study of the effect of the choice of renormalisation and factorisation scale 
choice on dijet predictions at both the Tevatron and the LHC, and the 
feasibility of including these datasets in a PDF fit is tested.

Before 2011, the only dijet data available over a range of rapidity was from D{\O} at 
the Tevatron. Studies into the comparison between data and theory were 
conducted \cite{d0-dijet-paper}, but inconsistencies in the scale uncertainty 
were found. The NLO calculations for the data set were performed using the 
average jet $p_T$ as the scale choice, and this was shown to exhibit strange 
behaviour at high rapidities. This is demonstrated in Fig. 
\ref{D0_data_ratios1}, where the predictions over (smoothed) data 
for 0.5, 1 and 2 times the scale 
choice are shown to cross over at high $y_{max}$ and mass,
and for %0.5p_T$ to become very low at high rapidity and jet mass. In order to 
understand the source of this behaviour, the kinematics of the process must be 
studied.

\subsection{Kinematics of Dijet Production}

\begin{figure}[h!]
\begin{center}
\includegraphics[width=\textwidth]{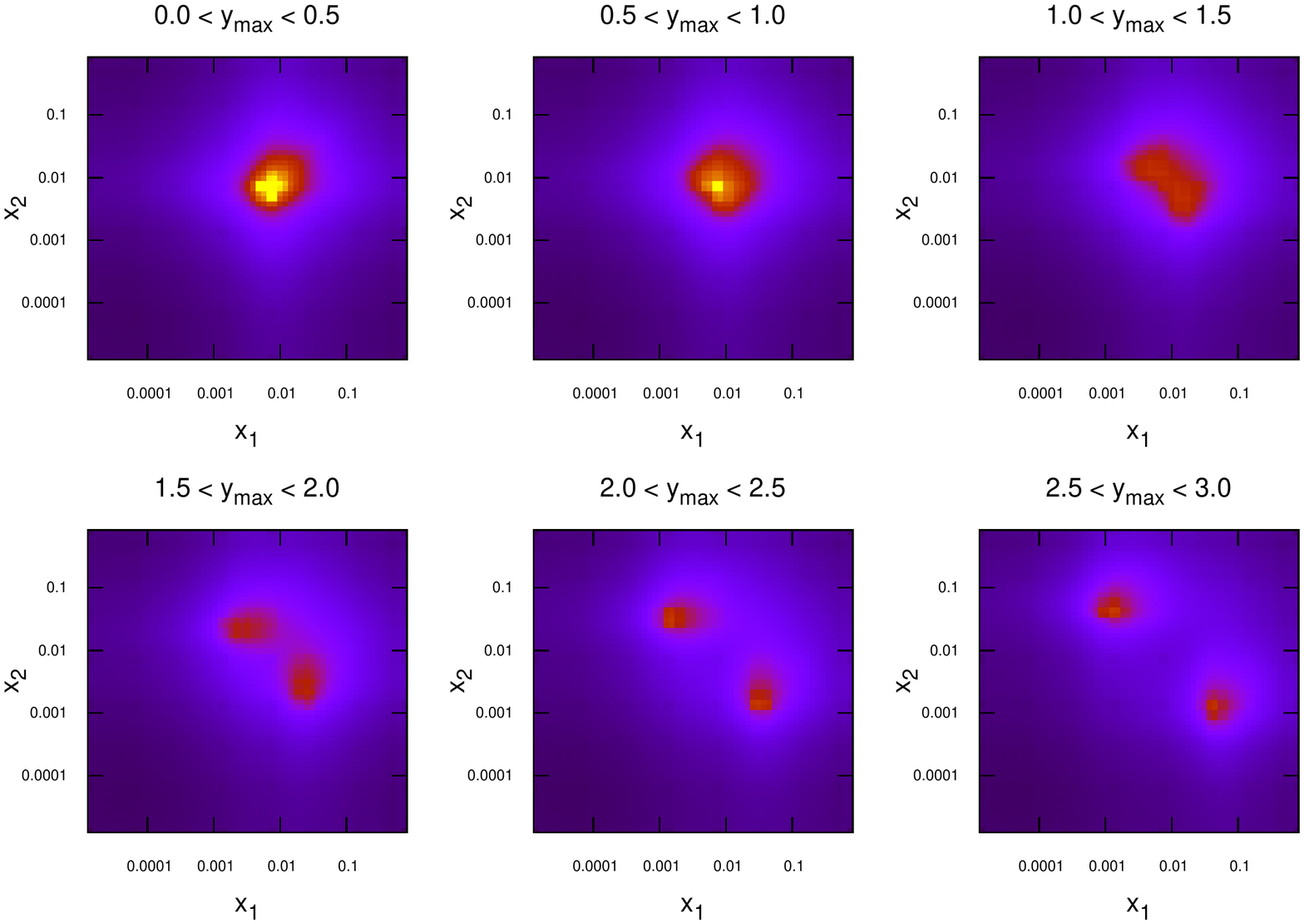}
\end{center}
\vspace{-0.8cm}
\caption{Values of $x_1$ and $x_2$ for each event generated in NLOJet++ for 
dijets at D{\O}}
\label{xreach3}
\end{figure}

\begin{figure}[h!]
\begin{center}
\includegraphics[width=\textwidth]{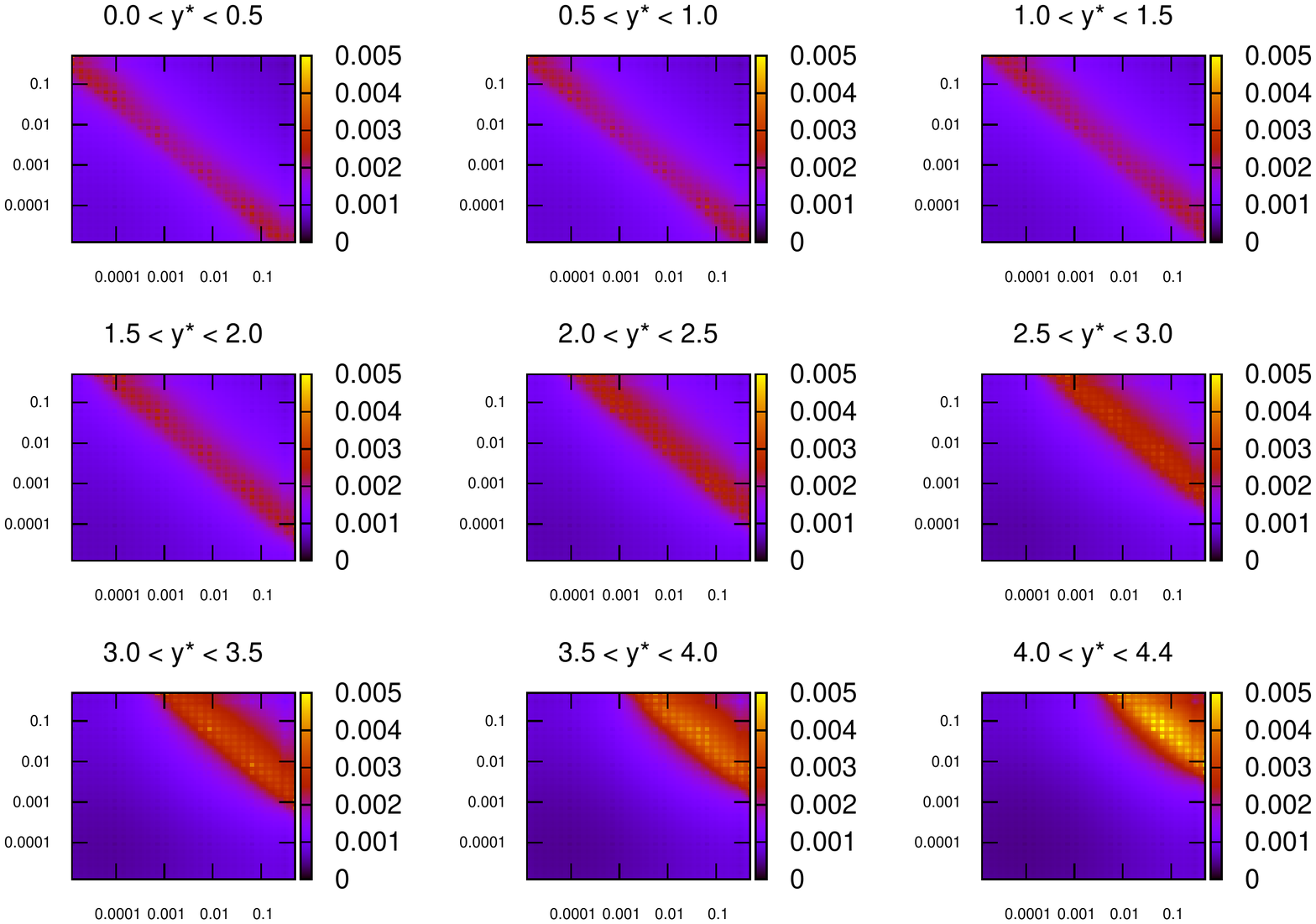}
\end{center}
\vspace{-0.8cm}
\caption{Values of $x_1$ and $x_2$ for each event generated in NLOJet++ for dijets at ATLAS}
\label{xreach4}
\end{figure}

The kinematics of the dijet production process are defined using the invariant 
mass of the dijet system, $M_{JJ}$, and the rapidity of each of the jets in the 
event. A double-differential cross section is constructed using bins in the 
dijet mass and a combination of the two rapidities. The flexibility in the 
latter leads to different possibilities for rapidity binning, and the D{\O} and 
ATLAS \cite{atlas-inc-paper} measurements use differing definitions. Where D{\O} 
uses $y_{max}$, the maximum rapidity of the two jets comprising the dijet pair, 
ATLAS chooses $y^*=(y_1-y_2)/2$, the difference between them.
This is the cause of the greatly differing $x$ distributions of 
Figs. \ref{xreach3} \& \ref{xreach4}. Using the maximum jet rapidity results 
in a similar pattern to inclusive jets, due to the fact that only the rapidity 
of one jet is considered. At high rapidities a single high-$x$ parton must 
combine with a single low-$x$ one, and low rapidities require equal values of 
$x$ in both partons.
Using the rapidity difference, however, allows a much wider range of parton 
momentum fractions to produce dijets in all $y^*$ bins. The observed shift 
towards high $x$ at high $y^*$ is in fact due to the fact that only high $M_{JJ}$ 
events are measured at these rapidities. These high $M_{JJ}$ events are also 
present in the other rapidity bins, however due to the power-like drop in cross 
section with dijet mass these events do not register in the respective plots 
and only the lowest mass bins can be seen.

\begin{figure}[h!]
\begin{center}
\subfigure[$y<0.5$]{\includegraphics[width=0.32\textwidth]{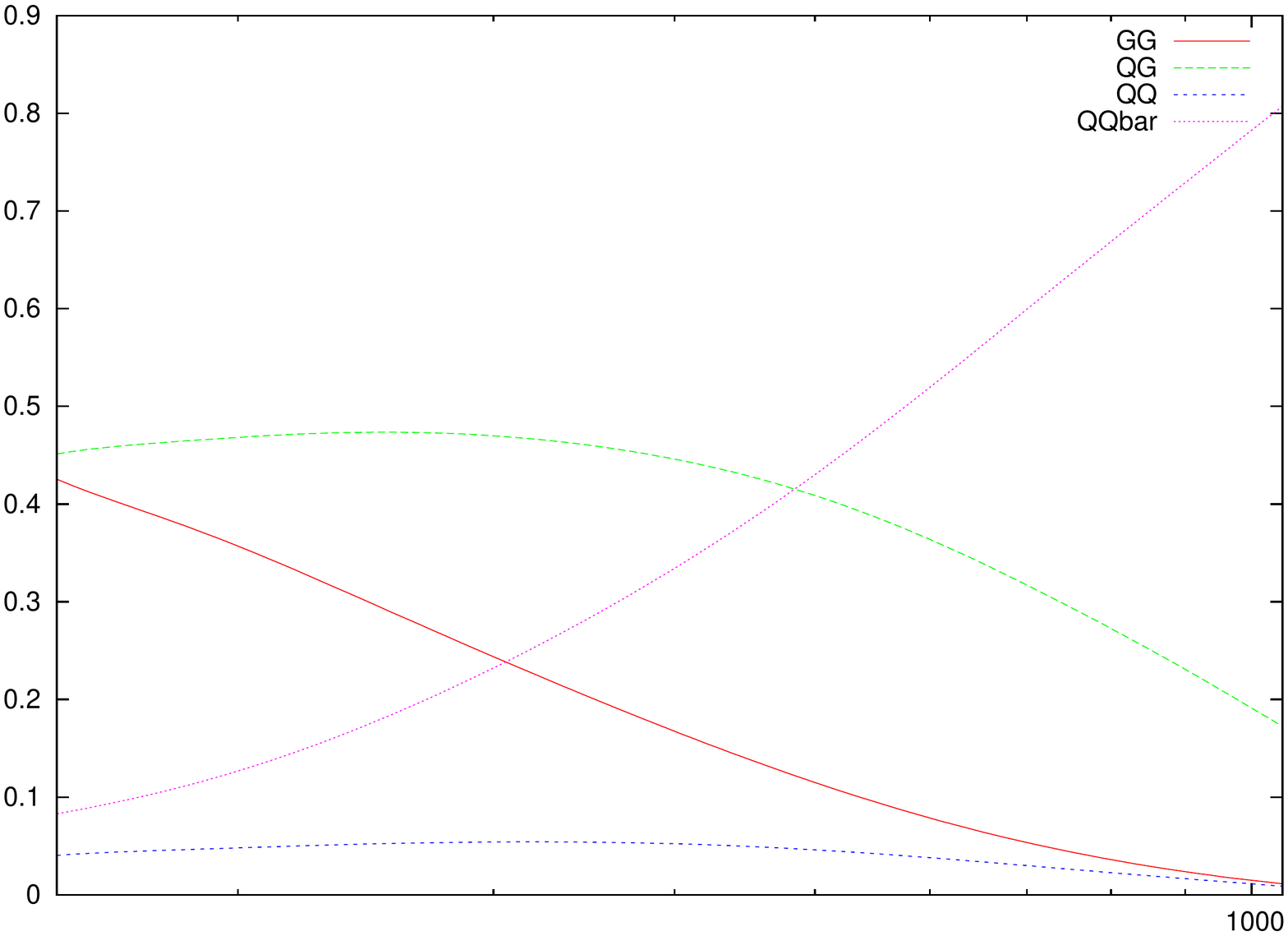}}
\subfigure[$0.5<y<1.0$]{\includegraphics[width=0.32\textwidth]{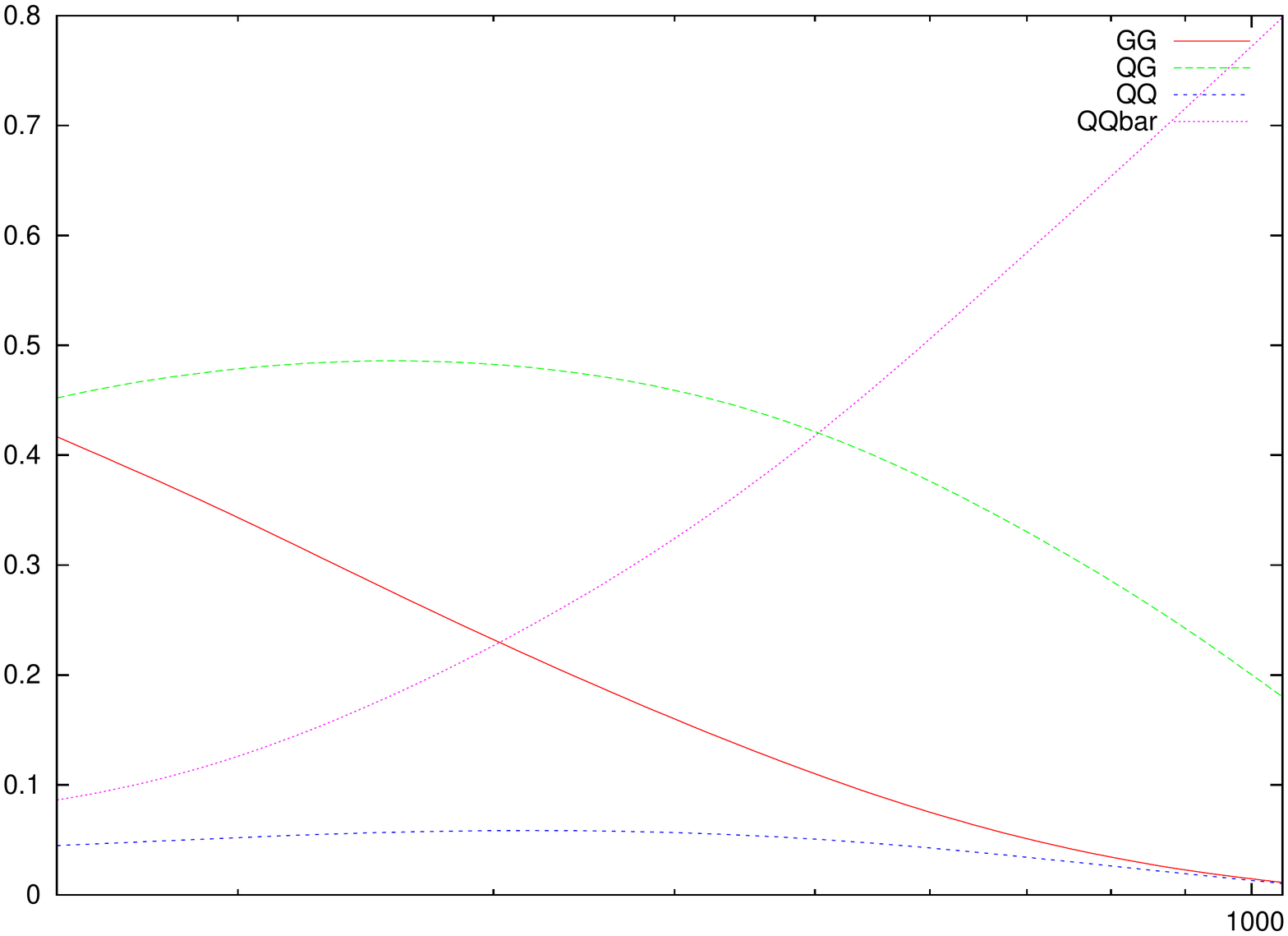}}
\subfigure[$1.0<y<1.5$]{\includegraphics[width=0.32\textwidth]{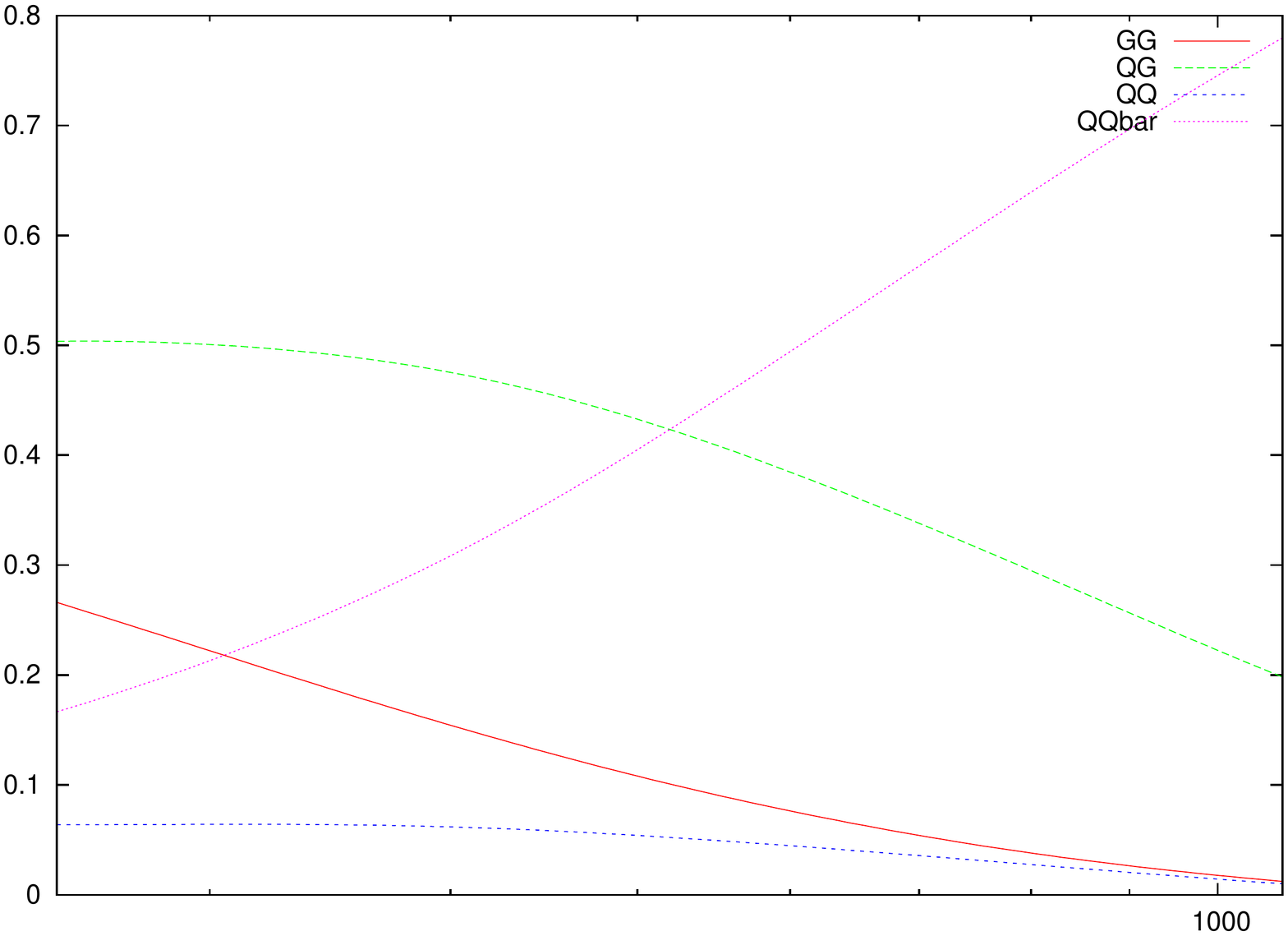}}
\subfigure[$1.5<y<2.0$]{\includegraphics[width=0.32\textwidth]{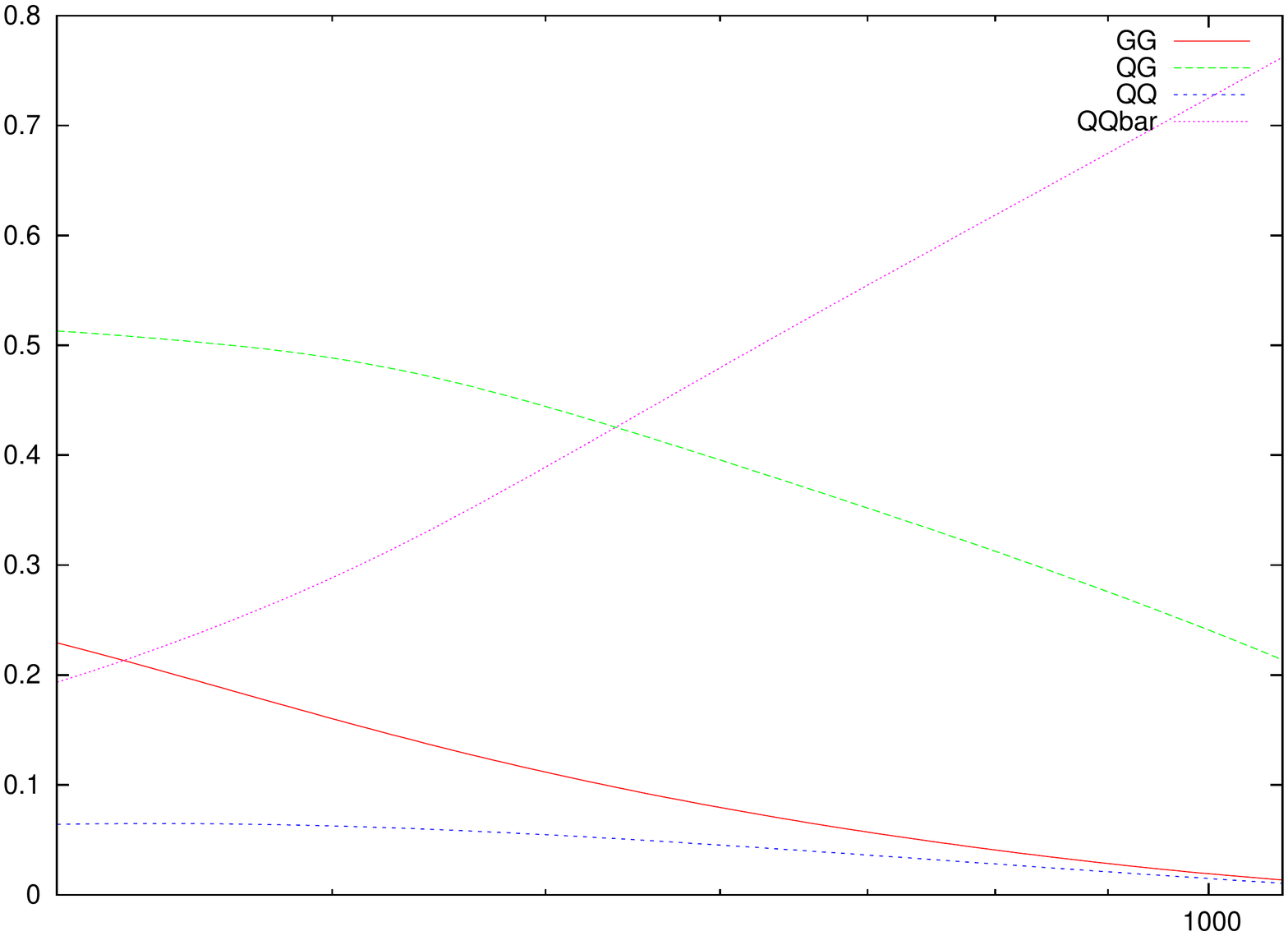}}
\subfigure[$2.0<y<2.5$]{\includegraphics[width=0.32\textwidth]{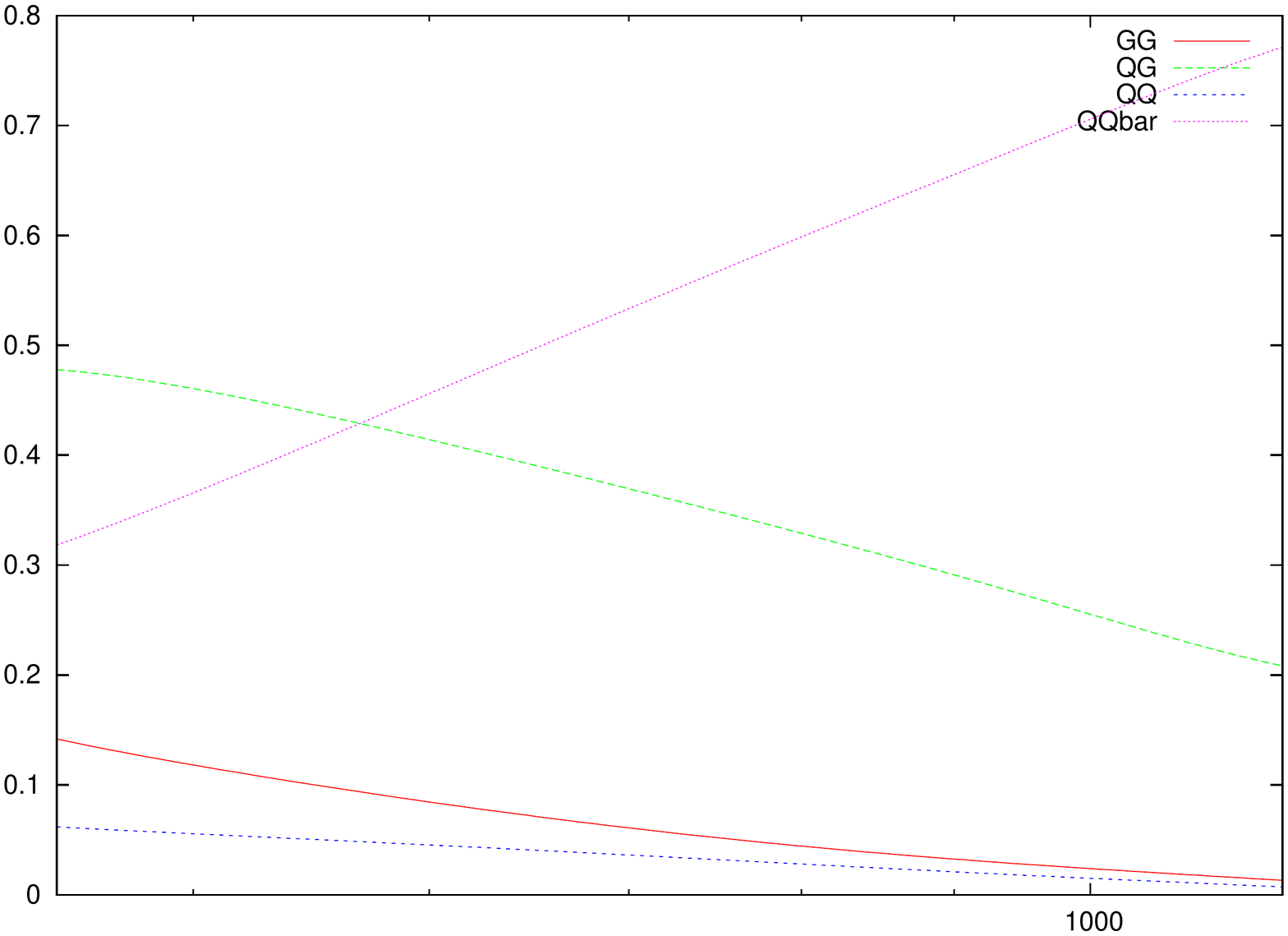}}
\subfigure[$2.5<y<3.0$]{\includegraphics[width=0.32\textwidth]{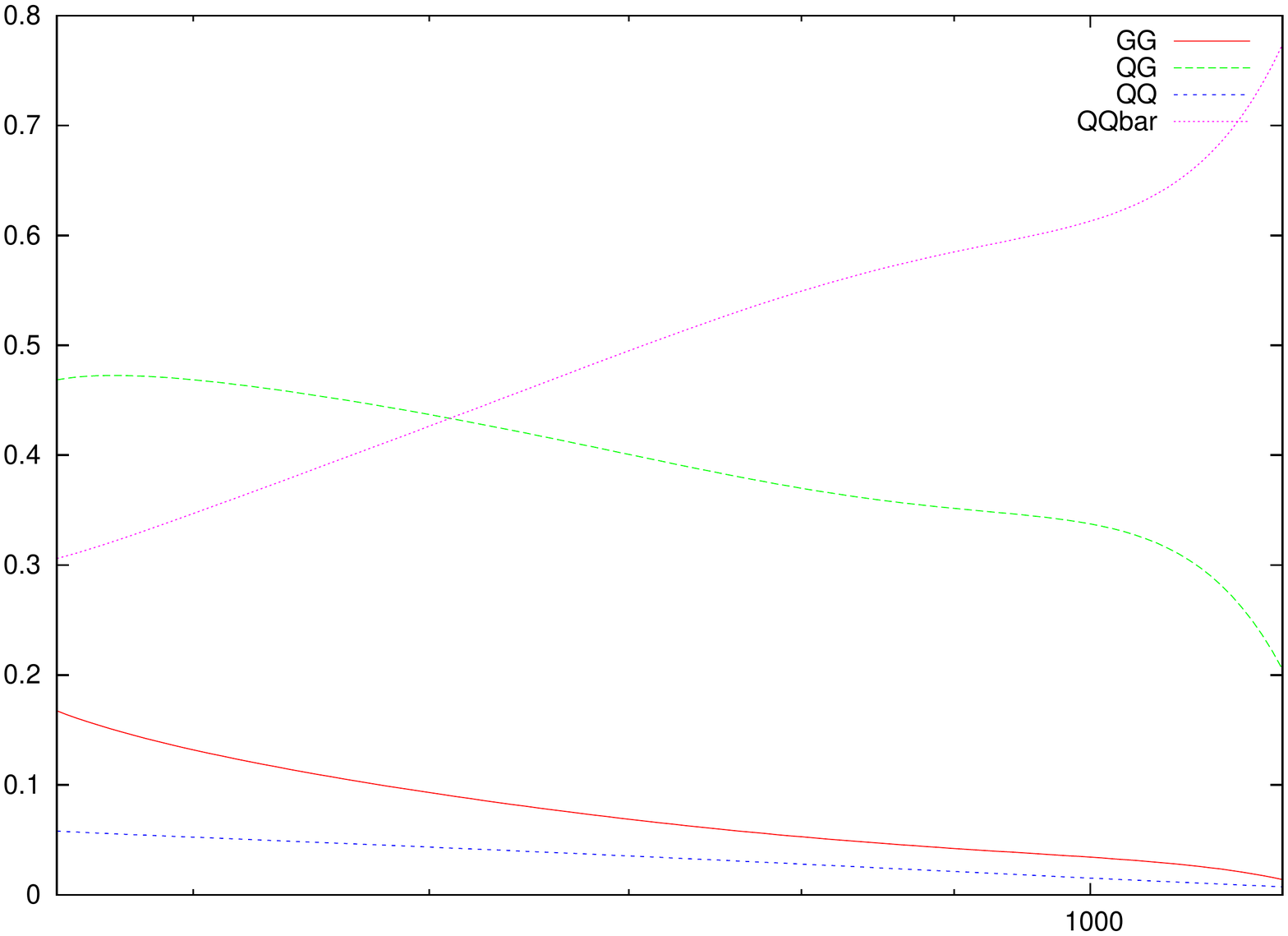}}
\end{center}
\vspace{-0.8cm}
\caption{Contributions of different initial-state parton combinations to the D{\O} dijet cross section calculation}
\label{partonsd}
\end{figure}

\begin{figure}[h!]
\begin{center}
\subfigure[$y<0.5$]{\includegraphics[width=0.32\textwidth]{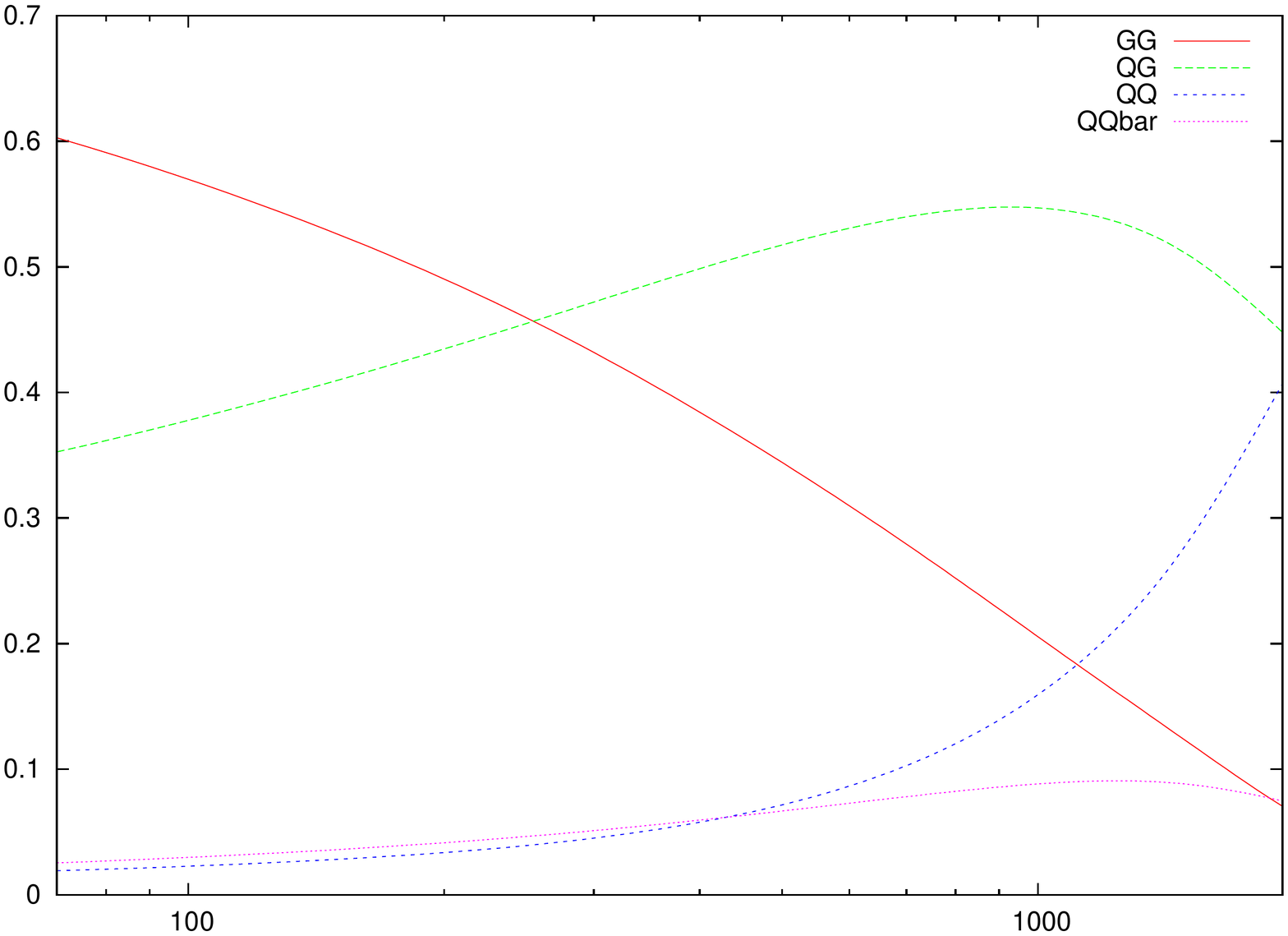}}
\subfigure[$0.5<y<1.0$]{\includegraphics[width=0.32\textwidth]{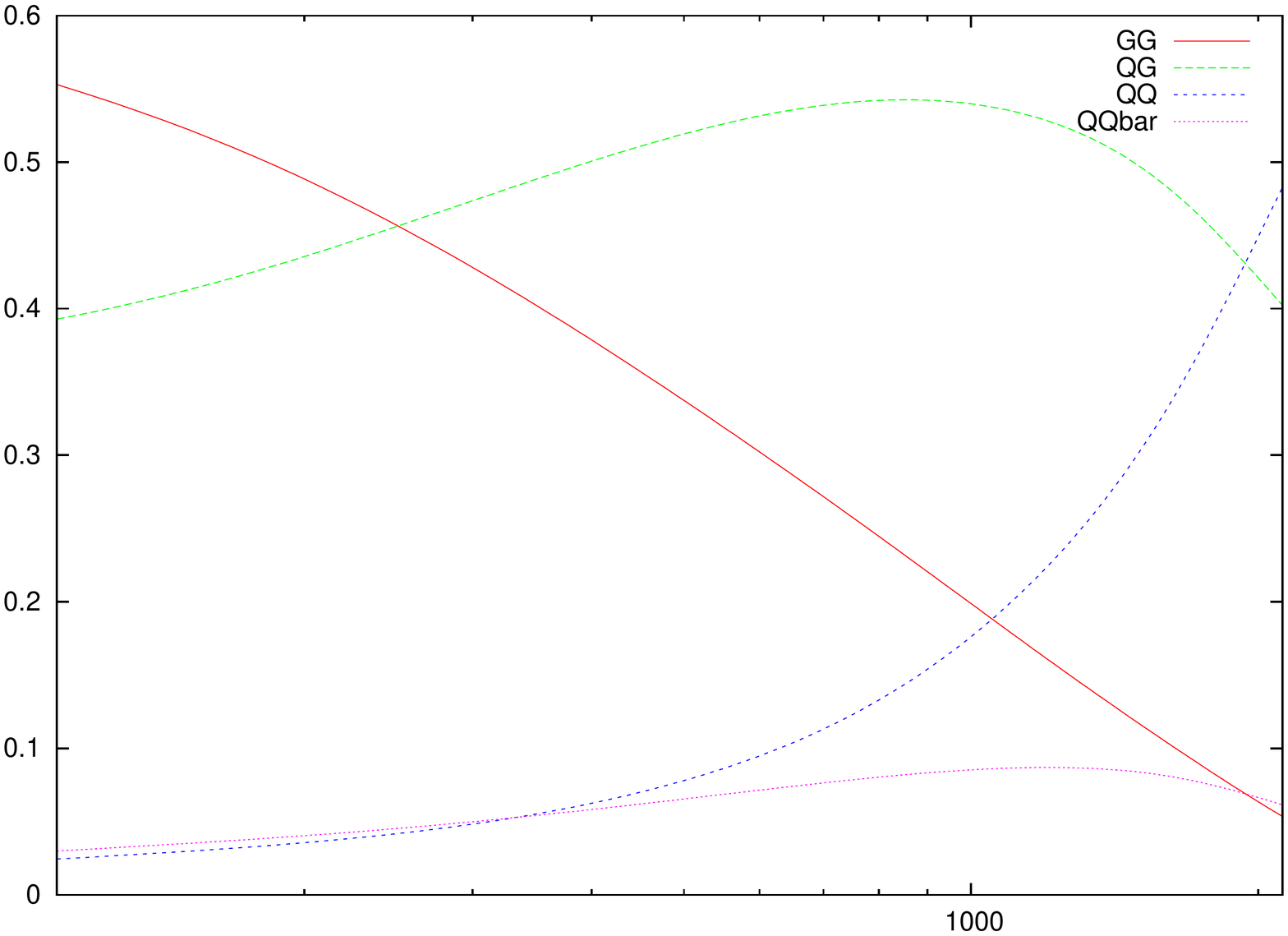}}
\subfigure[$1.0<y<1.5$]{\includegraphics[width=0.32\textwidth]{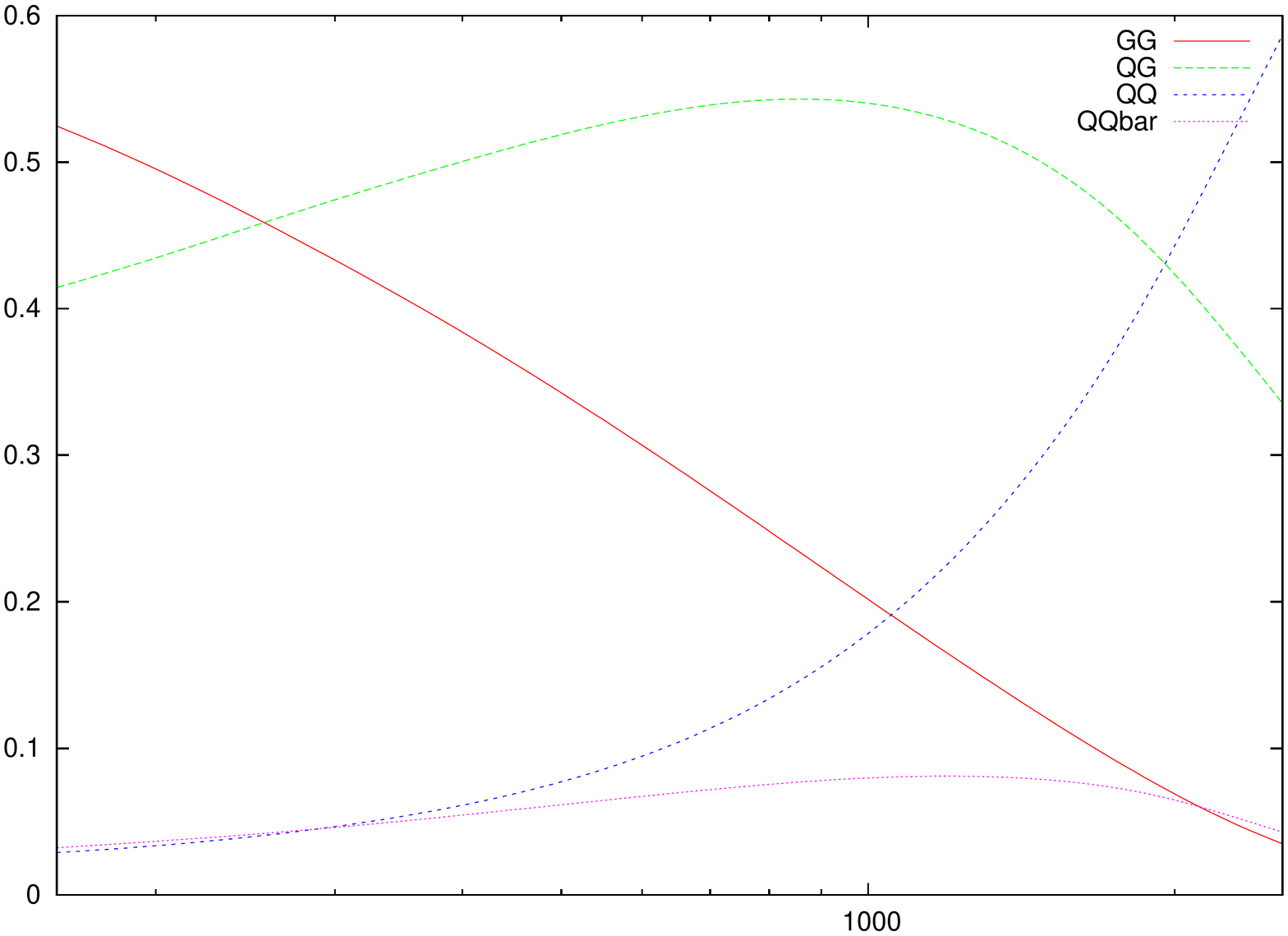}}
\subfigure[$1.5<y<2.0$]{\includegraphics[width=0.32\textwidth]{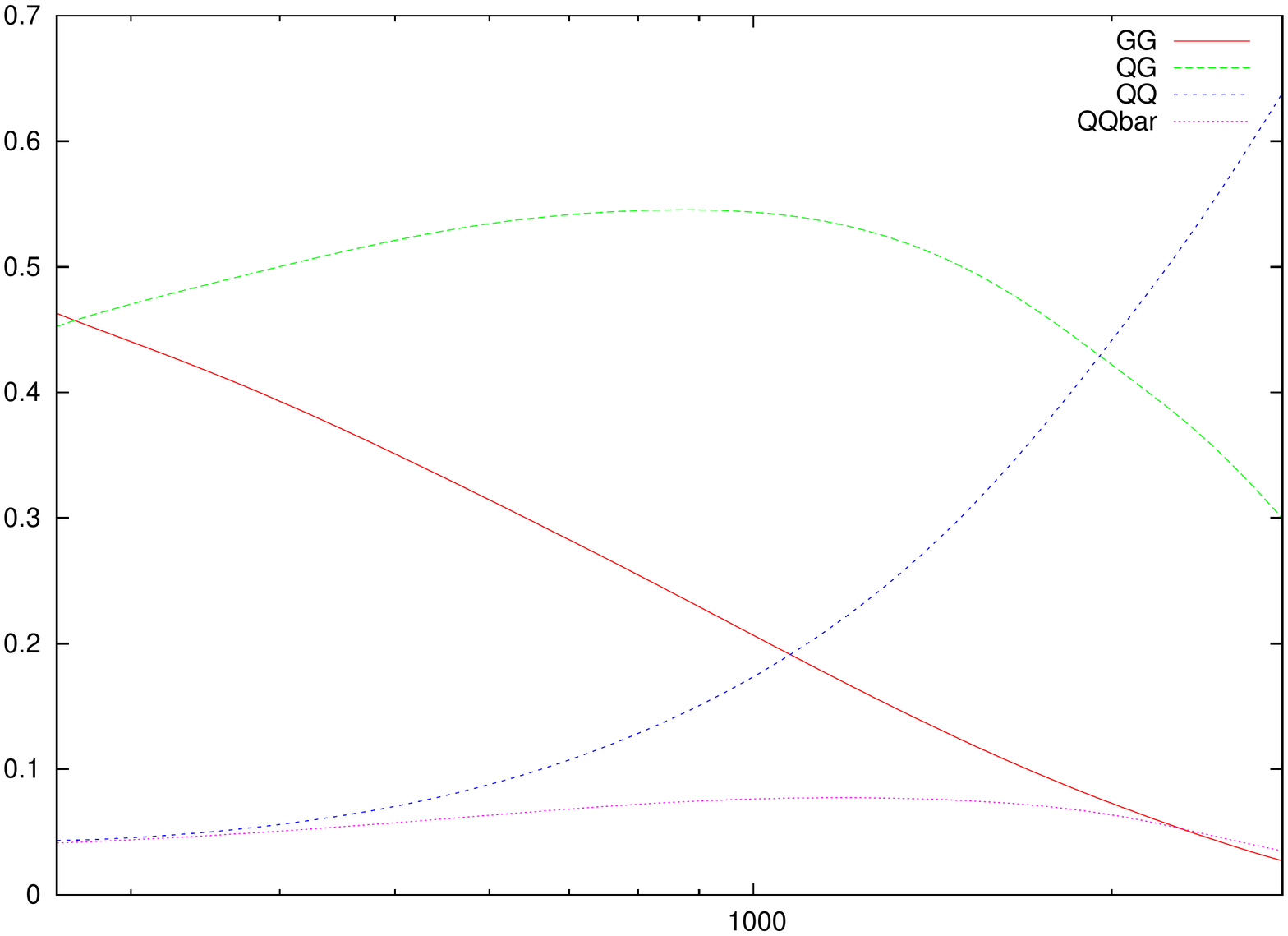}}
\subfigure[$2.0<y<2.5$]{\includegraphics[width=0.32\textwidth]{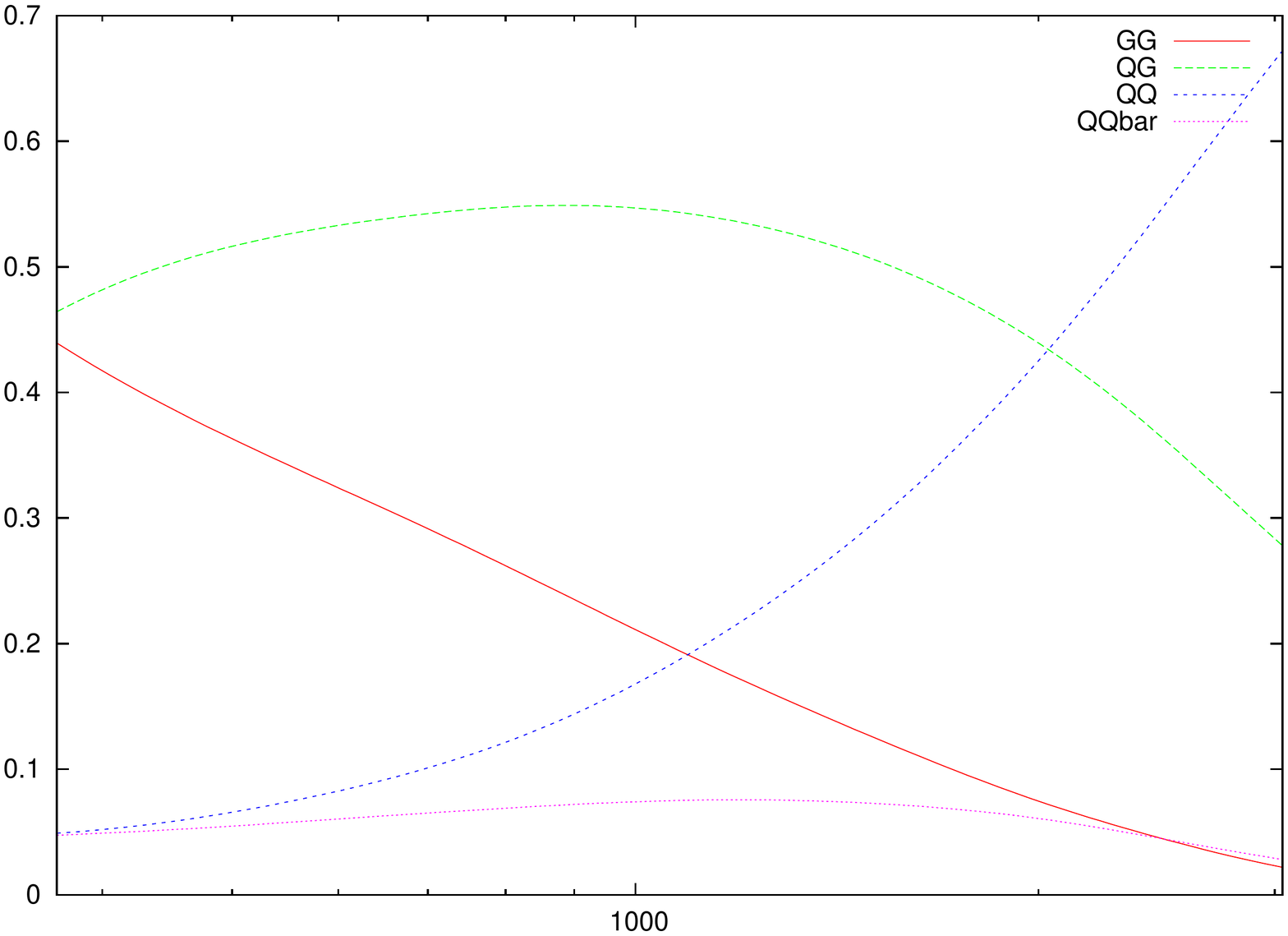}}
\subfigure[$2.5<y<3.0$]{\includegraphics[width=0.32\textwidth]{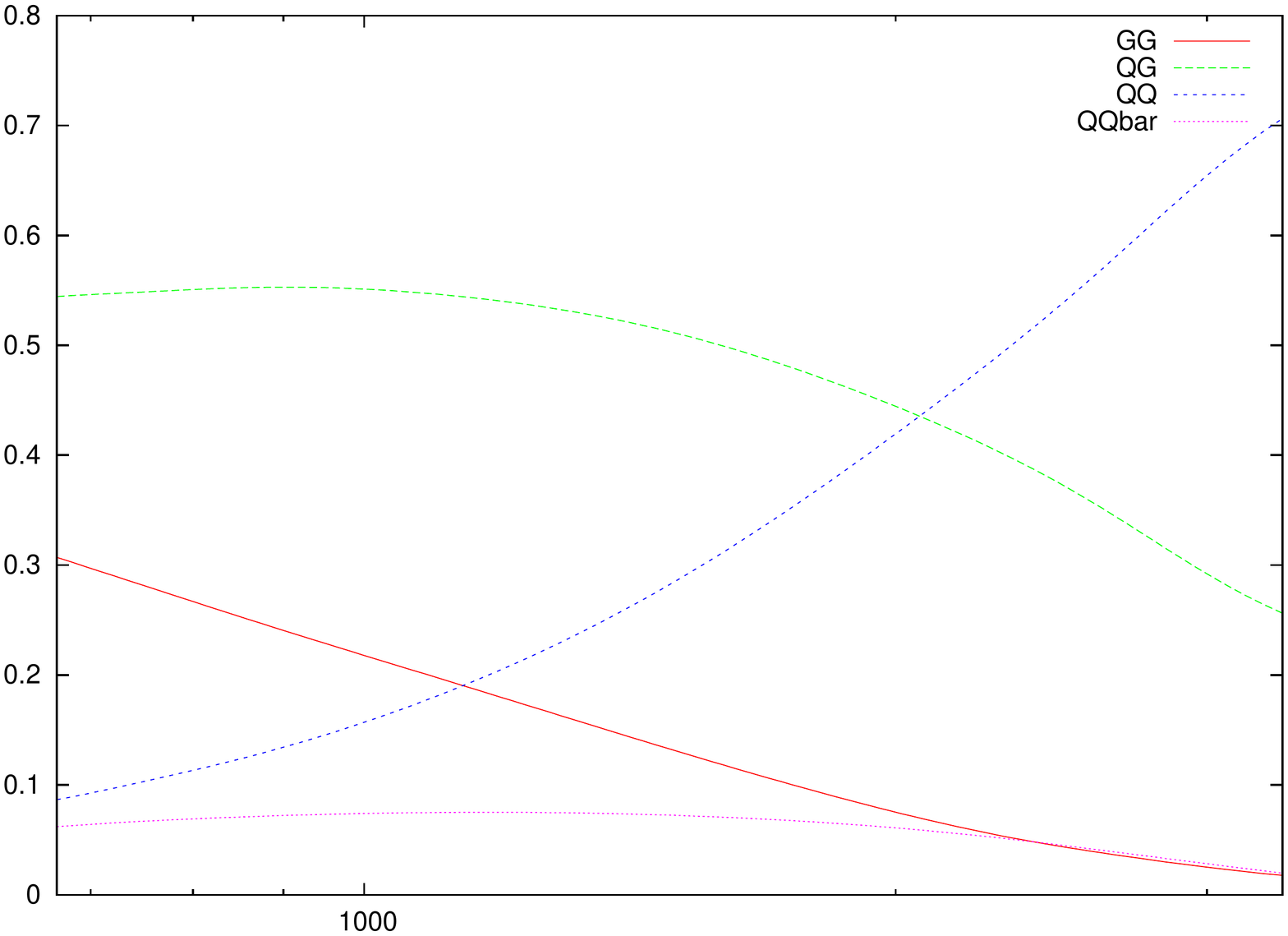}}
\subfigure[$3.0<y<3.5$]{\includegraphics[width=0.32\textwidth]{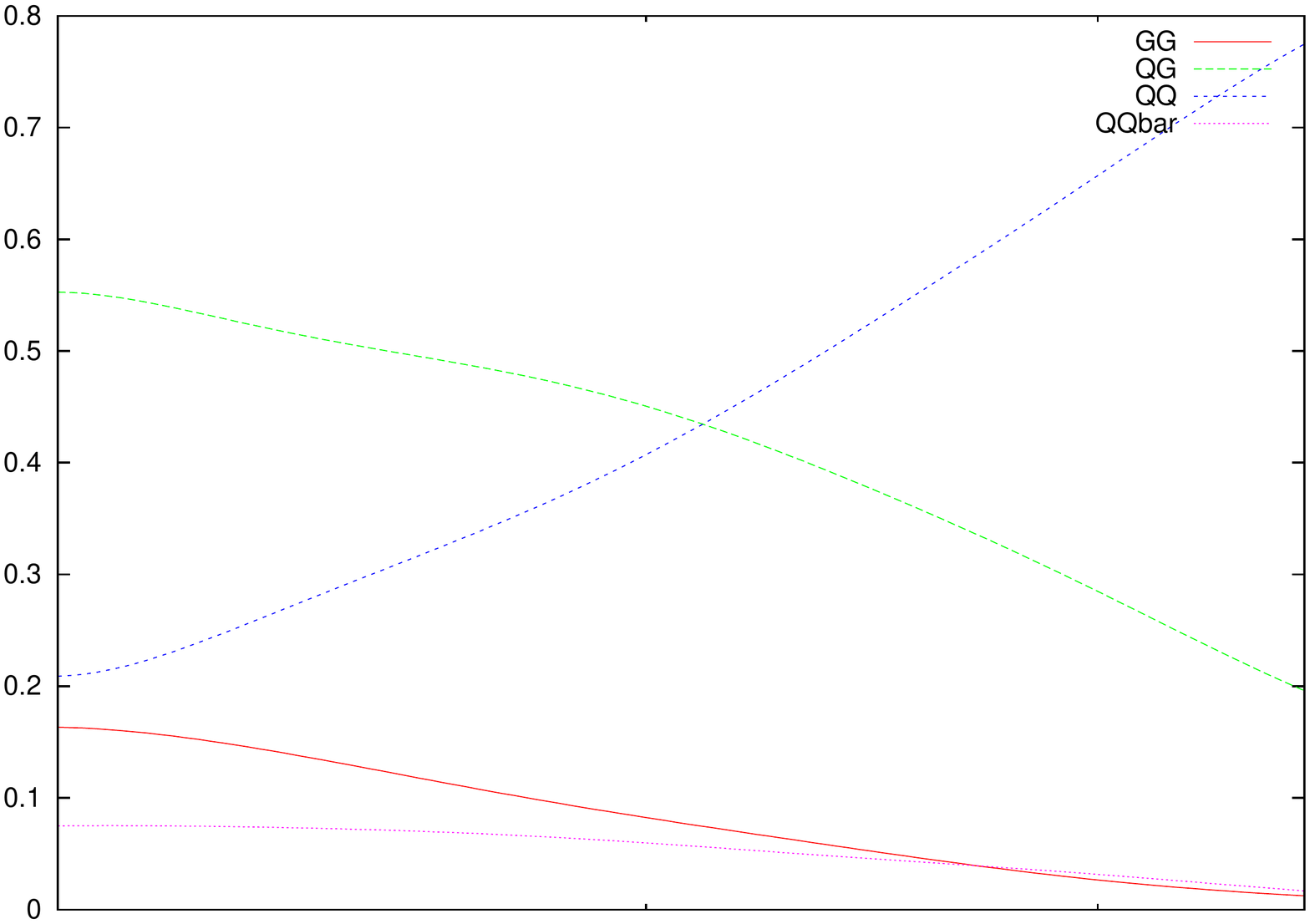}}
\subfigure[$3.5<y<4.0$]{\includegraphics[width=0.32\textwidth]{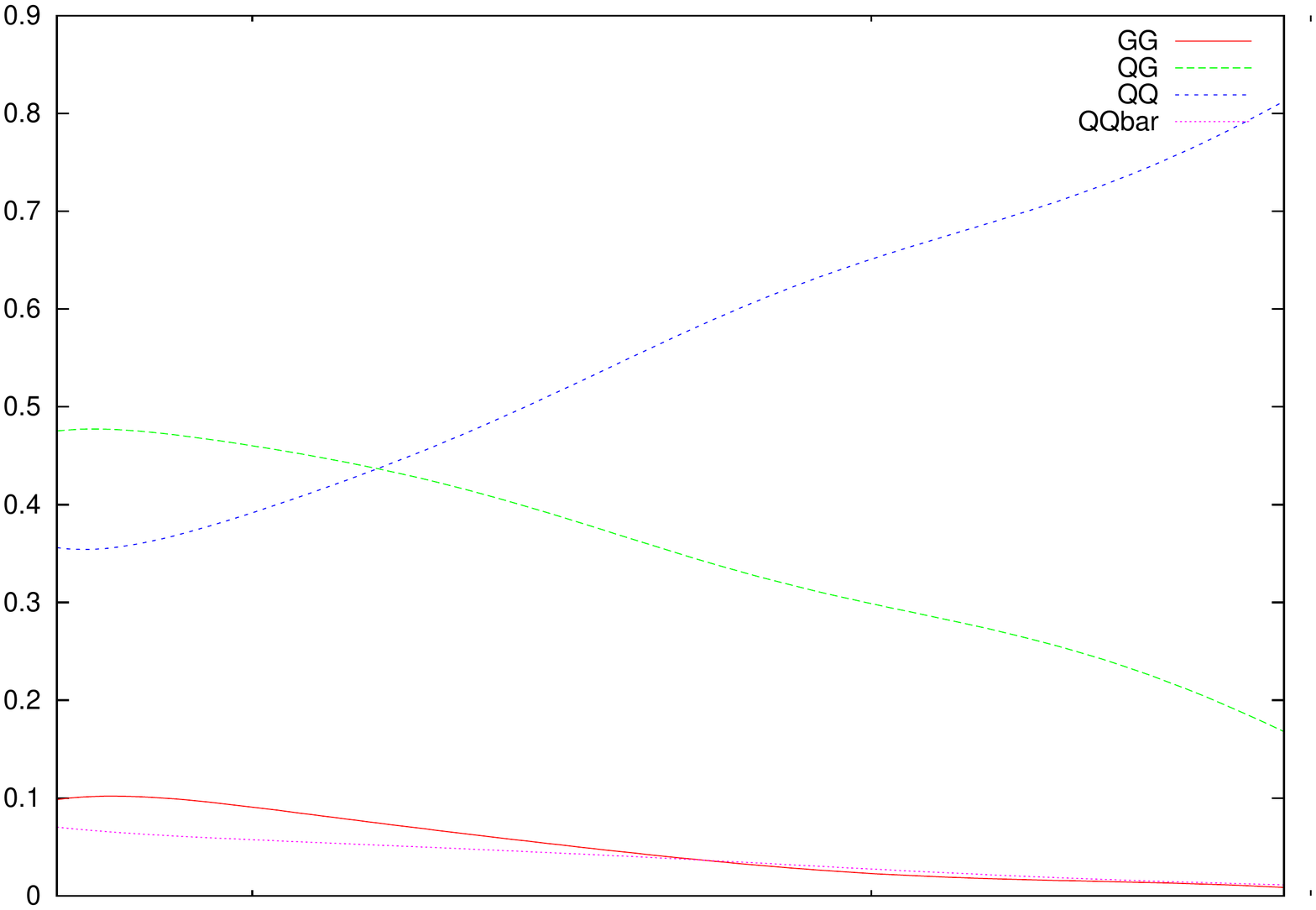}}
\subfigure[$4.0<y<4.5$]{\includegraphics[width=0.32\textwidth]{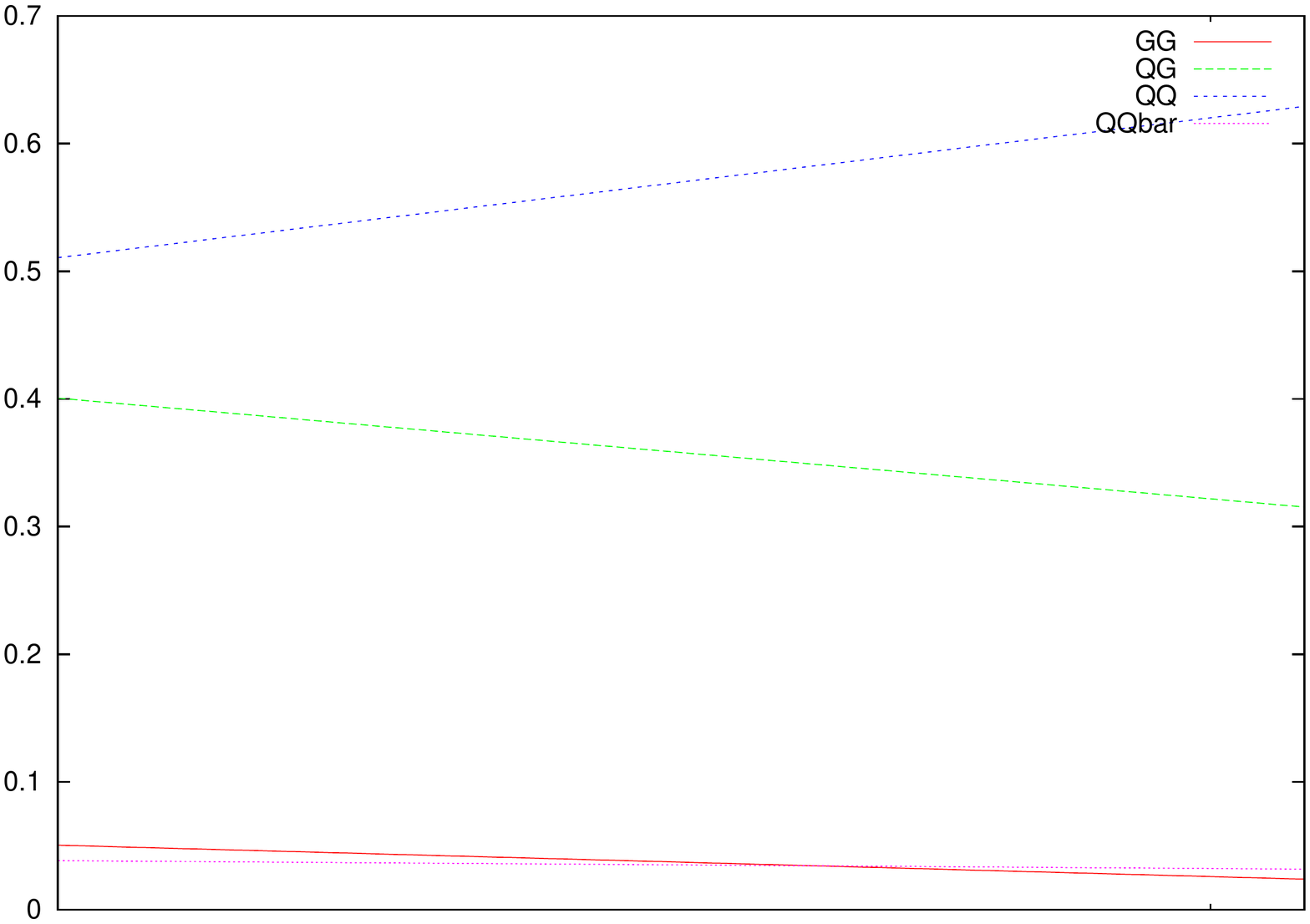}}
\end{center}
\vspace{-0.8cm}
\caption{Contributions of different initial-state parton combinations to the ATLAS dijet cross section calculation}
\label{partonsd2}
\end{figure}

The difference in the distributions of parton momenta leads to the question of 
which partons are being probed at different points in the phase space. Here we 
can begin to see the differences in the various datasets, especially when 
comparing to the relevant commensurate inclusive jet data. For the D{\O} dijet 
cross section in Fig. \ref{partonsd}, it is clear that the quark PDFs are in 
general the most important, with the $gg$ luminosity always below the $qg$, and 
mostly below the $q\bar{q}$ luminosities. 
For the ATLAS dijets, Fig. \ref{partonsd2} shows that for low rapidities, a 
similar behaviour to the corresponding inclusive jet plot is seen, with the 
gluon density dominating until the very high $M_{JJ}$ region. However, at higher 
rapidities, the requirement of two high-$x$ partons means the $qq$ luminosity 
becomes by far the most important. As a result, the dijet data for ATLAS should 
affect the quark densities far more than when using only the inclusive data.

\subsection{Scale Variations}

\begin{figure}[h!]
\centering
\includegraphics[width=0.8\textwidth]{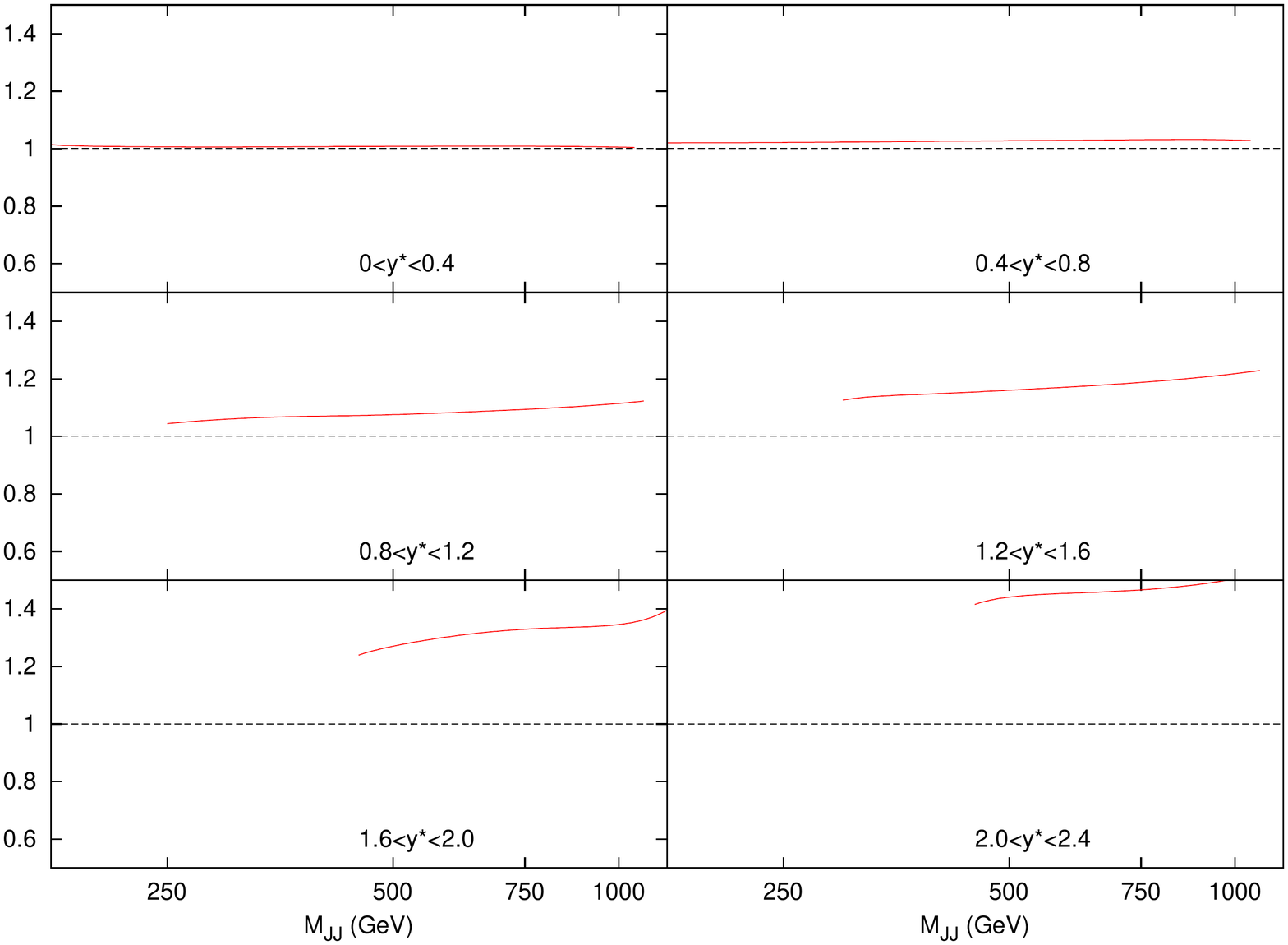}
\caption{Ratio of $M_{JJ}$ calculation to $2p_T^{av}$ calculation for D{\O} dijet calculation. The equivalence of the two scale choices at central rapidities is apparent, with large deviations for more forward jets. Both calculations are performed with NLOjet++.}
\label{D0_ptav_mjj}
\end{figure}

When considering dijet production, the choice of renormalisation and 
factorisation scales to include in the NLO calculation is not obvious. In 
general the behaviour of varying the scale on the full NLO calculation performed 
by NLOjet++ can be seen in the form of the differential cross section:

\begin{align*}
\frac{d^2\sigma}{dM_{JJ}dy}=&\left[\alpha_s^2(\mu_R)\sigma_{LO}+\alpha_s^3(\mu_R)\left(\sigma_{NLO}+2b_0\log\left(\frac{\mu_R}{M_{JJ}}\right)\sigma_{LO}-2\log\left(\frac{\mu_F}{M_{JJ}}\right)P_{ab}\otimes\sigma_{LO}\right)\right]  \\ &\otimes f_a(\mu_F)\otimes f_b(\mu_F)
\end{align*}

\noindent where the leading order and next to leading order cross sections, 
$\sigma_{LO}$ and $\sigma_{NLO}$ are computed using the matrix elements and 
evaluated at $\mu_R=\mu_F=M_{JJ}$, $b_0$ is the leading order QCD beta function 
coefficient, and $P_{ab}$ are the QCD splitting functions. The behaviour of this 
cross section under renormalisation scale variations is relatively simple, with 
only the running of $\alpha_s$ and a logarithm including this variable. The 
factorisation scale variations, however, are sensitive to the convolution with 
the PDFs, and so the particular $x$ values and partons probed in a particular 
event will affect the variations in $\mu_F$.

\begin{figure}[h!]
\centering
\subfigure[$M_{JJ}$]{\includegraphics[width=0.8\textwidth]{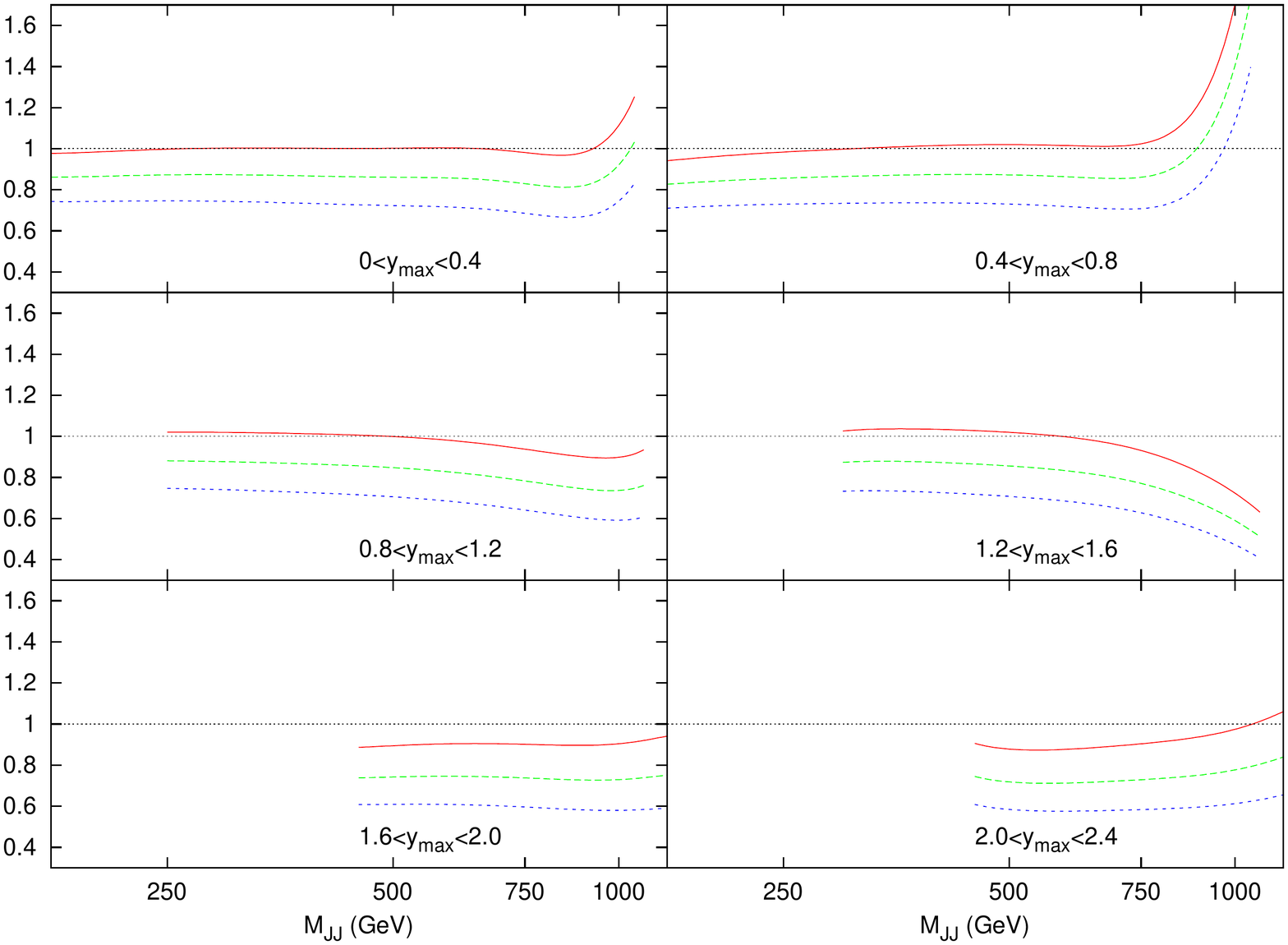}}
\vspace{-0.5cm}
\caption{Theory/Data ratio for D{\O} dijets, using multiples of $M_{JJ}$ as the choice of $\mu_R$ and $\mu_F$. The multiples are $0.5$ (red), $1.0$ (green) \& $2.0$ (blue)}
\label{D0_data_ratios2}
\end{figure}

Unlike inclusive jet production, in which the only physical scale involved in 
the events is the $p_T$ of the jet, dijet production has a number of possible 
choices of scale. The seemingly most obvious choice is the average $p_T$ of the 
two jets, however at high rapidities this can lead to problems due to the 
possible configuration of the event. A highly boosted hard scatter will have the 
same average $p_T$ as an unboosted soft scatter. Another variable which could 
be used as the scale choice is the dijet mass, $M_{JJ}$. This does not suffer 
from the same issues in event classification at high rapidities, though in this
case it is possible to have a very soft high rapidity scatter which still has
high $M_{JJ}$. At leading order, the mass is defined as:

\begin{equation}
M_{JJ}=2p_T\cosh(y^*)
\end{equation}

\noindent where $y^*=(y_{jet 1} - y_{jet 2})/2$ is half the rapidity difference 
of the final state jets making the dijet pair. At the limit $y^*=0$, for fully 
back-to-back jets, we have $M_{JJ}=2p_T$ as expected, and so the predictions 
using the two scale choices should agree. This is demonstrated in Fig. 
\ref{D0_ptav_mjj}, where the dijet cross section is calculated using both 
scales, and the ratio shown.

\begin{table}[h!]
\centering
\begin{tabular}{c c c c c}
\hline\hline
& $0.5*p_T^{av}$& $1.0*p_T^{av}$& $2.0*p_T^{av}$&\\
\hline
MSTW2008 NLO & 3.23 & 2.34 & 1.61&\\
\hline
\end{tabular}
\caption{$\chi^2$ values for D{\O} dijets}
\label{table:D0_chi1}
\end{table}

\begin{table}[h!]
\centering
\begin{tabular}{c c c c c}
\hline\hline
&$0.5*M_{JJ}$& $1.0*M_{JJ}$& $2.0*M_{JJ}$&\\
MSTW2008 NLO & 1.88 & 1.29 & 1.06&\\
\hline
\end{tabular}
\caption{$\chi^2$ values for D{\O} dijets}
\label{table:D0_chi2}
\end{table}

\begin{table}[h!]
\centering
\begin{tabular}{c c c c c}
\hline\hline
&$0.5*\frac{M_{JJ}}{2\cosh(0.7y^*)}$& $1.0*\frac{M_{JJ}}{2\cosh(0.7y^*)}$& $2.0*\frac{M_{JJ}}{2\cosh(0.7y^*)}$ &\\ 
\hline
MSTW2008 NLO & 3.06 & 2.15 & 1.44&\\
\hline
\end{tabular}
\caption{$\chi^2$ values for D{\O} dijets}
\label{table:D0_chi3}
\end{table}

Fig. \ref{D0_data_ratios2} (in comparison to the $p_T^{av}$ plot Fig. 
\ref{D0_data_ratios1}) demonstrates the apparent benefit of using dijet mass 
as the scale choice. In the case of $p_T^{av}$, although at low rapidity the 
prediction is stable and flat across all $M_{JJ}$, the predictions from 
different multiplicative factors of the scale begin to cross in the more forward 
bins. This has already been observed in \cite{d0-dijet-paper}, however other 
scale choices were not investigated.
In comparison, the theory/data ratio for the $M_{JJ}$ calculation is much more 
stable. The variation through multiplicative factors of the scale are constant 
throughout all rapidity bins, and the ratio remains generally flat.

\begin{table}[h!]
\centering
\begin{tabular}{c c c c c}
\hline\hline
& $p_T^{av}$& $2.0*p_T^{av}$& $4.0*p_T^{av}$&\\
\hline
MSTW2008 NLO & 6.66 & 1.94 & 1.91&\\
\hline
\end{tabular}
\caption{$\chi^2$ values for ATLAS dijets}
\label{table:ATLAS_chi1}
\end{table}`

\begin{table}[h!]
\centering
\begin{tabular}{c c c c c}
\hline\hline
&$0.5*M_{JJ}$& $1.0*M_{JJ}$& $2.0*M_{JJ}$&\\
MSTW2008 NLO &2.09  & 2.43 & 3.00&\\
\hline
\end{tabular}
\caption{$\chi^2$ values for ATLAS dijets}
\label{table:ATLAS_chi2}
\end{table}

\begin{table}[h!]
\centering
\begin{tabular}{c c c c c}
\hline\hline
&$0.5*\frac{M_{JJ}}{2\cosh(0.7y^*)}$& $1.0*\frac{M_{JJ}}{2\cosh(0.7y^*)}$& 2.0*$\frac{M_{JJ}}{2\cosh(0.7y^*)}$ &\\ 
\hline
MSTW2008 NLO & 2.59 & 2.27 & 2.11&\\
\hline
\end{tabular}
\caption{$\chi^2$ values for ATLAS dijets}
\label{table:ATLAS_chi3}
\end{table}

The $\chi^2$ values shown in Tables \ref{table:D0_chi1}-\ref{table:D0_chi3} 
confirm that the choice of $M_{JJ}$ provides the better fit to the D{\O} data. Also 
calculated is another choice of scale, namely multiples of 
$\frac{M_{JJ}}{2\cosh(0.7y^*)}$. This form of scale choice was suggested 
\cite{soper} as an empirical means to stabilise NLO corrections, and is almost 
equivalent to the choice $p_T\exp(0.3y^*)$ used by ATLAS \cite{atlas-inc-paper}. 
This choice allows the dependence on the dijet rapidity to be directly 
included. While it also provides an improvement on the $p_T^{av}$ calculation, 
it does not provide as good a fit for the D{\O} dijets as 
using simply $M_{JJ}$.

\begin{figure}[h!]
\includegraphics[width=\textwidth]{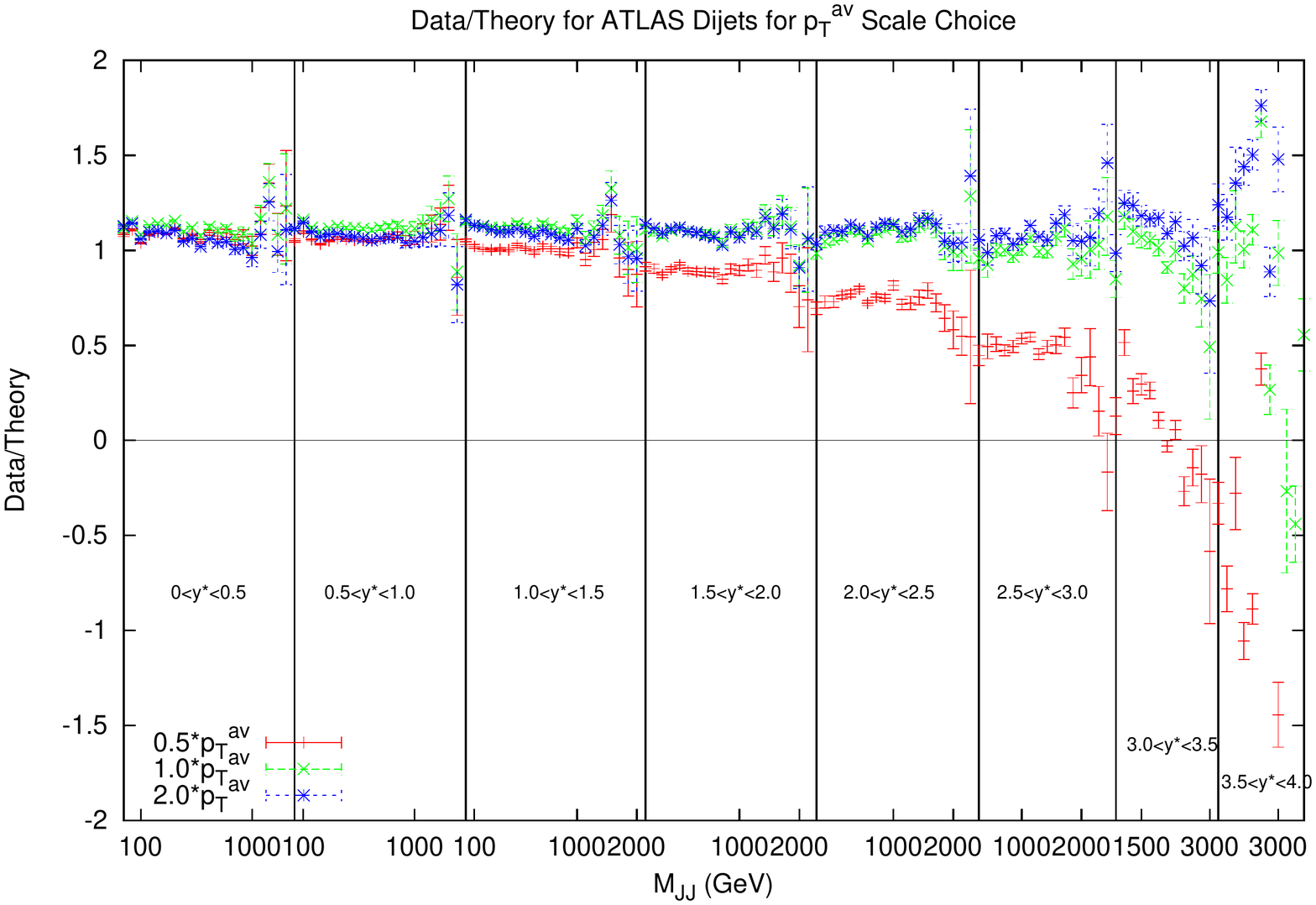}
\vspace{-0.8cm}
\caption{Ratio of data to theory for ATLAS dijets using 3 different multiples of $p_T^{av}$ as the scale choice. For the multiple of 1.0, the cross section becomes negative at high rapidity. This occurs much earlier for the lower multiple of 0.5.}
\label{ATLAS_dijets_ratio}
\end{figure}

The equivalent ATLAS results are now shown in 
Tables \ref{table:ATLAS_chi1}-\ref{table:ATLAS_chi3}. The tendency for the 
$p_T^{av}$ calculation to degrade at small multiplying factors is even more 
apparent here than with the D{\O} dijets, so much so that the $0.5p_T^{av}$ is 
not shown, and all values are multiplied by a further factor of 2. Even with 
this additional factor, the $1*p_T^{av}$ fit is terrible, and 
is due to the cross section calculation being negative in the high rapidity, 
high mass region, which can be seen in Fig. \ref{ATLAS_dijets_ratio}. This plot 
clarifies the issue with using $p_T^{av}$ that initially appeared in the D{\O} 
calculation, since it includes much higher rapidity and mass regions. It is 
clear that as higher rapidities are reached, the $p_T^{av}$ calculation 
dramatically falls off for low multiplying factors, to the point where it 
becomes negative for both the 0.5 and 1.0 factors. Despite this, once the 
multiplying factor is large enough, $p_T^{av}$ provides the best fit of the 
three choices, with $M_{JJ}$ in fact showing the worst fit of the three.

\begin{figure}[h!]
\includegraphics[width=0.49\textwidth]{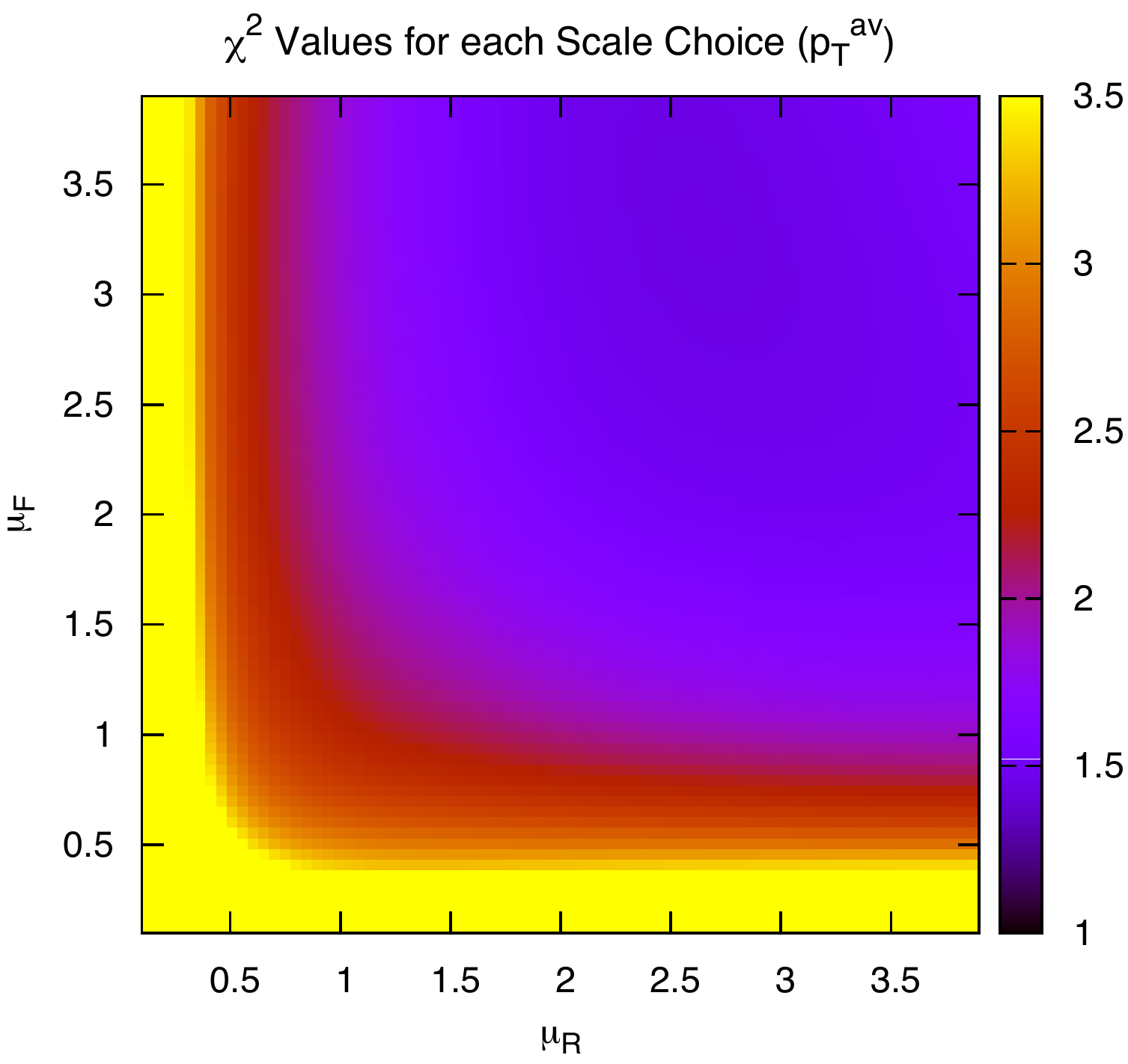}
\includegraphics[width=0.49\textwidth]{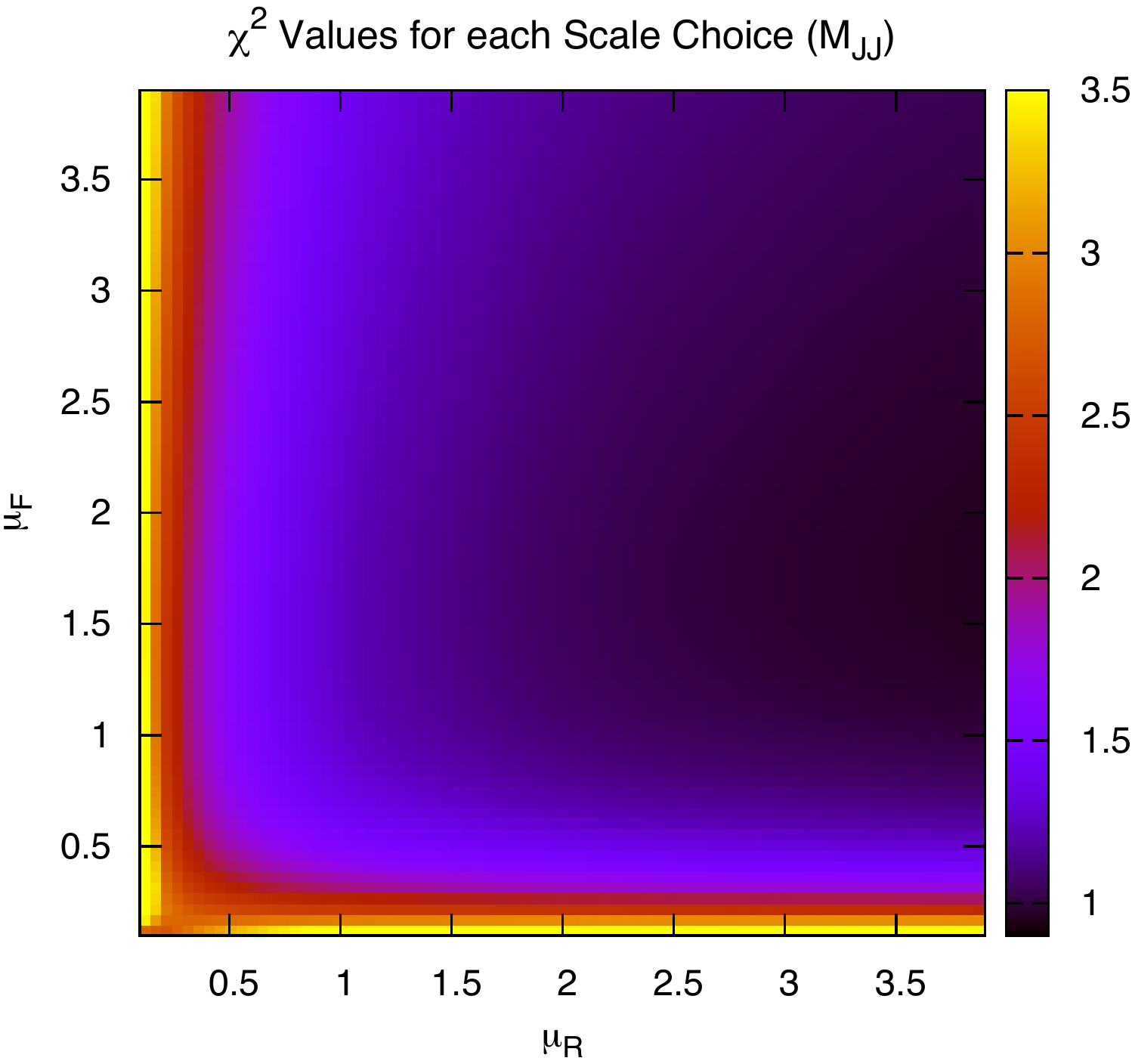}
\caption{$\chi^2$ per point for all values of multiplication factor for both $p_T^{av}$ and $M_{JJ}$ calculations for D{\O} dijets. The yellow area at low scales in the $p_T^{av}$ calculation is greatly off the scale, due to the calculation becoming negative in this region.}
\label{scale_comp_d0}
\end{figure}

\begin{figure}[h!]
\includegraphics[width=0.49\textwidth]{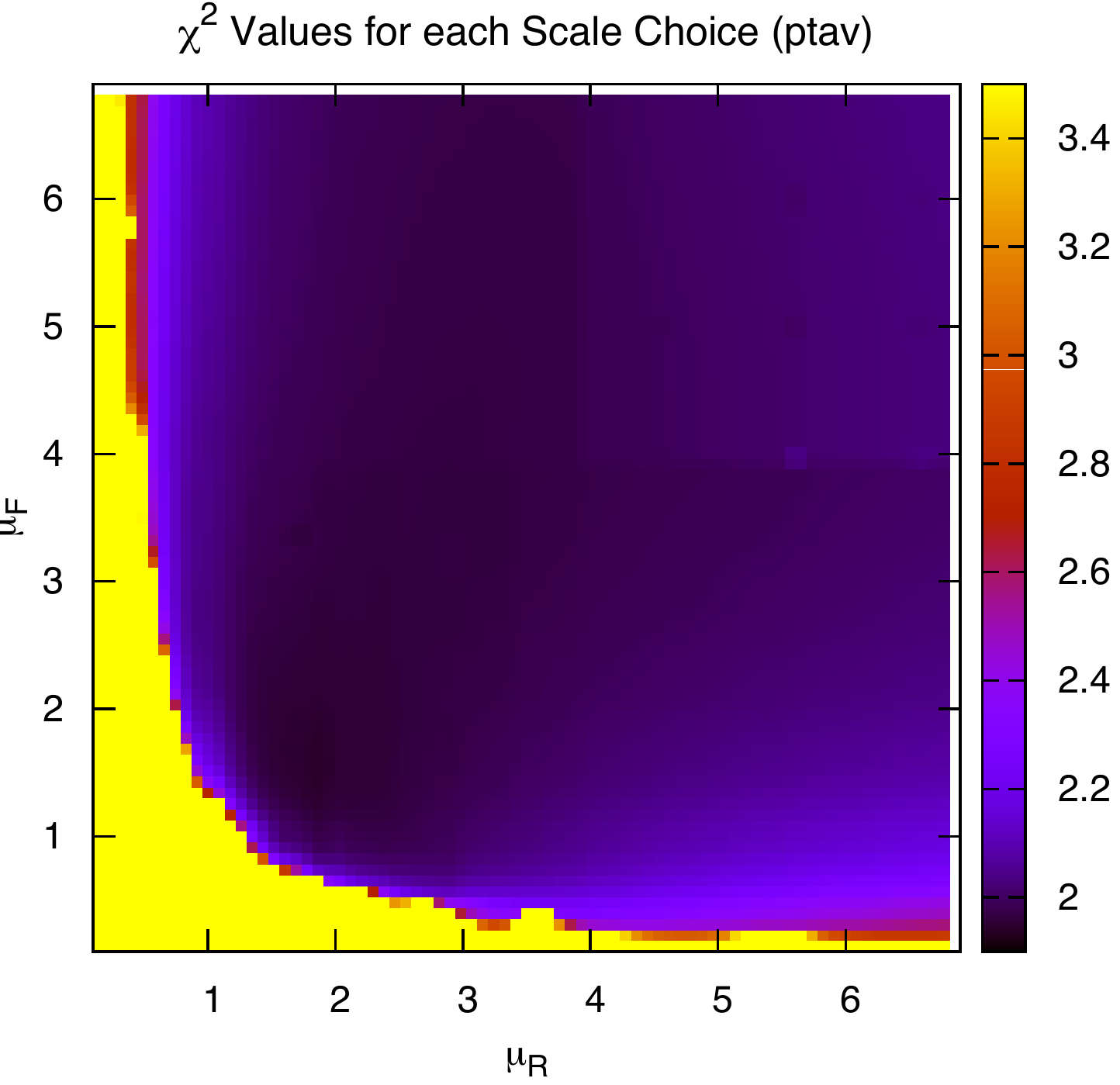}
\includegraphics[width=0.49\textwidth]{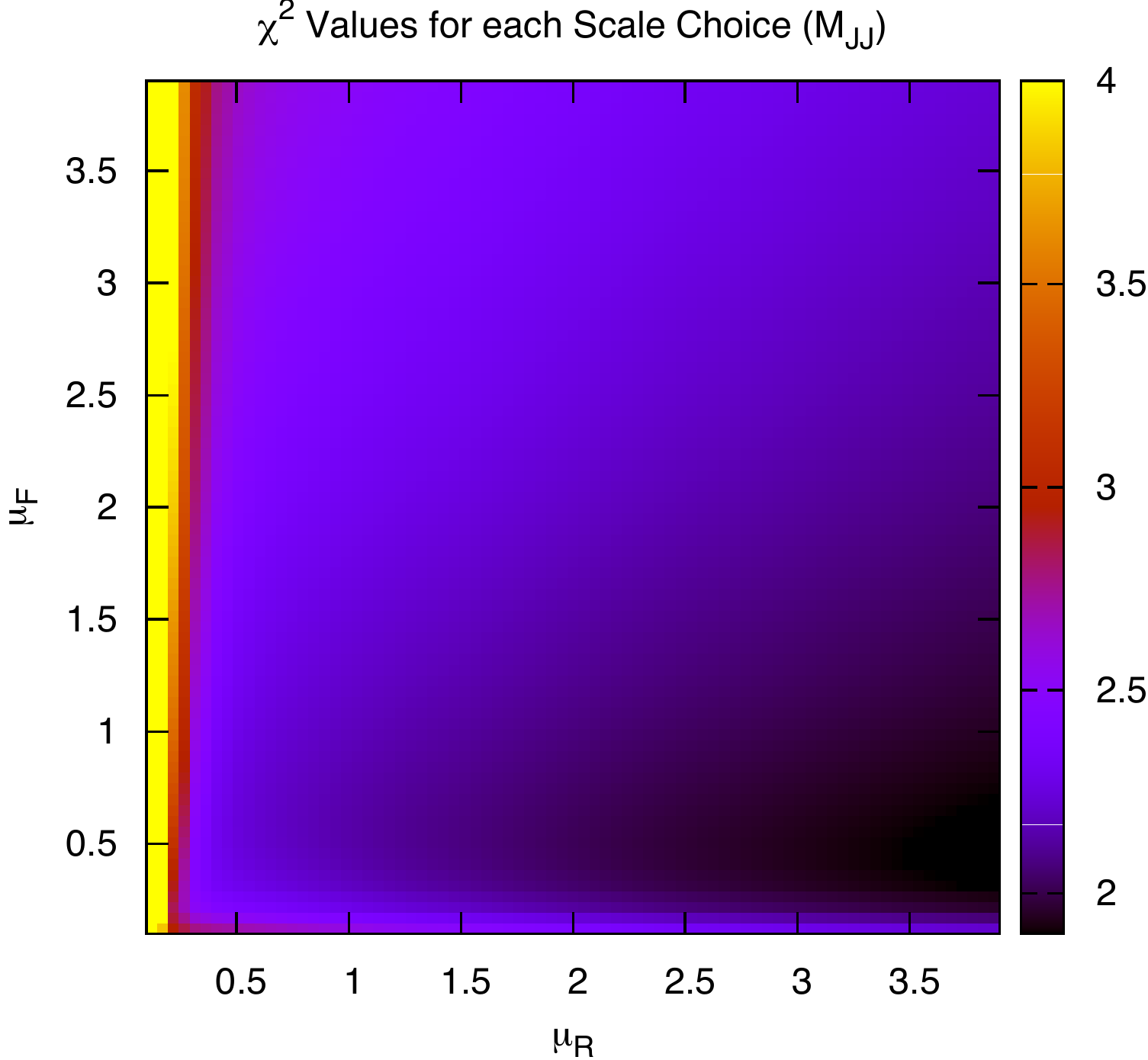}
\caption{$\chi^2$ per point for all values of multiplication factor for both $p_T^{av}$ and $M_{JJ}$ calculations. for ATLAS dijets The yellow area at low scales in the $p_T^{av}$ calculation is greatly off the scale, due to the calculation becoming negative in this region.}
\label{scale_comp_atlas}
\end{figure}

When considering the entire space of fits for any combination of 
$(\mu_R,\mu_F)$, using the dijet mass is again shown to be a more stable 
prediction that average $p_T$. Figs. \ref{scale_comp_d0} and 
\ref{scale_comp_atlas}, which show the fit quality for D{\O} and ATLAS 
respectively, more completely shows the degradation of the $p_T^{av}$ 
calculation at low values of scales. The yellow region, which for D{\O} covers 
the area in which either scale is below $0.5$, shows a rapid unbounded increase 
in $\chi^2$, deriving from the fact that the the cross section becomes 
increasingly negative as the scales approach 0. The fit becomes comparable in 
quality to the $M_{JJ}$ calculation at much higher choices of scale, however 
there is no clear minimum which can be identified as a stable choice. For ATLAS, 
the region of divergent $\chi^2$ is much larger for $p_T^{av}$, with normally 
sensible choices showing a very poor fit. Again, this is the result of the 
larger kinematic span of the ATLAS dijets exposing the failure of this 
calculation in the high rapidity, high mass region.

The $M_{JJ}$ calculation for both data sets shows a similar trend by increasing 
towards lower scale choices. However, due to the stability at high rapidities, 
the fit does not blow up in the same way as for $p_T^{av}$, and a lower $\chi^2$ 
is apparent across the entire parameter space. There is a much clearer minimum, 
although it occurs for unusually high values of $\mu_R$. This issue is 
discussed later and is shown to be related to the normalisation uncertainty.

\begin{figure}[h!]
\includegraphics[width=0.49\textwidth]{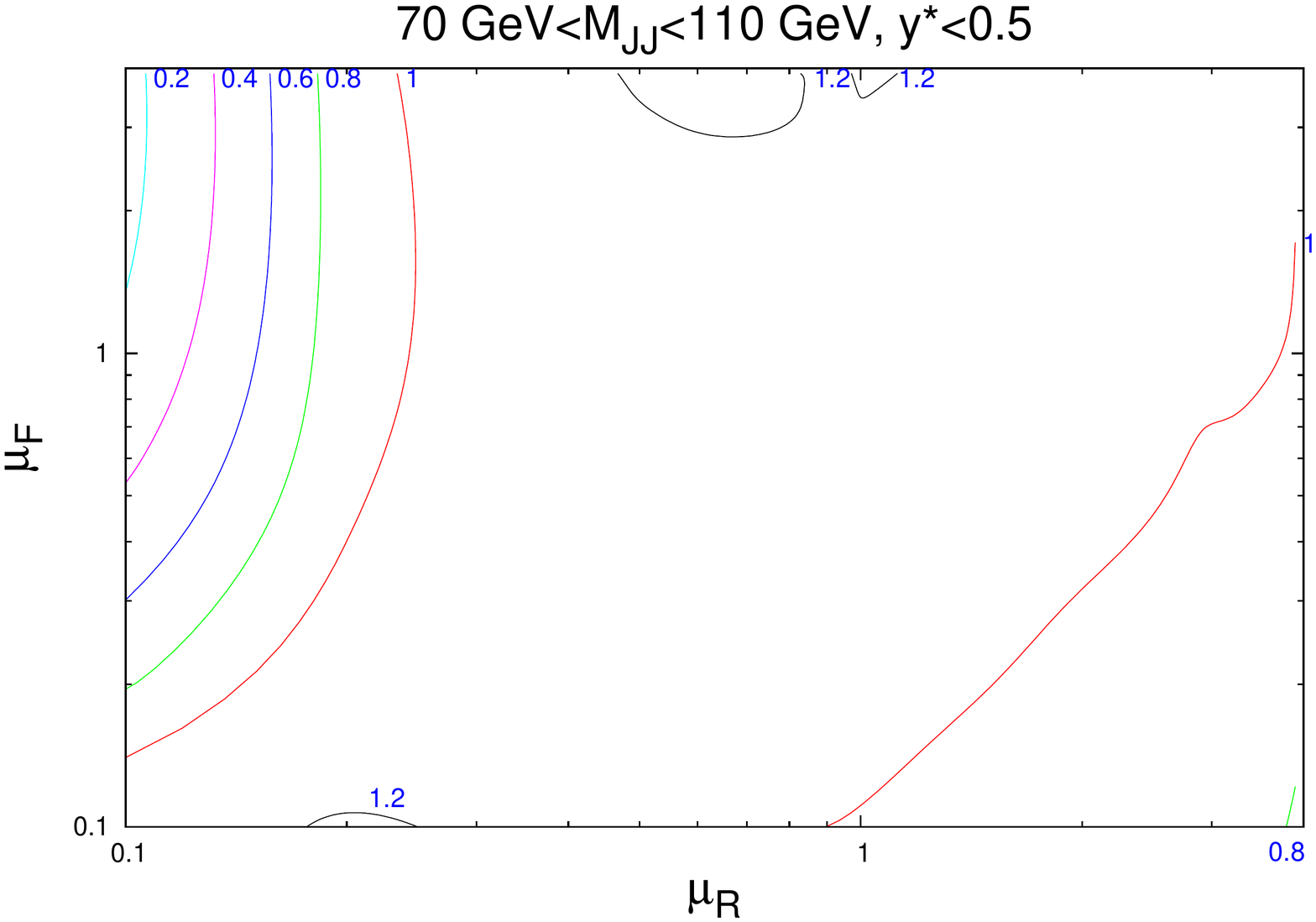}
\includegraphics[width=0.49\textwidth]{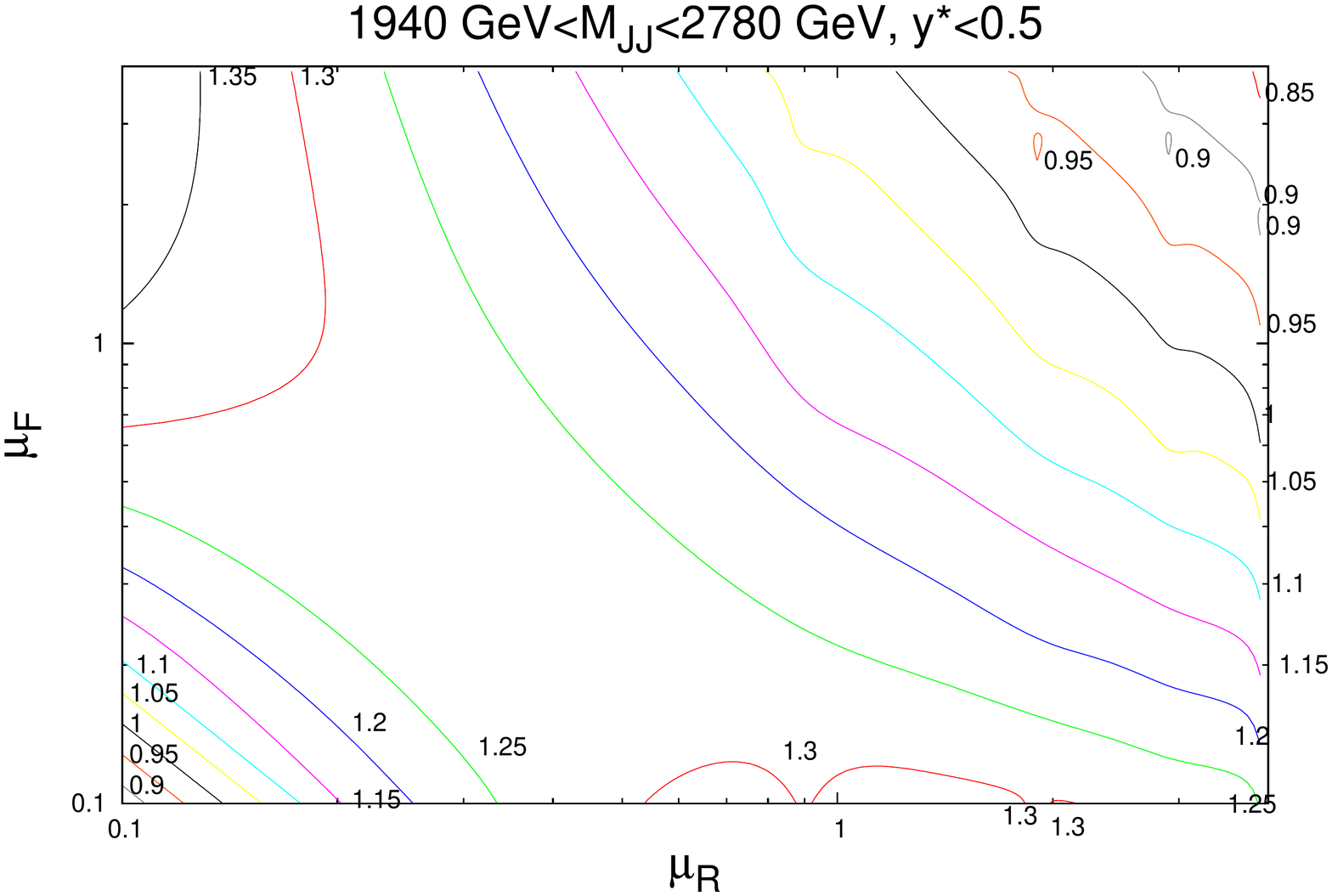}
\vspace{-0.8cm}
\caption{Comparison of scale variations for the (a) lowest and (b) highest $M_{JJ}$ bins in the $y^*<0.5$ rapidity bin of the ATLAS dijet calculation. The contour values are data/theory. The scales $\mu_R$ and $\mu_F$ are multiples of $M_{JJ}$}
\label{highlowvar}
\end{figure}

\begin{figure}[h!]
\subfigure{\includegraphics[width=0.49\textwidth]{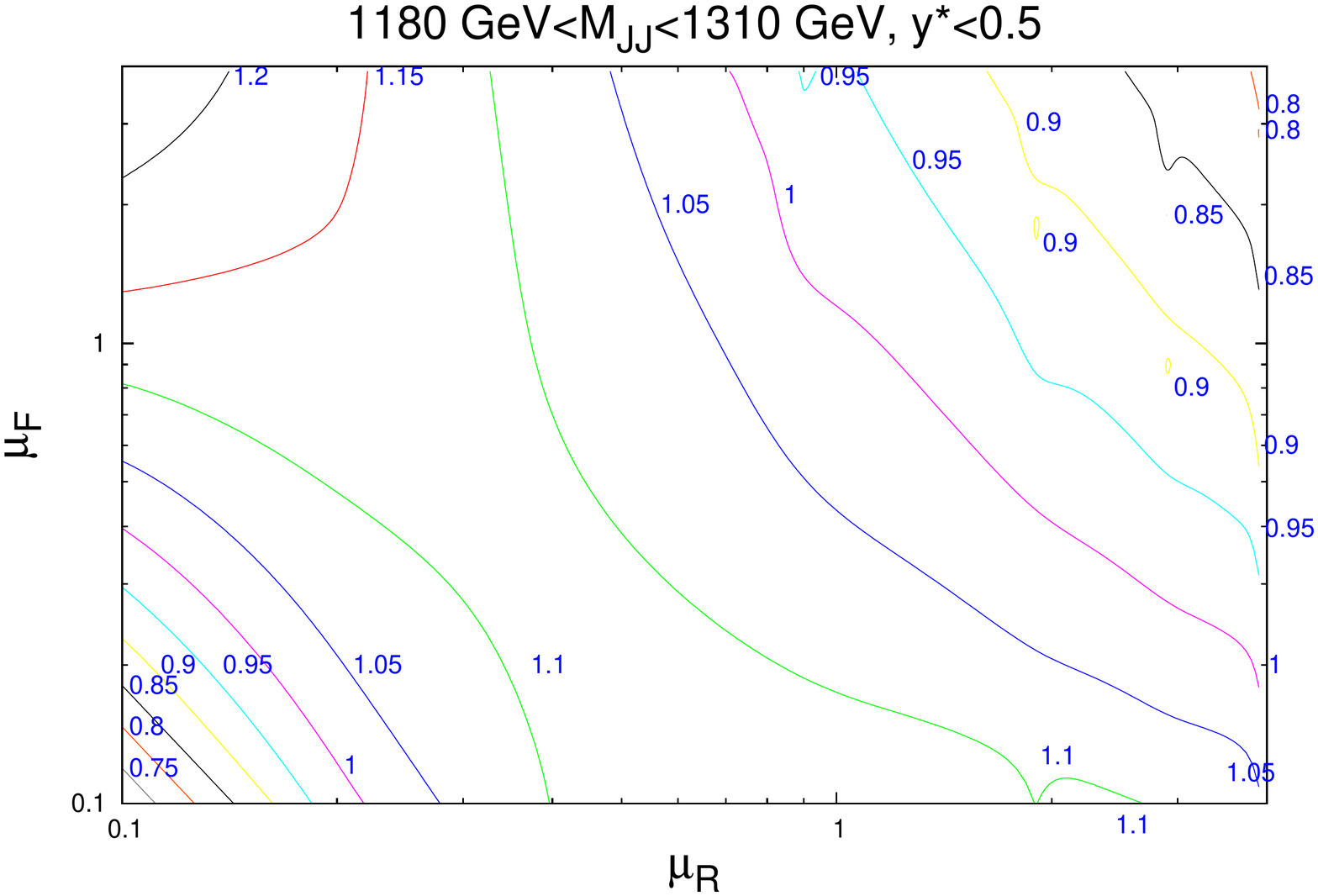}}
\subfigure{\includegraphics[width=0.49\textwidth]{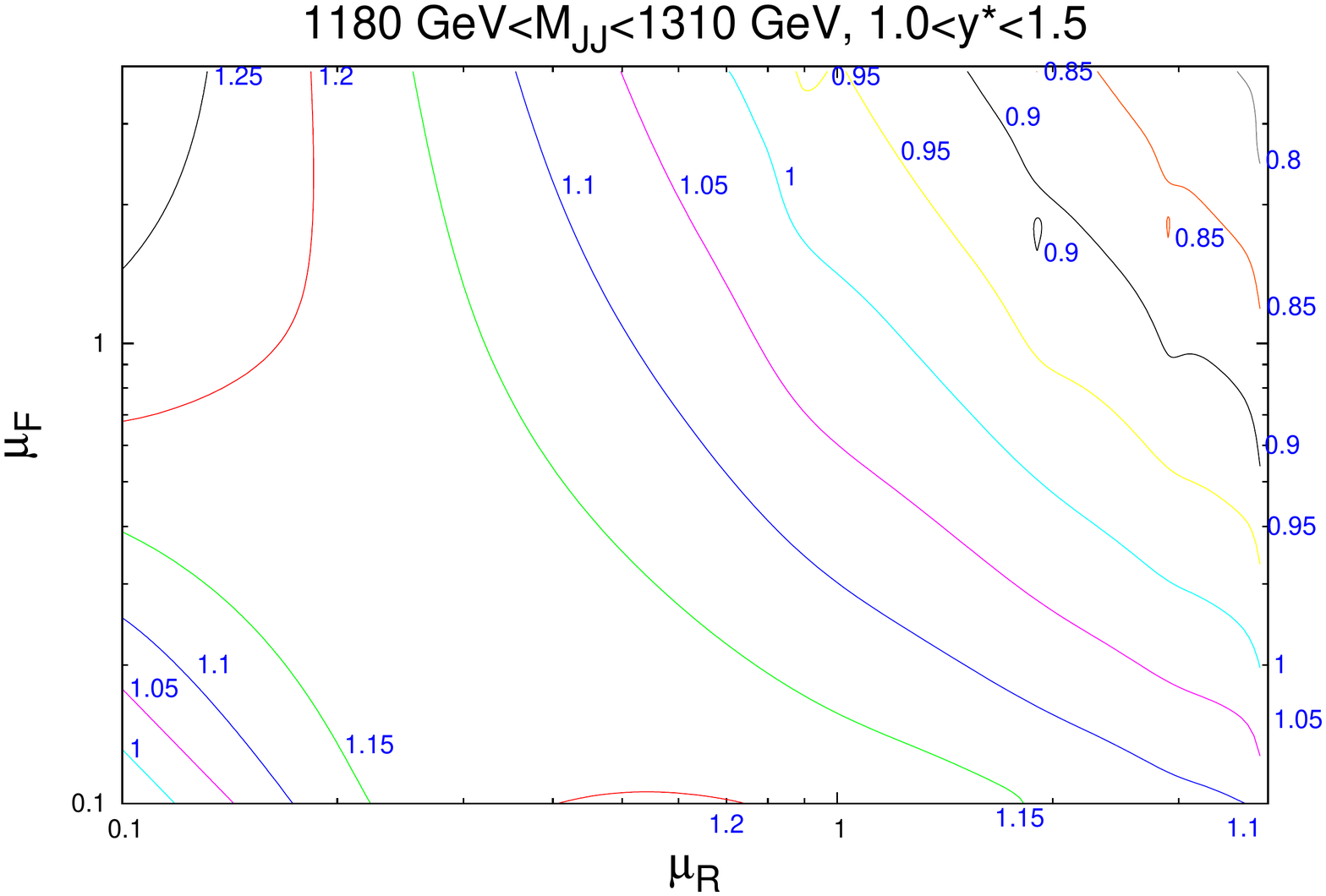}}
\subfigure{\includegraphics[width=0.49\textwidth]{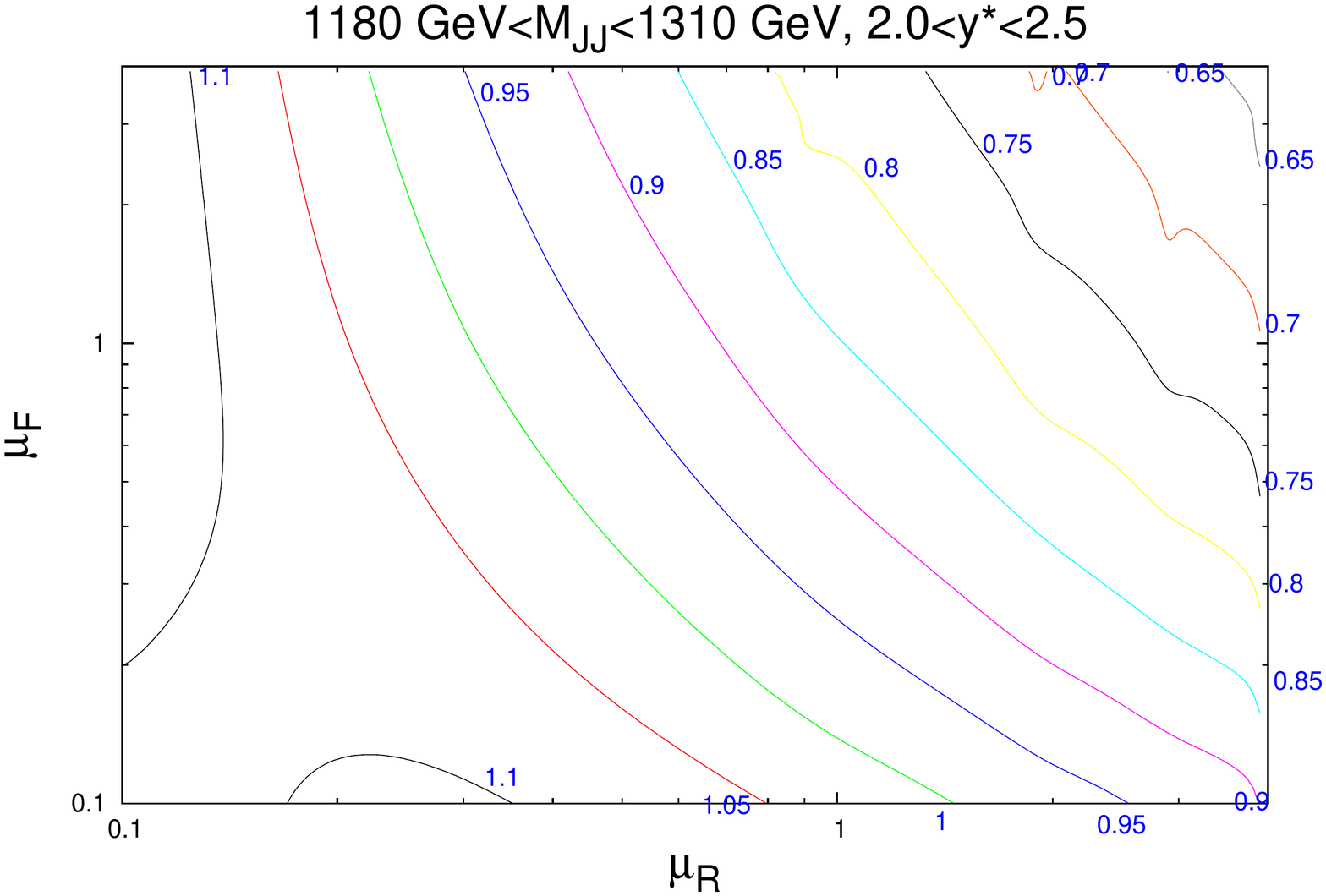}}
\subfigure{\includegraphics[width=0.49\textwidth]{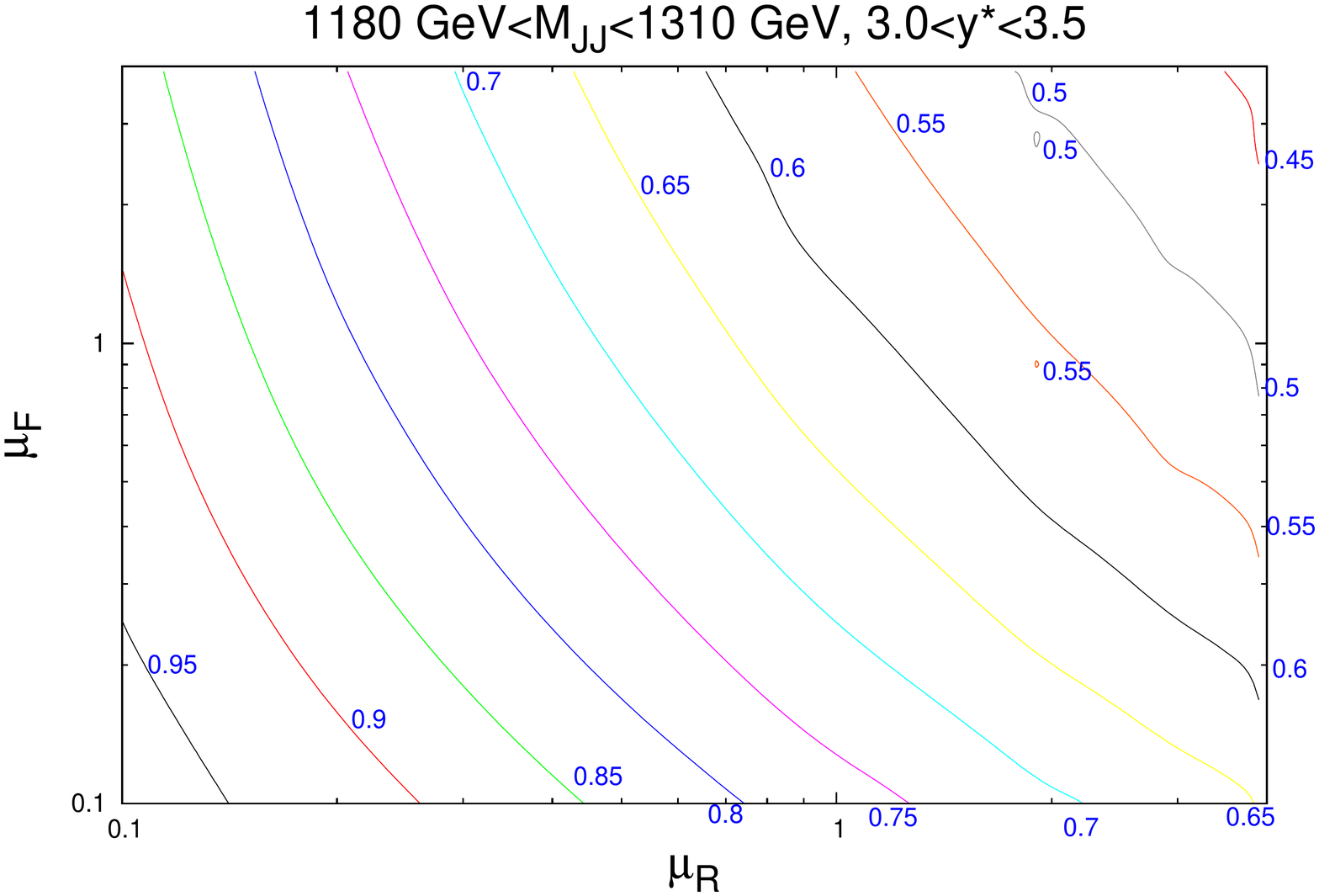}}
\vspace{-0.8cm}
\caption{Comparison of scale variations for different rapidity bins. The same $M_{JJ}$ range is used throughout.}
\label{samemjj}
\end{figure}

The nature of the effect of scales can be more deeply probed by studying 
individual cross-sections in finely defined regions of phase space. Whereas the 
previous discussion has focused on the fit to data of an entire data set, the 
following will study the variation of each point within that data set for each 
scale choice. As Figs. \ref{partonsd} \& \ref{partonsd2} have shown, the 
contributions from the individual PDFs depends greatly on the values of the 
kinematic variables, and so the variation of each point in the factorisation 
scale direction should change in a similar manner.
Fig. \ref{highlowvar} (similar to plots in \cite{huston}) demonstrates the scale 
variation of two single points in 
the kinematic phase space of the ATLAS data set. Both are in the lowest $y^*$ 
bin, however the first includes dijets with low mass (70-110~GeV) and the second 
includes those with high mass (1940-2780~GeV). The general behaviour is that 
of a stable saddle region in the central region, with data/theory decreasing 
away from the saddle along one axis and increasing along the other. The axes 
defining the saddle region, however, differ greatly between the two points. 
A smooth rotation anticlockwise is observed as the dijet mass is increased, 
resulting in the large rotation shown in the figure.
The dependence of this rotation on the kinematic variables is shown more clearly 
in Fig. \ref{samemjj}, where only the rapidity bin is changed. 
The $1.18~\TeV<M_{JJ}<1.31~\TeV$ bin is chosen for study as this is the bin 
appearing in the most rapidity bins. It is clear that the angle of the saddle 
point is dependent only on the dijet mass, however the overall behaviour is 
still affected by the rapidity. A migration towards lower scale choices is 
seen, such that at the highest rapidities, the saddle point disappears and 
the surface simply becomes a unidirectional slope. Ideally, the scale choice 
for a calculation would be the one which provides the most stable calculation, 
and hence would be within the saddle region for all of the points in the dataset. 

\begin{figure}[h!]
\includegraphics[width=0.49\textwidth]{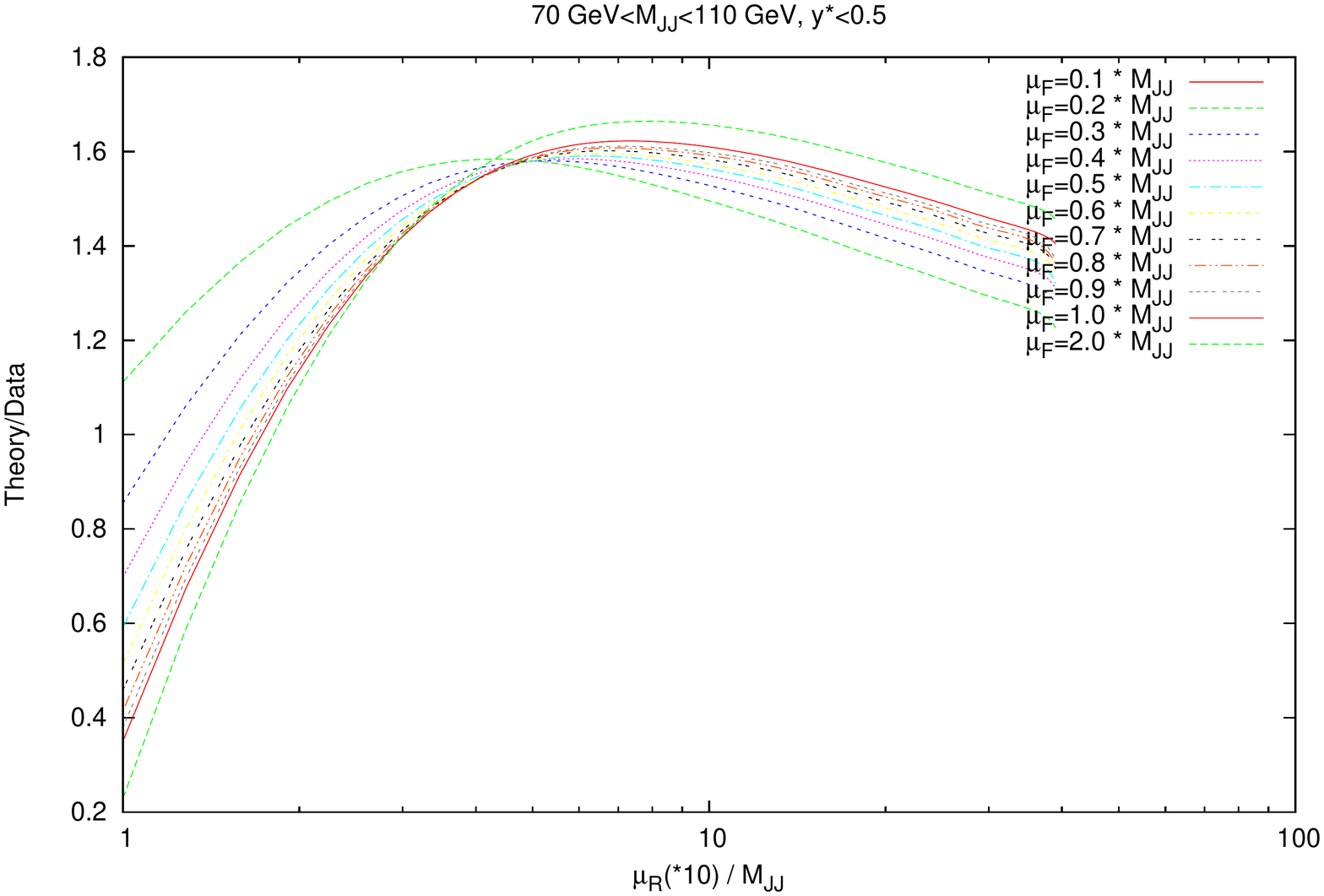}
\includegraphics[width=0.49\textwidth]{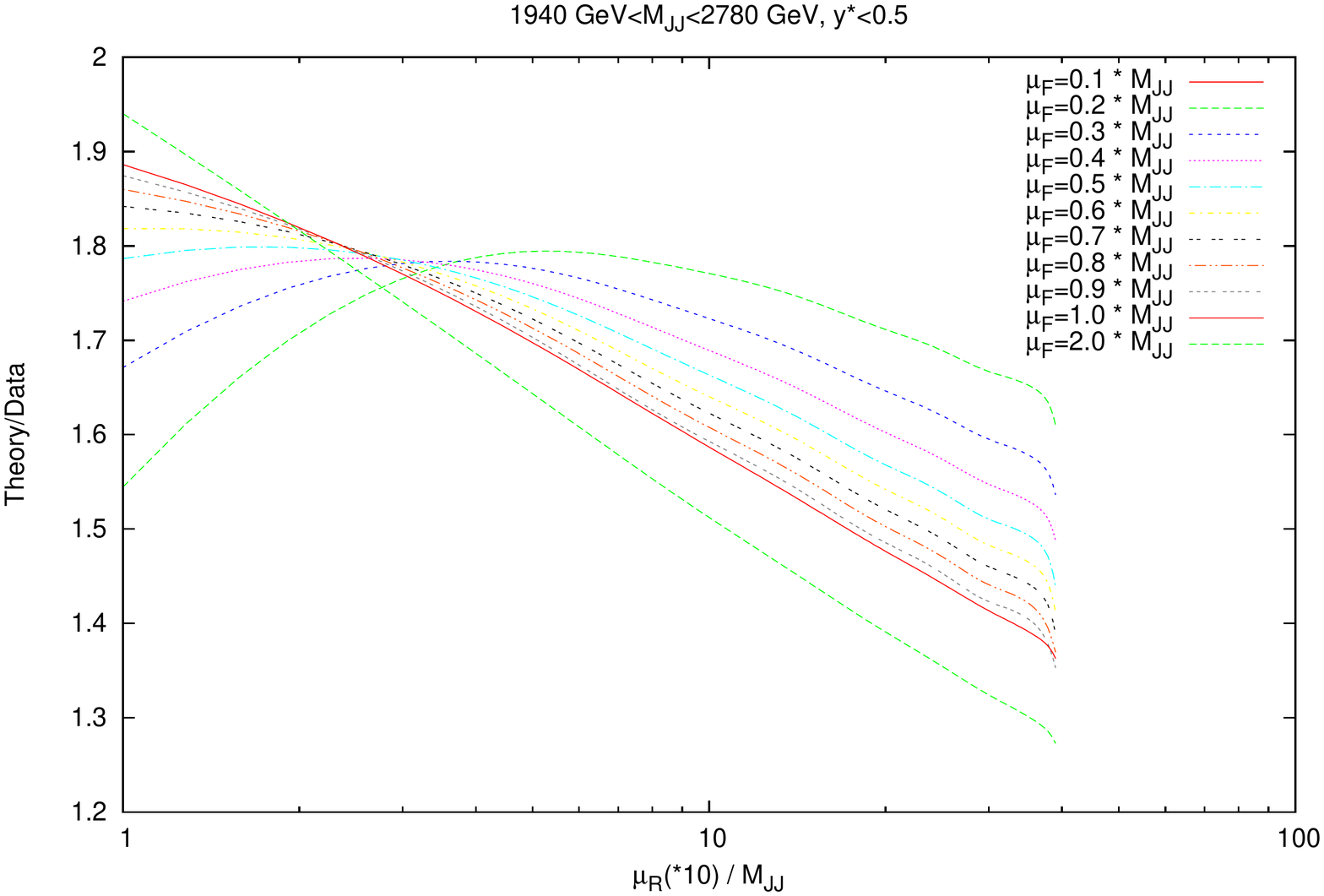}
\includegraphics[width=0.49\textwidth]{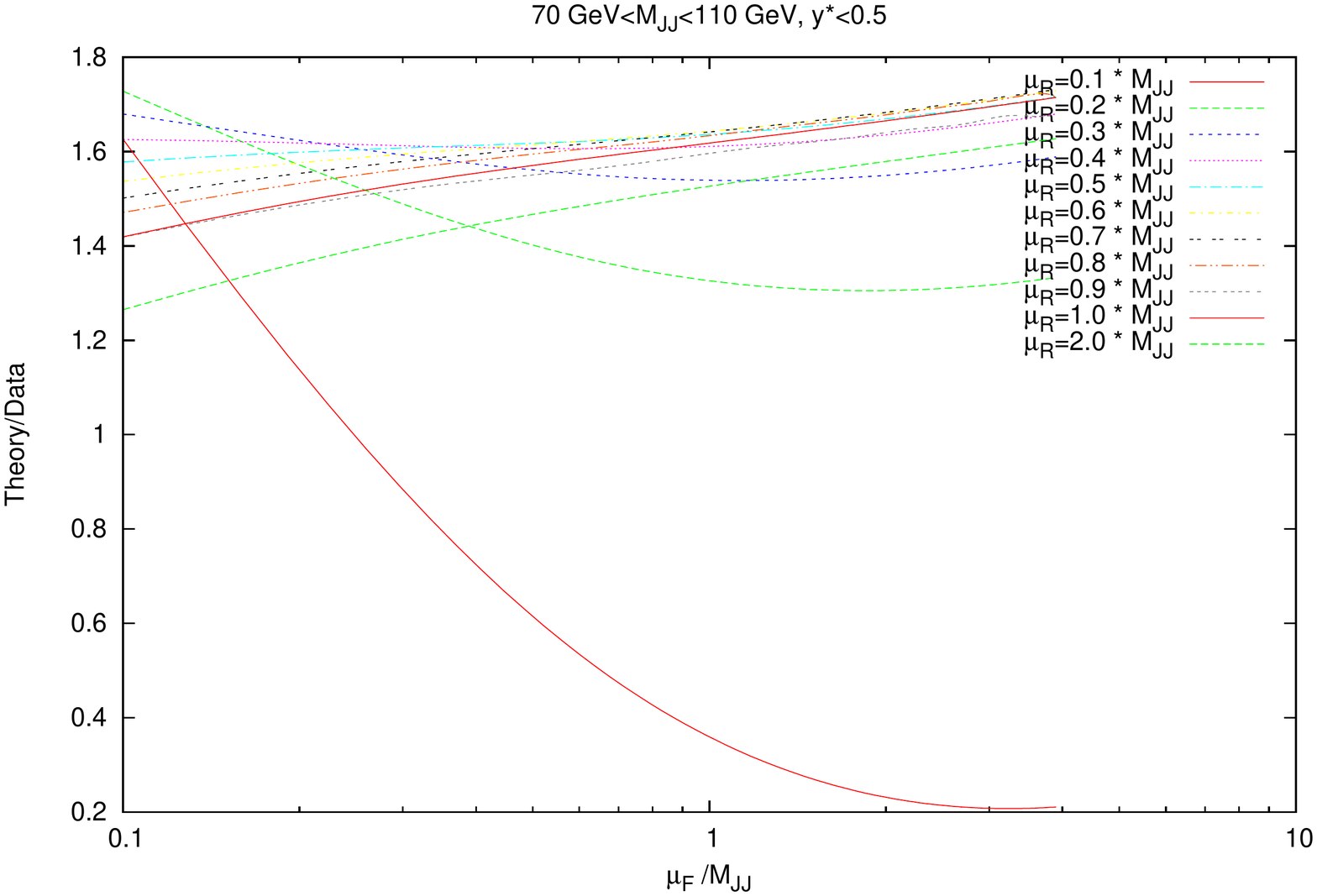}
\includegraphics[width=0.49\textwidth]{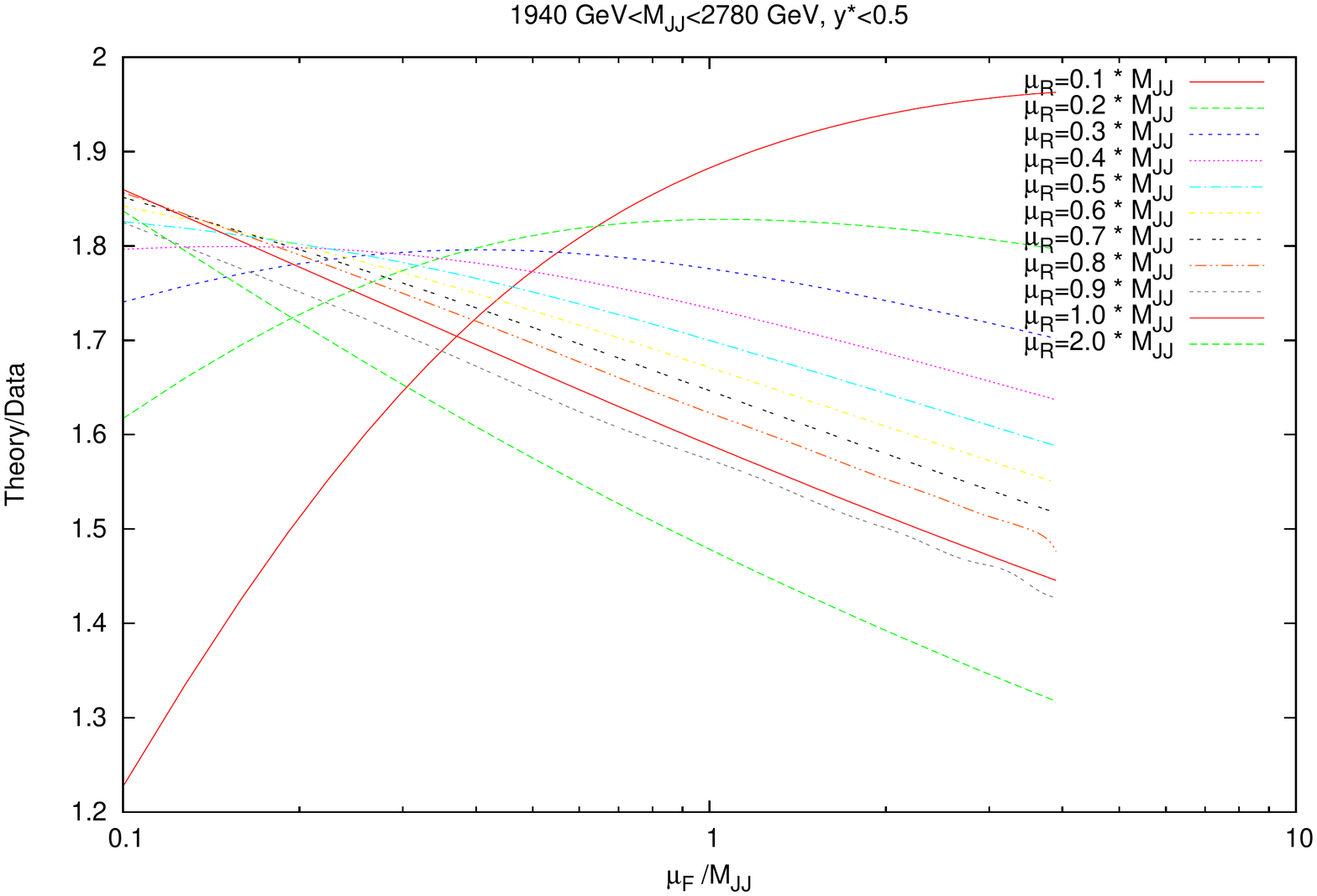}
\vspace{-0.8cm}
\caption{Plots demonstrating the variation of $\mu_R$ and $\mu_F$ independently.}
\label{slices}
\end{figure}

In order to understand the source of the observed behaviour, the variation in 
$\mu_R$ and $\mu_F$ are studied independently. Fig. \ref{slices} demonstrates 
this for two points, at low and high $M_{JJ}$ for the scale choice of $M_{JJ}$. 
The observed behaviour demonstrates that the rotation as a function of the 
dijet mass is governed by the factorisation scale changes. The renormalisation 
scale changes are similar at all values of $M_{JJ}$, with a smooth shape that 
changes little as the slices move through the factorisation scale range. The 
$\mu_F$ dependence, however, changes greatly with the dijet mass. In the first 
plot, with the lowest $M_{JJ}$ bin in the lowest rapidity bin shown, the 
factorisation scale dependence is roughly flat for all slices in $\mu_R$ except 
for the very lowest 2 $\mu_R$ choices. This is the cause of the vertical nature 
of the saddle point in the first plot in Fig. \ref{highlowvar}. In the second 
plot, at high $M_{JJ}$ in the lowest rapidity bin, the factorisation scale has a 
non-flat shape that depends greatly on the value of $\mu_R$ chosen. Because the 
variations in factorisation scale are now large, the saddle point in the second 
plot in Fig. \ref{highlowvar} is no longer vertical, and is rotated 
anticlockwise. The higher $\mu_F$ dependence at high $M_{JJ}$ can be understood 
through the $x$ values probed. In the high $M_{JJ}$ region, the high $x$ partons 
necessary for the events are evolved much more quickly than at medium/low $x$, 
and so a greater dependence on the factorisation scale is seen. The stability of 
the calculation, then, is dependent on the partons probed.

\begin{figure}[h!]
\includegraphics[width=0.49\textwidth]{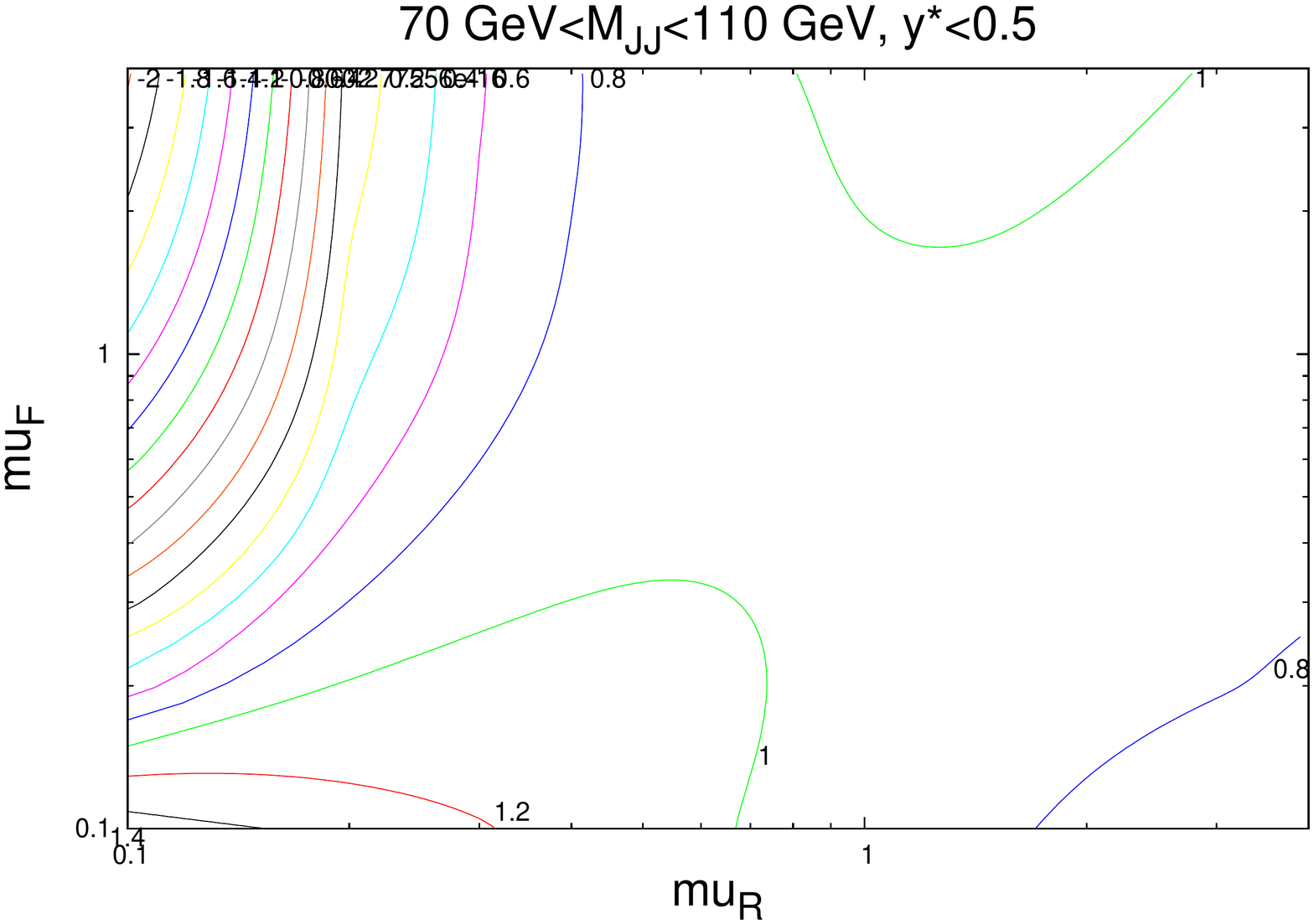}
\includegraphics[width=0.49\textwidth]{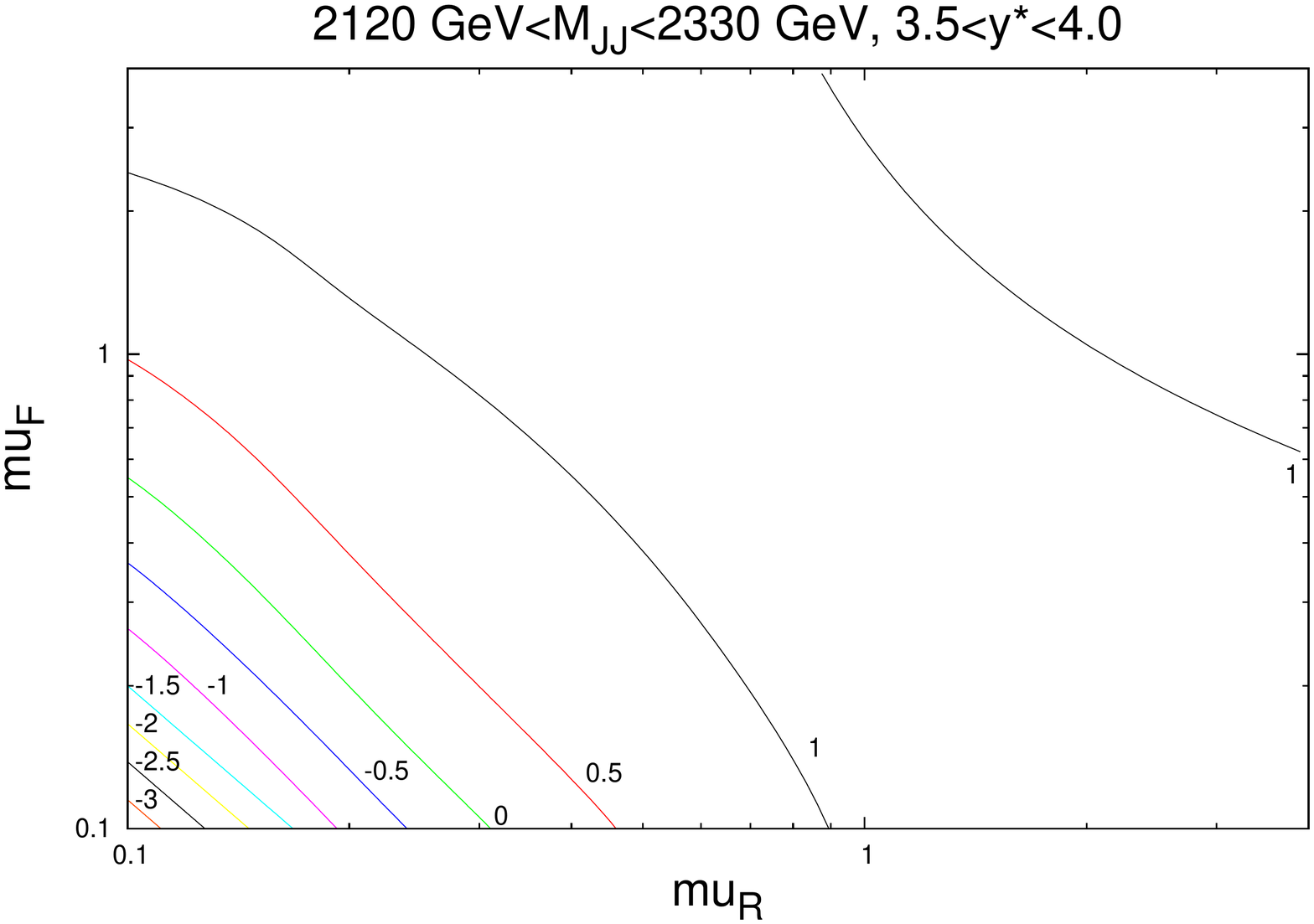}
\vspace{-0.8cm}
\caption{Scale variations for the scale choice $M_{JJ}/2\cosh(0.7y^*)$. Unlike when using $M_{JJ}$, the saddle point remains centrally located even in the high rapidity region. }
\label{mjjcoshy-scale}
\end{figure}

The need to choose a single scale for the entire calculation leads to the search 
for a choice where the saddle point is uniformly based at that choice. Since the 
calculation using $M_{JJ}$ as the kinematic scale choice seems to fail at higher 
rapidities, a function of $M_{JJ}$ and $y^*$ would be a logical choice to attempt 
to modulate this behaviour. The function $M_{JJ}/2\cosh(0.7y^*)$ is studied, 
which was shown in the previous section to improve the stability of the ATLAS 
calculation. The scale variations for this choice are shown in Fig. 
\ref{mjjcoshy-scale}, where even in the highest rapidity bin, the saddle point 
is located around the central scale choice. It is clear that for the ATLAS 
dataset, the phase space probed would prefer a scale choice including a 
rapidity term.

\subsection{Data Normalisation}

\begin{figure}[h!]
\begin{center}
\includegraphics[width=0.49\textwidth]{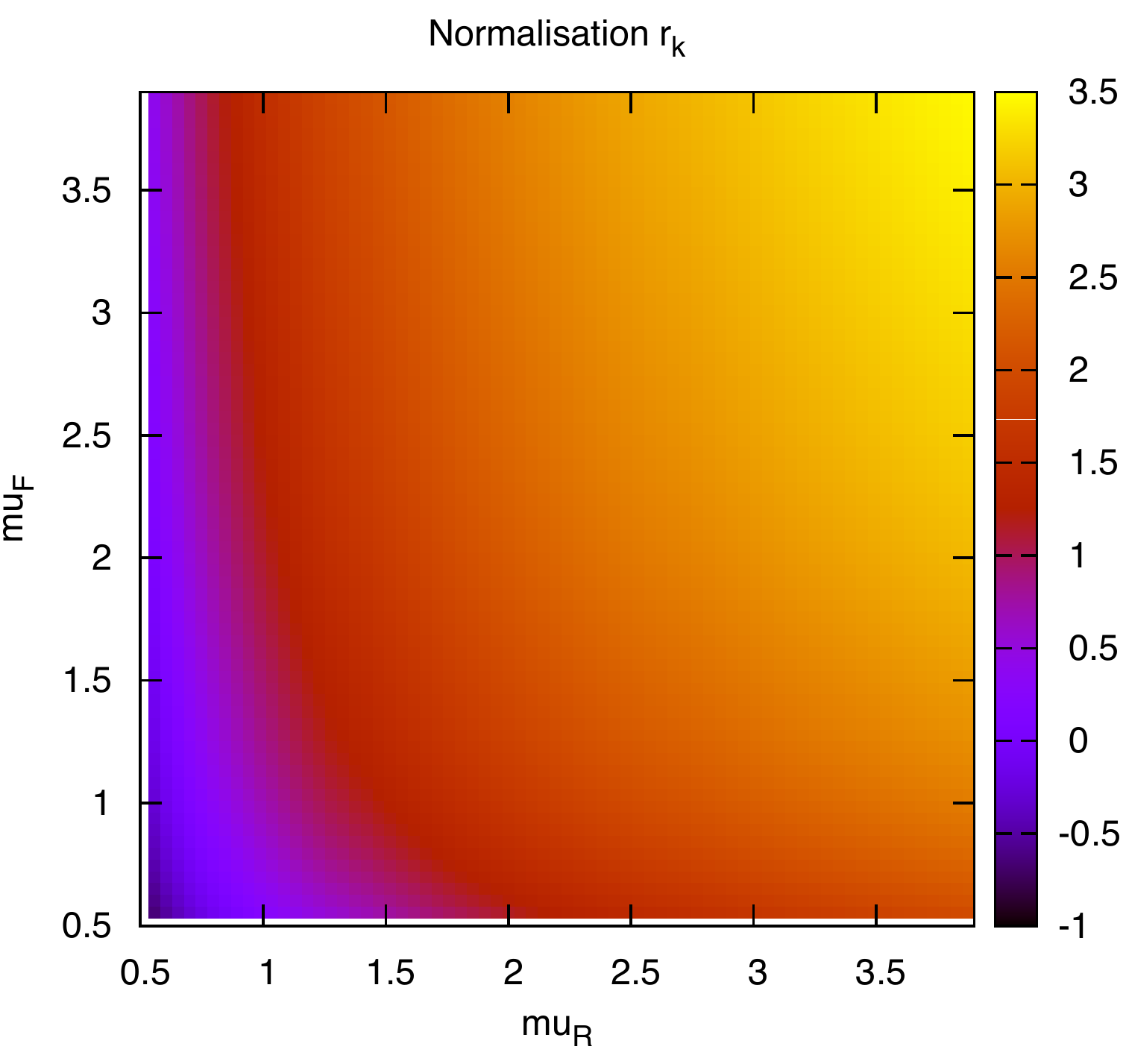}
\end{center}
\vspace{-0.8cm}
\caption{Value of the systematic shift associated with the normalisation uncertainty for each scale value (multiples of $M_{JJ}$)}
\label{norm}
\end{figure}

The treatment of normalisation errors on data sets has been a subject of 
previous discussion \cite{wattthorne}, and it is important to understand the 
effect they have on a fit. The only experimental source of the error is the 
luminosity uncertainty of the collider, and so it is correlated across all data 
sets produced at a single collider. 
For the Tevatron Run II data, the luminosity uncertainty is $6.1\%$, whilst the 
ATLAS 7~TeV run has a $3.4\%$ error. These provide the possibility for a 
theoretical prediction to move greatly up or down whilst incurring only a small 
penalty term in the $\chi^2$. Due to this effect, the MSTW 2008 PDFs include a 
more severe quartic penalty term for the normalisation.

\begin{figure}[h!]
\begin{center}
\includegraphics[width=0.49\textwidth]{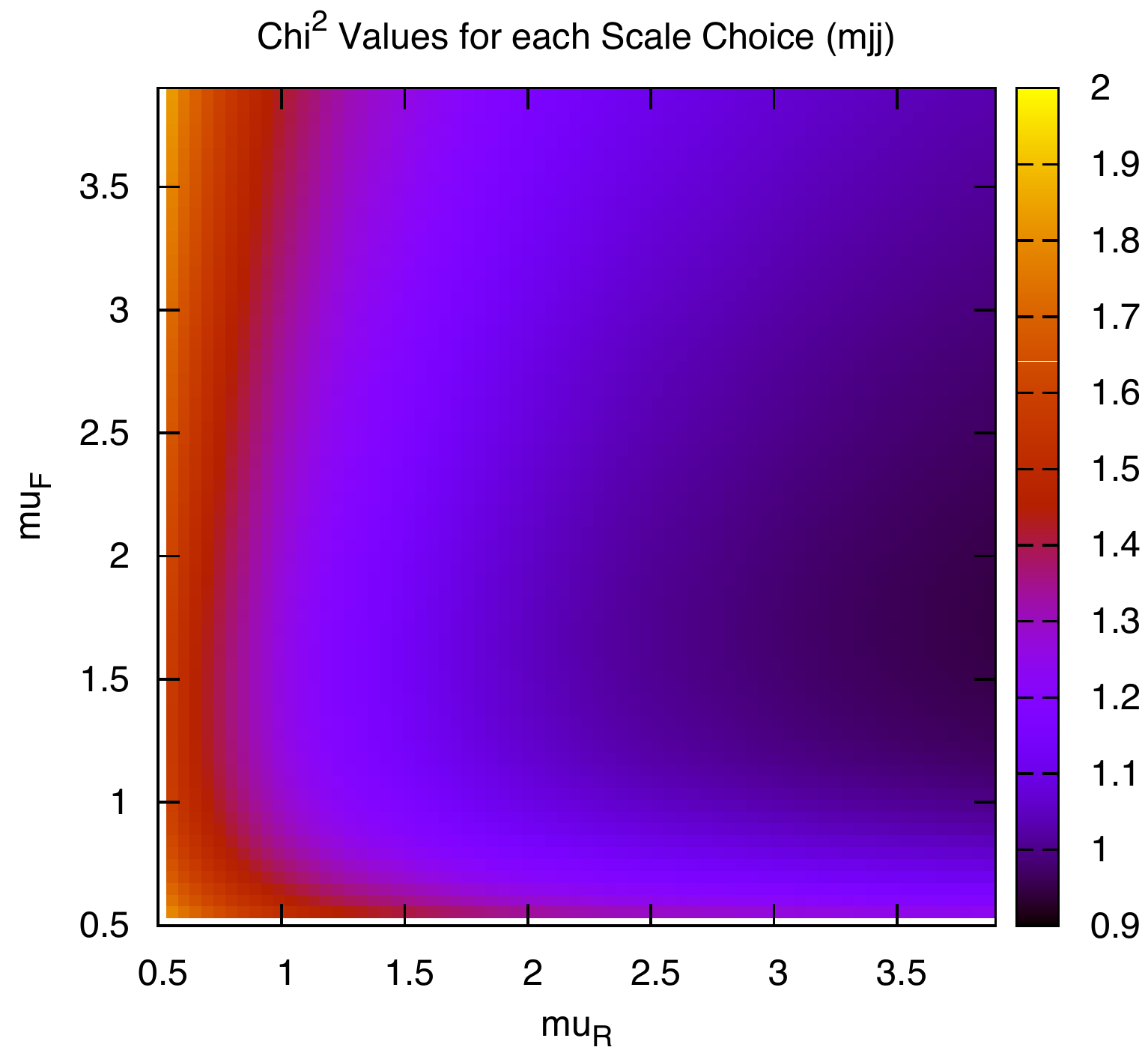}
\includegraphics[width=0.49\textwidth]{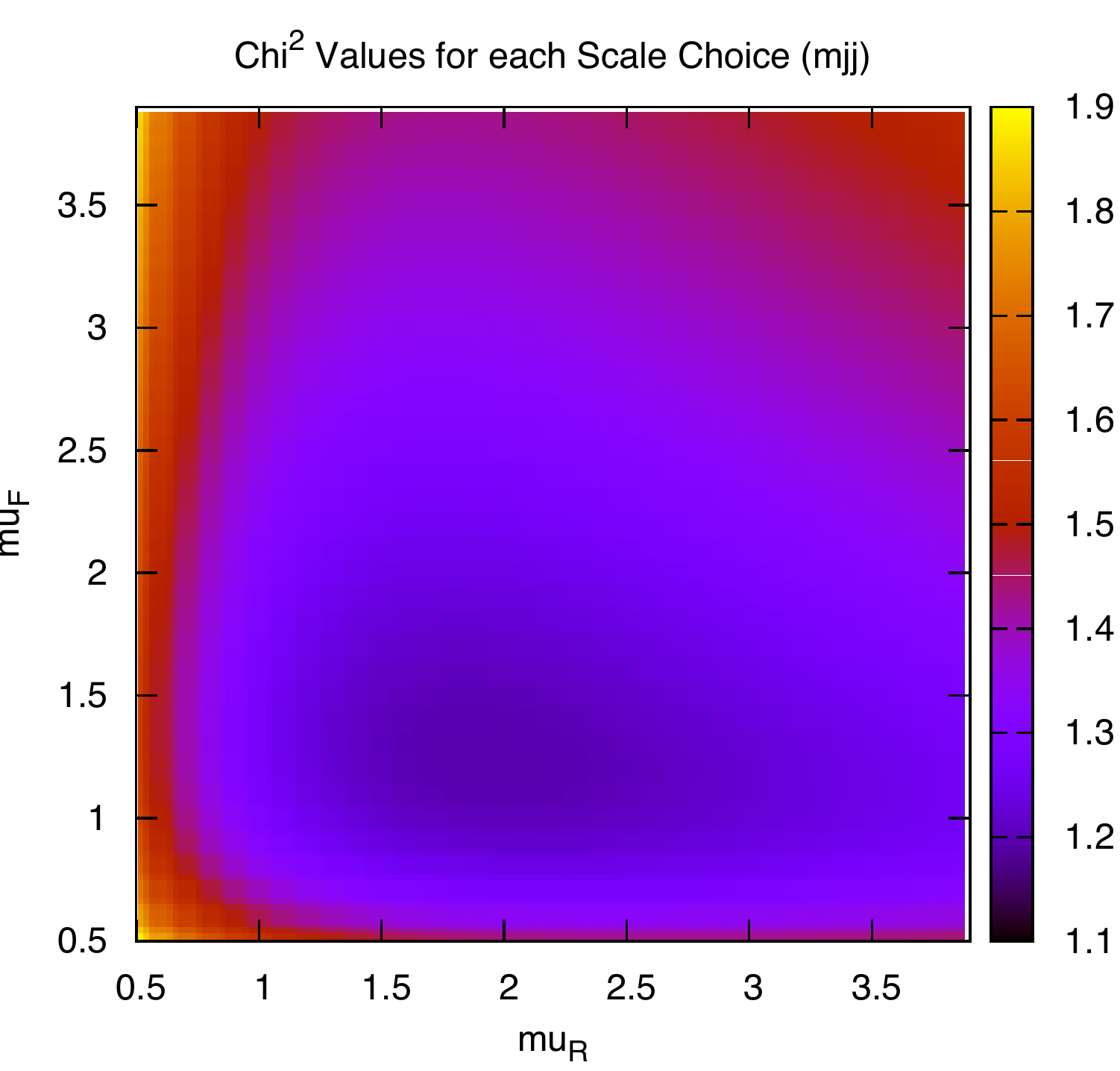}
\end{center}
\vspace{-0.8cm}
\caption{Goodness of fit for each combination of scales (multiples of $M_{JJ}$), first with and second without allowing the normalisation to move freely.}
\label{nonorm}
\end{figure}

When considering the best choice of scale variable for $D{\O}$ dijets, namely 
$M_{JJ}$, the best possible fit is obtained at very high values of 
renormalisation scale, 
as represented in Fig. \ref{nonorm}. However, if the normalisation $r_k$ of each 
fit is studied (where positive values of $r_k$ mean the data is normalised down),
it is clear that this minimum is obtained in a region where a $~2-3 \sigma$ 
shift is required, as can be seen in Fig. \ref{norm}. In fact, there is a very 
small area of the parameter space in which the normalisation parameter is moved 
less than $1 \sigma$, though this does include $\mu_{R,F}=M_{JJ}$.
The second plot in Fig \ref{nonorm} represents the same fit, but keeping the 
normalisation fixed. The minimum is now at a more sensible scale choice, at the 
cost of requiring a slightly higher $\chi^2$. Clearly equation \ref{chi2def} is 
inadequate for providing the most sensible fit, and a different treatment of 
the normalisation $r_k$ will ultimately be required. The difference in the 
normalisation treatments is most important for high values of the scales, where 
the calculation would naively appear to give the best fit. The effect is similar,
but rather less pronounced, for ATLAS dijet data.

\subsection{Effect on MSTW PDFs}

\begin{figure}[h!]
\begin{center}
\includegraphics[width=0.8\textwidth]{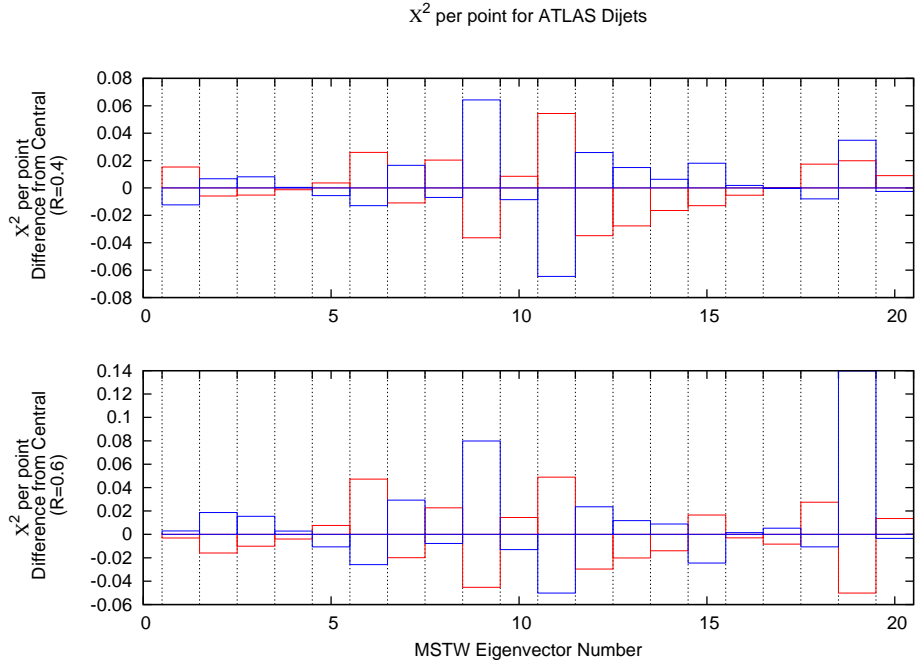}
\end{center}
\vspace{-0.8cm}
\caption{Deviations in fit quality from the MSTW 2008 NLO central value for each of the 20 eigenvector directions. Blue (red) bars indicate the positive (negative) direction of deviations in the eigenvector dimension.}
\label{evector-dojets}
\end{figure}

Fig. \ref{evector-dojets} shows the change in the $\chi^2$ for each eigenvector 
direction of the MSTW 2008 NLO set for the ATLAS dijet data, using 68\% 
confidence levels and a scale choice of $2p_T$. 
The plots show that, for the majority of the eigenvectors, 
a direction may be chosen in which the fit quality may improve, if only 
slightly. The eigenvector which contributes most significantly across the 
inclusive- and dijet data sets is number $9$, which is almost exclusively 
influenced by the gluon PDF. The other biggest contributors are influenced by a 
more mixed set of PDFs.

\begin{figure}[h!]
\begin{center}
\includegraphics[width=0.49\textwidth]{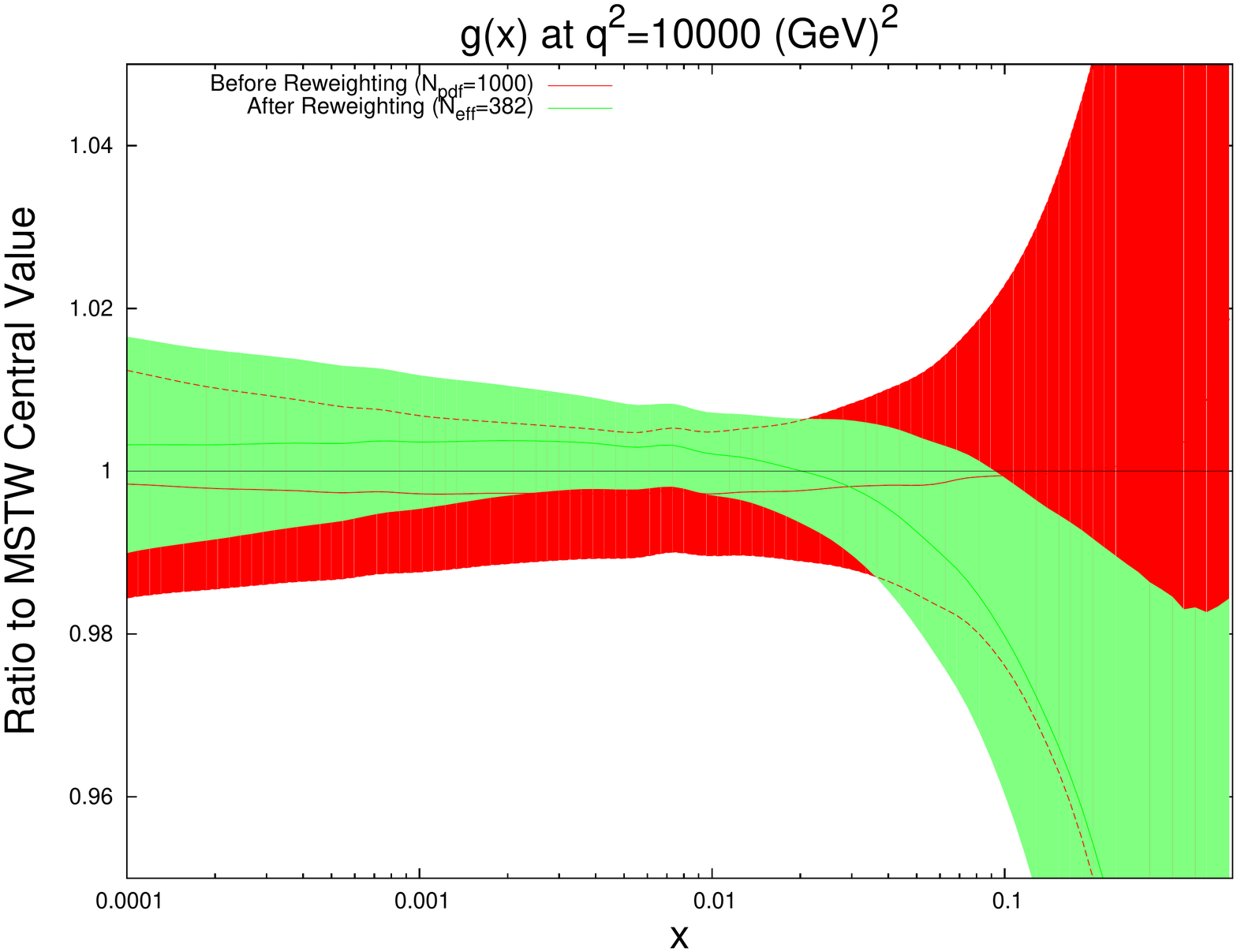}
\includegraphics[width=0.49\textwidth]{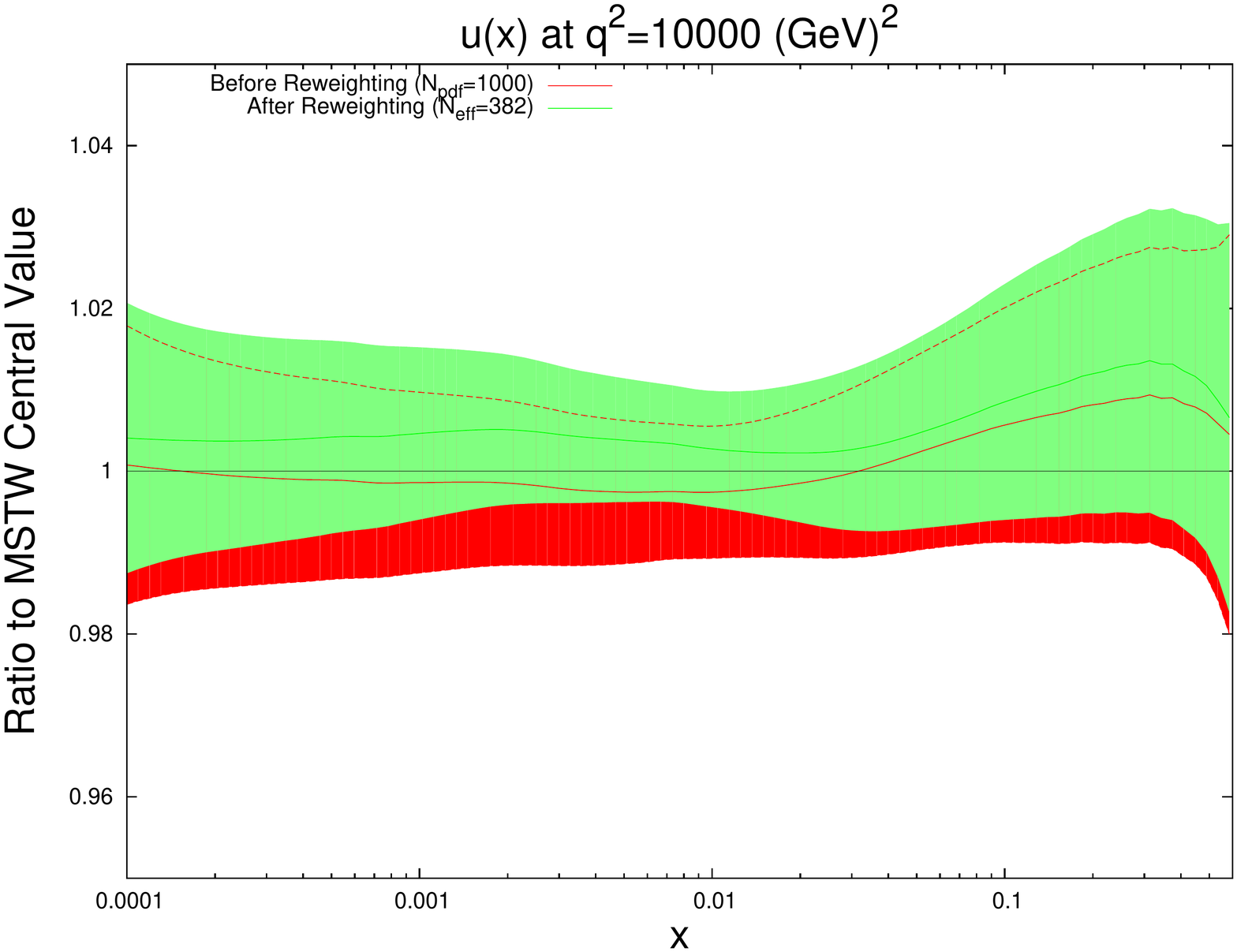}\\
\includegraphics[width=0.49\textwidth]{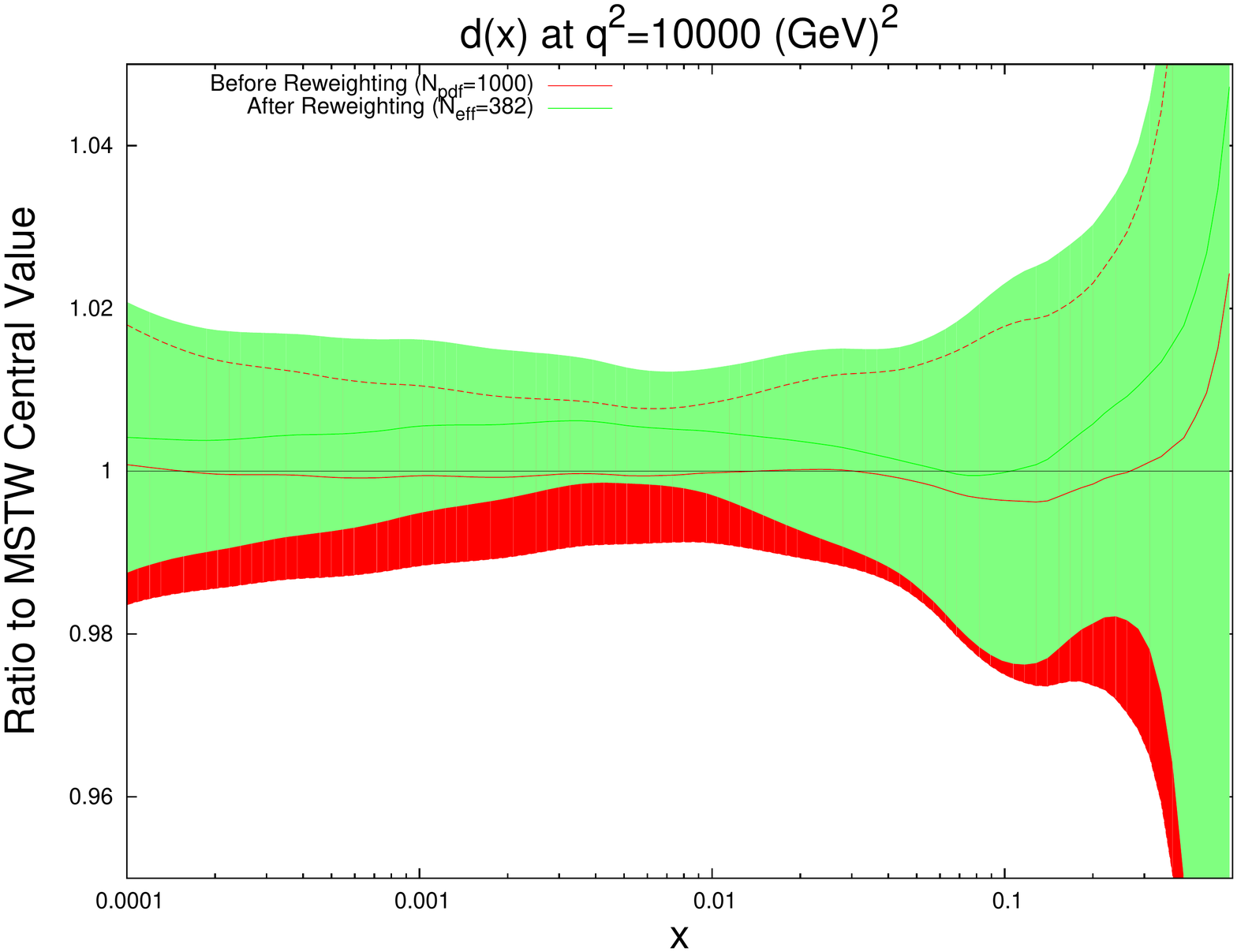}
\includegraphics[width=0.49\textwidth]{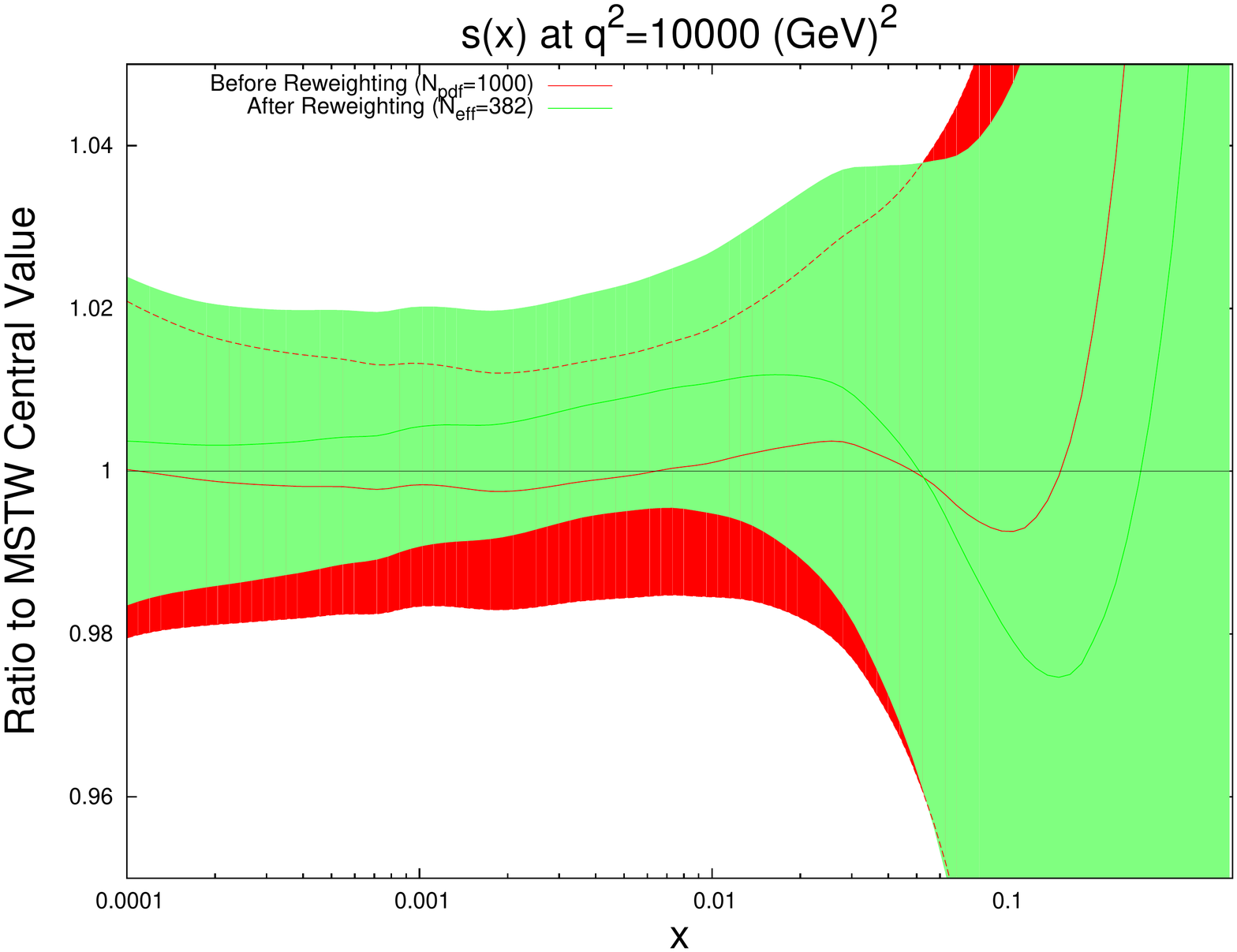}\\
\end{center}
\vspace{-0.8cm}
\caption{Effect of PDF reweighting on the gluon, up, down and strange distributions for D{\O} dijet data.}
\label{D0effect}
\end{figure}

\begin{figure}[h!]
\begin{center}
\includegraphics[width=0.49\textwidth]{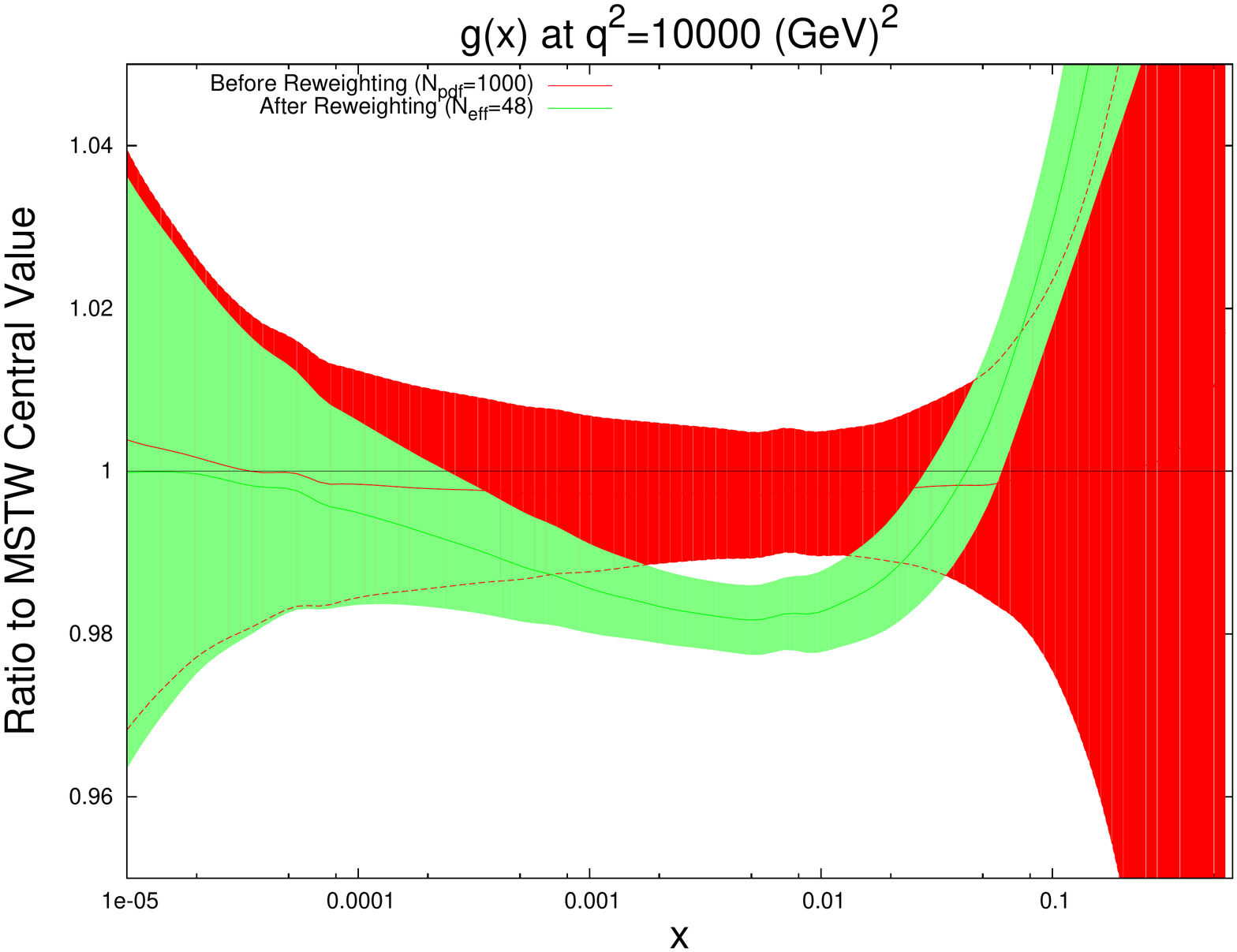}
\includegraphics[width=0.49\textwidth]{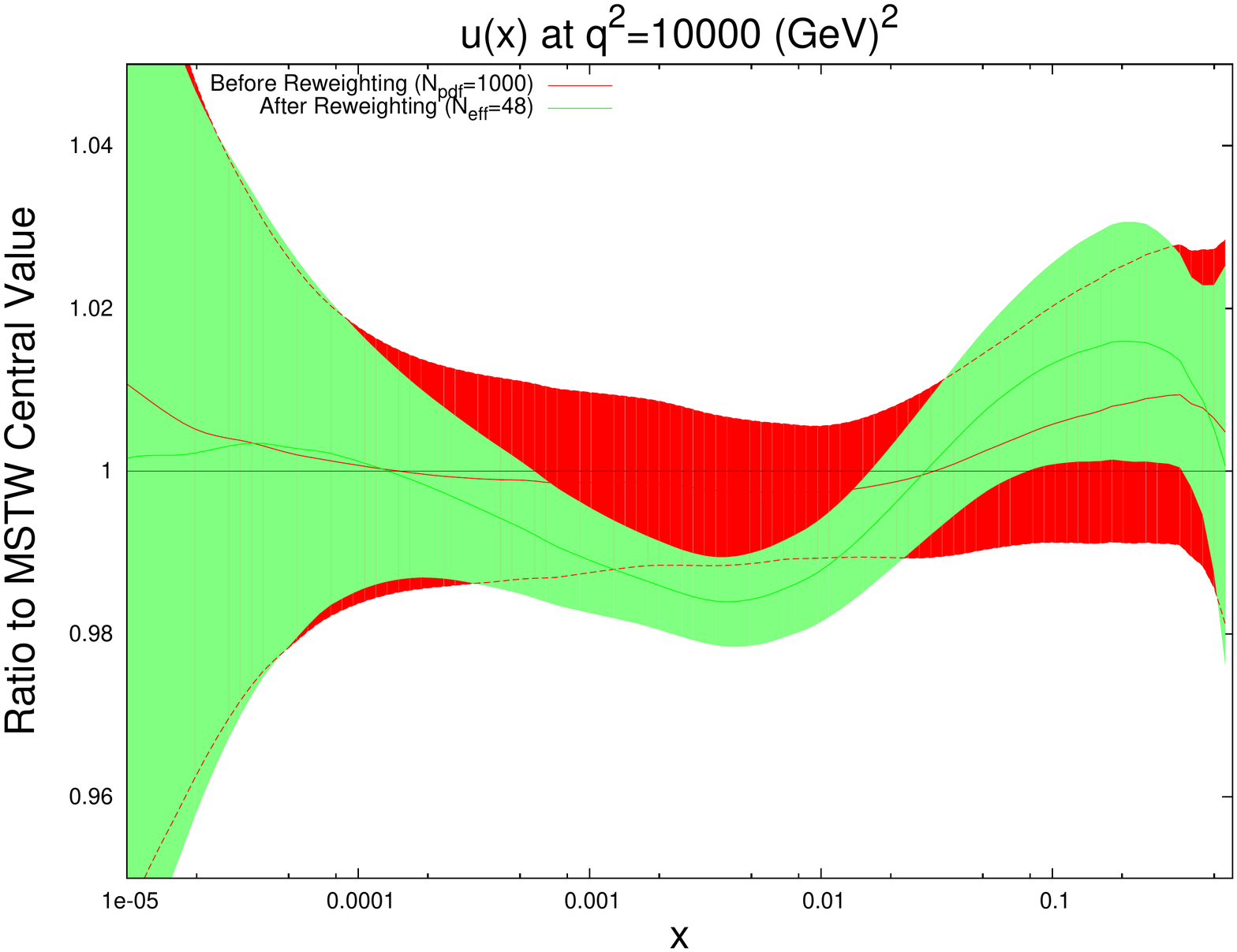}\\
\includegraphics[width=0.49\textwidth]{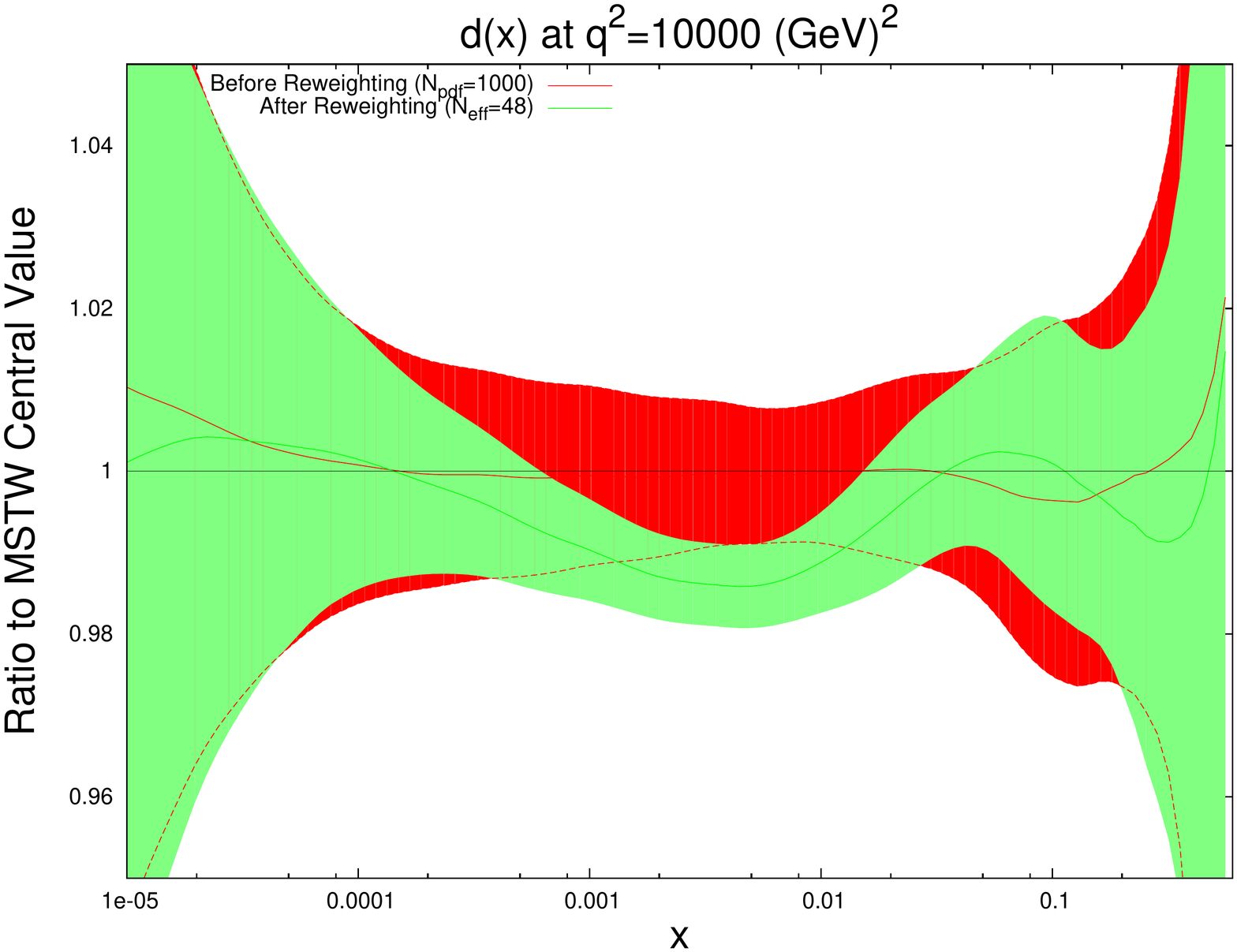}
\includegraphics[width=0.49\textwidth]{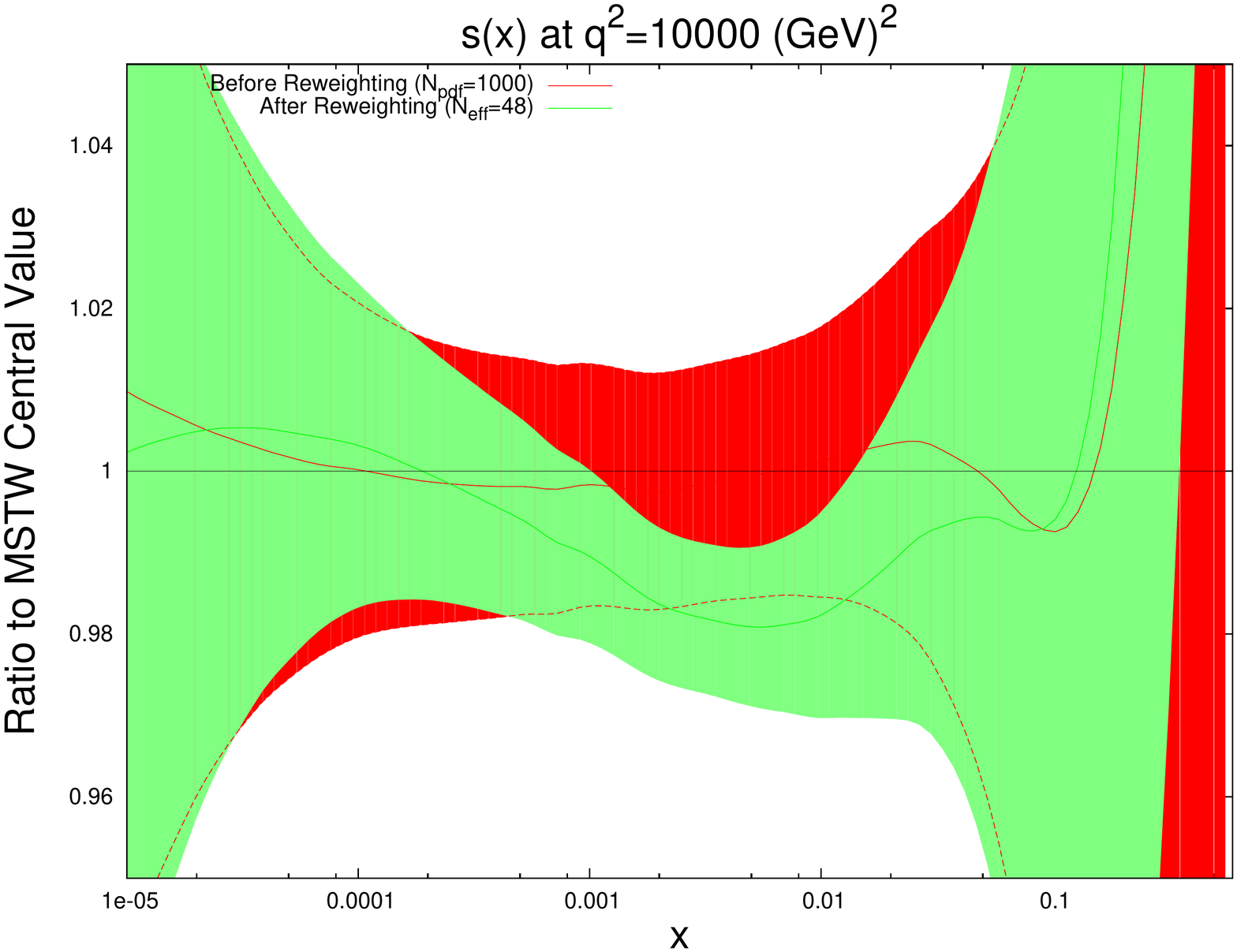}\\
\end{center}
\vspace{-0.8cm}
\caption{Effect of PDF reweighting on the gluon, up, down and strange distributions for ATLAS dijet data. The scale choice used is $M_{JJ}$}
\label{ATLASeffect}
\end{figure}

Next the reweighting procedure used in the previous section is repeated for the 
dijet datasets. The results for D{\O} dijets are shown in Fig. \ref{D0effect}. The 
scale choice used in the plots shown is $M_{JJ}$, however it was observed that a 
very similar effect was seen for the other two scale choices. Whilst the value 
of $N_{eff}$ changes from 382 in the shown plots to 166 for $p_T^{av}$ and 56 
for $M_{JJ}/0.7\cosh(y^*)$, the actual reweighted PDFs move in the same 
directions. All of the parton densities here are affected to some degree. 
Notable is the fact that there is a reasonable shift 
from the central values, especially for the gluon which also sees an improvement 
in the error band at the previous noted $x$ region. Given that the D{\O} inclusive 
jet data is included in the MSTW fit, this could be motivation to also attempt 
an inclusion of dijet data. The general trend of a larger gluon at low $x$ and 
lower at high $x$, along with slightly larger quark densities overall is similar 
to that of the ATLAS and CMS inclusive jet data shown in the previous Section.

%\begin{figure}[t!]
%\begin{center}
%\includegraphics[width=0.49\textwidth]{ATLAS/gluon-mjjcoshy.pdf}
%\includegraphics[width=0.49\textwidth]{ATLAS/up-mjjcoshy.pdf}
%\includegraphics[width=0.49\textwidth]{ATLAS/down-mjjcoshy.pdf}
%\includegraphics[width=0.49\textwidth]{ATLAS/strange-mjjcoshy.pdf}
%\end{center}
%\caption{Effect of PDF reweighting on the gluon, up, down and strange distributions for ATLAS dijet data. The scale choice used is $M_{JJ}/0.7\cosh(y^*)$}
%\label{ATLASeffect2}
%\end{figure}

\begin{figure}[h!]
\begin{center}
\includegraphics[width=0.49\textwidth]{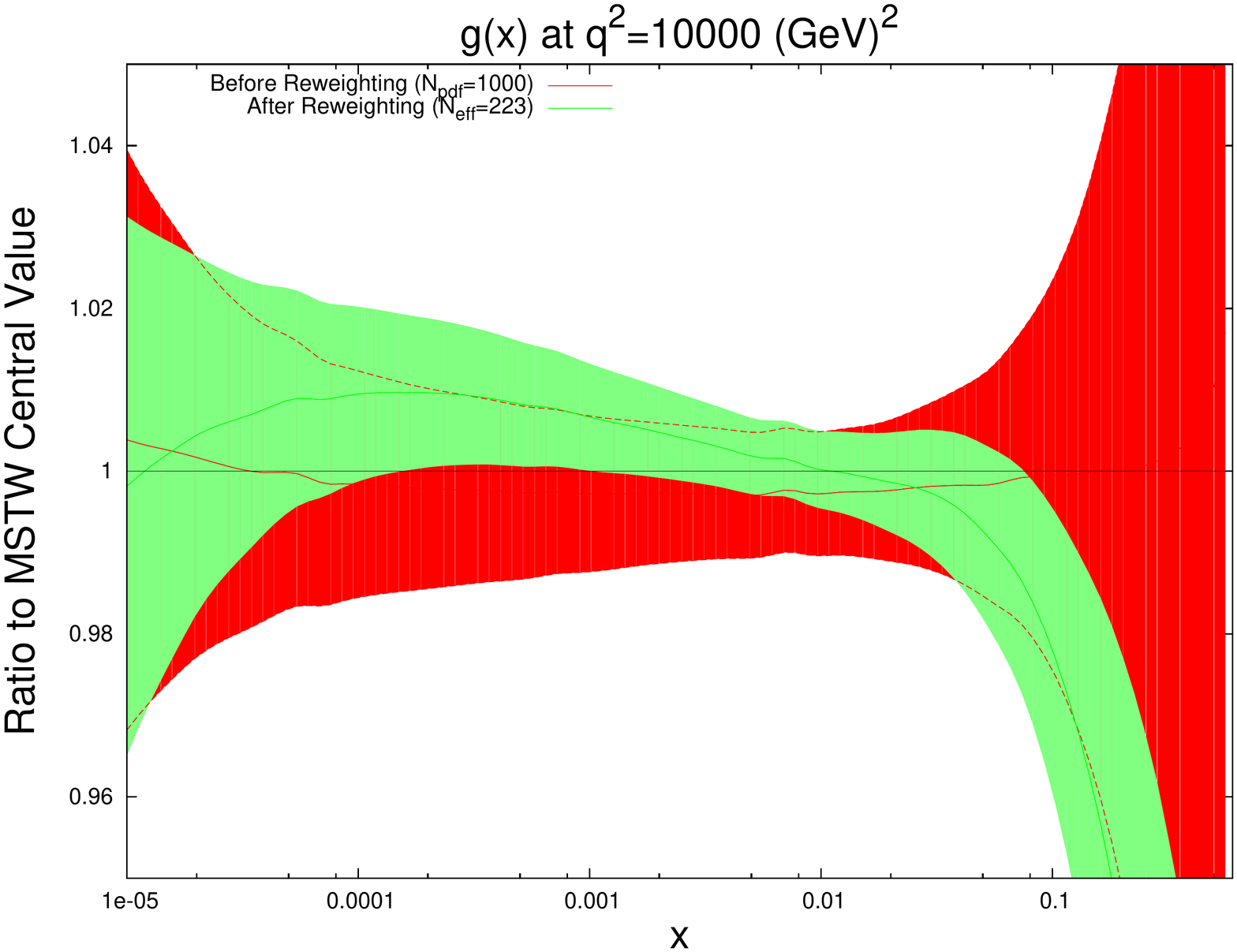}
\includegraphics[width=0.49\textwidth]{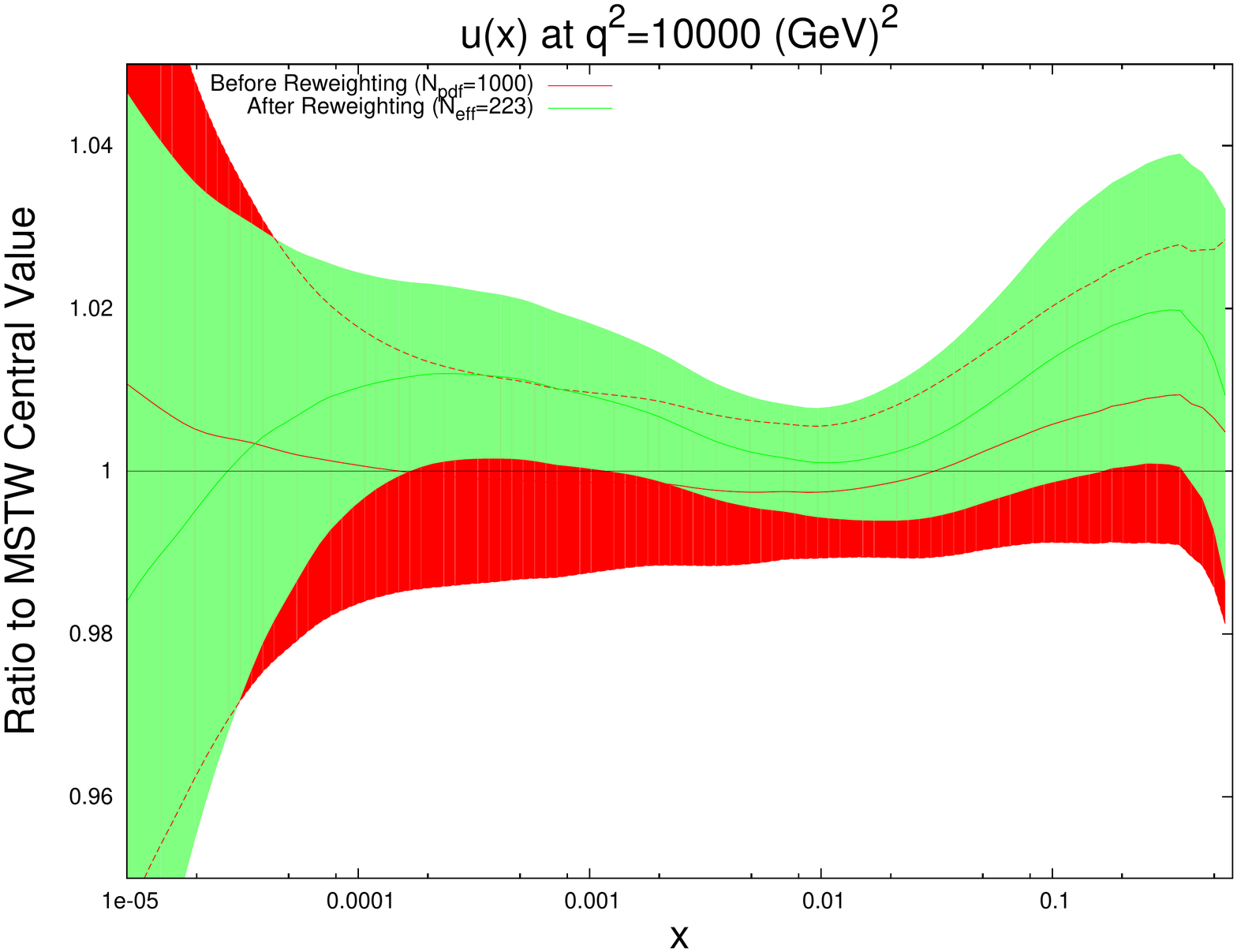}\\
\includegraphics[width=0.49\textwidth]{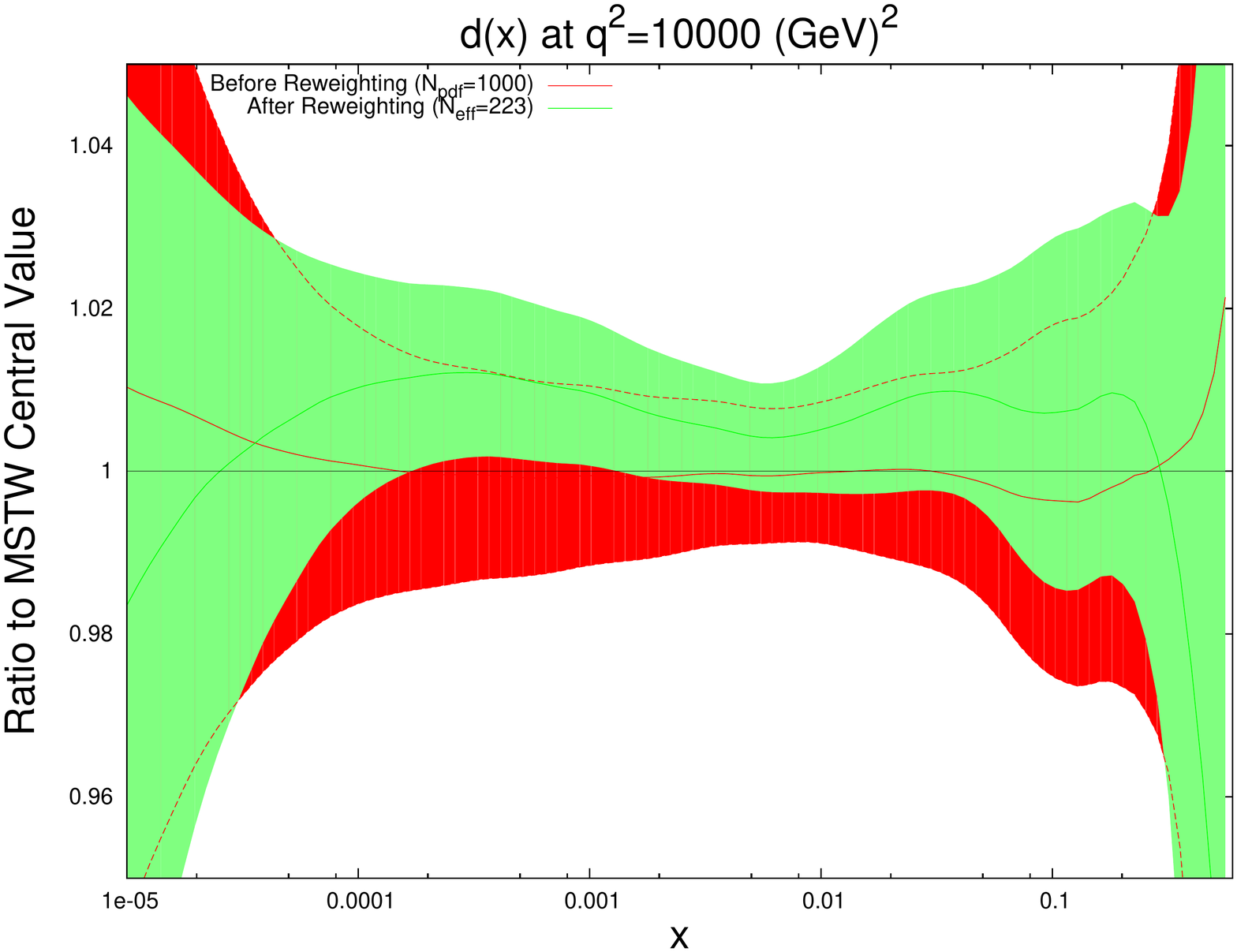}
\includegraphics[width=0.49\textwidth]{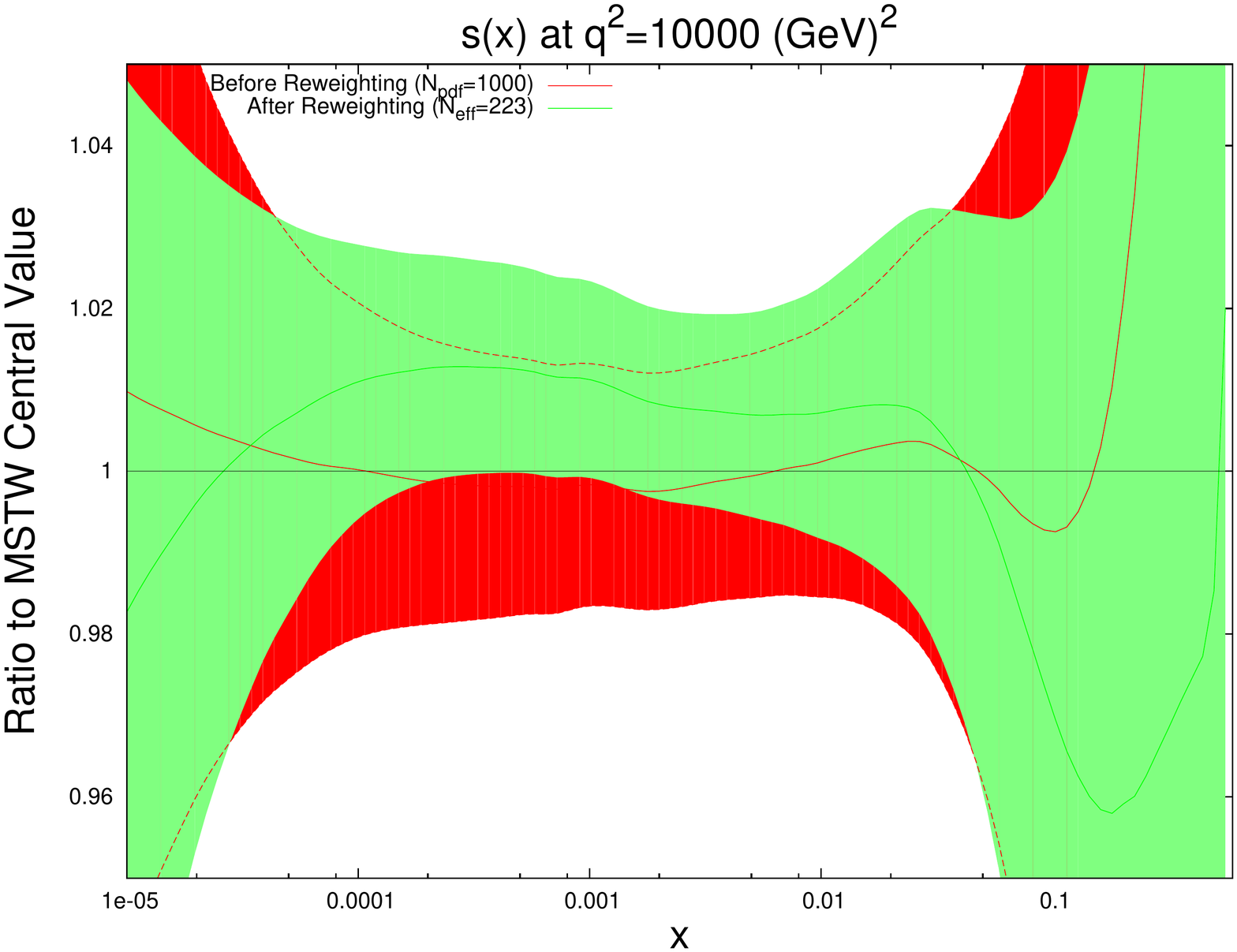}\\
\end{center}
\vspace{-0.8cm}
\caption{Effect of PDF reweighting on the gluon, up, down and strange distributions for ATLAS dijet data. The scale choice used is $2p_T^{av}$}
\label{ATLASeffect3}
\end{figure}

Next, the PDFs are reweighted using the ATLAS dijet data. This time, a 
difference in the PDF effect is observed between the different scale choices, 
which indicates a fundamental difference in the implied physics. For the 
choice of $M_{JJ}$, shown in Fig. \ref{ATLASeffect}, the gluon is moved well 
below its error band at moderate $x$ values, and above it at high-$x$. All of 
the quark PDFs are also significantly shifted with a reduction in error band 
size. For the other two scale choices, shown in Fig. \ref{ATLASeffect3} for 
$2p_T^{av}$ and very similar for $M_{JJ}/0.7\cosh(y^*)$, a less drastic and 
contradictory behaviour is seen, with the reweighted PDFs generally not 
moving outside of the error bands and the main effect being the softening of 
the gluon at high $x$. All of the reweighted PDFs give an improved fit to 
data from the standard MSTW predictions: $M_{JJ}$ changes from 2.30 to 1.95 
per point, whilst $2p_T^{av}$ moves from 1.98 to 1.90. However, the value of 
$N_{eff}$ is very low for the $M_{JJ}$ and $M_{JJ}/0.7\cosh(y^*)$ 
calculations, i.e. well under 100, and so the results should be considered 
with due care. Any value below 100 implies either that the original fit is 
very incompatible or the data is extremely constraining and hence 
that the reweighting is having a very large effect and is therefore not 
fully reliable. Without any clear preference for one of the particular 
scale choices it is difficult to conclude that the true effect on the PDFs 
from the ATLAS dijet data is within any PDF variation spanned by any of the 
scale choices. For the gluon this is then wider than the original uncertainty 
band. All choices seem to favour a slightly larger up quark distribution at 
high $x$, as does the D{\O} dijet data. Note that in each of these reweighting 
exercises the data normalisation moves by no more than one standard deviation.

\subsection{CMS Dijet Data}

\begin{figure}[h!]
\centering
\includegraphics[width=0.99\textwidth]{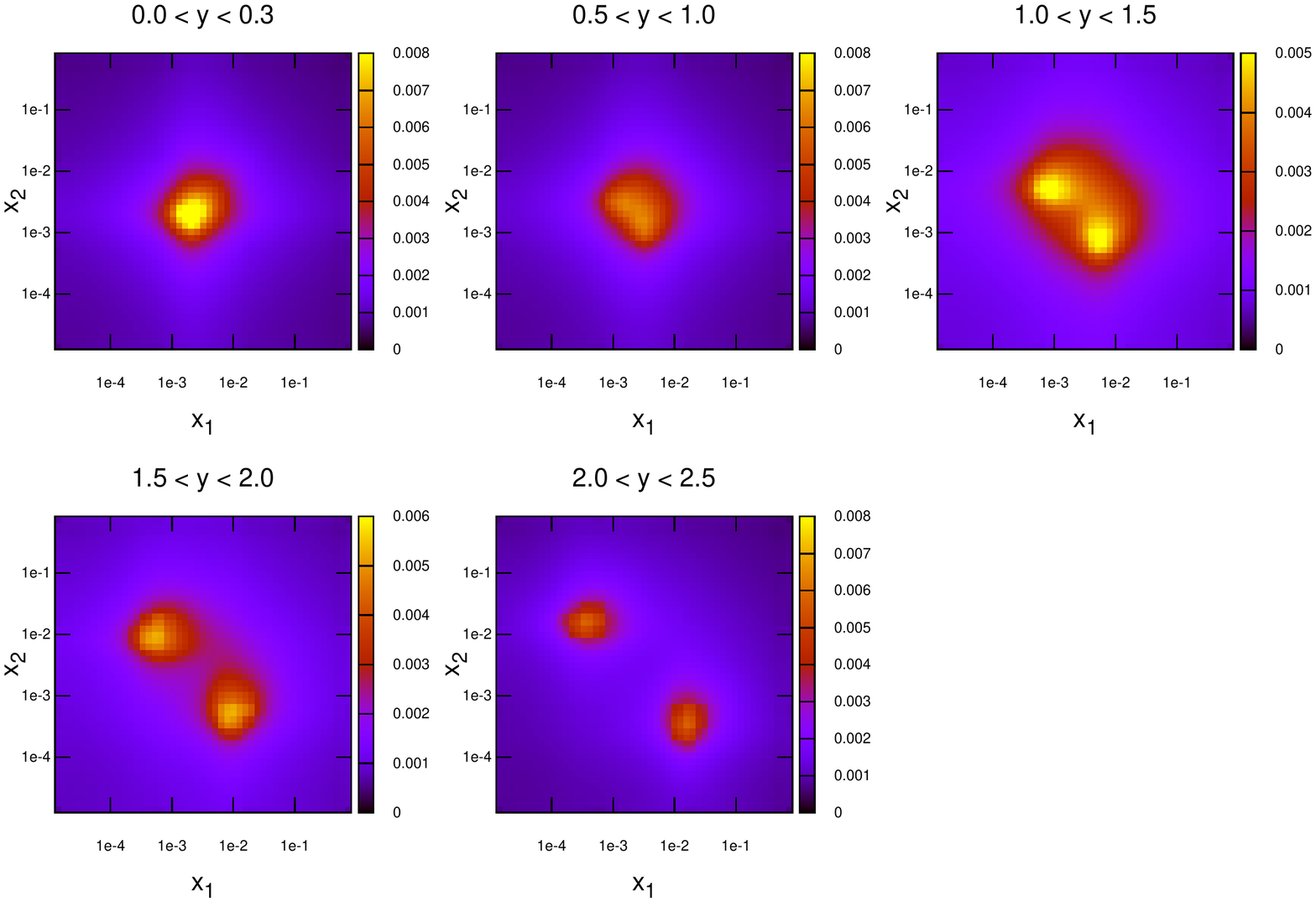}
\vspace{-0.8cm}
\caption{Values of $x_1$ and $x_2$ for each event generated in NLOJet++ for CMS dijets}
\label{cmsxes}
\end{figure}

Finally, as with the inclusive jets cross sections, the most recent and highest 
precision published dijet data has come from the CMS experiment \cite{cmsinc}. 
The data consists of 54 
points binned in $M_{JJ}$ and $y_{max}$. This is significant since it is the 
same rapidity binning as D{\O}, and different to ATLAS. Now any differences between 
the two approaches can be compared at the same collider. The $x$ distributions 
of NLOjet++ events generated for this data set are shown for each rapidity bin 
in Fig. \ref{cmsxes}. Due to the rapidity definition being the same as that at 
the Tevatron, the distribution resembles Fig. \ref{xreach3}, except with 
generally lower values of $x$ probed. Here, central dijets are probed at around 
$x\sim0.005$ for lowest $M_{JJ}$, with the highest rapidity dijets reaching $x\sim0.0001$.
The data in this case extends less far in rapidity, from $y_{max}=0$ to 
$y_{max}=2.5$, than the ATLAS data, which went all the way to $y^*=4.4$. 
Although the definitions are different, it must be true that the ATLAS data 
includes higher rapidity jets, since $y^*$ is defined as half the difference 
of the dijets' rapidities, and so the highest bin necessarily only includes 
two very high rapidity jets. Due to the low rapidity cut off, the issue of scale 
choice should be expected to not be as important, since all of the data is in 
the region where the $p_T^{av}$ choice behaved normally for ATLAS jets.

\begin{figure}[h!]
\centering
\includegraphics[width=0.99\textwidth]{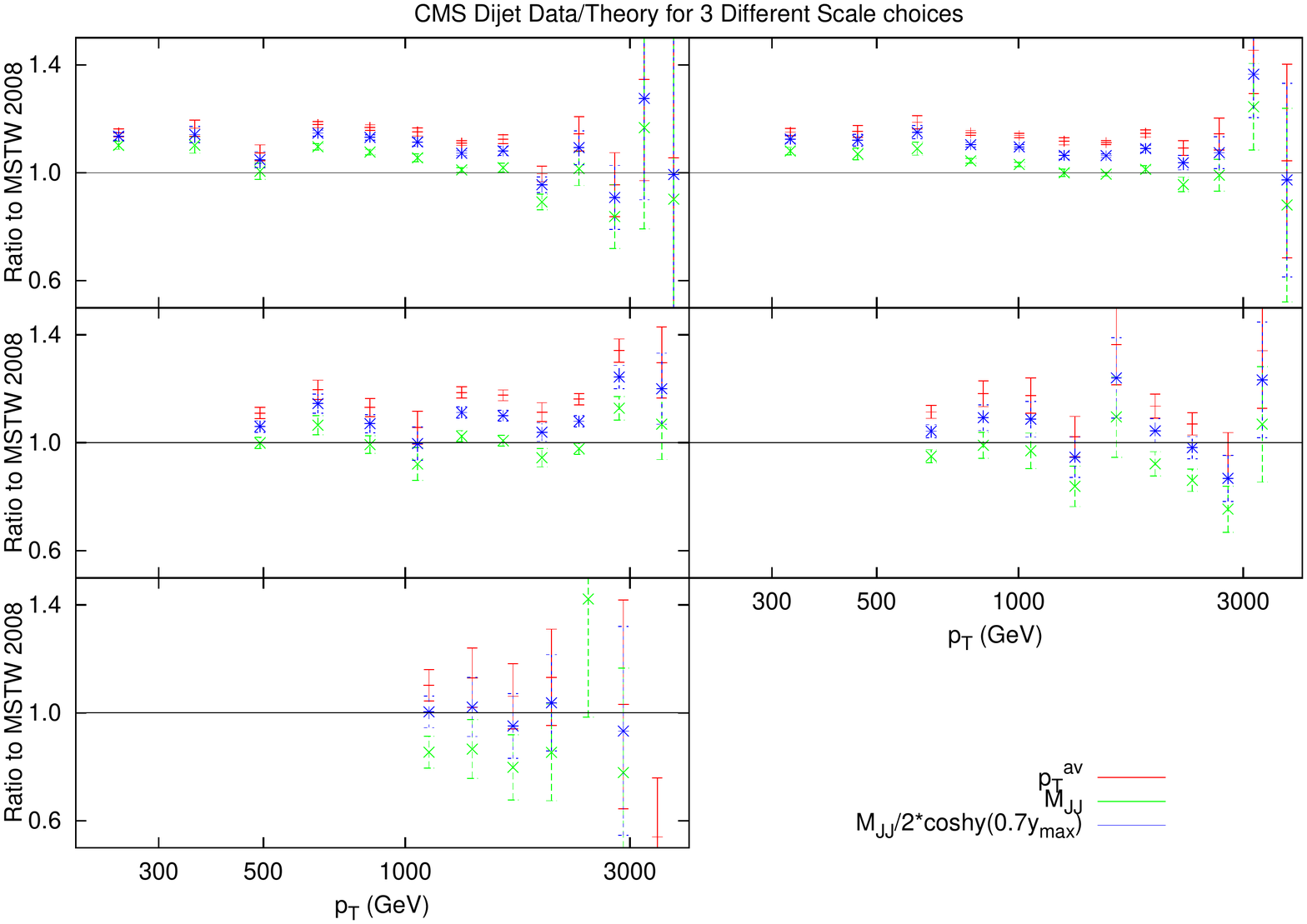}
\vspace{-0.8cm}
\caption{Ratio of data to theory for CMS dijets for all rapidity intervals. All 3 of the scale choices discussed are shown.}
\label{cmsratio}
\end{figure}

\begin{figure}[h!]
\centering
\includegraphics[width=0.5\textwidth]{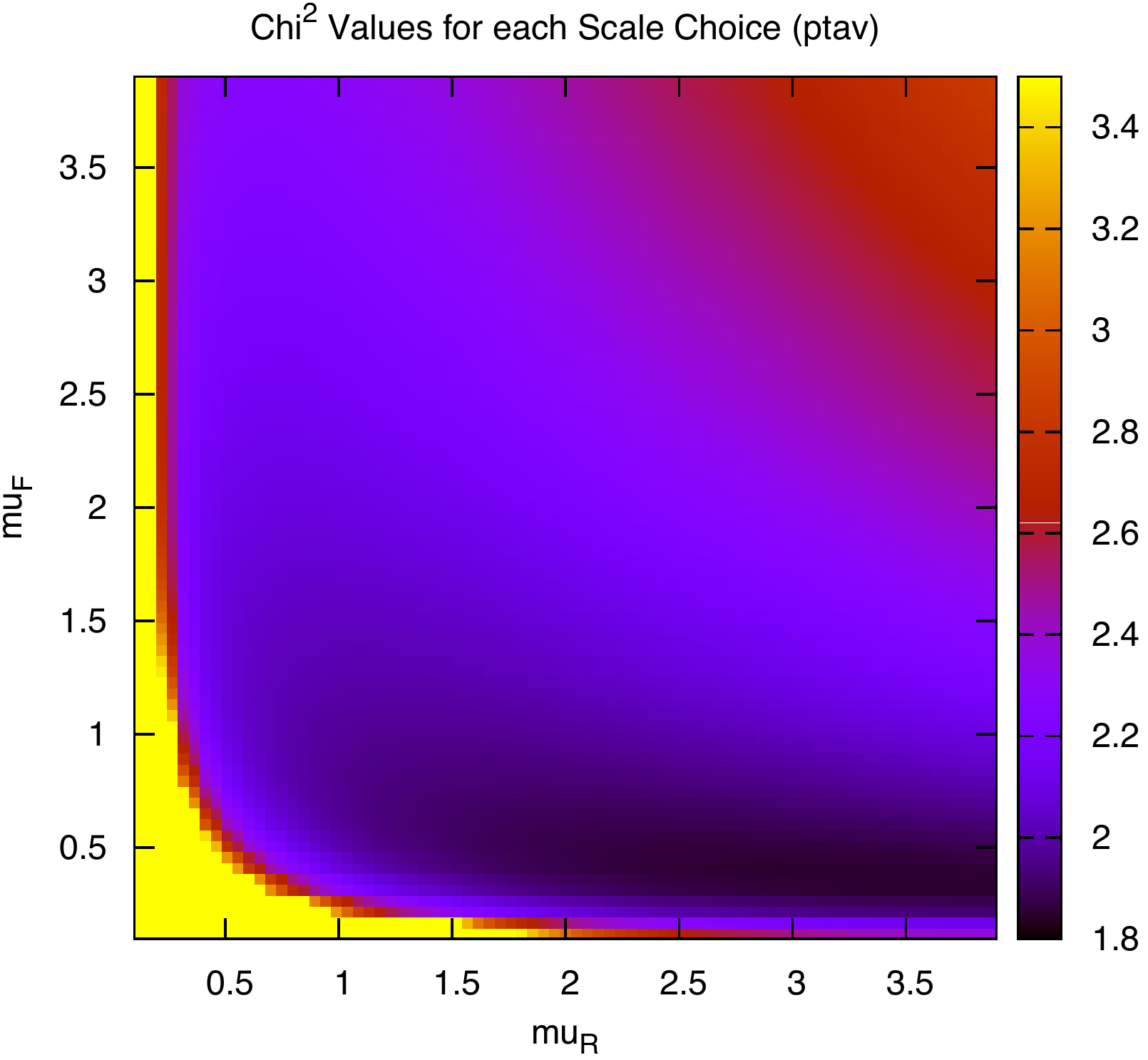}
\caption{$\chi^2$ value for every combination of $\mu_R$, $\mu_F$ for CMS dijets.}
\label{cmschi}
\end{figure}

The ratio of data to theory for the three scale choices is shown in Fig. 
\ref{cmsratio}. The scale variation has much less of an effect than for the 
ATLAS dijets, mostly due to the fact that the rapidity cut off is much lower, 
and the region where the most deviation occurred in the ATLAS dijets is 
avoided. The variation of the $\chi^2$ fit with the scales for the $p_T^{av}$ 
calculation is shown in Fig. \ref{cmschi}. Again, there is a region in the bottom 
left where the fit quality diverges exponentially, however this region is much 
smaller than for the ATLAS dijets, again because of the lack of the high-rapidity 
region, where the calculation is known to behave peculiarly. The results are 
summarized in Table \ref{cmspdfs}.

\begin{table}[h!]
\centering
\begin{tabular}{c c c c c}
\hline\hline
&$0.5*p_T^{av}$& $1.0*p_T^{av}$& $2.0*p_T^{av}$&\\
MSTW2008 NLO &2.76  & 1.97 & 2.18&\\
\hline
\end{tabular}
\caption{$\chi^2$ values for CMS dijets}
\label{cmspdfs}
\end{table}

\begin{figure}[h!]
\begin{center}
\includegraphics[width=0.49\textwidth]{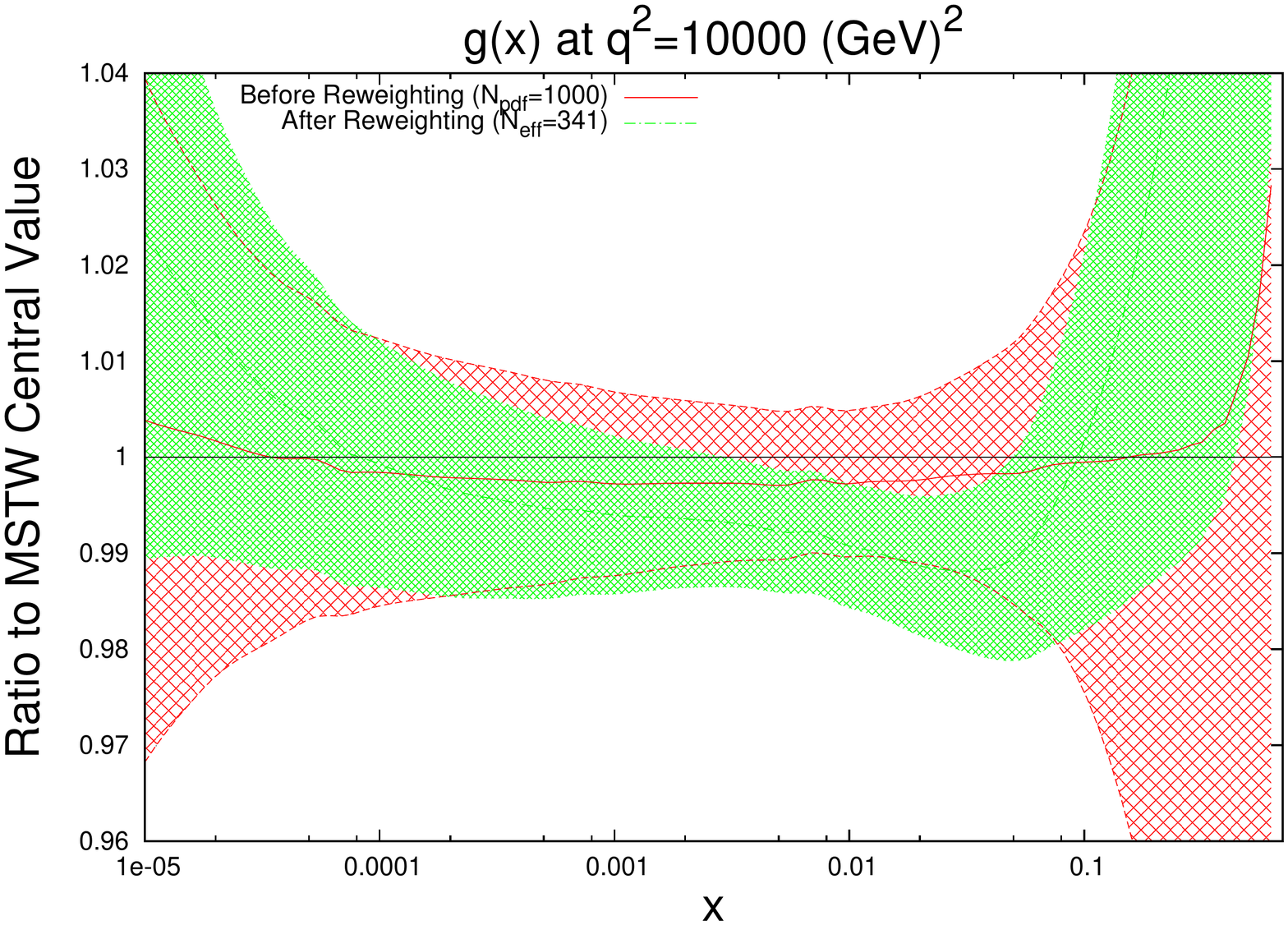}
\includegraphics[width=0.49\textwidth]{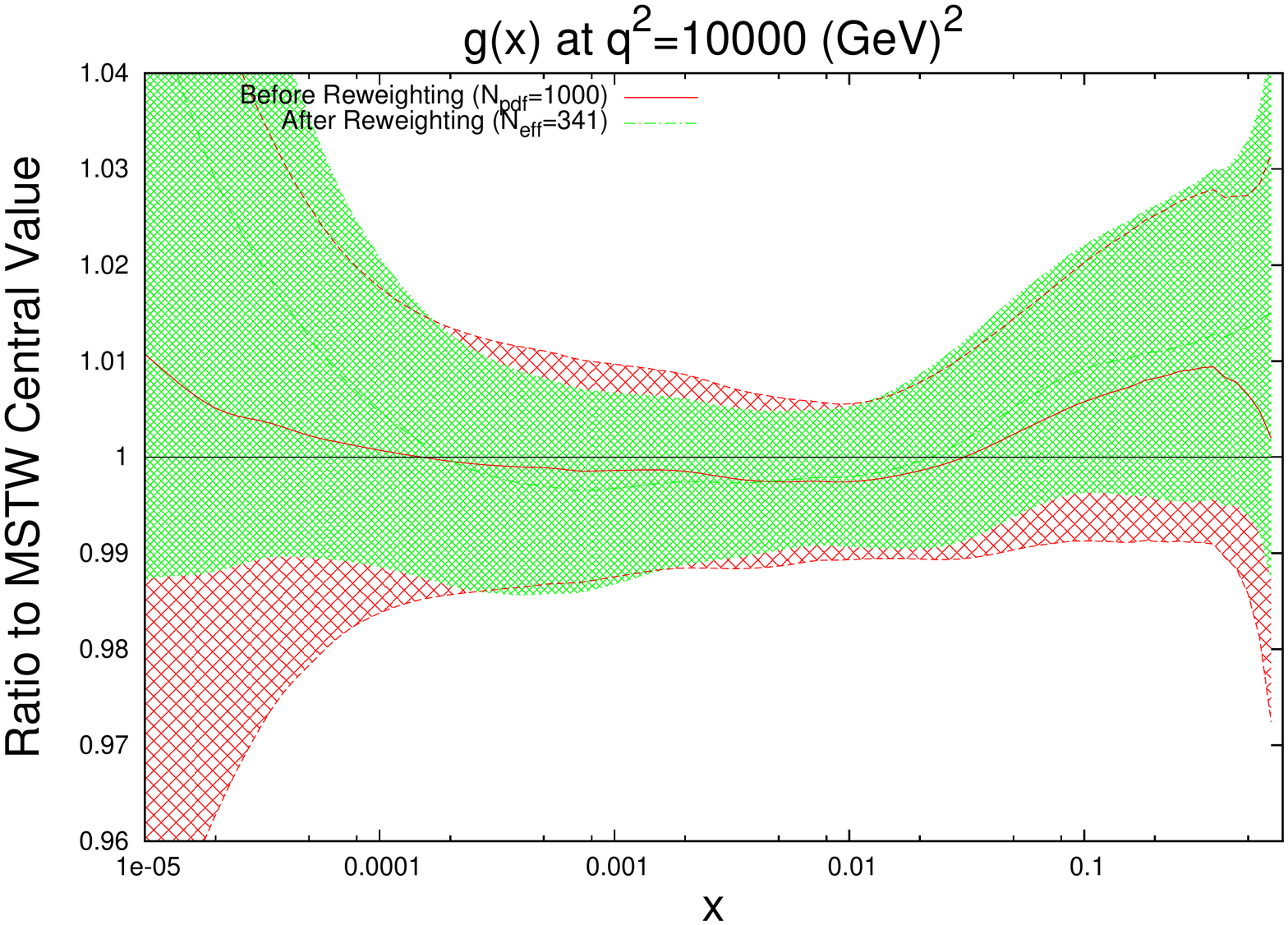}\\
\includegraphics[width=0.49\textwidth]{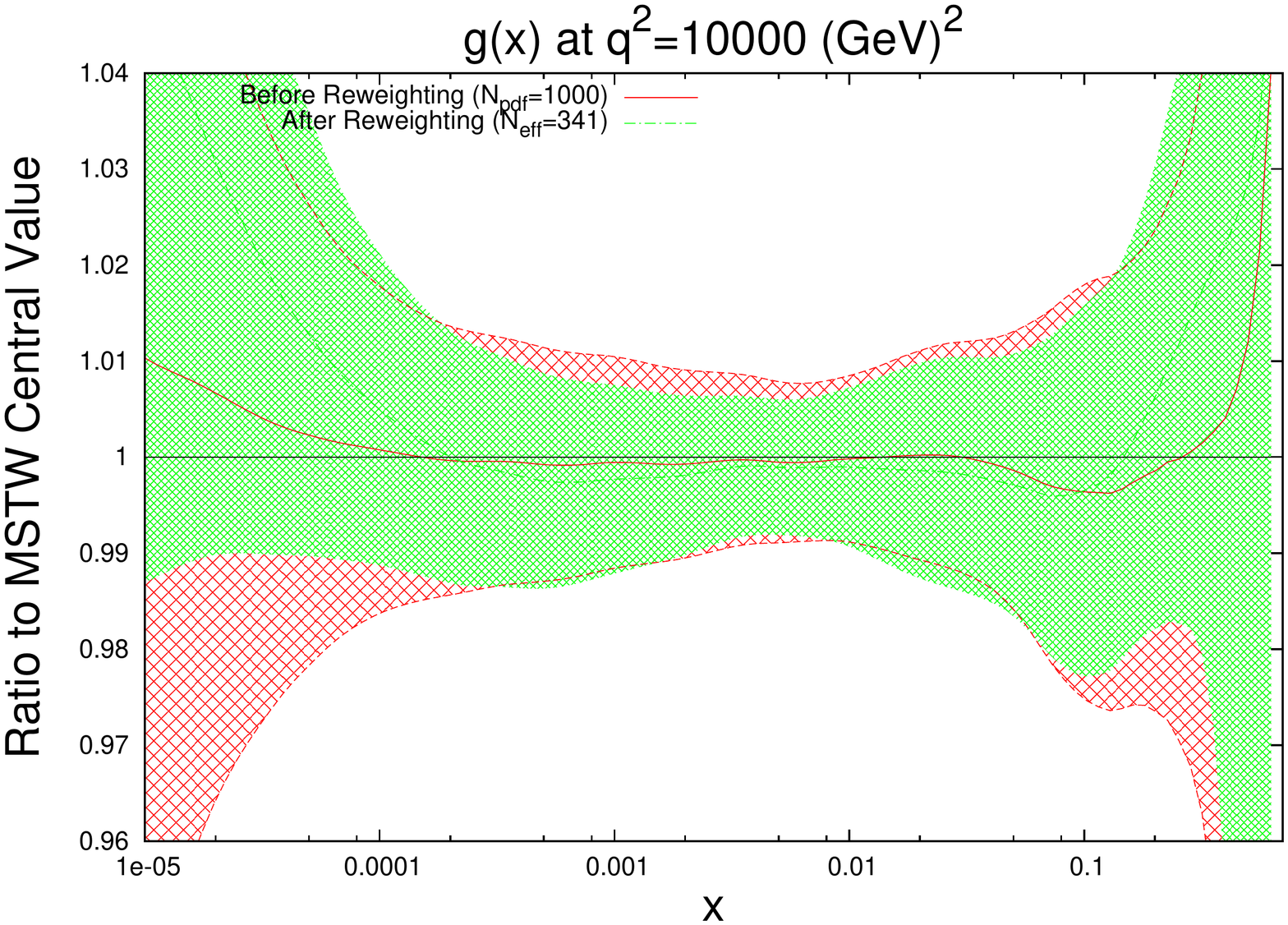}
\includegraphics[width=0.49\textwidth]{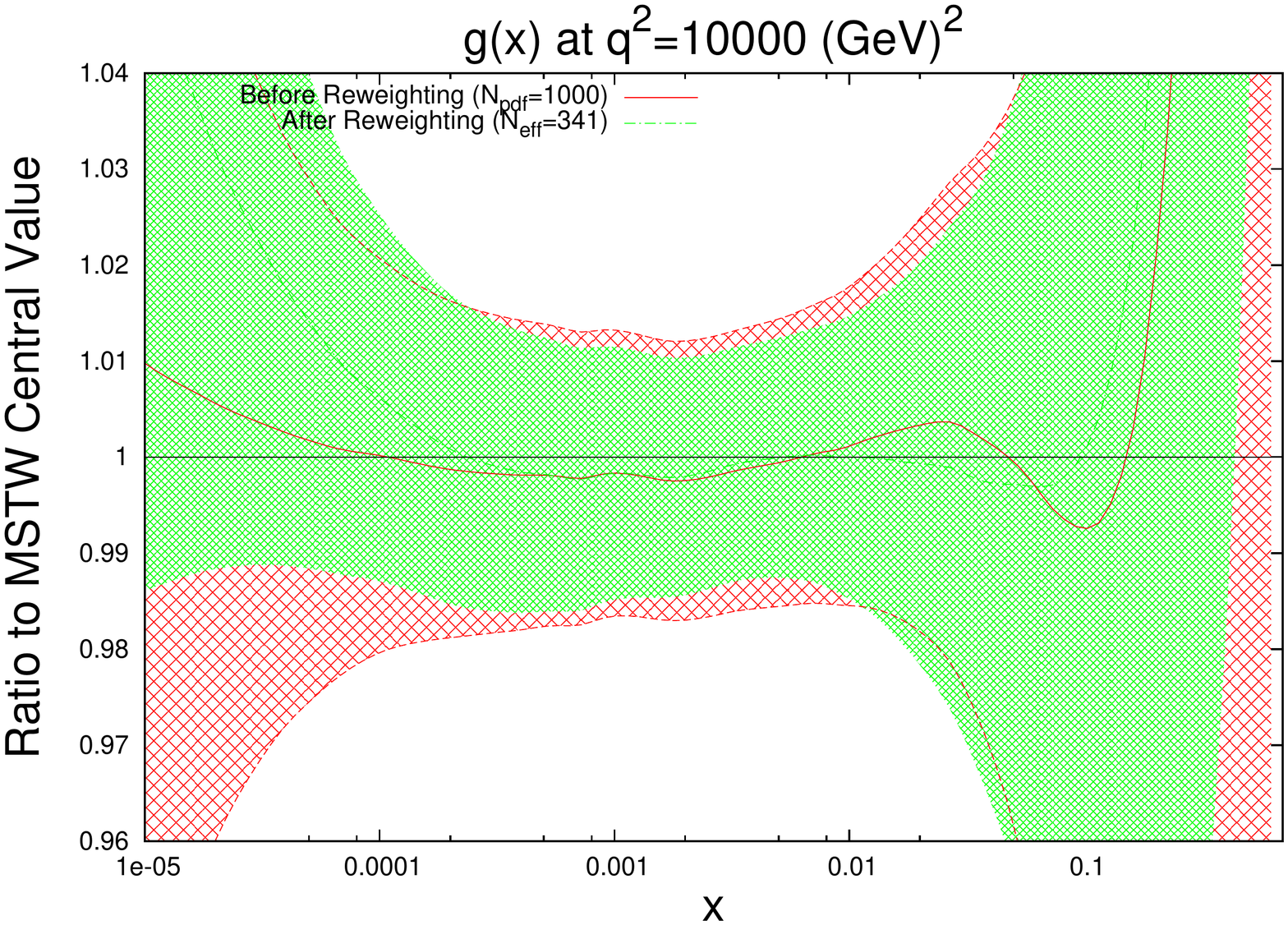}\\
\end{center}
\vspace{-0.8cm}
\caption{Effect of PDF reweighting on the gluon, up, down and strange distributions for CMS dijet data. The scale choice used is $p_T^{av}$}
\label{CMSeffect}
\end{figure}

The results of the PDF reweighting are shown in Fig. \ref{CMSeffect}. Only the 
plot for the scale choice $p^T_{av}$ are shown, since for this dataset the 3 
choices are all in general agreement, unlike for the ATLAS dijets. The shape 
of the reweighted gluon is similar to that of the $M_{JJ}$ ATLAS dijets, with 
a smaller gluon at moderate $x$ preferred, though the change is smaller than 
this, i.e. within the PDF uncertainty. This effect is in contradiction to the 
preferred gluon of the inclusive jet data, implying a conflict between the 
two datasets. However, it may be that higher order corrections beyond NLO 
in QCD (or electroweak corrections) 
do not have exactly the same shape dependence, and this could 
potentially remove, or reduce any tension between the constraints from 
inclusive and dijet data.

\section{Direct Inclusion of Inclusive data in PDF fits}
\label{chap:newpdfs}

In this section, new PDF sets are produced including the LHC inclusive jet data 
directly into PDF fits using the MSTW2008 framework. There was sufficient 
motivation from the eigenvector reweighting studies into the ATLAS combined 
2.76~TeV and 7~TeV data and the CMS data to justify this. In addition, this is 
an opportunity to further test the validity of the reweighting technique as a 
general method of quantifying the effect of a new data set on PDFs. Two fits 
are performed in this section, the first of which includes only the ATLAS 7~TeV 
and CMS inclusive data, both of which were calculated with FastNLO version 2. 
The second fit additionally includes the ATLAS combined data, which is 
calculated using APPLgrid and required further modifications to the fitting code.

\subsection{Fit With ATLAS 7~TeV and CMS Inclusive Jet Data}

\begin{table}[h!]
\begin{center}
\resizebox{12cm}{!} {
\begin{tabular}{| c | c | c | c |}
\hline
Data Set & MSTW2008 & MSTWCMS $\alpha_s$ Free & MSTWCMS $\alpha_s$ Fixed \\
\hline
BCDMS $\mu p$ $F_2$ \cite{data1}& 182/163& 172/163 & 182/163 \\
BCDMS  $\mu d$ $F_2$ \cite{data2}& 190/151&  188/151 &189/151\\
NMC  $\mu p$ $F_2$ \cite{data3}&121/123 &  122/123 &120/123\\
NMC  $\mu d$ $F_2$ \cite{data3}&102/123 &  103/123 &102/123\\
NMC  $\mu p/\mu d$ \cite{data5}&130/148 & 131/148  &130/148 \\
E665 $\mu p$ $F_2$ \cite{data6}&57/53 &  54/53 & 54/53\\
E665  $\mu d$ $F_2$ \cite{data6}& 53/53&  57/53 &57/53\\
SLAC  $\mu p$ $F_2$ \cite{data8,data9}&30/37 &  30/37 &30/37\\
SLAC  $\mu d$ $F_2$ \cite{data8,data9}& 30/38&  33/38 &30/38\\
NMC/BCDMS/SLAC $F_L$ \cite{data1,data3,data10}&38/41 &40/31 & 38/31\\
\hline
E866/NuSea pp DY \cite{data11}&228/184 & 227/184 & 229/184 \\
E866/NuSea pd/pp DY \cite{data12}& 14/15&  13/15&  14/15\\
\hline
NuTeV $\nu N$ $F_2$ \cite{data13}& 49/53&  50/53&  50/53\\
CHORUS $\nu N$ $F_2$ \cite{data14}& 26/42& 26/42 & 26/42\\
NuTev $\nu N$ $xF_3$ \cite{data13}& 40/45&  45/45&  40/45\\
CHORUS $\nu N$ $xF_3$ \cite{data14}&31/33 & 32/33& 31/33\\
CCFFR $\nu N\rightarrow\mu\mu X$ \cite{data17}& 66/86& 66/86 & 65/86\\
NuTeV $\nu N\rightarrow\mu\mu X$  \cite{data17}& 39/40& 39/40& 40/40\\
\hline
H1 MB 99 $e^+p$ NC \cite{data19}& 9/8& 9/8& 9/8\\
H1 MB 97 $e^+p$ NC \cite{data20}&42/64 & 43/64 & 44/64\\
H1 low $Q^2$ 96-97 $e^+p$ NC \cite{data20}&44/80 & 44/80& 45/80\\
H1 high $Q^2$ 98-99 $e^-p$ NC \cite{data22}& 122/126& 122/126& 120/126\\
H1 high $Q^2$ 99-00 $e^+p$ NC \cite{data23}& 131/147& 131/147& 128/147\\
ZEUS SVX 95 $e^+p$ NC \cite{data24}& 35/30& 35/30& 35/30\\
ZEUS 96-97 $e^+p$ NC \cite{data25}& 86/144& 86/144& 85/144\\
ZEUS 98-99 $e^-p$ NC \cite{data26}& 54/92 & 54/92 & 53/92\\
ZEUS 99-00 $e^+p$ NC \cite{data27}& 63/90& 63/90& 62/90\\
H1 99-00 $e^+p$ CC \cite{data28}& 29/28& 29/28& 29/28\\
ZEUS 99-00 $e^+p$ CC \cite{data29}& 38/30& 38/30& 38/30\\
H1/ZEUS $ep$ $F_2^{charm}$ \cite{data30a}-\cite{data30g}& 107/83& 106/83& 109/83\\
H1 99-00 $e^+p$ incl. jets \cite{data31}& 19/24& 17/24 & 18/24\\
ZEUS 96-97 $e^+p$ incl. jets \cite{data32}& 30/30& 29/30& 29/30\\
ZEUS 98-00 $e^\pm p$ incl. jets \cite{data33}& 17/30& 16/30& 16/30\\
\hline
D{\O} II $p\bar{p}$ incl. jets \cite{data34}& 114/110& 116/110& 115/110\\
CDF II $p\bar{p}$ incl. jets \cite{data35}& 56/76 & 60/76 & 58/76\\
CDF II $W\rightarrow l\nu$ asym.\cite{data36}&  29/22&  30/22&  29/22\\
D{\O} II $W\rightarrow l\nu$ asym. \cite{data37}&  25/10&  28/10&  26/10\\
D{\O} II Z rap. \cite{data38}& 19/28& 17/28& 19/28\\
CDF II Z rap. \cite{data39}& 49/29& 50/29& 50/29\\
\hline
ATLAS 7~TeV incl. jets (R=0.4) \cite{atlas-inc-paper}& (72/90)& 66/90& 70/90\\
CMS 7~TeV incl. jets \cite{cmsinc}& (180/133)& 163/133& 169/133\\
\hline
Total & 2795/2922 & 2781/2922 & 2786/2922 \\
\hline
\end{tabular}}
\end{center}
\caption{Table of $\chi^2$ values for each data set included in the fits for the standard MSTW 2008 NLO fit and the new NLO fits with ATLAS 7~TeV and CMS data. The ATLAS and CMS values are quoted for MSTW 2008 despite not being included in the fit. These are simply the $\chi^2$ values obtained when the fit code is run using the standard set without minimisation.}
\label{datasetnew}
\end{table}

In order to include the CMS data into an MSTW fit, the first necessary task was 
to modify the fit code to include FastNLO version 2 \cite{fastnlov2}. This new 
version allows more scale flexibility within the cross section calculation.  
The fit is performed allowing the same parameters to be free as in the standard 
MSTW2008 set. Initially, $\alpha_s(M_Z^2)$ was allowed to be free, and a reasonable 
improvement in the global fit from 2795 to 2781 over 2922 data points was 
obtained. This fit, however, included a decrease in $\alpha_s(M_Z^2)$ from 0.1202 to 
0.1189, which caused much of the improvement. Subsequently, in order to 
properly quantify the effect on just the PDFs, $\alpha_s(M_Z^2)$ was held fixed. 
This fit yielded a smaller improvement of only 8 points to 2786. 

\begin{figure}[h!]
\centering
\subfigure{\includegraphics[width=0.49\textwidth]{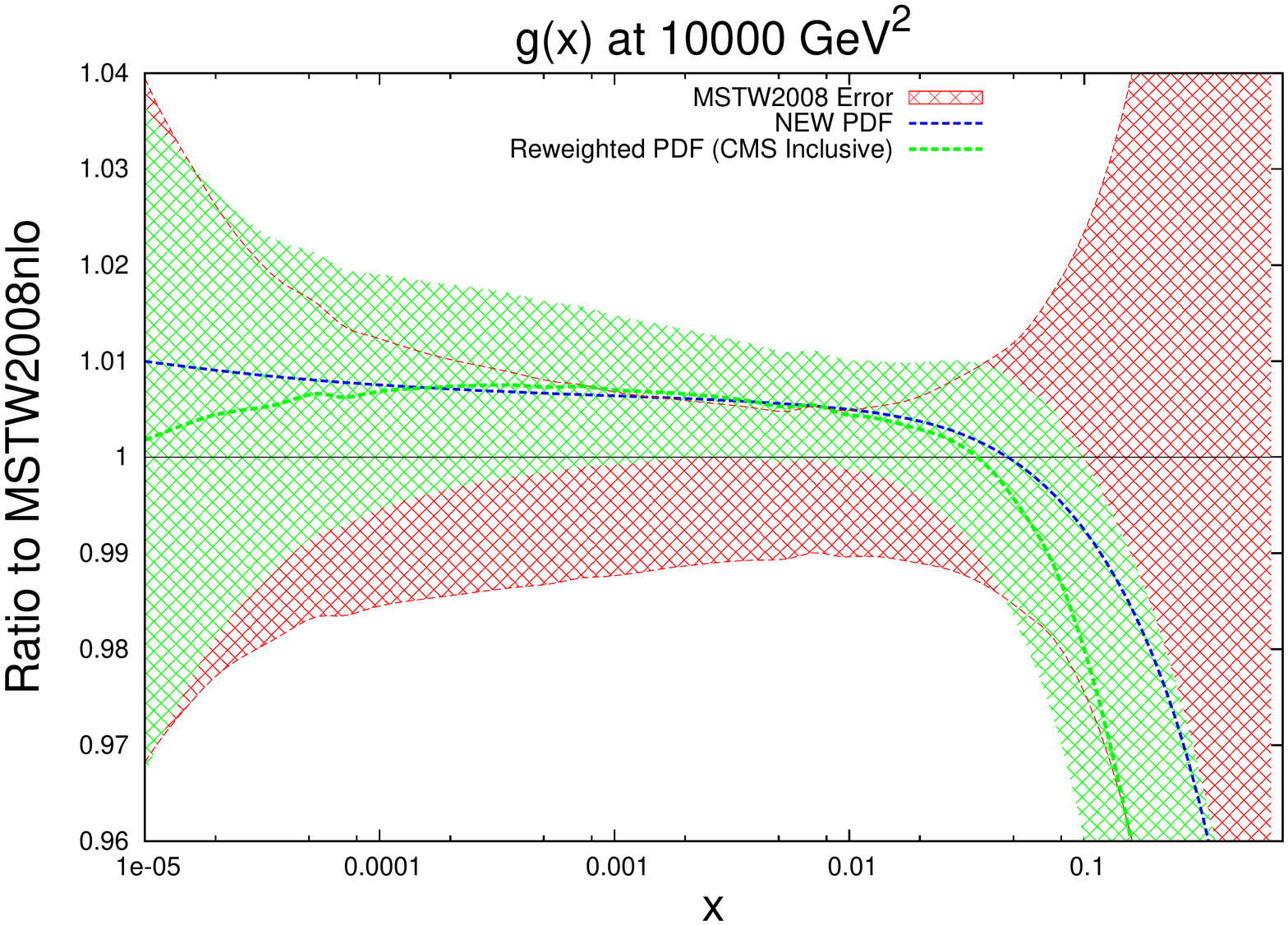}}
\subfigure{\includegraphics[width=0.49\textwidth]{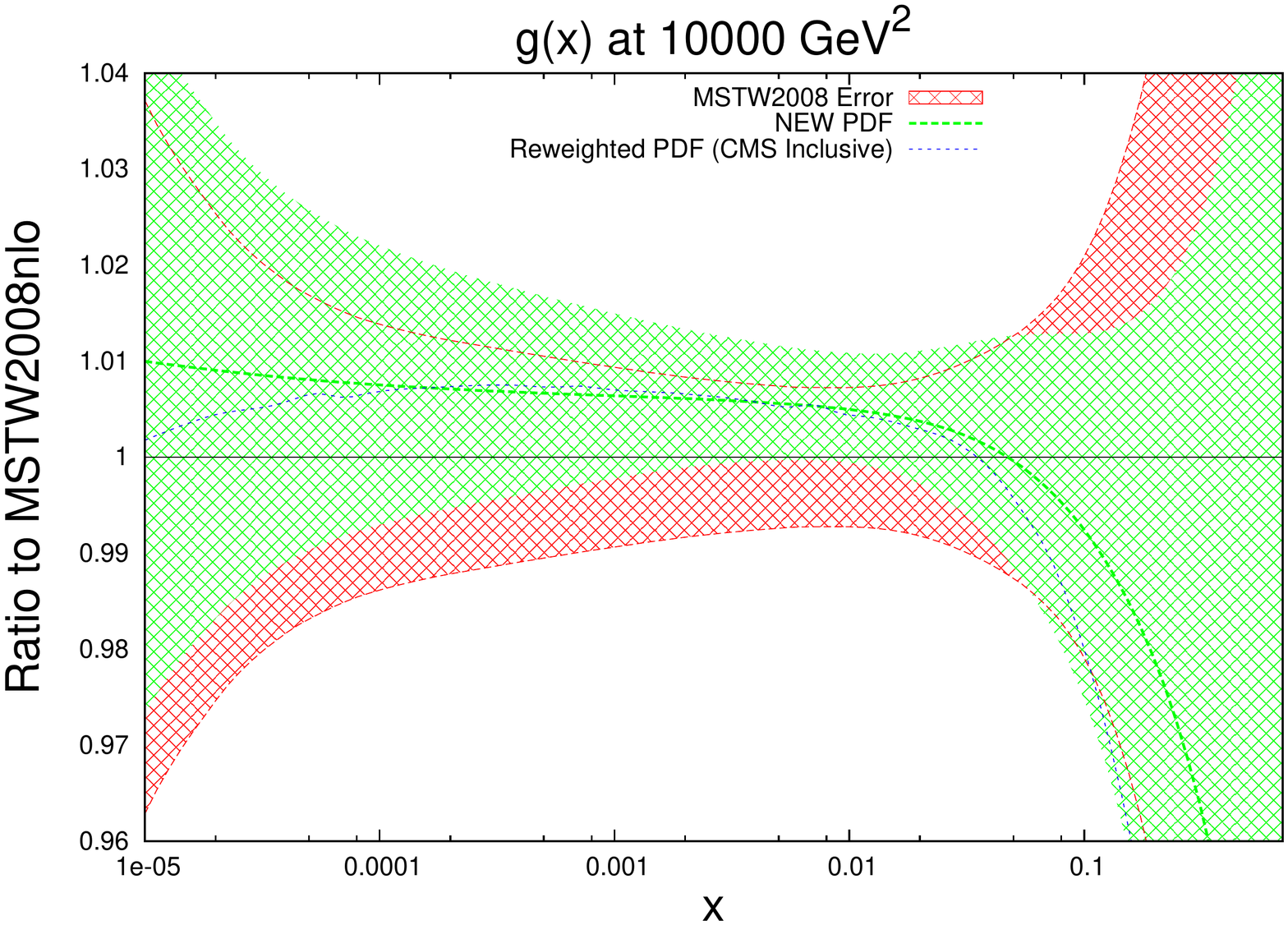}}
\vspace{-0.8cm}
\caption{Comparison of the gluon for standard MSTW fit, reweighted PDF (using CMS inclusive jets to reweight), and the new fit directly including the ATLAS \& CMS data. All 3 central values are shown on each plot; the first compares the error bands for MSTW against reweighting, and the second compares standard MSTW to the new fit.}
\label{newgluon}
\end{figure}

The effect on each data set included in the fit is shown in 
Table \ref{datasetnew}. The ATLAS and CMS $\chi^2$ values for MSTW2008 were 
first calculated using the fitting code by passing through the central value 
and bypassing the minimisation steps. Once they are included in the 
minimisation, a large improvement in the fit to CMS data is seen with a 
more modest improvement for the ATLAS data. The fact that both data sets prefer 
a smaller $\alpha_s(M_Z^2)$ is shown in the fact that the improvement is less 
pronounced when it is held fixed. In general, the fit to the various DIS data 
sets is left unchanged by both of the new fits. Interestingly, the Tevatron 
inclusive jet fits worsen very slightly with the inclusion of the LHC scenarios, 
although on the whole the Tevatron data remains also unchanged. The improvement 
of the global fit with $\alpha_s(M_Z^2)$ free can be understood through the stark 
improvement in the BCDMS proton $F_2$ measurement. This set returns to its 
original $\chi^2$ value once $\alpha_s(M_Z^2)$ is fixed. These PDFs will be named here 
MSTWCMS, due to the dominance of the CMS inclusive jet data on the improvement 
in fit quality.

\begin{figure}[h!]
\centering
\includegraphics[width=0.49\textwidth]{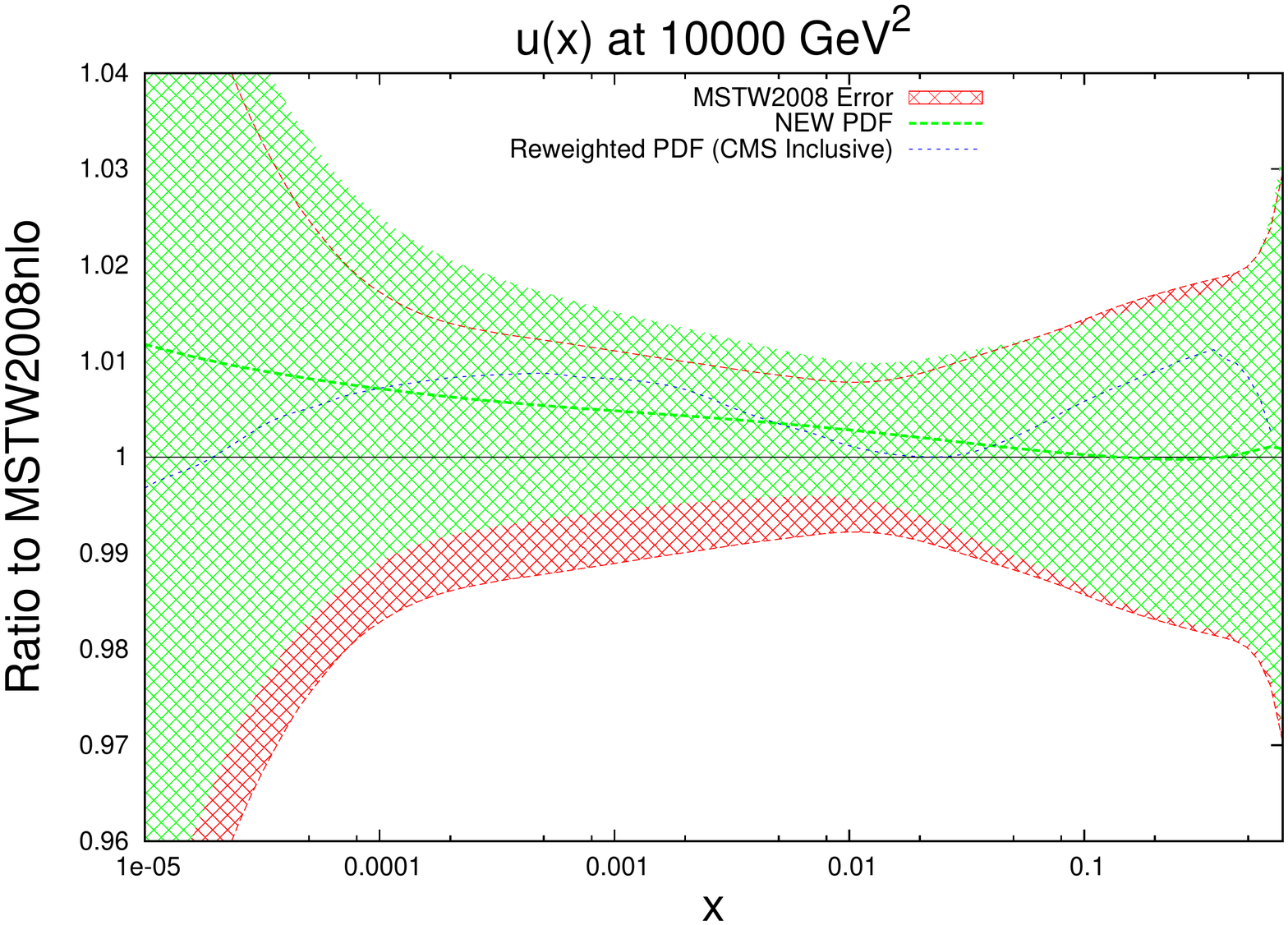}
\includegraphics[width=0.49\textwidth]{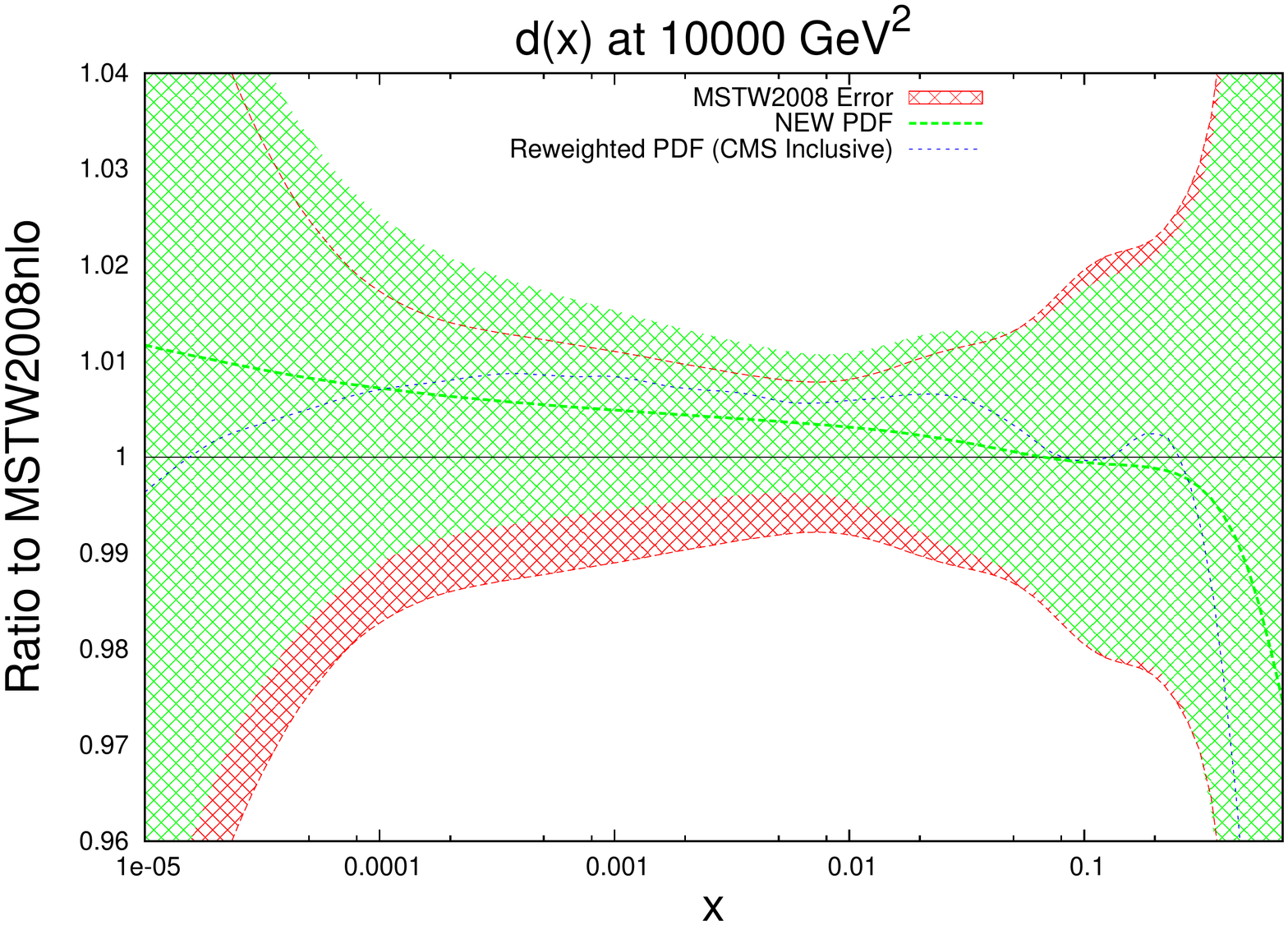}
\includegraphics[width=0.49\textwidth]{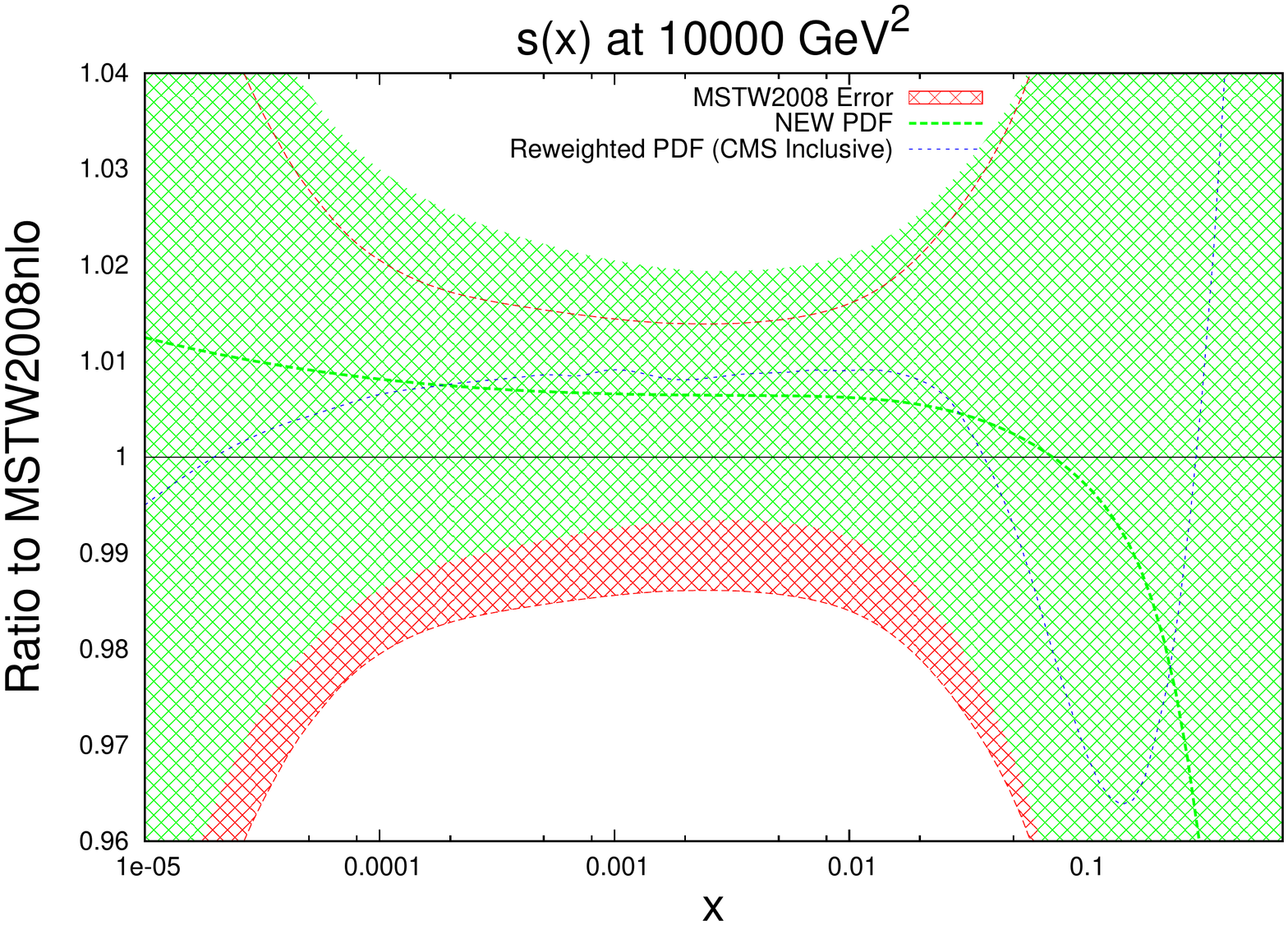}
\vspace{-0.8cm}
\caption{Ratio of the MSTWCMS quark distributions to MSTW2008. The central value of the reweighted PDF using CMS inclusive data is also shown for comparison.}
\label{newquarks}
\end{figure}

The new central PDF is shown in Fig. \ref{newgluon}, along with the reweighted 
PDF using the CMS inclusive data. The two error bands shown are the original 
MSTW2008 68\% confidence level, and the reweighted standard deviation of the 
randomly generated PDFs. It is clear that the new PDF requires a similar 
behaviour in the gluon as the reweighting technique. Whilst the two central 
lines to not exactly match, there is a trend for a $\sim$1\% increase in the 
gluon for much of the $x$ range, which turns into a rapidly decreasing gluon at 
around $x\sim0.1$. The error band of the reweighted PDF is in good agreement 
with that of the new fit for most values of $x$. The only region with 
disagreement is at high-$x$, where the reweighting technique appears to 
underestimate the error. Upon inspection of the top weighted PDFs used, all 
require a steeply falling gluon compared to MSTW2008, and so the standard 
deviation shows a strong grouping around this trend.

\begin{figure}[h!]
\includegraphics[width=0.9\textwidth]{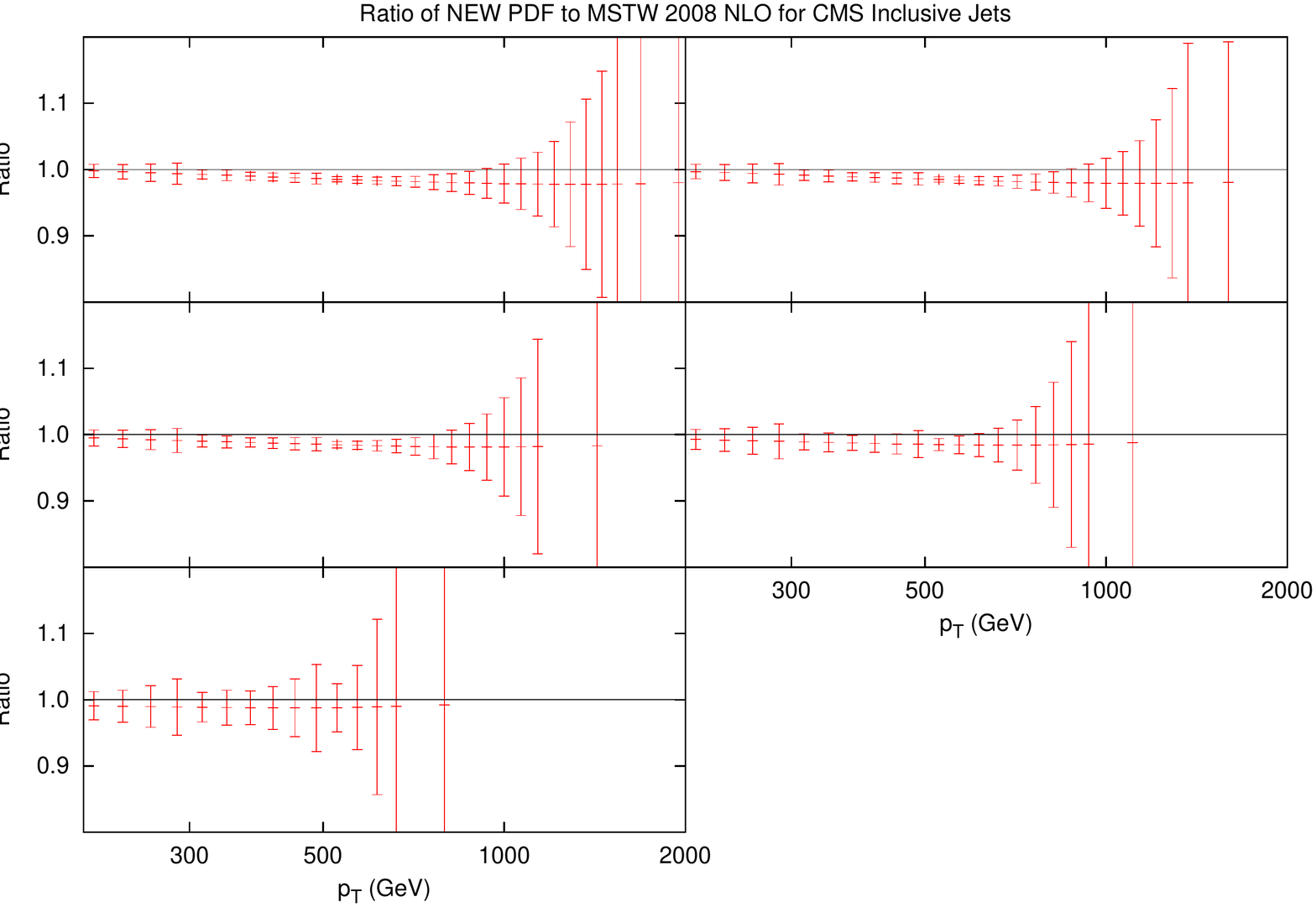}
\caption{Ratio of CMS inclusive jet cross section predictions for the new PDFs and the standard MSTW 2008 PDFs.}
\label{newcmsratio}
\end{figure}

The new quark PDFs are shown in Fig. \ref{newquarks}. These were shown to be 
important for the CMS inclusive jet data due to the $x$ values and resulting 
partons probed. The magnitude of change from MSTW2008 is comparable to the gluon,
lending further evidence for the importance of these data. Again, there is good 
agreement between the reweighting technique and the direct inclusion of data. 
The only significant disagreement is in the high-$x$ strange distribution where 
the uncertainties are very large.

The new prediction for CMS inclusive jets is shown in Fig. \ref{newcmsratio}. 
There is no change in shape between the new prediction and the MSTW2008 
prediction, and most points lie within the experimental error bars. However a 
systematic downward shift of $\sim$ 1\% is seen across most higher $p_T$ 
data points. For 
lower rapidity bins, where the experimental error is smallest, this shift 
brings some points out of agreement with the MSTW2008 
prediction, and this is where 
the largest change in $\chi^2$ originates. 

\subsection{Eigenvectors}

\begin{figure}
\centering
\includegraphics[width=0.8\textwidth]{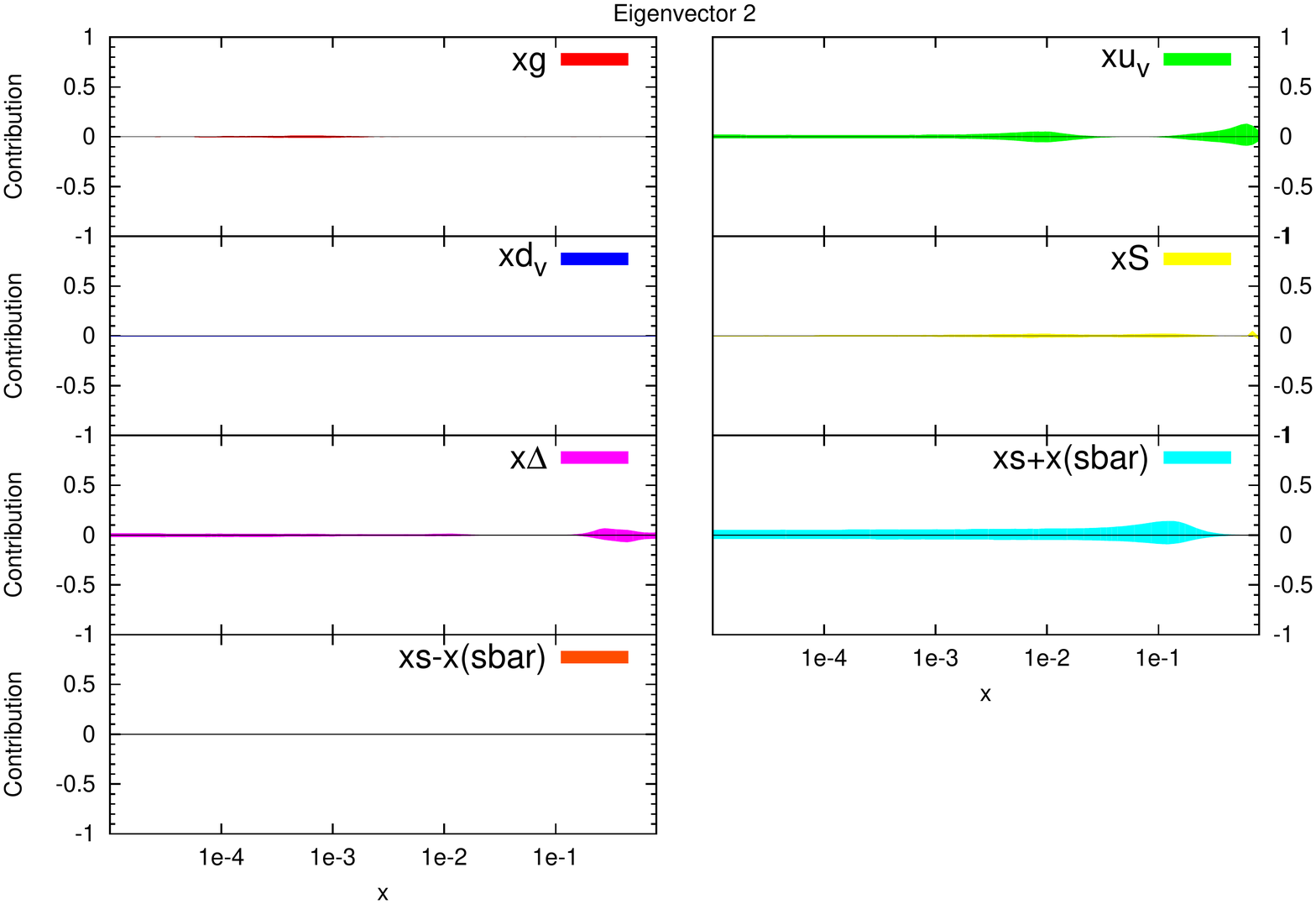}
\includegraphics[width=0.8\textwidth]{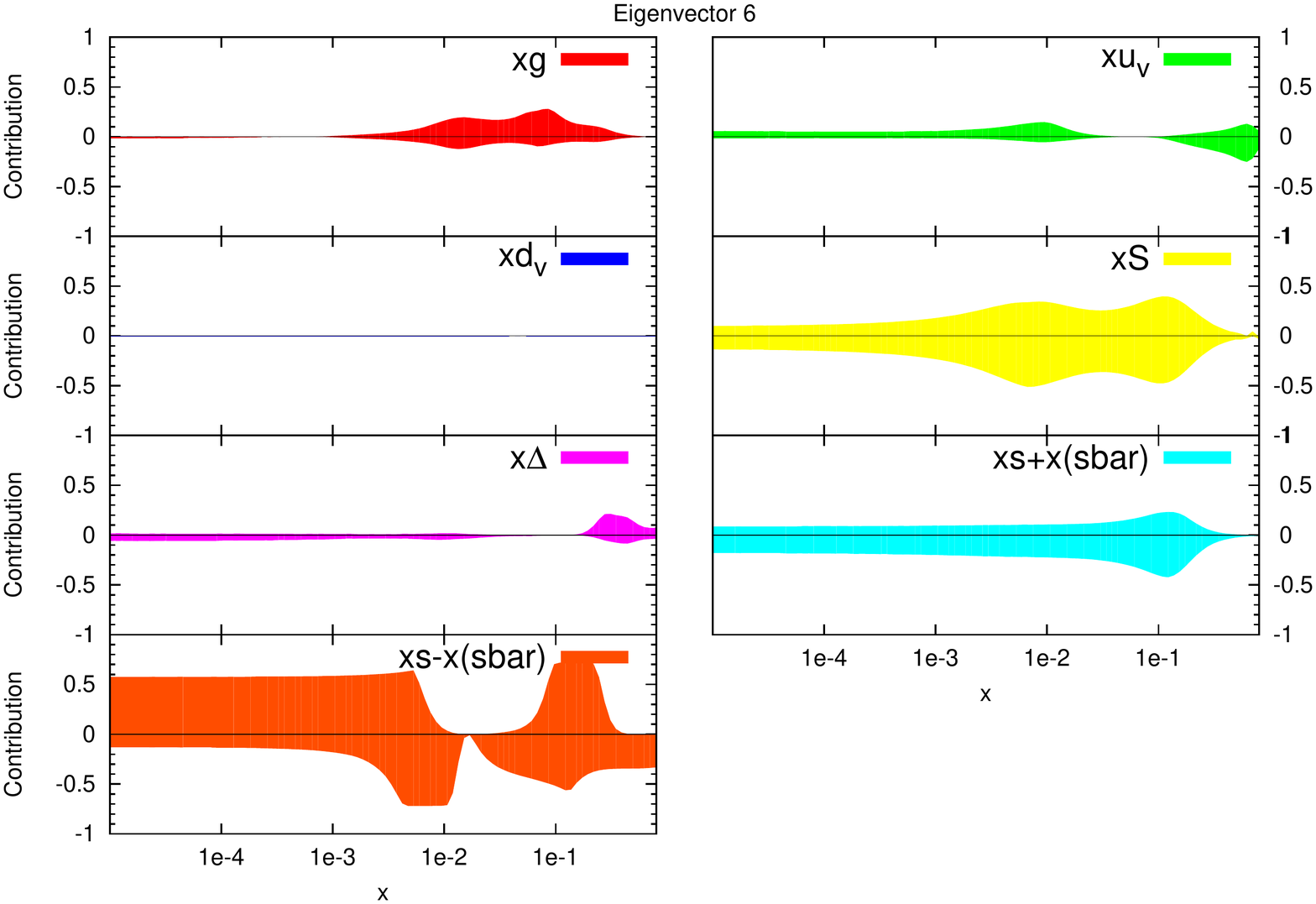}
\vspace{-0.8cm}
\caption{Fractional contribution to the uncertainty on major distributions from each eigenvector. Eigenvector 2 \& 6 shown.}
\label{eig1-2}
\end{figure}

\begin{figure}
\centering
\includegraphics[width=0.8\textwidth]{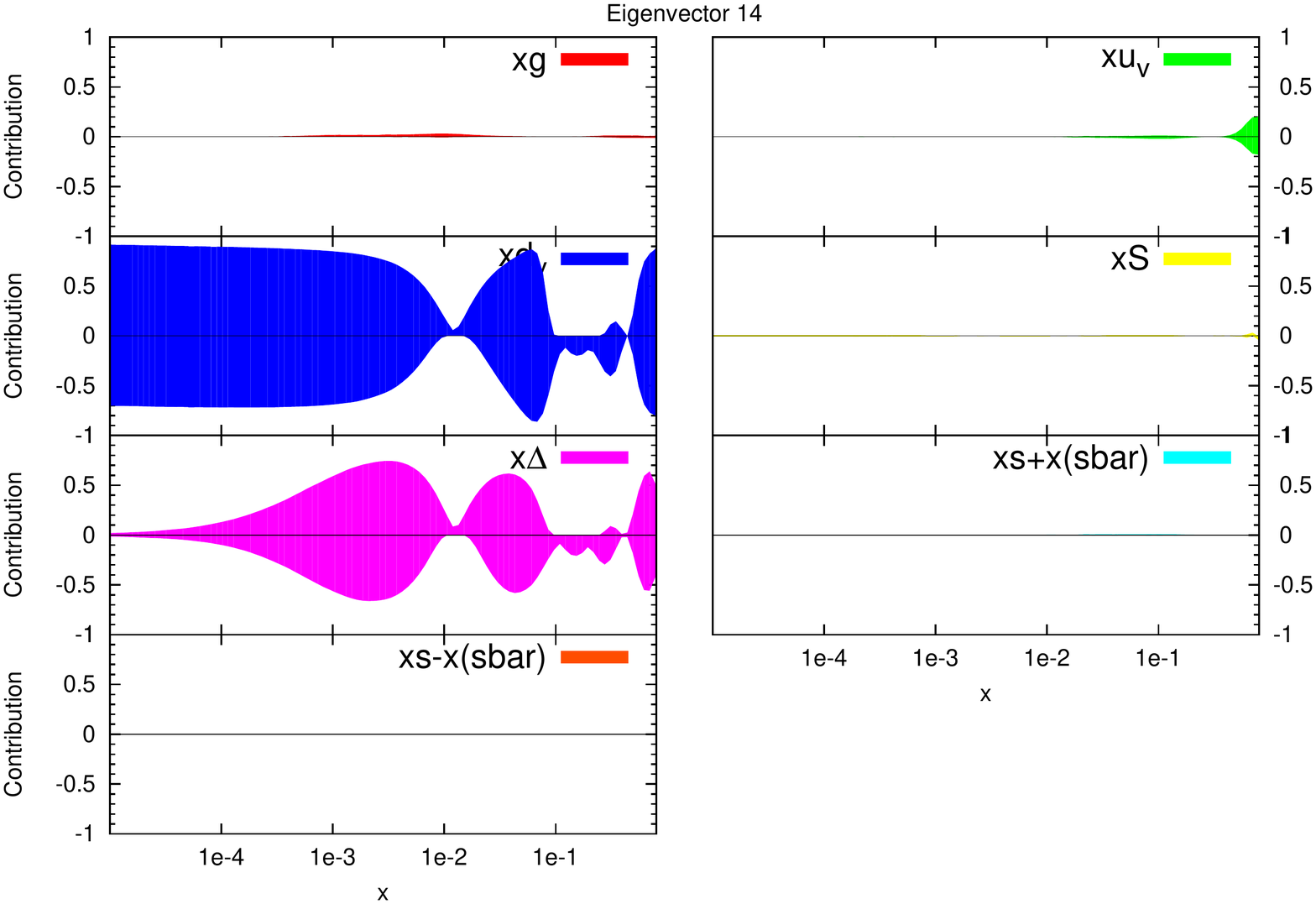}
\includegraphics[width=0.8\textwidth]{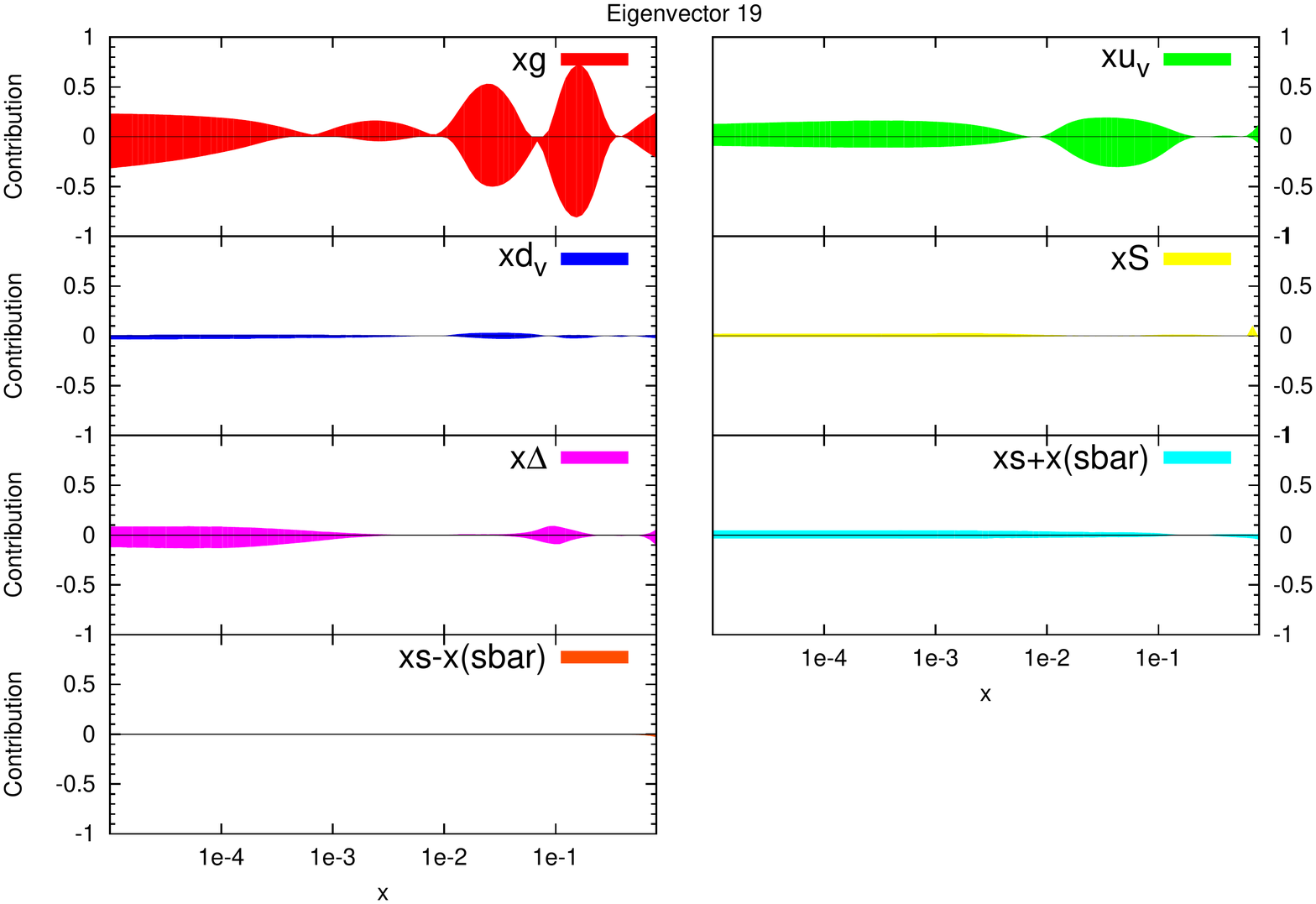}
\vspace{-0.8cm}
\caption{Fractional contribution to the uncertainty on major distributions from each eigenvector. Eigenvector 14 \& 19 shown.}
\label{eig3-4}
\end{figure}

The eigenvectors for the new fit are calculated in the same manner as the usual 
MSTW global fits. There are again 20 eigenvectors due to the same parameters 
being free, however the dependence of each eigenvector on the underlying 
parameters and data sets has changed. The fractional contribution to the 
total uncertainty on selected distributions from some eigenvectors is shown in 
Figs. \ref{eig1-2} and \ref{eig3-4}. These can be interpreted as the sensitivity 
to the underlying PDFs of each eigenvector, and can be compared to the 
equivalent plots for the MSTW2008 fit presented in \cite{MSTW}.

The CMS data itself directly constrains eigenvector 19 in this set. This can 
be seen to be almost entirely dependent on the gluon, although the up valence 
quark is also affected. Both distributions are most sensitive to this 
eigenvector in the higher $x$ region, which is consistent with the conclusions 
of the reweighting study, where the gluon and quark distributions were shifted 
the most at high $x$ after reweighting to the CMS data.

\begin{figure}[h!]
\includegraphics[width=0.49\textwidth]{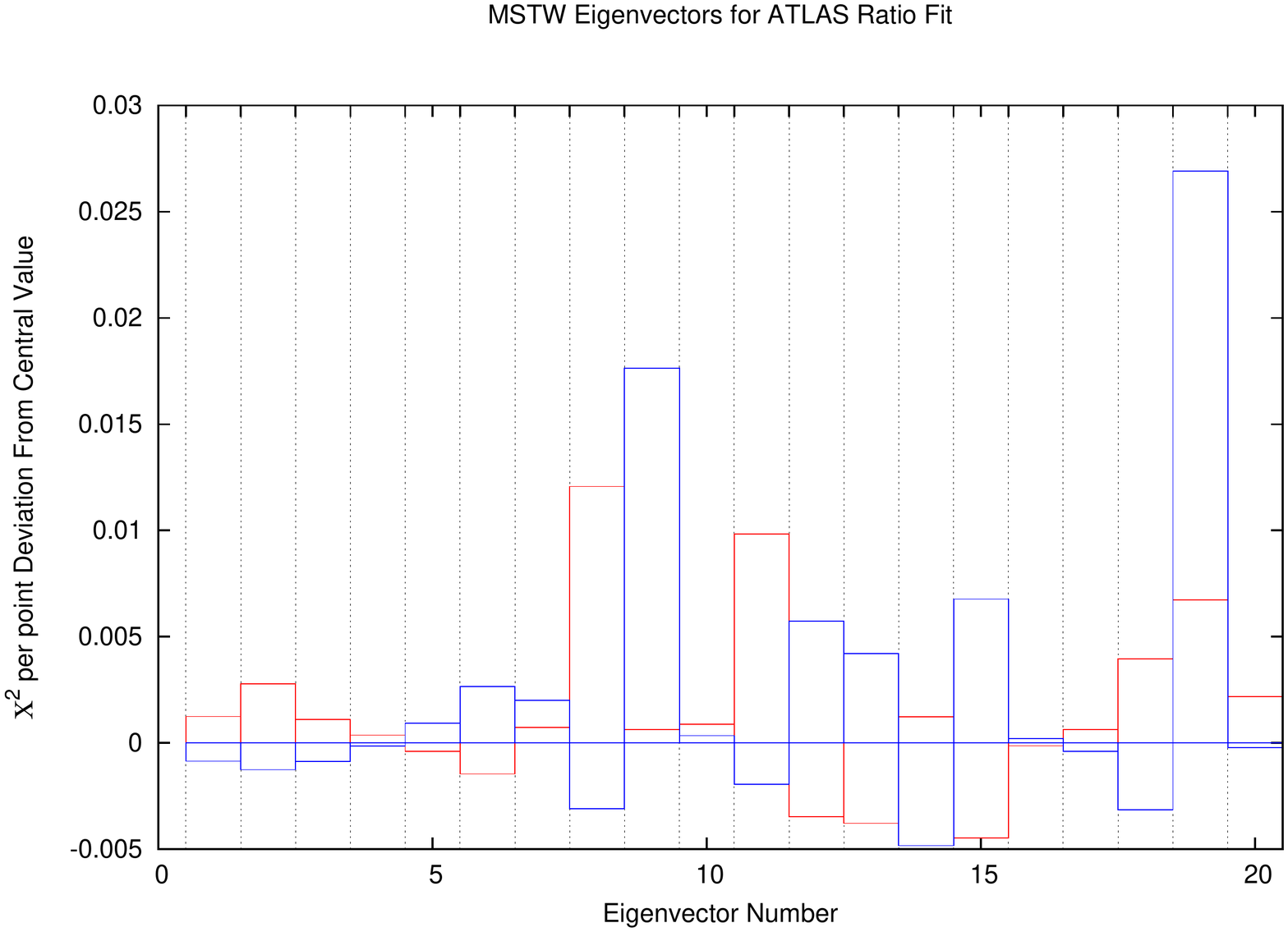}
\includegraphics[width=0.49\textwidth]{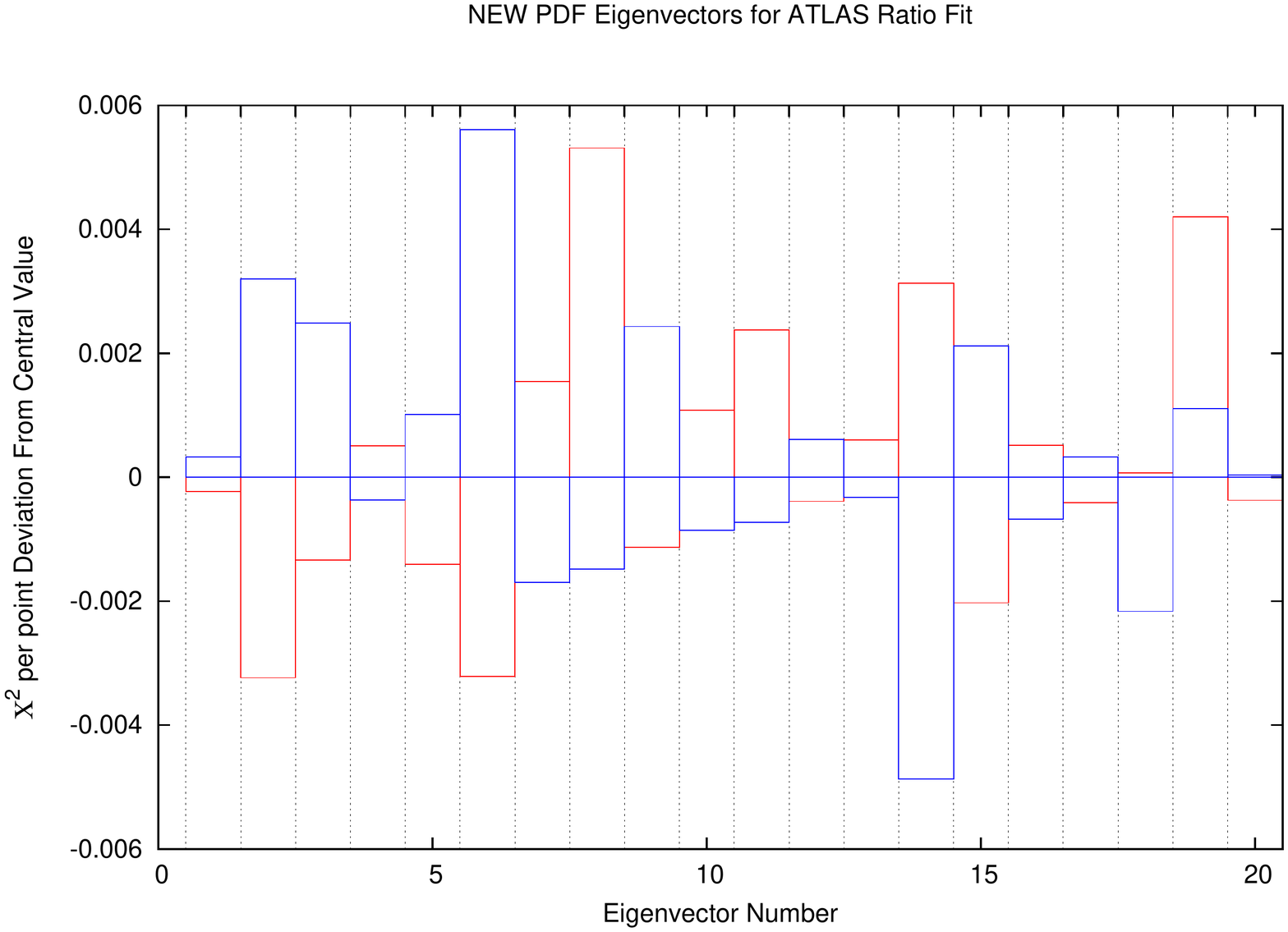}
\vspace{-0.8cm}
\caption{Change in fit quality to the ATLAS combined 2.76~TeV and 7~TeV cross sections from the MSTW2008 (left) and MSTWCMS (right) central values for each eigenvector in the respective fits.}
\label{neweigs}
\end{figure}

The change in fit quality to the ATLAS inclusive jet combined data for each of 
the new eigenvectors is shown in Fig. \ref{neweigs} alongside the corresponding 
plot for MSTW2008. There is more dependence on the eigenvectors of the MSTW2008 
set, and large increases in $\chi^2$ can be obtained for many eigenvectors. The 
new eigenvectors do not produce this dramatic reduction in fit quality, implying 
a better agreement with the data. Despite this, there are still many 
eigenvectors which can improve the fit to a reasonable degree. The largest are 
eigenvectors 2, 6 and 14, shown in the figures. There are no longer any
eigenvectors which give nearly such a large deterioration in fit quality, which
shows that the CMS data has already provided much of the constraint possible 
from the combined ATLAS data in a completely compatible manner.
Hence, the new PDF can then be said to provide a better fit to ATLAS combined 
data, with some scope still for further improvement.

\subsection{Reweighting of the New PDFs}

\begin{figure}[h!]
\includegraphics[width=0.9\textwidth]{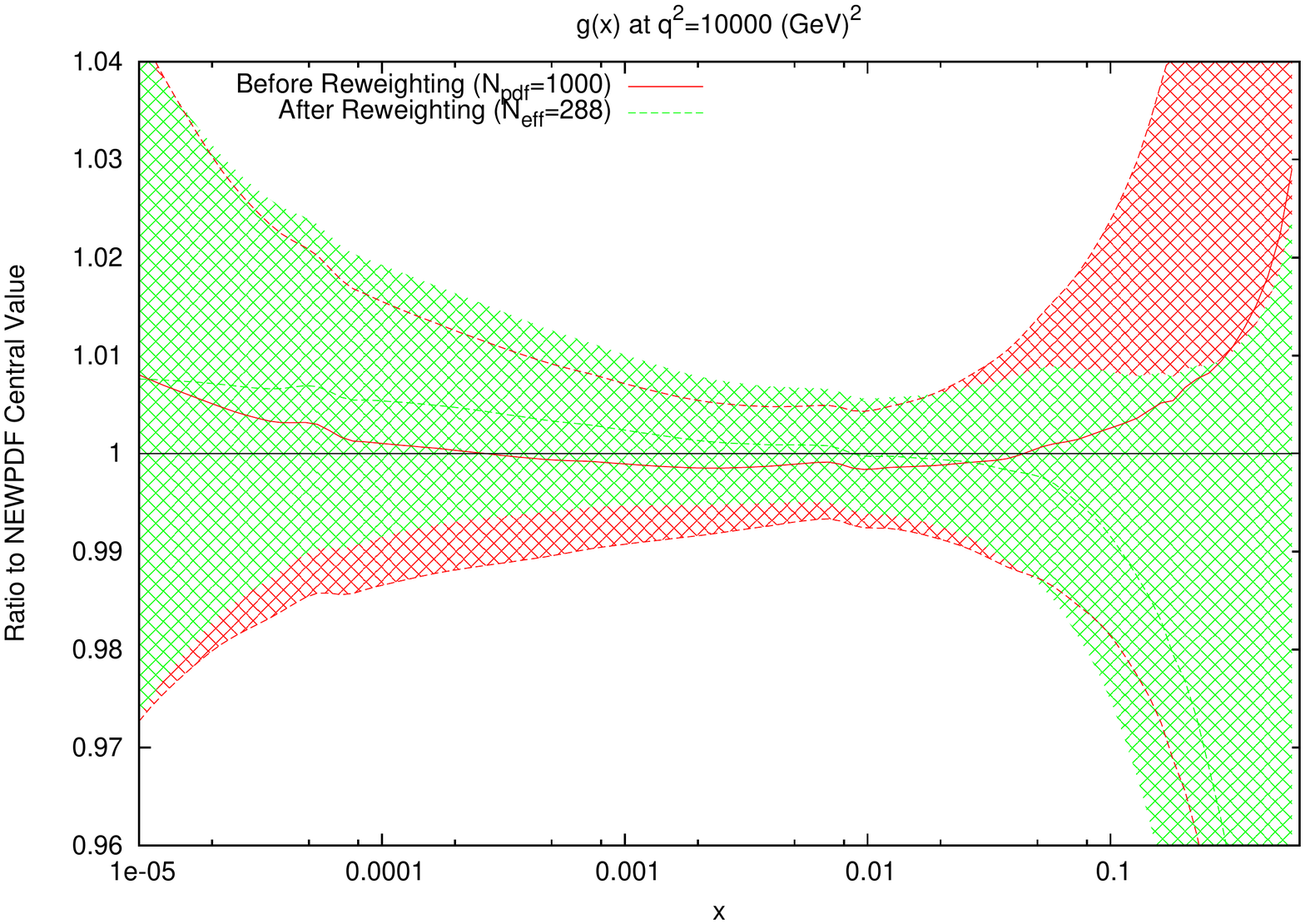}
\vspace{-0.8cm}
\caption{Reweighting of the new gluon PDF using ATLAS combined data.}
\label{ratiogluon}
\end{figure}

An important study which can now be performed is to reweight the new PDFs, which 
will again check the compatibility of the method with the standard fitting 
procedure. By using the new central value and eigenvectors, the $\chi^2$ for 
ATLAS combined jets is calculated for 1000 PDFs randomly generated in the 
eigenvector space. The distribution can then be compared to that of the PDFs 
randomly distributed in the standard MSTW eigenvector space.
The observed effect is shown in Fig. \ref{ratiogluon}. The cuts previously 
discussed are used along with the additive treatment of systematic errors. 
There is still a shift required of the gluon under reweighting. 
This can be interpreted as further evidence that the 7~TeV ATLAS inclusive data 
has little effect on the PDFs, and the combined data including the 2.76~TeV 
set must be used.

\begin{figure}[h!]
\includegraphics[width=0.9\textwidth]{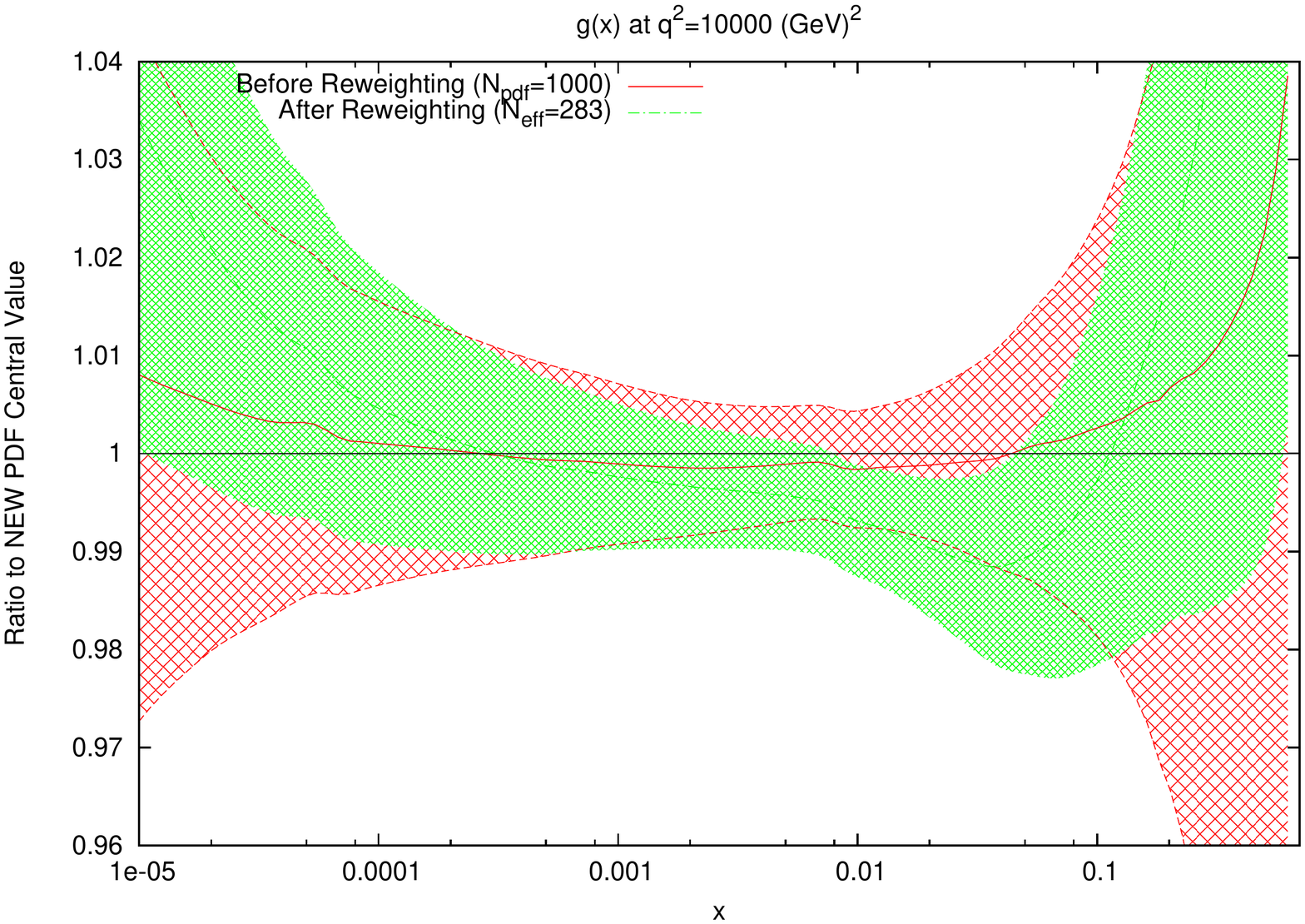}
\vspace{-0.8cm}
\caption{Reweighting of the new gluon PDF using CMS dijet data ($p_T^{av}$ scale choice)}
\label{ratiodijet}
\end{figure}

Finally the dijet cross sections are studied using the new PDFs. After the 
previous studies which showed that in general the dijet data sets require a 
different shift in the PDFs to the equivalent inclusive jet data, this is the 
ideal test of compatibility between the data types.
Fig. \ref{ratiodijet} demonstrates the effect of the CMS dijet data on the new 
PDFs. There is in fact very little difference between the shape of the 
reweighted gluon with respect to the new PDF as that with respect to the 
MSTW2008 set. In fact, the slight reduction in the error band for 
the new PDF causes the reweighted central value to be marginally outside of 
the error band for a small $x$ range. The trend is still opposing the inclusive 
jet data, with a smaller gluon required at moderate $x$, and a larger gluon at 
low $x$. The reweighted PDF has a $\chi^2$ of 1.77 per point, compared to the 
unweighted central value which is 2.02 per point. Both of these values are 
larger than the 1.67 per point which is the value after reweighting to the 
MSTW2008 PDFs, which implies that the new PDFs are in fact slightly worse at 
describing the CMS dijet data, despite the corresponding inclusive jet data 
being newly included in these sets.

\subsection{$\Delta\chi^2=1$ Treatment}

\begin{figure}[h!]
\subfigure{\includegraphics[width=0.49\textwidth]{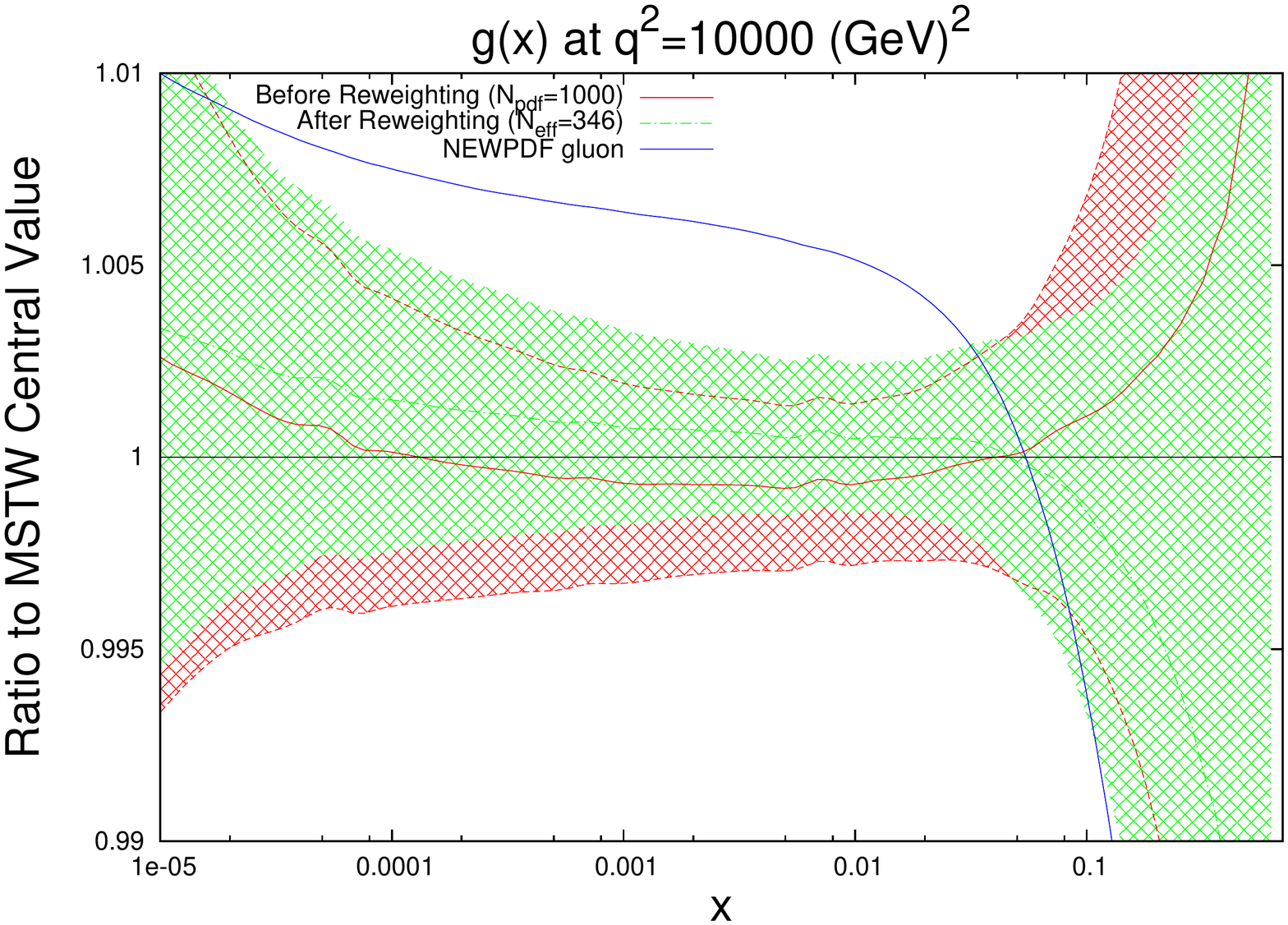}}
\subfigure{\includegraphics[width=0.49\textwidth]{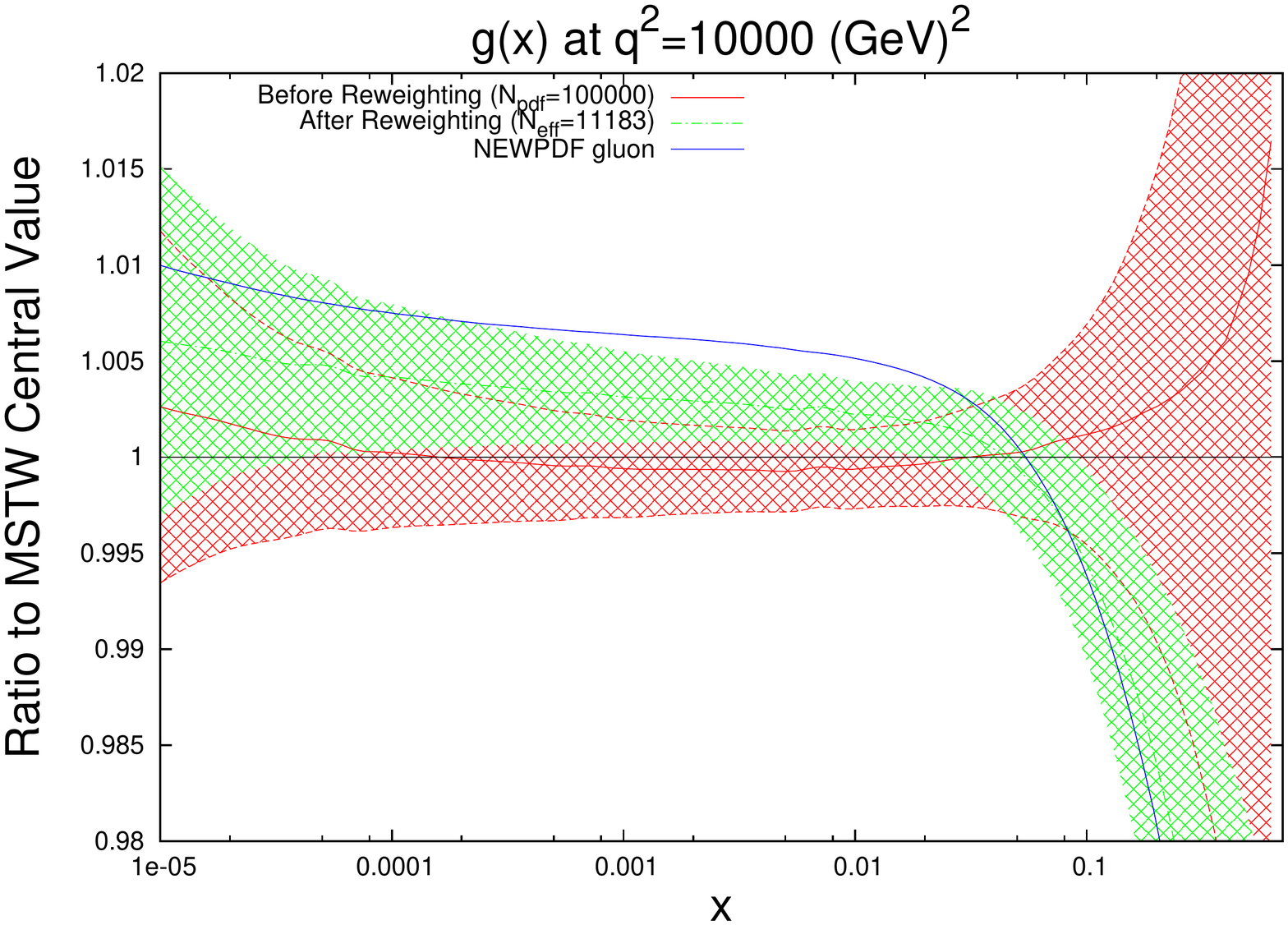}}
\vspace{-0.8cm}
\caption{Reweighted gluon using CMS inclusive data, the $\Delta\chi^2=1$ 
error treatment and 1000 PDFs (left) or 100,000 PDFs (right). 
The reweighting formula is a pure exponential.}
\label{deltachi}
\end{figure}

Until now, the reweighting procedure used in this article and in the previous 
MSTW study have used the standard MSTW eigenvectors, which are defined using 
dynamical tolerance levels. In this procedure, some eigenvectors are allowed 
to move further from the global minimum in $\chi^2$ than others, depending on 
the deterioration in the fit quality to individual data sets 
in the relevant direction. This practice is similar to that used in the 
CTEQ/CT PDF determination, and though the NNPDF determination of PDFs uses a 
very different approach to determine the uncertainties, where a particular 
$\Delta \chi^2$ is difficult to identify, the PDF uncertainties from MSTW and 
NNPDF (and CT10), are very similar, see e.g \cite{Watt:2011kp,Ball:2012wy}. 
However, when reweighting using the eigenvectors, it may be interesting to  
to instead consider the use of a set tolerance of $\Delta\chi^2=1$ in each 
direction, i.e. the conventional ``textbook'' choice.

The reweighted gluon using this technique is shown with the gluon of the new 
fit PDF in the left of Fig. \ref{deltachi}. 
Here we also test the hypothesis that when 
using the ``textbook'' method for uncertainty determination the appropriate 
reweighting function is a pure exponential. In fact the results do not seem 
strongly dependent on the reweighting function used. 
Whilst the reweighting had previously agreed 
well with the required shift for the new PDF, there is clear disagreement 
here. The new PDF is well outside the $1 \sigma$ error band. This can be 
explained simply by an inability for the the random PDFs to be generated in 
the required range. Given that on average, the dynamic tolerance level for 
the eigenvectors in the MSTW2008 fit are approximately 3 to 4, by rescaling 
to a value of 1, we can assume that all error bands and fluctuations 
will be reduced by a 
factor of 3 or 4.

\begin{figure}[h!]
\subfigure{\includegraphics[width=0.49\textwidth]{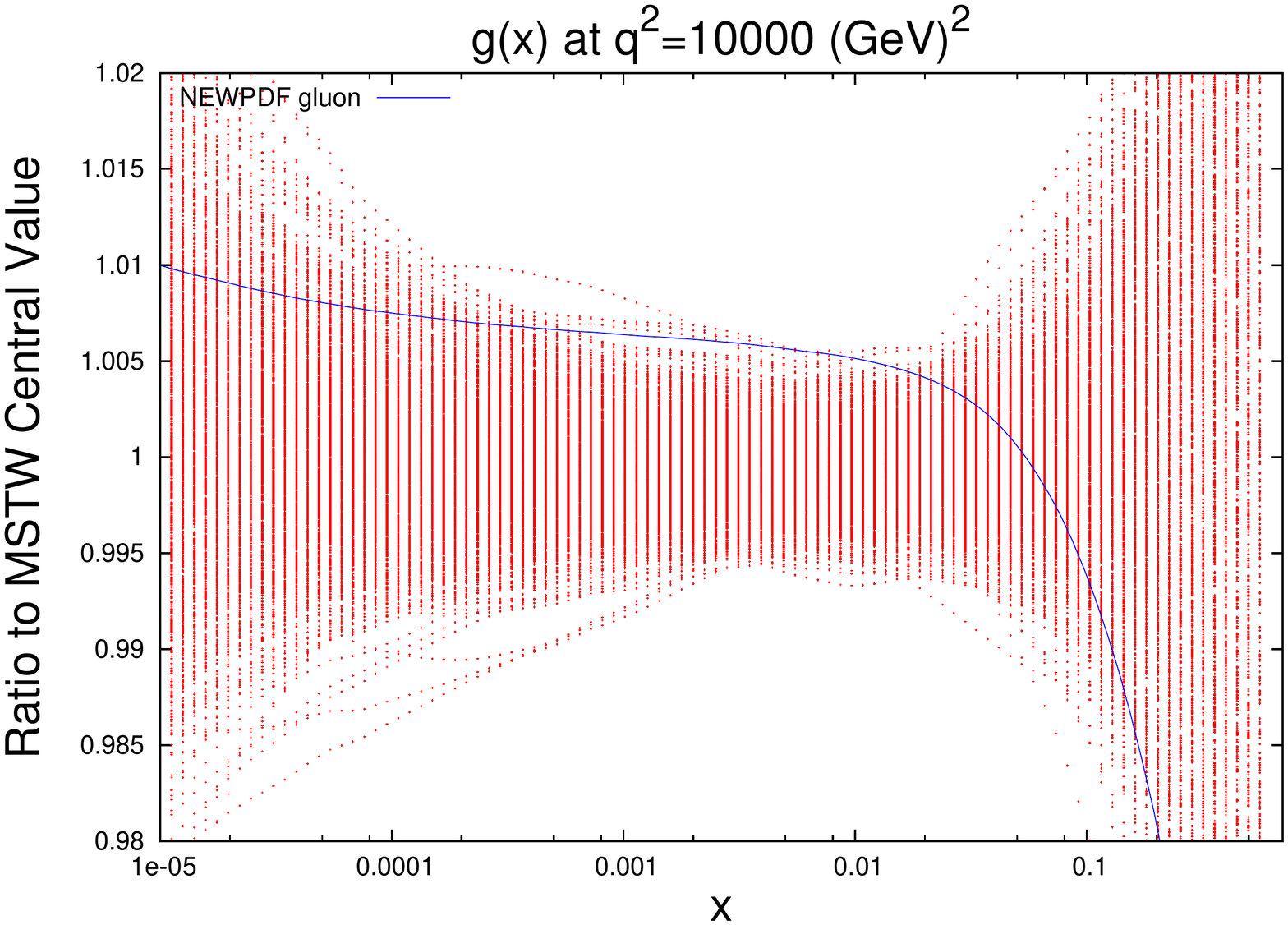}}
\subfigure{\includegraphics[width=0.49\textwidth]{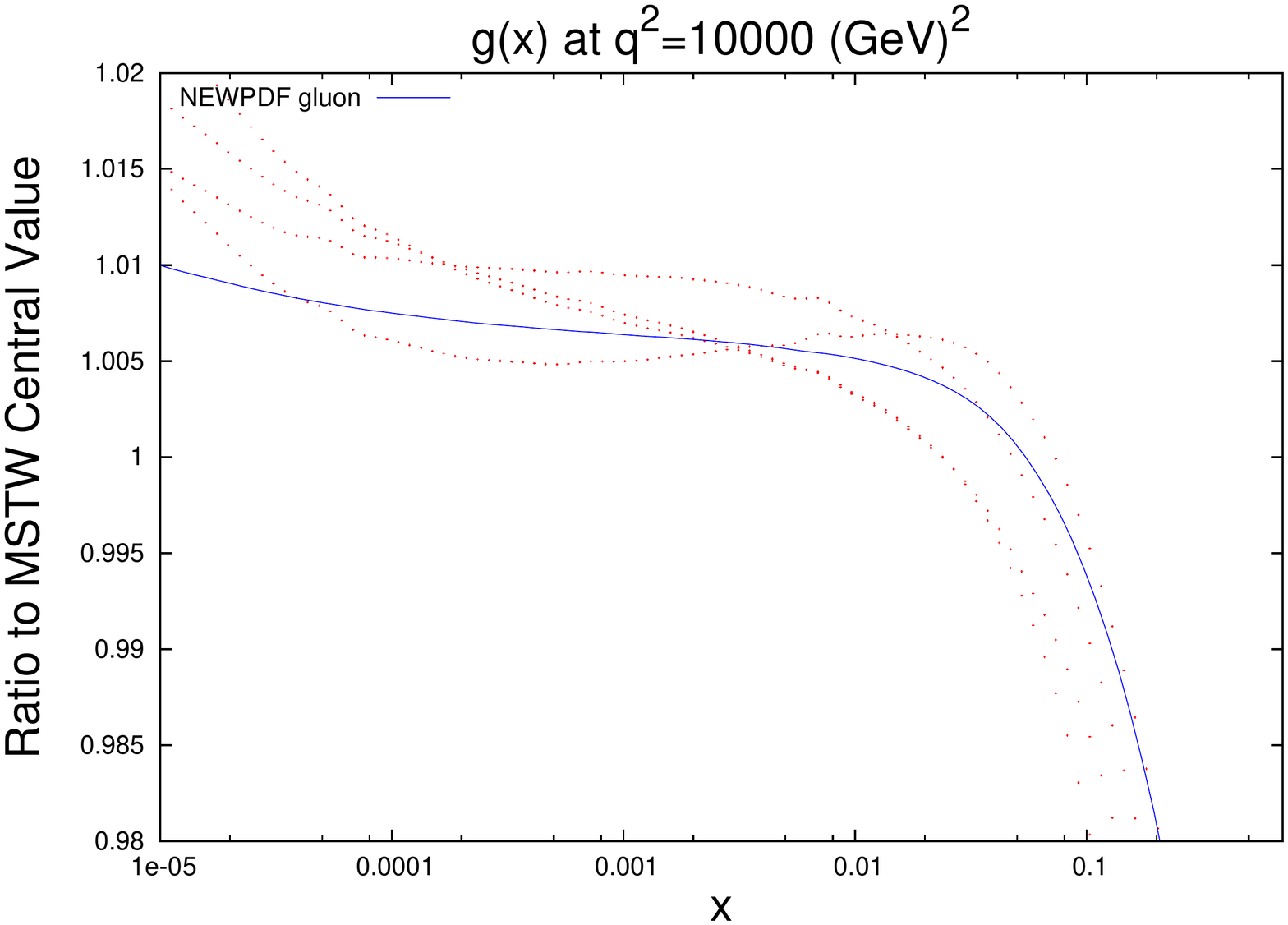}}
\vspace{-0.8cm}
\caption{Plot of the 1000 randomly distributed PDFs under the $\Delta\chi^2=1$ prescription (left) and the highest weighted of 100,000 randomly distributed 
PDFs under the $\Delta\chi^2=1$ prescription (right).}
\label{topweights}
\end{figure}

The 1000 randomly distributed PDFs using the $\Delta\chi^2=1$ method are 
shown in the left of Fig. \ref{topweights}, and demonstrate this 
inability to replicate 
the new PDF gluon. Whilst a very small handful extend to the required upward 
shift, these are drowned out by the vast majority which, whilst weighted 
lightly, contribute the most to the reweighted PDF. Indeed, very few of 
these 1000 PDF sets give a gluon distribution which is similar to that 
required by the full global fit including the CMS jet data (let alone also a 
set of quark distributions of exactly the correct shape). This is not that 
surprising since the deterioration of the other data in the fit is a 
few units, and the new gluon is 2-3$\sigma$ from the MSTW2008 gluon if 
$\Delta\chi^2=1$ is used as the uncertainty criterion. Hence, 1000 random 
PDFs sets is very likely not enough to produce a significant number near 
the best fit, and hence to provide a correct reweighting procedure.

%\begin{figure}[h!]
%\includegraphics[width=0.9\textwidth]{top_weights.pdf}
%\vspace{-0.8cm}
%\caption{{Plot of the highest weighted of 100,000 randomly distributed PDFs 
%under the $\Delta\chi^2=1$ prescription}}
%\label{topweightsa}
%\end{figure} 

%\begin{figure}[h!]
%\includegraphics[width=0.9\textwidth]{gluon.pdf}
%\vspace{-0.8cm}
%\caption{Reweighted gluon using CMS inclusive data and the $\Delta\chi^2=1$ error treatment, and 100,000 PDFs. The reweighting formula is a pure exponential.}
%\label{deltachia}
%\end{figure} 

Hence, we repeat the exercise using  100,000 PDF sets.In the right of Fig. 
\ref{topweights} we show the highest weighted of these PDF sets. Even with 
this number of random PDFs only a small number have a gluon of very nearly 
the ideal shape. In the right of 
Fig. \ref{deltachi} we show the reweighted gluon using 
100,000 PDF sets. Clearly this is different from that with 1000 sets and 
is much nearer to that obtained with the full fit. It does not appear as 
though even this number of sets has led to convergence, but it is 
impractical to generate an even larger number of PDFs. However, we can 
make two conclusions from this study. If the true PDF modification is well 
outside what is defined to be the uncertainty band of the PDF the reweighting 
procedure becomes very inefficient. We also conclude that when using the 
conventional MSTW uncertainty prescription the CMS inclusive jet data is 
compatible with the MSTW2008 PDFs at about the one $\sigma$ level, and hence 
has a significant, but not dramatic effect on new PDFs, 
whereas using the ``textbook''
uncertainty determination the CMS data is quite distinctly incompatible 
with the MSTW2008 set, and by inference with some of the data used in 
the PDF determination. Similar size changes in PDFs and in $\chi^2$ have 
frequently been observed when adding new data sets to the PDF fit, but the 
reweighting procedure allows us to illustrate the results using this 
particular new set in a new manner.

\subsection{Direct Inclusion of ATLAS 2.76~TeV + 7~TeV Data}

\begin{table}
\begin{center}
\resizebox{15cm}{!} {
\begin{tabular}{| c | c |c| c |}
\hline
Data Set & MSTW2008 & MSTWATLAScomb $\alpha_s$ Free & MSTWATLAScomb $\alpha_s$ Fixed\\
\hline
BCDMS $\mu p$ $F_2$& 182/163& 170/163 &182/163  \\
BCDMS  $\mu d$ $F_2$& 190/151& 189/151& 190/151\\
NMC  $\mu p$ $F_2$&121/123 & 123/123& 119/123 \\
NMC  $\mu d$ $F_2$&102/123 &  103/123&101/123 \\
NMC  $\mu p/\mu d$&130/148 &131/148& 129/148  \\
E665 $\mu p$ $F_2$&57/53 &53/53&  54/53 \\
E665  $\mu d$ $F_2$& 53/53&57/53&  57/53\\
SLAC  $\mu p$ $F_2$&30/37&30/37 &  30/37\\
SLAC  $\mu d$ $F_2$& 30/38&33/38&  29/38 \\
NMC/BCDMS/SLAC $F_L$&38/41&40/31 &38/31 \\
\hline
E866/NuSea pp DY&228/184 &227/184& 228/184 \\
E866/NuSea pd/pp DY& 14/15& 13/15& 14/15\\
\hline
NuTeV $\nu N$ $F_2$& 49/53& 50/53& 50/53\\
CHORUS $\nu N$ $F_2$ & 26/42&26/42& 26/42 \\
NuTev $\nu N$ $xF_3$& 40/45& 45/45& 40/45\\
CHORUS $\nu N$ $xF_3$ &31/33 &32/33& 31/33\\
CCFFR $\nu N\rightarrow\mu\mu X$ & 66/86&67/86& 65/86 \\
NuTeV $\nu N\rightarrow\mu\mu X$  & 39/40&49/40& 40/40\\
\hline
H1 MB 99 $e^+p$ NC& 9/8&9/8& 9/8\\
H1 MB 97 $e^+p$ NC&42/64 &42/64& 44/64\\
H1 low $Q^2$ 96-97 $e^+p$ NC&44/80 &44/80& 45/80\\
H1 high $Q^2$ 98-99 $e^-p$ NC& 122/126& 122/126&119/126\\
H1 high $Q^2$ 99-00 $e^+p$ NC& 131/147&132/147& 127/147\\
ZEUS SVX 95 $e^+p$ NC& 35/30& 35/30&35/30\\
ZEUS 96-97 $e^+p$ NC& 86/144&86/144& 85/144\\
ZEUS 98-99 $e^-p$ NC& 54/92 &54/92& 54/92 \\
ZEUS 99-00 $e^+p$ NC& 63/90& 63/90&62/90\\
H1 99-00 $e^+p$ CC& 29/28&29/38& 29/28\\
ZEUS 99-00 $e^+p$ CC& 38/30& 38/30&38/30\\
H1/ZEUS $ep$ $F_2^{charm}$& 107/83&105/83& 109/83\\
H1 99-00 $e^+p$ incl. jets& 19/24& 16/24&19/24\\
ZEUS 96-97 $e^+p$ incl. jets& 30/30&29/30& 29/30\\
ZEUS 98-00 $e^\pm p$ incl. jets& 17/30& 16/30&17/30\\
\hline
D{\O} II $p\bar{p}$ incl. jets & 114/110&116/110& 116/110\\
CDF II $p\bar{p}$ incl. jets & 56/76 &63/76& 58/76 \\
CDF II $W\rightarrow l\nu$ asym.&  29/22&29/22&  29/22\\
D{\O} II $W\rightarrow l\nu$ asym.&  25/10& 28/10& 25/10\\
D{\O} II Z rap. & 19/28& 18/28&19/28\\
CDF II Z rap.& 49/29& 49/29&50/29\\
\hline
ATLAS 2.76~TeV and 7~TeV incl. jets (R=0.4)& (159/114)&144/114 &155/114\\
CMS 7~TeV incl. jets& (180/133)& 161/133&166/133\\
\hline
Total & 2882/2946& 2862/2946 & 2869/2946 \\
\hline
\end{tabular}
}
\end{center}
\caption{Table of $\chi^2$ values for each data set included in the fits for the standard MSTW 2008 NLO fit and the new NLO fits with CMS and ATLAS combined 2.76~TeV and 7~TeV data. The ATLAS and CMS values are quoted for MSTW2008 despite not being included in the fit. These are simply the $\chi^2$ values obtained when the fit code is run using the standard set without minimisation.}
\label{atratfit}
\end{table}

The final new fit performed in this study is to include the ATLAS 2.76~TeV data 
in conjunction with the already present 7~TeV data. Whilst FastNLO tables for 
the 7~TeV data are available, this is not the case for the 2.76~TeV data. This 
presents the opportunity to interface APPLgrid, which did not exist at the time 
of the MSTW2008 fit, into the MSTW fitting code. 
Due to the fact that the MSTW code uses by default additive errors, the 
stringent cuts on the ATLAS ratio data discussed previously were applied to the 
data set in the fit. APPLgrid grids were used for both of the ATLAS cross 
sections, and FastNLO was kept for all of the other jet cross sections, 
including the CMS inclusive jet data introduced in the previous section.

\begin{figure}
\centering
\subfigure{\includegraphics[width=0.49\textwidth]{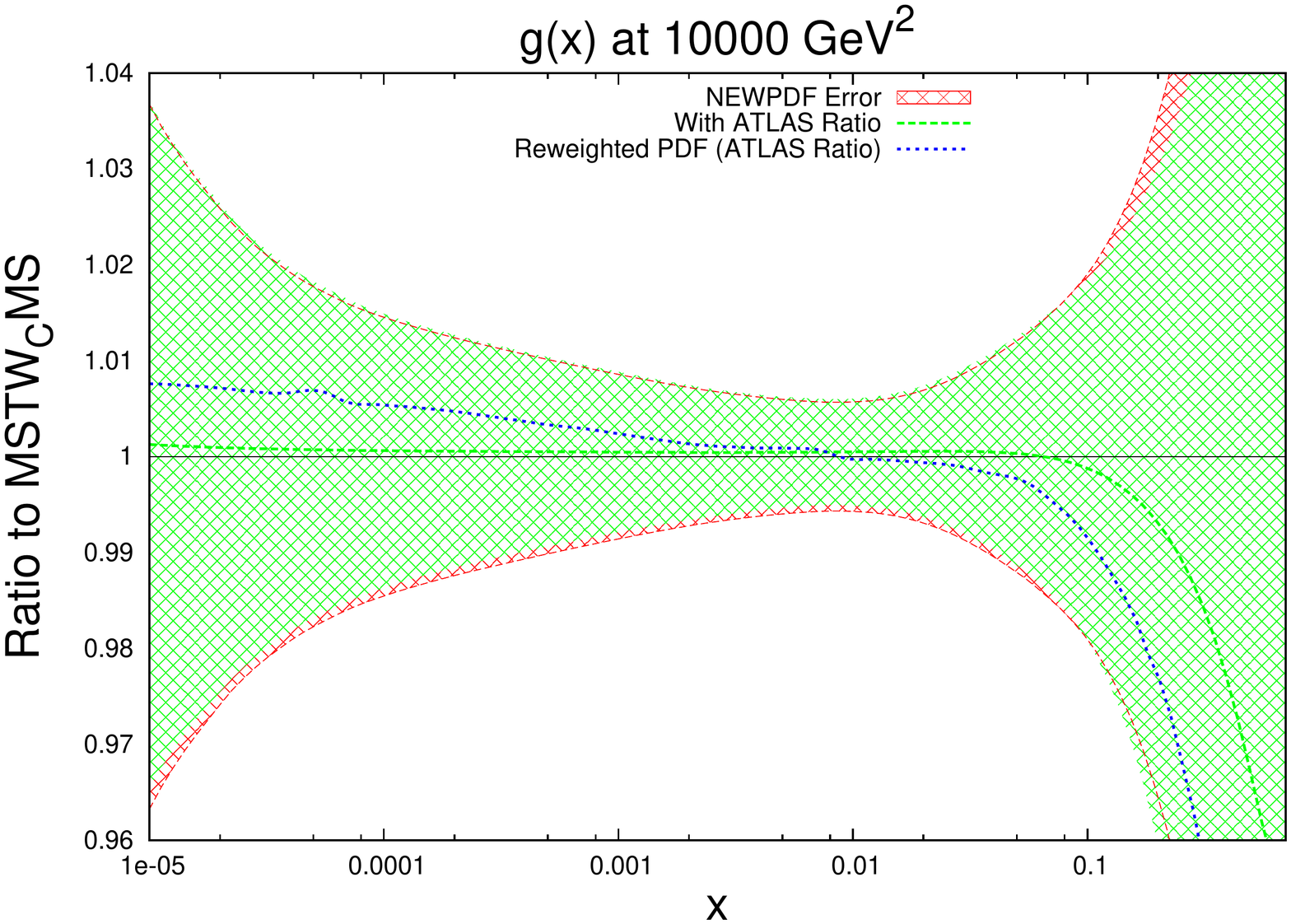}}
\subfigure{\includegraphics[width=0.49\textwidth]{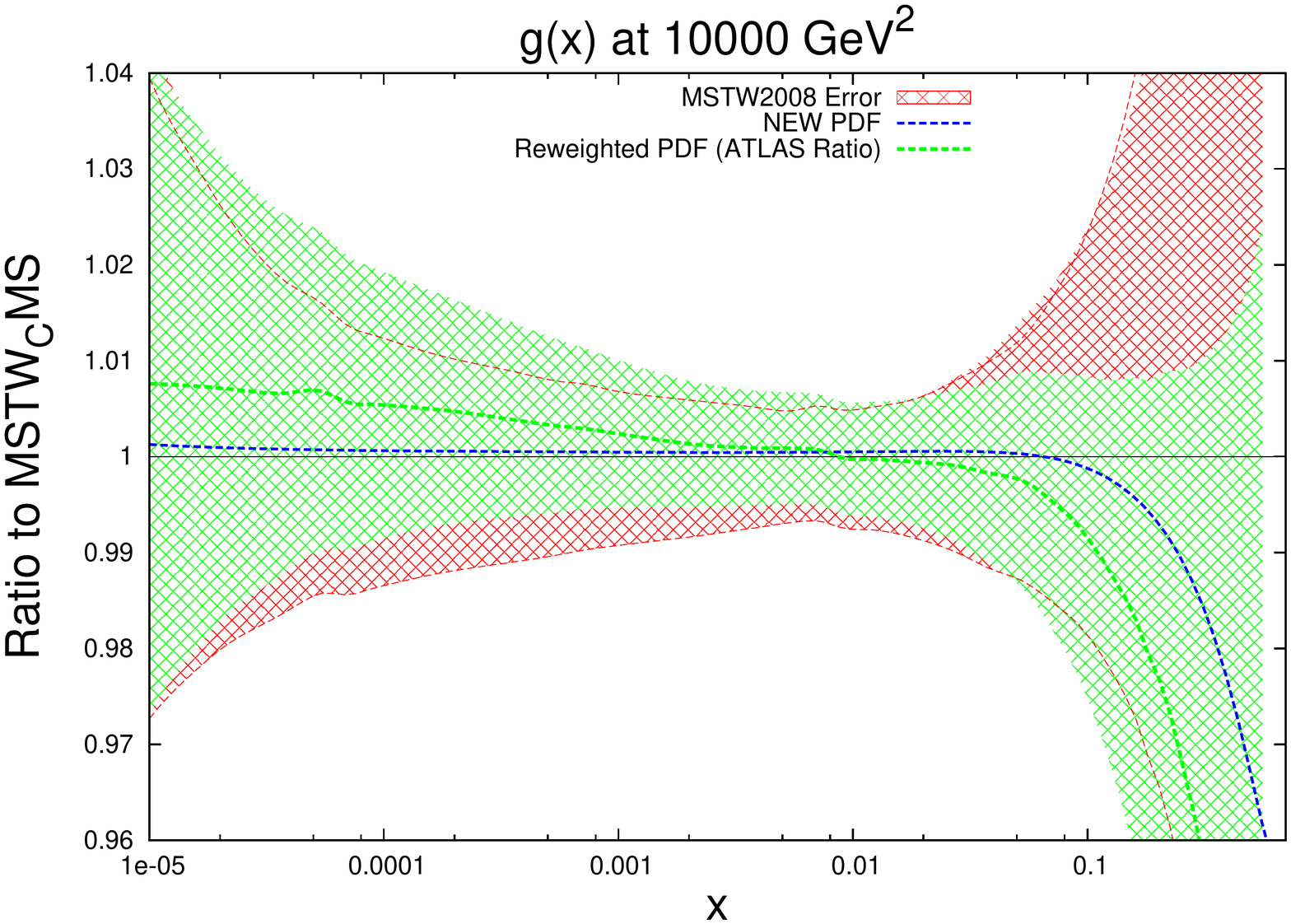}}
\vspace{-0.8cm}
\caption{Comparison of the gluon for the CMS fit, reweighted PDF (using ATLAS Ratio jets to reweight), and the new fit directly including the ATLAS Ratio \& CMS data. The central values are the same on both plots; however the first plot shows the new PDF's error band in green, whilst the second shows the reweighed PDF's error band in green.}
\label{newgluonatrat}
\end{figure}

Again, $\alpha_s(M_Z^2)$ is allowed to initially go free, yielding an improvement of 
20 fit points, and yielding a new value of $\alpha_s(M_Z^2)=0.1187$. Most 
notable in 
this fit, shown as the second column in Table \ref{atratfit}, is the very 
significant improvement in the ATLAS combined jet data. This improvement mostly 
goes away after holding $\alpha_s(M_Z^2)$ fixed at its MSTW2008 value. The total 
improvement in $\chi^2$ in this case from the MSTW2008 NLO fit is 13 points, and 
so is better than the previous fit which only included the CMS and ATLAS 7~TeV 
data. The majority of the improvement is again caused by the CMS data which 
reduces by 14 points. The ATLAS combined data improves by 4 points, better than 
the 2 by the ATLAS 7~TeV data in the previous fit, however there are 20 more 
points in the combined data set. This fit will be named here as 
MSTWATLAScomb, due to the additional inclusion of the combined ATLAS data. The 
fact that the CMS data improves more in this fit than the last demonstrates the 
excellent compatibility between the two LHC data sets. The ATLAS combined data, 
whilst only improving a small amount, is clearly having an additional affect 
on the global fit in the same direction preferred by the CMS data.
The increase in global $\chi^2$ and the most constraining eigenvector is shown in Table. \ref{chi2eig}. This is similar to the fit adding the CMS data alone, 
though the nominal order of the eigenvectors is altered slightly due to small changes in the size of the eigenvalues with the additional ATLAS jet data.

Finally, the reweighting procedure can be once again tested against the direct 
inclusion of a new data set. This is achieved by reweighting the new MSTWCMS 
PDFs using the ATLAS combined data. When comparing this to the change in the 
gluon by moving from MSTWCMS to MSTWATLAScomb, the results should agree if the 
two methods are consistent. The results for the gluon are shown in Fig.
\ref{newgluonatrat}. The agreement between the two methods is not as obvious 
as in the previous case with the inclusion of the CMS data, however the general 
trends are comparable, and both agree within their respective error bands. The 
MSTWATLAScomb fit is almost identical to MSTWCMS for most of the $x$ range, with 
the only divergence coming at high $x$ where the uncertainties are highest. This 
is testament to the dominance of the CMS data in both fits. The ATLAS combined 
data has limited effect on its own when additionally added to the CMS fit. The 
left plot in Fig. \ref{newgluonatrat} shows that there is a small improvement 
in the error band of the PDFs when including the ATLAS ratio data, and so 
there is a benefit to including both data sets simultaneously.

\begin{table}
\begin{center}
\begin{tabular}{| c | c |l| c |l|}
\hline
eigenvector & $+$ direction & most constraining & $-$ direction & most constraining\\
number & $\sqrt{\Delta \chi^2}$ & data set & $\sqrt{\Delta \chi^2}$ & data set \\
\hline
1 & 4.30 & Zeus ep 95-00 $\sigma_r^{NC}$ & 3.40  & H1 ep 97-00 $\sigma_r^{NC}$ \\
2 & 3.90 & NuTeV  $\nu N \to \mu\mu X$ & 3.50  & NMC $\mu d \,\,F_2$ \\
3 & 2.20 &  CCFR  $\nu N \to \mu\mu X$ & 1.30  &  NuTeV  $\nu N \to \mu\mu X$ \\
4 & 3.50 & NMC $\mu n/p  \,\,F_2$  & 2.30  & E866/NuSea $pd/pp$ DY   \\
5 & 2.20 & NuTeV $\mu N \,\, xF_3$ & 1.55 & NuTeV  $\nu N \to \mu\mu X$   \\
6 & 4.35 & H1 ep 97-00 $\sigma_r^{NC}$ & 3.00  &  NuTeV  $\nu N \to \mu\mu X$ \\
7 & 2.05 & D{\O} II $W \to l \nu$ asym. & 2.80  &  BCDMS $\mu d F_2$ \\
8 & 4.90 & NuTeV $\mu N \,\,F_2$ & 1.90  & BCDMS $\mu p \,\,F_2$ \\
9 & 5.00 & Zeus ep 95-00 $\sigma_r^{NC}$ & 3.90  & H1 ep 97-00 $\sigma_r^{NC}$ \\
10 & 2.95 & D{\O} II $W \to l \nu$ asym. & 3.25  & SLAC $\mu p \,\,F_2$ \\
11 & 4.80 & CDF $p\bar p \to$ jets & 4.05  & H1 ep 97-00 $\sigma_r^{NC}$ \\
12 & 5.45 & NuTeV  $\nu N \to \mu\mu X$ & 3.10  & E866/NuSea $pd/pp$ DY \\
13 & 1.40 & NuTeV  $\nu N \to \mu\mu X$ & 3.35  & E866/NuSea $pp$ DY \\
14 & 3.60 & NMC $\mu d \,\,F_2$ & 3.50  & NMC $\mu n/p  \,\,F_2$   \\
15 & 2.40 & H1 ep 97-00 $\sigma_r^{NC}$ & 3.80  & NuTeV $\mu N \,\,F_2$ \\
16 & 2.05 & CCFR  $\nu N \to \mu\mu X$ & 1.10  &   E866/NuSea $pd/pp$ DY \\
17 & 1.60 & E866/NuSea $pd/pp$ DY & 2.70  & NuTeV  $\nu N \to \mu\mu X$ \\
18 & 2.15 & D{\O} II $W \to l \nu$ asym. & 1.80  & E866/NuSea $pd/pp$ DY \\
19 & 2.80 & H1 ep 97-00 $\sigma_r^{NC}$ & 4.30  & CMS $pp \to$ jets \\
20 & 5.30 & NuTeV  $\nu N \to \mu\mu X$ & 1.95  & NuTeV  $\nu N \to \mu\mu X$  \\
\hline
\end{tabular}
\end{center}
\caption{Table of $\Delta \chi^2$ values for $68\%$ confidence level uncertainty for
each eigenvector and the most constraining data sets for the new NLO fits with CMS and ATLAS combined 2.76~TeV and 7~TeV data.}
\label{chi2eig}
\end{table}

\section{NNLO PDFs}

When considering NNLO PDFs, it is strictly necessary to use NNLO matrix 
elements for the theoretical predictions. For hadron-hadron inclusive jet 
cross sections these calculations have to date not been produced, and so 
approximations must be utilised to obtain the theoretical cross sections. The 
approximation used in the MSTW2008 analysis for Tevatron inclusive jets is 
based on the calculation by Kidonakis and Owens \cite{kidonakis}. This 
calculation produces a threshold resummation which is based around the 
assumption that the parton-parton scattering phase space is restricted to the 
threshold region of $x_T=2p_T/\sqrt{s}\sim 1$, due to the rapid decrease in 
PDFs at high $x$. The corrections are provided within the FastNLO framework, 
and so have been included for the use of Tevatron inclusive jet data in NNLO 
fits. These corrections have recently been reproduced in \cite{Kumar:2013hia}. 
In this latter article a comparison is also made between the threshold 
approximation and the full NLO result, observing that best agreement is for 
low values of  cone radius $R \sim 0.3-0.4$. This is suggests that
the threshold approximation of the NNLO corrections may be a little low 
for the $R$ values more like $R=0.7$ used at the Tevatron, though this 
is not definite, and as discussed more below, these corrections are quite 
small. A more detailed threshold calculation has also recently been performed 
in \cite{deFlorian:2013qia}, where good agreement is seen between the 
threshold approximation and the full NLO calculation at high $p_T$
independent of $R$.
However, despite a large $R$ dependence at NLO, the further correction from 
NLO to NNLO shows much less dependence, and is often of order $15\%$ if the 
scale choice is $p_T$. Unfortunately these results are not yet in a form 
which can easily be incorporated in a PDF fit. 

As a check on reliability of NNLO results 
we have rerun the NNLO MSTW08 fit with the threshold 
corrections multiplied by quite an extreme factor of two. This results in 
a lowering of $\alpha_s(M_Z^2)$ by about 0.001 and a slightly higher gluon PDF 
at low $x$ and slightly smaller gluon at high $x$, with changes about one 
sigma 
or less. Hence, the change is not dramatic, and actually rather similar to 
the changes seen at NLO in this article which are 
induced by the LHC jet data. The fit
quality does deteriorate, particularly for D{\O} data, but more due to details 
of shape rather than normalisation, i.e. the threshold corrections are 
unlikely  to be exactly the correct shape in $p_T$, particularly at 
low $p_T$ values, but the gluon distribution probed here is already very
well constrained by HERA DIS data. Similarly, removing the threshold factor 
entirely and performing an NNLO fit (the default procedure used by some 
groups  in NNLO fits) results in a raising of $\alpha_s(M_Z^2)$ by about 
0.001 and a slightly lower gluon PDF at low $x$ and slightly larger gluon
at high $x$, 
with changes about one sigma or less. Simply using a constant $K$-factor of 
$15\%$ changes the fit quality by only about one unit, and both PDFs and 
$\alpha_S$ change by much less than one standard deviation.

\begin{figure}[h!]
\centering
\includegraphics[width=0.9\textwidth]{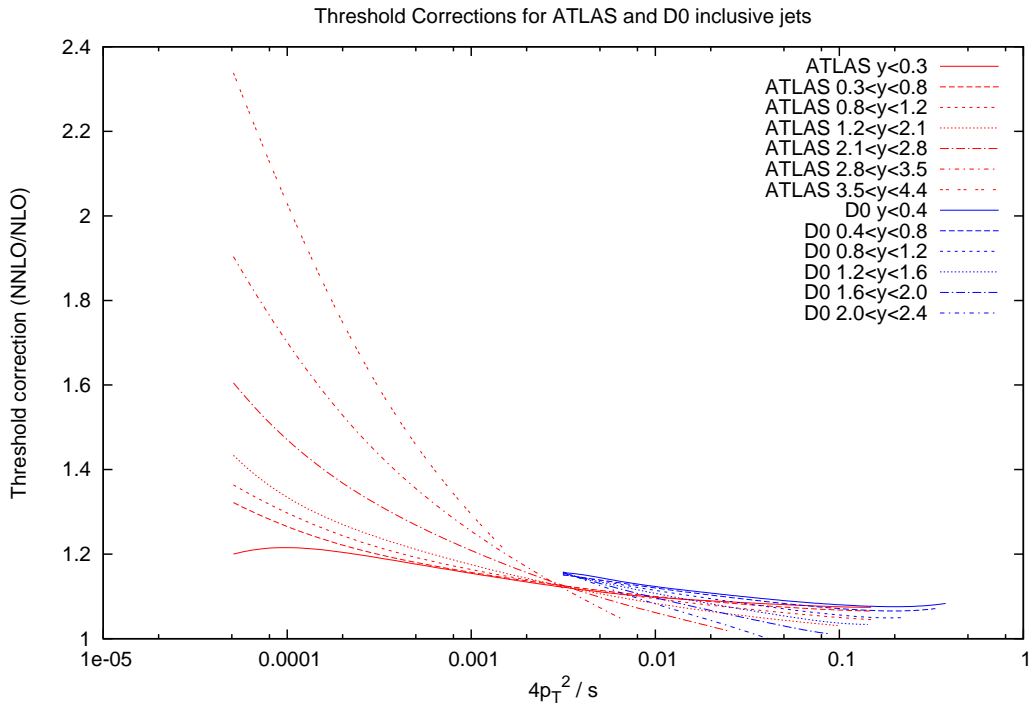}
\vspace{-0.8cm}
\caption{Comparison of NNLO threshold corrections for ATLAS and D{\O} inclusive jets as a function of $x_T^2=4p_T^2/s$.}
\label{thres}
\end{figure}

In order to help facilitate the inclusion of the LHC jet data into an NNLO 
fit the threshold corrections 
have also now been calculated for the new data and implemented in FastNLO for 
ATLAS data. The results are shown in Fig. \ref{thres}, where the ATLAS data is 
presented alongside that of D{\O} (similar plots appear in \cite{Kumar:2013hia}, but none extending to such low $p_T$ values or showing a range in 
rapidity values, and in \cite{deFlorian:2013qia}). 
The main point of note is that 
the LHC phase space spans a region which extends much further from the 
threshold region than the Tevatron. The Tevatron threshold corrections 
maintain a small correction of approximately 
$\sigma_{NNLO}\sim 1.1 - 1.2 \sigma_{NLO}$ across the majority of the phase space. 
However, this correction clearly increases away from threshold. The 
corresponding ATLAS calculation demonstrates that this trend continues 
even further, and although the central jets maintain a reasonable correction 
throughout, the forward jet corrections become very large with decreasing 
$x_T$.

It is clear that for LHC jets it will be necessary to include the full NNLO 
matrix elements in order to perform a full NNLO fit. Although not yet fully 
performed, the gluon-gluon process has been calculated by Gehrmann de Ridder 
et. al \cite{gggp,Currie:2013dwa}. These calculations have shown a NNLO 
results of between 1.1 and 1.3 times the NLO prediction across all jet $p_T$ 
values, and suggest that the threshold corrections indeed are not applicable 
to all LHC scenarios, i.e. at $p_T$ values such that one is very far from 
threshold. For inclusive jet cross sections at the Tevatron the threshold 
corrections seem reasonable when compared to the full NNLO corrections so far 
known, certainly when one considers the very large 
systematic uncertainties, including luminosity, which allow the data to move 
relative to theory.

\begin{table}[h]
\centering
\begin{tabular}{|c |c| c|c|}
\hline\hline
 & No Cuts & HERAPDF Cuts & Additive Errors\\
\hline
MSTW 2008 & 1.32 & 0.927 & 1.44 \\
\hline
\end{tabular}
\caption{$\chi^2$ per point for ATLAS combined data, both with and without $p_T$ cuts. The third column uses additive errors and has two additional anomalous points cut. The NNLO PDF set is used.}
\label{table:chi_ratio_nnlo}
\end{table}

\begin{table}[h]
\centering
\begin{tabular}{c c}
\hline\hline
Scale & $p_T$ \\
\hline
MSTW 2008 & 1.37\\
\hline
\end{tabular}
\caption{$\chi^2$ per point (133 points) for NNLO PDFs for CMS inclusive jets.}
\label{table:chi_cms2}
\end{table}

In order to give some indication of how NNLO PDFs perform for LHC jet cross 
sections the NNLO PDF predictions for the ATLAS jet cross 
sections are calculated using just the NLO QCD cross section.
The fit value with NNLO PDFs and NNLO coupling is shown in 
Table \ref{table:chi_ratio_nnlo}. The NNLO MSTW set describes the data after 
the low $p_T$ cuts are applied well. Again the additive error treatment with 
the additional cuts is shown in the third column.We also repeat this 
exercise for the CMS inclusive jet data, and the results are in 
Table \ref{table:chi_cms2}. Again the fit quality is very good. We have also 
tried, as a rough experiment, to compare the prediction using NNLO PDFs and
a very approximate NNLO K-factor based on the results in  
\cite{gggp,Currie:2013dwa}. This causes the fit quality to deteriorate quite 
significantly 
if not dramatically. A refit of the PDFs results in a fit quality similar to 
that obtained in NLO fits (though slightly worse), 
little change in NNLO PDFs (though with a trend 
similar to the changes the LHC jet data induce in NLO PDFs) and a reduction 
in the NNLO $\alpha_S(M_Z^2)$ value extracted of order 0.001. However, since 
the gluon-gluon initiated contribution is not dominant, particularly at high 
$p_T$, it is difficult to draw strong conclusions beyond the fact that the 
still large systematic uncertainties on jet data at the LHC will likely allow 
fairly good quality fits for something similar to the current PDF and 
$\alpha_S$ values unless the full NNLO corrections are somewhat larger than 
seems likely.

\section{Conclusions}

The data which has been measured during the first run of the 
LHC at 7~TeV is our first look at QCD in a new energy regime, and so 
the jet data is an important test of our current knowledge of PDFs. 
The conclusion from these first data sets is that the MSTW2008 PDFs hold up 
well in this regime, since none of the inclusive jet data from either 
ATLAS or CMS has required a PDF to move outside its 1$\sigma$ error band.
The earliest released measurement was the least discerning for PDFs; the 
ATLAS inclusive jet cross section at 7~TeV using 36 $pb^{-1}$ of luminosity 
was inevitably dominated by systematics uncertainties. The fit quality 
obtained using MSTW2008 PDFs is very good and any variation 
in physics parameters used is incapable of improving the fit in any 
significant way. The lack of constraint due to large systematic uncertainties 
is significantly improved by the inclusion of a simultaneous measurement at 
center of mass energy 2.76~TeV. The cancellation of systematic effects 
associated with jet energy scale provides a more suitable environment for 
testing PDFs. In this measurement, too, a good fit is found for MSTW2008 
PDFs. The potential impact of the data was investigated using the PDF 
reweighting procedure.  
Although the data prefers a larger low-$x$ and softer high-$x$ gluon, 
these movements are still entirely within the error bands. A significant 
improvement in error is seen for the gluon across all $x$, which implies 
that, if included in a new fit, this data could provide more accurate PDFs 
for the LHC era. The published CMS inclusive data at 7~TeV is also analysed. 
With much higher luminosity than the ATLAS data, this is currently the 
published measurement with the most potential for PDF effects. Again a 
reasonable fit is found for MSTW2008, although the $\chi^2$ per point is 
higher than the ATLAS fit. Due to the kinematics of the measurement, more 
focus is given to the quark densities for this set, and some reduction in the 
error bands is seen for all flavours. Again, including this data into a new 
fit would appear to provide PDFs with some improved constraints.

A detailed study into hadron-hadron dijet cross sections in relation to PDFs 
has been also been presented. The instability of the calculation observed at 
the Tevatron using the scale choice of $p_T^{av}$ is explained by the 
behaviour of the kinematics at high rapidities. Calculations using other 
scale choices involving the dijet mass do not exhibit these problems, and so 
potentially provide a more reliable estimate of the scale uncertainty.
For ATLAS dijets, the instability is even more clear for the $p_T^{av}$ 
calculation, with a very poor fit for low values of the scale multiplier 
quickly becoming an excellent fit for higher, unrealistic values. A study of 
the behaviour of the individual data points under scale variations 
demonstrates a saddle point structure which is centred around the central 
scale choice for low rapidity bins, and which can become a constantly 
decreasing plane at higher rapidities. The best scale choice to maintain 
the stability of each bin under scale variations is something similar to 
$M_{JJ}/0.7\cosh(y^*)$, as suggested many years ago \cite{soper}.
The best fit to data is clearly obtained for scale choices similar 
to $\mu=M_{JJ}$ for Tevatron data, whereas choices with less rapidity 
dependence are preferred by the ATLAS data. The difference is perhaps   
related to the fact that one is a proton-antiproton collider and the 
other a proton-proton collider, so different PDF combinations are probed
even after one takes account of the different collider energies.
The reweighting procedure 
has been conducted for D{\O}, ATLAS and CMS dijet data, 
and in general the resulting 
preferred PDF depends upon the scale choice used. This is not an ideal 
situation, since the physics cannot depend on an unphysical mathematical 
property of the calculation. However, for the CMS dijet cross section, 
which does not extend to very large rapidity, an 
agreement is reached between the scale choices, which is for a 
slightly smaller gluon 
across much of the $x$ range, except very high $x$, 
with the largest change at moderate $x \sim 0.05$ values. 
This also agrees with one of the scale choices for ATLAS dijets. This result 
is notable due to it being the opposite effect required to describe the 
ATLAS and CMS inclusive jet data, implying a possible conflict between the 
two datasets, or different forms of higher order QCD corrections (the shape 
of the NNLO inclusive and dijet corrections so far available 
\cite{gggp,Currie:2013dwa} does not appear to be identical).

For the first time, LHC jet data has also been included directly in the 
framework of the MSTW PDF fit. The data sets included represent the highest 
precision inclusive jet cross sections from both ATLAS and CMS to date. Two 
fits were initially performed with the new data, including the CMS inclusive 
jet data and ATLAS 7~TeV inclusive jet data: one allowing all standard MSTW 
parameters to be free, and one with $\alpha_s(M_Z^2)$ fixed to the 
MSTW2008 value. 
The entirely free set showed a significant reduction in global $\chi^2$, 
although much of this was due to a shift in $\alpha_s(M_Z^2)$ which significantly 
improved some fixed target data. With $\alpha_s(M_Z^2)$ fixed the fit again 
improved, although to a lesser extent, with the majority of improvement 
coming from the fit to the new data sets. The improvement was dominated by 
the CMS data due to the previously noted issue of the large ATLAS systematic 
errors. With the new central values and eigenvectors, the reweighting 
procedure was applied to study the change in the effect after the addition of 
the ATLAS 2.76~TeV data set. This was shown to still have an effect on the 
gluon, with a similar but less pronounced shape than was seen when 
reweighting the MSTW2008 set with the same data. Dijet data was shown again 
to have a different effect on the PDFs to the corresponding inclusive data, 
which is further evidence for their value in a future global fit.
The ATLAS combined data, having shown an effect through reweighting, was then 
included in a second fit along with the CMS data. This further improved the 
global fit, with the fit to CMS showing a similar improvement to the first 
set, and an additional improvement from the ATLAS data itself. This 
demonstrated an excellent agreement between the ATLAS and CMS data sets, 
which had already been observed through the reweighting technique.

\section*{Acknowledgements}

We would like to thank A. D. Martin, W. J. Stirling  and G. Watt for
numerous discussions on PDFs. We would also like to thank A.M. Cooper-Sarkar, 
and M. Wing for discussions on jets at the LHC, D. Britzger and M. Sutton 
for discussions on fastNLO and APPLgrid, and W. Vogelsang for communication 
on NNLO approximations using threshold corrections. This work is
supported partly by the London Centre for Terauniverse Studies (LCTS),
using funding
from the European Research Council via the Advanced Investigator Grant 267352.
RST would also like to thank the IPPP, Durham, for the award of a Research Associateship. We would like to thank the Science and Technology 
Facilities Council (STFC) for support.


\begin{thebibliography}{99}

\bibitem{cdfjet} CDF Collaboration, Phys.Rev.D75:09 (2006), [ arXiv:hep-ex/0701051].
\bibitem{d0jet} D{\O} Collaboration,  Phys.Rev.D85:05 (2006), [arXiv:1110.3771].
\bibitem{MSTW} A. D. Martin W.J. Stirling, R.S. Thorne, G. Watt, Eur.Phys.J. C63 (2009) 189-285, [arXiv:0901.0002].
\bibitem{nnpdfpaper} 
 R.~D.~Ball, V.~Bertone, S.~Carrazza, C.~S.~Deans, L.~Del Debbio, S.~Forte, A.~Guffanti and N.~P.~Hartland {\it et al.},
  %``Parton distributions with LHC data,''
  Nucl.\ Phys.\ B {\bf 867} (2013) 244
  [arXiv:1207.1303 [hep-ph]].
\bibitem{herapdf} A. M. Cooper-Sarkar, PoS DIS2010:023, (2010) [arXiv:1006.4471].
\bibitem{ct10} M. Guzzi \emph{et al}, SMU-HEP-10-11, [arXiv:1101.0561].
\bibitem{abm}
  S.~Alekhin, J.~Bluemlein and S.~Moch,
  %``The ABM parton distributions tuned to LHC data,''
  arXiv:1310.3059 [hep-ph].
  %%CITATION = ARXIV:1310.3059;%%
\bibitem{d0-dijet-paper} D{\O} Collaboration,  Phys.Lett.B693:531-538 (2010), [arXiv:1002.4594].
\bibitem{wattthorne} R.S. Thorne, G. Watt, JHEP 1108:100 (2011), [arXiv:1106.5789].
%\cite{Alekhin:2012ig}
\bibitem{Alekhin:2012ig}
  S.~Alekhin, J.~Blumlein and S.~Moch,
  %``Parton Distribution Functions and Benchmark Cross Sections at NNLO,''
  Phys.\ Rev.\ D {\bf 86} (2012) 054009
  [arXiv:1202.2281 [hep-ph]].
\bibitem{ewcorr} K. Mishra \emph{et al}, Snowmass QCD Working Group (2013) [ arXiv:1308.1430].
\bibitem{cheb} A.D. Martin, A.J.Th.M. Mathijssen, W.J. Stirling, R.S. Thorne, B.J.A. Watt, G. Watt, Eur. Phys. 
J. C 73:2318 (2013), [arXiv:1211.1215].
\bibitem{mstwhera} R.S. Thorne, A.D. Martin, W.J. Stirling, G. Watt, PoS DIS2010:052, (2010), [arXiv:1006.2753].
\bibitem{robert1} R.S. Thorne, Phys.Rev. D86:074017 (2012), [arXiv:1201.6180]. 
\bibitem{atlas-inc-paper} ATLAS Collaboration, Phys.Rev. D86 (2012) 014022, [arXiv:1112.6297].
\bibitem{nlojet} Z.Nagy Phys. Rev. D68, 094002 (2003), [arXiv:0307268].
\bibitem{nlojet2} Z. Nagy, Phys. Rev. Lett.88, 122003 (2002), [arXiv:0110315].
\bibitem{fastnlo} T. Kluge, K. Rabbertz, M. Wobisch, in proceedings "14th International Workshop on Deep Inelastic Scattering" (2006) [arXiv:hep-ph/0609285].
\bibitem{fastnlov2} D.Britzger,  K. Rabbertz, F. Stober, M. Wobisch, Proceedings of the XX International Workshop on Deep Inelastic Scattering, University of Bonn, 26-30th March 2012 [arXiv:1208.3641].
\bibitem{geile}W. T. Giele and S. Keller, Phys.Rev. D58 094023 (1998) , [hep-ph/9803393].
\bibitem{nnrew1} NNPDF Collaboration, R. D. Ball \emph{et al}, Nucl.Phys. B849 112Ð143 (2011), [arXiv:1012.0836].
\bibitem{nnrew2} NNPDF Collaboration, R. D. Ball \emph{et al}, Nucl.Phys. B855 (2012) 608Ð638, [arXiv:1108.1758].
\bibitem{reweight} G. Watt, R.S. Thorne, JHEP 1208:052 (2012), [ arXiv:1205.4024]. 
\bibitem{dilorenzi} F. De Lorenzi, CERN-THESIS-2011-237 (2011).
\bibitem{Sato:2013ika}
  N.~Sato, J.~F.~Owens and H.~Prosper,
  %``Bayesian Reweighting for Global Fits,''
  arXiv:1310.1089 [hep-ph].
%\cite{Paukkunen:2014zia}
\bibitem{Paukkunen:2014zia}
  H.~Paukkunen and P.~Zurita,
  %``PDF reweighting in the Hessian matrix approach,''
  arXiv:1402.6623 [hep-ph].
  %%CITATION = ARXIV:1402.6623;%%


\bibitem{ratiopaper} ATLAS Collaboration, EPJC 73:2509 (2013) [ arXiv:1304.4739].
\bibitem{applgrid} T. Carli \emph{et al}, Eur Phys J C 66 (2010) 503 , [arXiv:0911.2985].
\bibitem{cmsinc} CMS Collaboration, Phys. Rev. D 87 112002 (2013) [arXiv:1212.6660].

\bibitem{soper} S.D. Ellis,	 Z.Kunszt, D.E. Soper,Phys. Rev. Lett. 69, 1496Ð1499 (1992). 

\bibitem{huston} J.Huston - https://indico.cern.ch/contributionDisplay.py?contribId=3\&confId=226756.

\bibitem{data1} A. C. Benvenuti et al. [BCDMS Collaboration], Phys. Lett. B 223 (1989) 485.
\bibitem{data2} A. C. Benvenuti et al. [BCDMS Collaboration], Phys. Lett. B 237 (1990) 592.
\bibitem{data3} M. Arneodo et al. [New Muon Collaboration], Nucl. Phys. B 483 (1997) 3 [arXiv:hep- ph/9610231].
\bibitem{data5} M. Arneodo et al. [New Muon Collaboration], Nucl. Phys. B 487 (1997) 3 [arXiv:hep- ex/9611022].
\bibitem{data6} M. R. Adams et al. [E665 Collaboration], Phys. Rev. D 54 (1996) 3006.
\bibitem{data8} L. W. Whitlow, E. M. Riordan, S. Dasu, S. Rock and A. Bodek, Phys. Lett. B 282 (1992)
475. 
\bibitem{data9} L. W. Whitlow, Ph.D. thesis, Stanford University, 1990, SLAC-0357.

\bibitem{data10} L. W. Whitlow, S. Rock, A. Bodek, E. M. Riordan and S. Dasu, Phys. Lett. B 250 (1990) 193.
\bibitem{data11} J. C. Webb, Ph.D. thesis, New Mexico State University, 2002, arXiv:hep-ex/0301031; Paul E. Reimer, private communication (for the radiative corrections).
\bibitem{data12} R. S. Towell et al. [FNAL E866/NuSea Collaboration], Phys. Rev. D 64 (2001) 052002 [arXiv:hep-ex/0103030]
\bibitem{data13} M. Tzanov et al. [NuTeV Collaboration], Phys. Rev. D 74 (2006) 012008 [arXiv:hep- ex/0509010]
\bibitem{data14} G. Onengut et al. [CHORUS Collaboration], Phys. Lett. B 632 (2006) 65.
\bibitem{data17} M. Goncharov et al. [NuTeV Collaboration], Phys. Rev. D 64 (2001) 112006 [arXiv:hep-
ex/0102049].
\bibitem{data19} E. M. Lobodzinska [H1 Collaboration], arXiv:hep-ph/0311180.
\bibitem{data20} C. Adloff et al. [H1 Collaboration], Eur. Phys. J. C 21 (2001) 33 [arXiv:hep-ex/0012053].
\bibitem{data22} C. Adloff et al. [H1 Collaboration], Eur. Phys. J. C 19 (2001) 269 [arXiv:hep-ex/0012052].
\bibitem{data23} C. Adloff et al. [H1 Collaboration], Eur. Phys. J. C 30 (2003) 1 [arXiv:hep-ex/0304003].
\bibitem{data24} J. Breitweg et al. [ZEUS Collaboration], Eur. Phys. J. C 7 (1999) 609 [arXiv:hep- ex/9809005].
\bibitem{data25} S. Chekanov et al. [ZEUS Collaboration],  Eur. Phys. J. C 21 (2001) 443 [arXiv:hep-ex/0105090].
\bibitem{data26} S. Chekanov et al. [ZEUS Collaboration], Eur. Phys. J. C 28 (2003) 175 [arXiv:hep-ex/0208040].
\bibitem{data27} S. Chekanov et al. [ZEUS Collaboration], Phys. Rev. D 70 (2004) 052001 [arXiv:hep- ex/0401003].
\bibitem{data28} C. Adloff et al. [H1 Collaboration], Eur. Phys. J. C 30 (2003) 1 [arXiv:hep-ex/0304003].
\bibitem{data29} S. Chekanov et al. [ZEUS Collaboration], Eur. Phys. J. C 32 (2003) 1 [arXiv:hep- ex/0307043].
\bibitem{data30a} C. Adloff et al. [H1 Collaboration], Z. Phys. C 72 (1996) 593 [arXiv:hep-ex/9607012].
\bibitem{data30b} C. Adloff et al. [H1 Collaboration], Phys. Lett. B 528 (2002) 199 [arXiv:hep-ex/0108039].
\bibitem{data30c}  A. Aktas et al. [H1 Collaboration], Eur. Phys. J. C 45 (2006) 23 [arXiv:hep-ex/0507081].
\bibitem{data30d} A. Aktas et al. [H1 Collaboration], Eur. Phys. J. C 40 (2005) 349 [arXiv:hep-ex/0411046].
\bibitem{data30e} J. Breitweg et al. [ZEUS Collaboration], Eur. Phys. J. C 12 (2000) 35 [arXiv:hep- ex/9908012].
\bibitem{data30f} S. Chekanov et al. [ZEUS Collaboration], Phys. Rev. D 69 (2004) 012004 [arXiv:hep- ex/0308068].
\bibitem{data30g} S. Chekanov et al. [ZEUS Collaboration], JHEP 0707 (2007) 074 [arXiv:0704.3562 [hep- ex]].
\bibitem{data31} A. Aktas et al. [H1 Collaboration], Phys. Lett. B 653 (2007) 134 [arXiv:0706.3722 [hep- ex]].
\bibitem{data32} S. Chekanov et al. [ZEUS Collaboration], Phys. Lett. B 547 (2002) 164 [arXiv:hep- ex/0208037].
\bibitem{data33} S. Chekanov et al. [ZEUS Collaboration], Nucl. Phys. B 765 (2007) 1 [arXiv:hep- ex/0608048].
\bibitem{data34} V. M. Abazov et al. [D{\O} Collaboration], Phys. Rev. Lett. 101 (2008) 062001 [arXiv:0802.2400 [hep-ex]].
\bibitem{data35} A. Abulencia et al. [CDF - Run II Collaboration], Phys. Rev. D 75 (2007) 092006 [Erratum- ibid. D 75 (2007) 119901] [arXiv:hep-ex/0701051].
\bibitem{data36} D. Acosta et al. [CDF Collaboration], Phys. Rev. D 71 (2005) 051104 [arXiv:hep- ex/0501023].
\bibitem{data37} V. M. Abazov et al. [D{\O} Collaboration], Phys. Rev. D 77 (2008) 011106 [arXiv:0709.4254 [hep-ex]].
\bibitem{data38} V. M. Abazov et al. [D{\O} Collaboration], Phys. Rev. D 76 (2007) 012003 [arXiv:hep- ex/0702025].
\bibitem{data39} J. Han et al. [CDF Collaboration], Òd?/dy distribution of DrellÐYan dielectron pairs,Ó Public Note, May 2008, http://www-cdf.fnal.gov/physics/ewk/2008/dszdy/.



%\cite{Watt:2011kp}
\bibitem{Watt:2011kp}
  G.~Watt,
  %``Parton distribution function dependence of benchmark Standard Model total cross sections at the 7 TeV LHC,''
  JHEP {\bf 1109} (2011) 069
  [arXiv:1106.5788 [hep-ph]].

%\cite{Ball:2012wy}
\bibitem{Ball:2012wy}
  R.~D.~Ball, S.~Carrazza, L.~Del Debbio, S.~Forte, J.~Gao, N.~Hartland, J.~Huston and P.~Nadolsky {\it et al.},
  %``Parton Distribution Benchmarking with LHC Data,''
  JHEP {\bf 1304} (2013) 125
  [arXiv:1211.5142 [hep-ph]].


\bibitem{kidonakis}Nikolaos Kidonakis, J. F. Owens, Phys. Rev. D 63, 054019 (2001)
%\cite{Kumar:2013hia}
\bibitem{Kumar:2013hia}
  M.~C~Kumar and S.~-O.~Moch,
  %``Phenomenology of threshold corrections for inclusive jet production at hadron colliders,''
  arXiv:1309.5311 [hep-ph].

%\cite{deFlorian:2013qia}
\bibitem{deFlorian:2013qia}
  D.~de Florian, P.~Hinderer, A.~Mukherjee, F.~Ringer and W.~Vogelsang,
  %``Approximate next-to-next-to-leading order corrections to hadronic jet production,''
  arXiv:1310.7192 [hep-ph].
  %%CITATION = ARXIV:1310.7192;%%
  %1 citations counted in INSPIRE as of 27 Nov 2013


\bibitem{gggp}A. Gehrmann-De Ridder \emph{et al} , Phys. Rev. Lett. 110, 162003 (2013)

%\cite{Currie:2013dwa}
\bibitem{Currie:2013dwa}
  J.~Currie, A.~Gehrmann-De Ridder, E.~W.~N.~Glover and J.~Pires,
  %``NNLO QCD corrections to jet production at hadron colliders from gluon scattering,''
  arXiv:1310.3993 [hep-ph].
  %%CITATION = ARXIV:1310.3993;%%


\end{thebibliography}
\end{document}